\documentclass[
reprint,
superscriptaddress,
nofootinbib,showkeys,
amsmath,amssymb,
aps, prd,
floatfix,
]{revtex4-2}
\usepackage{xcolor}
\usepackage{appendix}
\usepackage{graphicx}
\usepackage{dcolumn}
\usepackage{bm}
\usepackage[colorlinks=true]{hyperref}
\hypersetup{
     colorlinks=true,
     linkcolor=blue,
     filecolor=blue,
     citecolor = red,      
     urlcolor=purple,
     }
\usepackage{booktabs}
\usepackage{xspace}
\usepackage[doipre={DOI:~}]{uri}
\usepackage{natbib}
\usepackage{aas_macros}

\newcommand{\Ms}{{\rm M}_\odot}
\newcommand{\bpop}{\textsc{B-pop}\xspace}

\newcommand{\gwfish}{\textsc{GWFish}\xspace}
\newcommand{\dragonii}{\textsc{Dragon-II}\xspace}

\newcommand{\mobse}{\textsc{Mobse}\xspace}
\newcommand{\sevn}{\textsc{SEvN}\xspace}

\newcommand{\der}{{\rm d}}
\newcommand{\yrgpc}{~{\rm yr}^{-1}~{\rm Gpc}^{-3}}
\newcommand{\Rate}{\mathcal{R}}

\newcommand{\orcidicon}[1]{\href{https://orcid.org/#1}{\includegraphics[width=11pt]{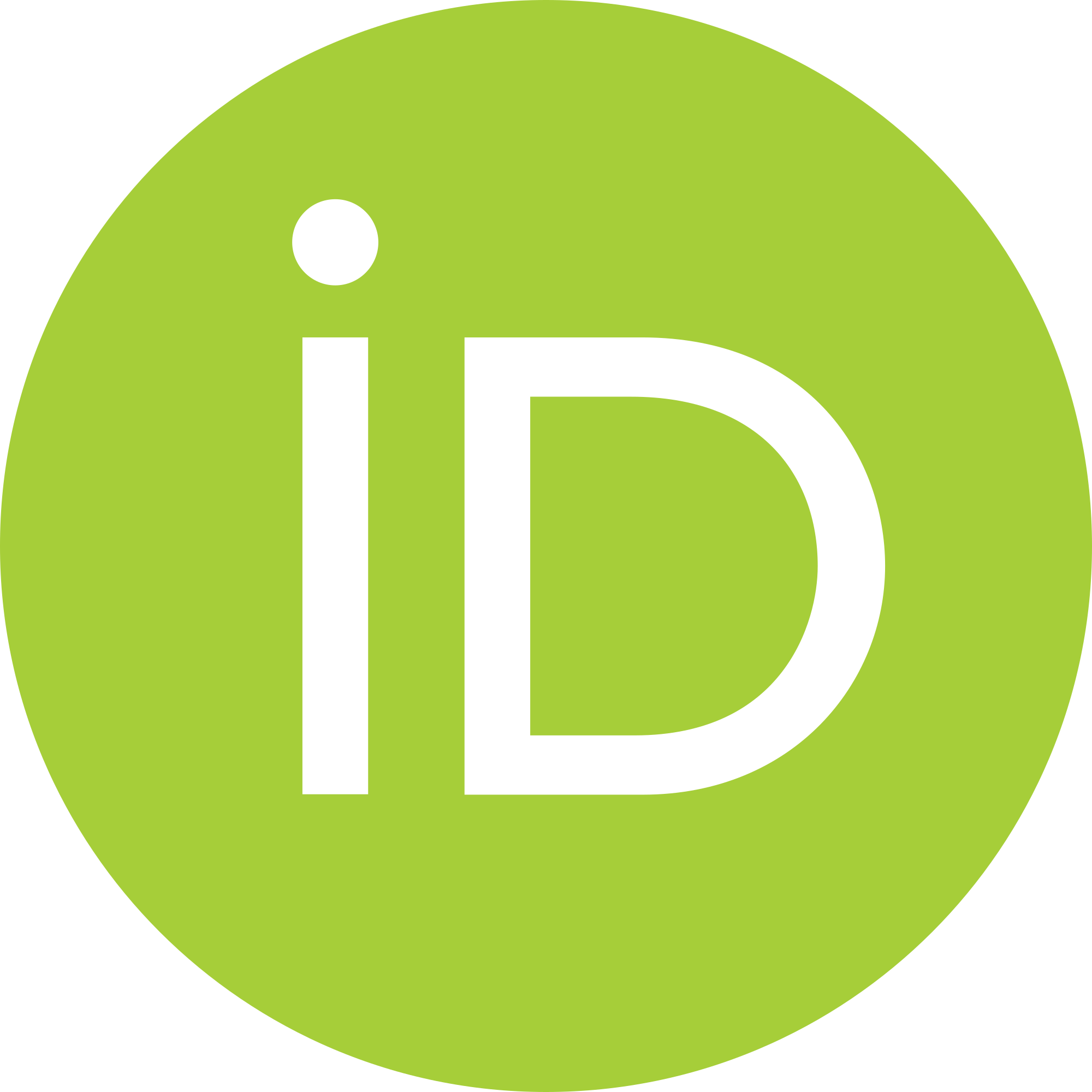}}}
\newcommand{\orcid}[1]{\href{https://orcid.org/#1}{\protect\orcidicon{#1}}}

\begin{document}

\preprint{APS/PRD}

\title{Isolated or Dynamical? Tracing Black Hole Binary Formation through the Population of Gravitational-Wave Sources}

\author{Manuel Arca Sedda\orcid{0000-0002-3987-0519}}
\email{manuel.arcasedda@gssi.it}
\affiliation{Gran Sasso Science Institute, Via F. Crispi 7, L'Aquila, I-67100, Italy}
\affiliation{INFN - Laboratori Nazionali del Gran Sasso, I-67100 Assergi, Italy}
\affiliation{INAF - Osservatorio Astronomico di Roma, I-00040 Monte Porzio Catone (Rome), Italy}

\author{Lavinia Paiella\orcid{0009-0001-7605-991X}} 
\affiliation{Gran Sasso Science Institute, Via F. Crispi 7, L'Aquila, I-67100, Italy}
\affiliation{INFN - Laboratori Nazionali del Gran Sasso, I-67100 Assergi, Italy}
\affiliation{INAF - Osservatorio Astronomico di Roma, I-00040 Monte Porzio Catone (Rome), Italy}

\author{Cristiano Ugolini}
\affiliation{Gran Sasso Science Institute, Via F. Crispi 7, L'Aquila, I-67100, Italy}
\affiliation{INFN - Laboratori Nazionali del Gran Sasso, I-67100 Assergi, Italy}
\affiliation{INAF - Osservatorio Astronomico di Roma, I-00040 Monte Porzio Catone (Rome), Italy}

\author{Filippo Santoliquido}
\affiliation{Gran Sasso Science Institute, Via F. Crispi 7, L'Aquila, I-67100, Italy}
\affiliation{INFN - Laboratori Nazionali del Gran Sasso, I-67100 Assergi, Italy}

\author{Benedetta Mestichelli \orcid{0009-0002-1705-4729}}
\affiliation{Gran Sasso Science Institute, Via F. Crispi 7, L'Aquila, I-67100, Italy}
\affiliation{INFN - Laboratori Nazionali del Gran Sasso, I-67100 Assergi, Italy}
\affiliation{INAF - Osservatorio Astronomico di Roma, I-00040 Monte Porzio Catone (Rome), Italy}
\affiliation{Institut f\"ur Theoretische Astrophysik, Zentrum f\"ur Astronomie, Universit\"at Heidelberg, Albert Ueberle Str. 2, D-69120 Heidelberg, Germany}

\author{Ilaria Usai}
\affiliation{Gran Sasso Science Institute, Via F. Crispi 7, L'Aquila, I-67100, Italy}
\affiliation{Dipartimento di Fisica, Università degli Studi di Cagliari,
Cittadella Universitaria, 09042 Monserrato (CA), Italy}
\affiliation{INFN - Sezione di Cagliari,
Cittadella Universitaria, 09042 Monserrato (CA), Italy}

\author{Filippo Simonato\orcid{0009-0008-5500-4160}}
\affiliation{Gran Sasso Science Institute, Via F. Crispi 7, L'Aquila, I-67100, Italy}
\affiliation{INFN - Laboratori Nazionali del Gran Sasso, I-67100 Assergi, Italy}
\author{Marica Branchesi}
\affiliation{Gran Sasso Science Institute, Via F. Crispi 7, L'Aquila, I-67100, Italy}
\affiliation{INFN - Laboratori Nazionali del Gran Sasso, I-67100 Assergi, Italy}
\affiliation{INAF - Osservatorio Astronomico d’Abruzzo, Via Maggini, I-64100 Teramo, Italy}

\date{\today}

\begin{abstract}
The population of binary black hole (BBH) mergers observed by the LIGO–Virgo–KAGRA (LVK) collaboration offers a window into the cosmic evolution of compact binaries and the physical mechanisms that drive their formation. We employ the semi-analytic population-synthesis code B-POP to model BBHs assembled through isolated binary evolution and dynamical interactions in young, globular, and nuclear star clusters. Our framework self-consistently incorporates star formation history, metallicity evolution, and single and binary stellar evolution to quantify their impact on the observable properties of the BBH population and on the relative contribution of distinct formation channels. Our models are characterized by a merger rate, $\mathcal{R} = 17.5$–$24.1~\mathrm{Gpc}^{-3}~\mathrm{yr}^{-1}$, broadly consistent with LVK constraints. Moreover, the predicted distributions of primary mass, mass ratio, and effective inspiral spin parameter are compatible with those inferred from current LVK observations. Our primary-mass distribution is dominated by isolated binaries at $m_1 \lesssim 20\,\Ms$, while dynamically assembled first- and higher-generation mergers dominate at larger masses. As a consequence, the sub-population of mergers with $m_1 \gtrsim 45\,\Ms$ exhibits a nearly flat mass-ratio distribution and distinctive spin properties. We leverage our models to explore how: (i) the fraction of stars in isolated binaries and the fraction of stellar mass bound in clusters regulate the overall merger rate; (ii) common-envelope physics shapes the primary-mass distribution and its redshift evolution; (iii) the inclusion of stellar-collision products enhances the formation of higher-generation mergers; and (iv) the natal spin distribution of stellar-mass black holes influences the effective spin of merging BBHs. Using our models to assess possible origins of selected GW events, we illustrate how the complexity of the underlying astrophysical processes can hinder the possibility to draw definitive conclusions.
\end{abstract}


\maketitle

\section{Introduction}

With the conclusion of first part of the fourth observation campaign (O4a) run by the LIGO-Virgo-Kagra (LVK) collaboration \citep{2010CQGra..27h4006H,2013PhRvD..88d3007A,2015CQGra..32g4001L,2015CQGra..32b4001A,2018LRR....21....3A,2020LRR....23....3A,2021PTEP.2021eA101A}, the gravitational-wave transient catalogue (GWTC-4) counts $165$ robust detections\footnote{Here, we consider all sources with a parameter $p_{\rm astro} > 0.5$. Note that the number of detections which also have a false-alarm rate $<1$ yr$^{-1}$ is 151. Unless otherwise stated, the LVK data utilised to perform comparisons with our models are publicly available at \doi{10.5281/zenodo.16911562}.} of coalescing black hole binaries (BBHs) \citep{2025arXiv250818082T}.  

Despite the ever-growing number of detected BBHs, understanding their origin still remains an unsolved problem of GW astrophysics, primarily due to the many degeneracies and uncertainties affecting the underlying stellar physics of BH formation \citep[e.g.][]{2022PhR...955....1M, 2022Galax..10...76S}. From the theoretical viewpoint, the population of detected BBHs likely belongs to a heterogeneous mixture of binaries formed through different formation channels \citep{2019MNRAS.482.2991A, 2020ApJ...894..133A, 2020AeA...635A..97B, 2021ApJ...910..152Z,2022MNRAS.511.5797M,2023MNRAS.520.5259A}. Broadly speaking, the origin of BBHs can be categorized into two main scenarios: the isolated formation scenario \citep{2012ApJ...759...52D, 2020ApJ...905L..15B, 2002ApJ...572..407B, 2023MNRAS.524..426I, 2019MNRAS.485..889S, 2016A&A...594A..97B, 2019ApJ...882..121S, 2019ApJ...887...53F,2023arXiv230401288G}, in which BBH mergers form from the binary stellar evolution of two stars paired at birth, and the dynamical formation scenario \citep{1993ApJ...418..147L, 2010MNRAS.407.1946D, 2010MNRAS.402..371B,  2015PhRvL.115e1101R, 2016ApJ...831..187A, 2017MNRAS.464L..36A, 2018MNRAS.473..909B, 2019MNRAS.487.2947D, 2019PhRvD.100d3027R, 2021ApJ...908..194T, 2022A&A...665A..20B,  2022MNRAS.513.4527C, 2022MNRAS.517.2953T, 2024MNRAS.528.5140A, 2025MNRAS.538..639B}, in which black holes (BHs) meet and pair up via dynamical interactions in star clusters and galactic nuclei. 

Several quantities inferred from LVK detections could be used to untangle signatures of different formation scenarios, such as the local merger rate density (MRD), $\mathcal{R}_{\rm LVK} = (14–26)\yrgpc$, the mass distribution of binary primaries, the possible evolution of BBH masses and merger rate with redshift, the spin distribution. Additionally, exceptional BBH sources may hint at very specific formation scenarios. Among others, the most massive BBH mergers ever detected, like GW190521 \citep{2020PhRvL.125j1102A} and GW231123 \citep{2025arXiv250708219T}, hardly reconcile with the expectations of isolated stellar evolution \citep[but see][]{2025ApJ...995L..76P,2025arXiv250801135T}, possibly being byproducts of dynamical processes in dense stellar environments and galactic nuclei \citep{2025ApJ...994L..37K,  2025arXiv250808558B,2025arXiv250908298L, paiella_letter_2025}, or deriving from an even more exotic scenario \citep[e.g. the merger of primordial BHs,][]{2025PhRvD.112h1306Y,2025arXiv250809965D}. 

In the last few years, several studies utilised phenomenological models to infer properties of BBH mergers at population-level and to assess their statistical significance, either regarding the BBH masses \citep{2017ApJ...851L..25F,2018ApJ...856..173T,2019MNRAS.484.4216R,2021ApJ...913L..19T}, spins \citep{2017PhRvD..96b3012T,2019PhRvD.100d3012W,2020ApJ...895..128M,2021PhRvD.104h3010R,2023PhRvD.108j3009G}, or the possible combination of BBH features \citep{2020PhRvD.102l3022R,2023ApJ...946...16E,2024PhRvX..14b1005C,2025PhRvD.112b3531A,2025arXiv250909123A,2023arXiv230401288G,2025ApJ...991...17R,2025CQGra..42v5008K,2024ApJ...962...69F,2025ApJ...991...17R}.
From the numerical point of view, semi-analytical population synthesis codes represent powerful tools to simulate statistically relevant populations of BBH mergers to compare with observations. Generally, these tools are tailored to represent one scenario at a time, either isolated \citep{2023MNRAS.524..426I, 2008ApJS..174..223B, 2002MNRAS.329..897H} or dynamical \citep{2020MNRAS.492.2936A, 2024PhRvD.110d3023K, 2021ApJ...908..194T, 2021Symm...13.1678M}. Among others \citep{2019MNRAS.486.5008A,2021Symm...13.1678M,2022arXiv221010055K}, \bpop \citep{2019MNRAS.482.2991A,2020ApJ...894..133A,2023MNRAS.520.5259A} is a semi-analytical tool that simultaneously simulates populations of BBH mergers forming via isolated binary (IB) stellar evolution or dynamical encounters in young (YC), globular (GC), and nuclear clusters (NCs). 
This allows \bpop to simulate a synthetic Universe that naturally contains a mixture of BBH mergers originating in different environments and through different formation scenarios. Comparing observations with simulated BBH populations, constructed by varying the cosmic star formation history of different environments or the stellar evolution paradigm, can help placing constraints on the mutual contribution of different formation channels to the population of observed BBH mergers. 

In this work, we leverage on \bpop to perform "forward modeling" and predict the population of BBH mergers in synthetic Universe models. Each model is defined by a set of parameters, including the cosmic star formation history of galaxies and clusters, the metallicity evolution with redshift, stellar evolution of single and binary stars, and the structural properties of star clusters.
Comparing models and observational constraints on BBH population properties, we explore the impact of model parameters on BBH observables, such as the MRD, the distribution of masses and spins, and the evolution of such quantities with redshift. We show that, despite the many uncertainties, a fiducial model based on astrophysically motivated initial conditions naturally leads to a population of BBHs with properties broadly consistent with those inferred from the sample of observed GW sources. We also attempt to constrain the origin of some GW events, showing how the complexity of astrophysical models hinders the possibly to conclusively assign a specific formation scenario for single sources. 

The paper is organized as follows: Section~\ref{sec:method} describes the method adopted and the features of investigated models; Section~\ref{sec:res} presents the main results describing the main features of our mock population of BBHs, the comparison with GWTC-4 observational constraints, and the impact of the adopted initial conditions; and Section~\ref{sec:end} summarizes the conclusions of the work. Additionally, a comprehensive description of the code is available in Appendix~\ref{app:method}.

\section{Method}
\label{sec:method}

\bpop \citep[Binary merger POPulations][]{2019MNRAS.482.2991A, 2020ApJ...894..133A,2023MNRAS.520.5259A} is a semi-analytical code for the simulation of BBH mergers forming from either IBs or dynamical interactions in YCs, GCs, and NCs\footnote{The code is available at \url{https://github.com/marcasedda/BPOP}}. The code implements several recipes to initialize the cosmic star formation, metallicity, star cluster masses and sizes and their long-term evolution, natal mass, spin, and kick of stellar BHs, properties of dynamically formed binaries, mass of BHs formed from stellar collisions and interactions. 
Moreover, \bpop handles multiple generation mergers in star clusters using numerical relativity fitting formulae to calculate the merger remnant mass and spin
\citep{2017PhRvD..95f4024J} and gravitational recoil \citep{2007PhRvL..98w1102C,2007PhRvL..98i1101G,2008PhRvD..77d4028L,2012PhRvD..85h4015L}. Hence, \bpop creates a synthetic Universe filled with an heterogeneous population of BBH mergers formed in different environments. 
We refer the interested reader to \citep{2020ApJ...894..133A, 2023MNRAS.520.5259A} for further details about the code workflow and features. More details about the different features of the code are described in Appendix~\ref{app:method}.

Recently, we upgraded \bpop to include state-of-the-art stellar evolution tracks for single and binary stars, star cluster evolution dynamical evolution, a reliable redshift-dependent metallicity distribution of BH progenitors, and the possible formation, through stellar collisions and mergers, of BHs with masses beyond the limit set by ordinary single and binary stellar evolution. A dedicated Python library processes \bpop catalogs to generate MRD profiles for different environments and a mock catalog of sources for a given time interval.

\subsection{Investigated models}
To properly model a cosmic population of merging BBHs, we need several ingredients to describe: a) the stellar evolution paradigm adopted to link stellar progenitors and their remnants, and b) the cosmological evolution of the environments in which they form and evolve. 

While deferring to Appendix \ref{appA} a detailed discussion about the structure and features of \bpop, we briefly summarize here the main properties of our fiducial synthetic Universe and its variations.

\subsubsection{Black holes natal masses and spins}
The natal mass of BHs is drawn from catalogs of BHs forming from single and binary stars generated with the \sevn population synthesis code \citep{2015MNRAS.451.4086S, 2019MNRAS.485..889S,2023MNRAS.524..426I}. We initialize stellar binaries following the fiducial model described in \cite[][see section 2.1-2.3 and Table 5]{2023MNRAS.524..426I}. To find a balance between the completeness of our models and computational demand, we only vary the $\alpha_{\rm CE}$ parameter, which regulates the fraction of orbital energy converted into kinetic energy during the common envelope phase. In this work, we assume $\alpha_{\rm CE} = (1,~5)$, i.e. values that bracket the maximum and minimum values of the local merger rate evaluated for IB mergers only \citep[see Fig. 22 in][]{2023MNRAS.524..426I}.

We generate $2\times 10^8$ binaries and single stars divided into 12 values of the metallicity in the range $Z = 10^{-4}-0.03$. While BBH orbital properties of isolated binaries are provided by \sevn, the orbital properties of dynamical binaries are calculated according to the environment in which they form. We assume that a certain fraction, $f_{\rm mix}$, of BBH mergers involve at least one component whose progenitor was previously in a primordial (or original) binary, i.e. a binary star that was already paired at cluster birth. The remaining $1-f_{\rm mix}$ binaries, instead, are assumed to involve only BHs formed from single stars \citep[see also][]{2023MNRAS.520.5259A}. An important consequence of this choice is the possibility to take into account also BHs produced in binary mergers events, which can attain masses within or above the upper mass gap, i.e. the range of masses where the onset of pair-instability supernova (PISN) rips off the stellar progenitor and leaves no remnant \citep{2021ApJ...912L..31W}. Upper mass gap BHs formed this way may crucially determine the properties of BBHs with a dynamical origin \citep[see e.g.][]{2020MNRAS.497.1043D,2020ApJ...903...45K,paiella_letter_2025}. In this work, we assume $f_{\rm mix} = 0,~0.5$, or $1$, i.e. either none, half, or all BHs have a progenitor that evolved in a binary. Additionally, BHs in the upper mass gap can form from a dynamical merger among initially unrelated stars. We leverage the results from direct $N$-body simulations presented in \cite{2020MNRAS.497.1043D} (hereafter DC19), and implement a metallicity-dependent function to calculate the fraction of BBHs with at least one dynamically-formed component with mass in the upper mass gap. We extract the mass of these BHs in the range $(50-100)\,\Ms$, assuming a flat distribution. The latter assumption represents a reasonable balance between the impact of the adopted Kroupa initial mass function \cite{2001MNRAS.322..231K}, which favors lighter BH progenitors, and the fact that heavier BH progenitors have larger cross sections, thus a higher probability to interact and pair up. 
As evidenced in several recent works \citep{2021ApJ...908L..29G,2021MNRAS.501.5257R,2023MNRAS.526..429A,2024MNRAS.531.3770R,2026arXiv260107917R}, multiple stellar collisions and star-BH interactions can also trigger the formation of intermediate-mass black holes (IMBHs), i.e. objects with a mass $>100\,\Ms$. To include this further feature, in \bpop it can be assumed that a fraction $f_{\rm IMBH}$ of all clusters with given properties can form a very massive star (VMS) via repeated collisions. In this work, we assume that only a fraction $f_{\rm IMBH}=0.2$ of clusters with a mass $>5,000\,\Ms$ and a density $>10^5\,\Ms$ pc$^{-3}$ can form a VMS, as suggested by the works referenced above. The terminal VMS mass is calculated as $m_{\rm vms} = \max(2\times10^4\, \Ms, 0.02\, m_{\rm cl})$, following the numerical results described in \cite{2024MNRAS.531.3770R} (hereafter R24). We assume that the VMS directly collapses to an IMBH if its mass exceeds $m_{\rm vms} > 270\,\Ms$, and calculate the IMBH mass as $m_{\rm IMBH}\simeq0.9 \, m_{\rm vms}$ \citep{2017MNRAS.470.4739S}. 
More details about BHs natal masses and the adopted stellar evolution paradigm are provided in Appendix~\ref{appA1}.

BH natal spins are assigned depending on the BH formation history \citep[following][]{2020AeA...635A..97B}. In particular, we assign a negligible spin to BHs formed from single stars and to first-born BHs in binaries  \citep{2019ApJ...881L...1F}, while we assume a flat spin distribution limited between 0 and 1 for all second-born BHs in binary stars and BHs formed from stellar collisions (see Appendix~\ref{appA} for further details). We refer to this model as to B20.
Additionally, we perform a run in which all BHs have a natal spin extracted from a Maxwellian distribution with $\sigma_\chi = 0.2$ , following our previous papers \citep{2020ApJ...894..133A,2023MNRAS.520.5259A}, to mimick a distribution skewed toward values $>0.1$ as suggested by LVK data analysis \citep{2023PhRvX..13a1048A,2025arXiv250818083T}. Hereafter, we refer to this as the LVK model. 
Finally, we model spins orientation differently depending on the formaiton channel. For dynamical binaries, we assume randomly oriented spin vectors. For isolated binaries, we adopt a distribution for the polar angles $\theta_{1,2}$, i.e. the angles between each spin and the binary angular momentum, such that up to $\sim 55\%$ of isolated BBHs feature a relative difference between $\der\theta/\theta < 0.05$. More details about BH natal spins and spin alignment are provided in Appendix~\ref{appA2}.

In the following sections, we will explore the properties of BBH spins through global quantities, such as the binary effective spin parameter, i.e. a combination of the projections of spin vectors ($\vec{\chi_i}$) on the binary angular momentum ($\vec{L}$) \citep{2008PhRvD..78d4021R,2010PhRvD..82f4016S,2011PhRvL.106x1101A}
\begin{equation}
    \chi_{\rm eff} = \frac{\left(m_1 \vec{\chi}_1 + m_2 \vec{\chi}_2\right)\cdot \vec{L} }{m_1+m_2},
\end{equation}
and the precession spin parameter, defined as \citep{2015PhRvD..91b4043S,2021PhRvD.103f4067G}
\begin{equation}
    \chi_p = {\rm max} \left(\chi_1 \sin\theta_1,  q\frac{4q-3}{4-3q}\chi_2 \sin\theta_2\right).
\end{equation}

\subsubsection{Star cluster evolution and black hole dynamics}

A star cluster in \bpop is uniquely identified by its initial mass and half-mass radius. These two quantities are used to calculate the cluster structural properties: core radius, density, velocity dispersion, and escape velocity. Cluster evolution is modeled to follow three main phases: i) initial contraction driven by two-body relaxation and mass-segregation, ii) core-collapse, iii) rebound and expansion. The formulation is based on the \dragonii $N$-body simulations \citep{2024MNRAS.528.5119A} and semi-analytical results by \citep{2023MNRAS.522.5340G}, and is described in detail in Appendix~\ref{appA3}. 

In \bpop, the dynamical evolution of the cluster directly affect the formation of BBHs in star clusters. In fact, dynamical BBHs in \bpop form through either three-body or binary--single scatterings, and shrink via stellar hardening and GW emission down to a point where they either merge in the cluster or are ejected, possibly merging afterwards. The final properties of merger remnants are calculated according to numerical relativity fitting formulae \citep{2017PhRvD..95f4024J}, following our previous works \citep{2020ApJ...894..133A,2023MNRAS.520.5259A}. If the relativistic kick is smaller than the cluster escape velocity, the remnant is retained, possibly forming new binaries and undergoing further ``hierarchical'' mergers. 
Since cluster properties crucially affect the timescales over which BBHs evolve, in \bpop we incorporate the cluster evolution into the set of equations regulating BBH formation and evolution, as shown in Section~\ref{appA4}.

\subsubsection{Star formation history and metallicity evolution}
To infer the MRDs and compile mock catalogs, we convolve our population with the star formation rate (SFR) of different environments, ensuring that the sum of all formation rates from different channels equals the cosmic star formation rate (CSFR), which we assume as 
\begin{equation}
\psi_{\rm MF17}(z) = \frac{0.01 (1+z)^{2.6} }{ 1 + \left[(1+z)/{3.2}\right]^{6.2}} ~\Ms\, {\rm yr}^{-1}\, {\rm Mpc}^{-3},
\label{eq:sfr}
\end{equation} 
following \cite[][hereafter MF17]{2017ApJ...840...39M}. 
In the fiducial model, we assume this functional form to model the formation history of IBs, YCs, and NCs. However, we scale the SFR by a factor that represents the amount of stellar material contributing to a specific channel. In the case of IBs, we assume that only a fraction $f_{\rm IB}$ of the whole SFR forms stellar binaries, and we set $f_{\rm IB} = 0.4$, following \cite{2023MNRAS.524..426I}. Similarly, we assume that a fraction $f_{\rm YC} = 0.01$ of stellar material builds YCs \citep{2008MNRAS.390..759B}, and that a fraction $f_{\rm NC} = 0.0005$ of stellar mass eventually resides in NCs \citep{2020A&ARv..28....4N}. For GCs, instead, we assume a redshift-dependent Gaussian distribution
\begin{equation}
    \psi_{\rm EB19}(z) = B_{\rm GC} \exp\left[-\frac{(z - z_{\rm GC})^2}{(2\sigma_{\rm GC})^2}\right],
    \label{eq:EB19}
\end{equation}
Based on the recent work of \cite{2019MNRAS.482.4528E}. In the fiducial model, we assume a normalizing factor $B_{\rm GC} = 1.2\times10^{-4}\,\Ms~$yr$^{-1}$ Mpc$^{-3}$, mean redshift value $z_{\rm GC}=4.5$, and dispersion $\sigma_{\rm GC}=2$, following \citep{2019MNRAS.482.4528E}. We refer to this model as EB19. Alternatively, we assume $B_{\rm GC} = 2\times10^{-4}\,\Ms~$yr$^{-1}$ Mpc$^{-3}$, $z_{\rm GC}=3.2$, $\sigma_{\rm GC}=1.5$, following \citep{2022MNRAS.511.5797M}. We refer to this model as M20. 

We run an additional model in which we adopt the same Gaussian distribution for NCs too, assuming in this case $B_{\rm NC} = 5\times10^{-5}\,\Ms~$yr$^{-1}$ Mpc$^{-3}$, $z_{\rm NC}=z_{\rm GC}$, and $\sigma_{\rm NC}=\sigma_{\rm GC}$. Note that this assumption relies on the dry-merger formation scenario for NCs, by which they formed --- at least partly --- from repeated mergers of massive clusters formed in the inner regions of the host galaxy \citep[e.g.][]{1975ApJ...196..407T, 1993ApJ...415..616C, 2014ApJ...785...71G,2014MNRAS.444.3738A}. 

More details about the treatment of the CSFR in \bpop are described and discussed in Appendix~\ref{appA5}.

The metallicity evolution is modeled assuming it follows a log-normal distribution around a mean value, modeled as 
\begin{equation}
     \left\langle{\rm Log}  \left(Z/{\rm Z}_\odot\right) \right\rangle = \alpha + \beta z^\gamma + A(Z),
\end{equation}
where we adopt $\alpha=0.153,~ \beta = -0.074$, and $\gamma = 1.34$, following \cite{2017ApJ...840...39M,2020AeA...635A..97B}, and $A(Z)=\ln(10)\,\sigma_Z^2/2$ \citep{2022MNRAS.511.5797M} for the MF17 star formation model. For the EB19 star formation model, instead, we follow \citep{2016MNRAS.456.2140M} and calculate the metallicity as
\begin{equation}
{\rm Log}\left(\frac{Z}{{\rm Z}_\odot}\right) = 0.4{\rm Log}\left(\frac{M}{\Ms}\right) + 0.67 \exp\left(-0.5 z\right) - 5.04.
\end{equation}
Further details about the metallicity-redshift relation are provided in Appendix~\ref{appA6}.

The main features of our fiducial model and its variations are summarized in Table~\ref{tab1}.

\begin{table*}[]
    \centering
    \caption{Main properties of investigated models. Column 1: model name. Column 2: common envelope parameter. Column 3-4: fraction of stars in isolated binaries and fraction of stellar mass bound in young clusters. Column 5: fraction of BHs in clusters formed from progenitors born in a binary. Column 6-7: model adopted for dynamically formed upper-mass gap BHs and IMBH seeds. Column 8: model adopted for natal spins. Column 9-12: star formation history model assigned to isolated binaries and dynamical binaries in young, nuclear, and globular clusters. }
    \setlength{\tabcolsep}{3pt} 
    \renewcommand{\arraystretch}{1.3}
    \begin{tabular}{c|c|cc|c|cc|c|cccc}    
    \hline\hline
     {\bf Model} & ${\bf \alpha_{\rm CE}}$ & ${\bf f_{\rm IB}}$ & ${\bf f_{\rm YC}}$ & ${\bf f_{\rm mix}}$ & {\bf Upper}  & {\bf IMBH} &  {\bf Spin} & ${\bf \psi_{\rm IB}}$ &${\bf \psi_{\rm YC}}$ &${\bf \psi_{\rm NC}}$ &${\bf \psi_{\rm GC}}$ \\
     {\bf name}     & & & & & {\bf mass gap} & {\bf seeds} & & \\
    \hline 
    \multicolumn{12}{c}{{\bf Fiducial}}\\[3pt]
    \hline  
    F    & 1 & 0.4 & 0.01 & 0.5 & DC19 & R24 & B20 & MF17 & MF17 & MF17 & EB19 \\
    \hline 
    \multicolumn{12}{c}{{\bf Primordial binary fraction}}\\[3pt]
    \hline 
    Fb0  & 1 & 0.4 & 0.01 & 0   & DC19 & R24 & B20 & MF17 & MF17 & MF17 & EB19 \\
    Fb1  & 1 & 0.4 & 0.01 & 1   & DC19 & R24 & B20 & MF17 & MF17 & MF17 & EB19 \\
    \hline     
    \multicolumn{12}{c}{{\bf Common Envelope}}\\[3pt]
    \hline 
    F5   & 5 & 0.4 & 0.01 & 0.5 & DC19 & R24 & B20 & MF17 & MF17 & MF17 & EB19 \\
    \hline 
    \multicolumn{12}{c}{{\bf Star formation history}}\\[3pt]
    \hline
    Fe  & 1 & 0.4 & 0.01 & 0.5 & DC19 & R24 & B20 & MF17 & MF17 & MF17 & M20   \\
    Fb  & 1 & 0.4 & 0.01 & 0.5 & DC19 & R24 & B20 & MF17 & MF17 & EB19 & EB19  \\
    \hline 
    \multicolumn{12}{c}{{\bf BH natal spins}}\\[3pt]
    \hline 
    Fmxl & 1 & 0.4 & 0.01 & 0.5 & DC19 & R24 & LVK & MF17 & MF17 & MF17 & EB19 \\
    \hline
    \multicolumn{12}{c}{{\bf Seeds}}\\[3pt]
    \hline 
    Fgp  & 1 & 0.4 & 0.01 & 0.5 & DC19 &  -  & B20 & MF17 & MF17 & MF17 & EB19 \\
    Fst  & 1 & 0.4 & 0.01 & 0.5 &   -  &  -  & B20 & MF17 & MF17 & MF17 & EB19 \\
    \bottomrule             
    \end{tabular}
    \label{tab1}
\end{table*}

\subsection{Merger rates and mock sources}
Following previous works, we define the cosmic MRD for each formation channel as \cite[e.g.][]{2020ApJ...898..152S,2021MNRAS.507.5224B}
\begin{align}
    \mathcal{R}_i(z)  = & \int_{Z_{\rm min}}^{Z_{\rm max}} \int_z^{z_{\rm max}} \der Z \, \der z' \, \frac{\der t_{\rm lb}(z')}{\der z'}\nonumber \\
    & \psi_i(z') \, p_i(z',Z) \, \eta(Z) \, \dot{\mathcal{F}}_i(z',z,Z)    
    \label{rz}
\end{align}
where $\der t_{\rm lb}(z')/\der z' = \left[(1+z) \, H_0 \, \left[(1+z)^3\,\Omega_m + \Omega_\Lambda\right]^{1/2}\right]^{-1}$ is the cosmological time element; $\psi_i(z')$ is the SFR calculated at the binaries' formation redshift, $z'$; $p_i(Z,z')$ is the probability that a star formed at $z'$ has metallicity $Z$; $\eta(Z)$ is the BBH merger efficiency at the metallicity $Z$; $\dot{\mathcal{F}}_i(z',z,Z)$ is the number of mergers forming at redshift $z'$ and merging at redshift $z$ in environments with metallicity $Z$, per time unit. We adopt cosmological parameters from Planck 2018 results \citep{2020A&A...641A...6P}.

For each $i$-th formation channel and metallicity value, we integrate the corresponding MRD ($\mathcal{R}_i(z',Z)$) in different redshift bins, taking into account the cosmological volume and the dilation time, such that the number of events per year is given by \citep[e.g.][]{2018ApJ...863L..41F}
\begin{equation}
    \der\Gamma_i(z,Z) = \int_{z_i}^{z_{i+1}} \mathcal{R}_i(z',Z)\, \frac{\der V(z')}{(1+z')\,\der z'}\, \der z',
\end{equation}
where $\mathcal{R}_i(z',Z)$ encodes the astrophysical properties of our model, $\der V(z')/\der z'$ is the element of comoving volume, and the term $(1+z')^{-1}$ accounts for time dilation. The number of sources for each channel forming at a redshift $z$ is thus given by $N_{\rm gw,i}(z,Z) = \der\Gamma_i(z,Z)\,T$, where $T$ is the considered time-span. In the following, we assume $T=10$ yr of events. This procedure allows us to create a synthetic population of BBH mergers for each formation channel and metallicity bin, and assess the contribution of different environments to the cosmic population of merging BHs \footnote{All simulated data are available at \url{10.5281/zenodo.19115567}}.

\section{\label{sec:res} Results}

\subsection{Black hole binary mergers in different environments: general properties}

\begin{figure*}
    \centering
    \includegraphics[width=0.95\columnwidth]{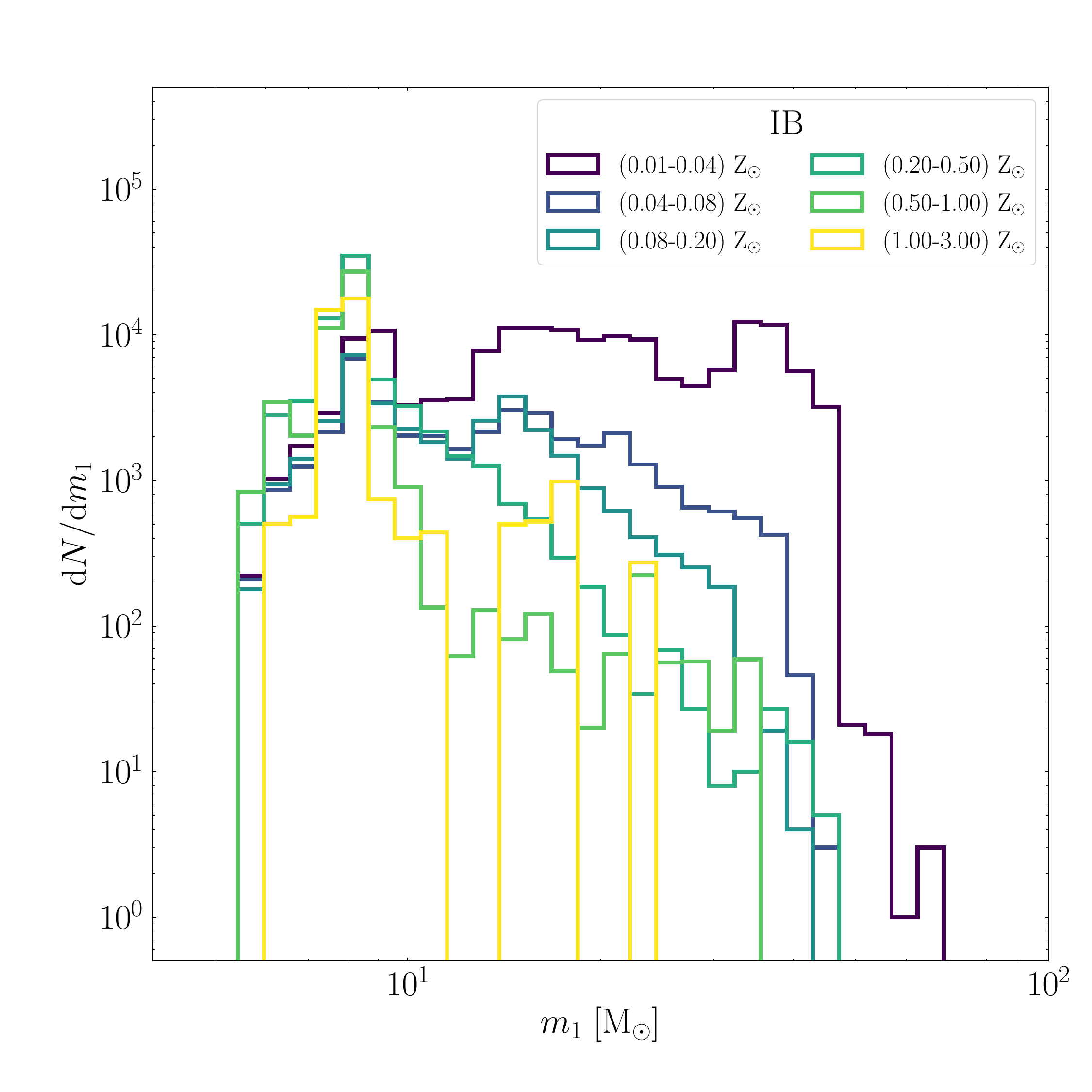}
    \includegraphics[width=0.95\columnwidth]{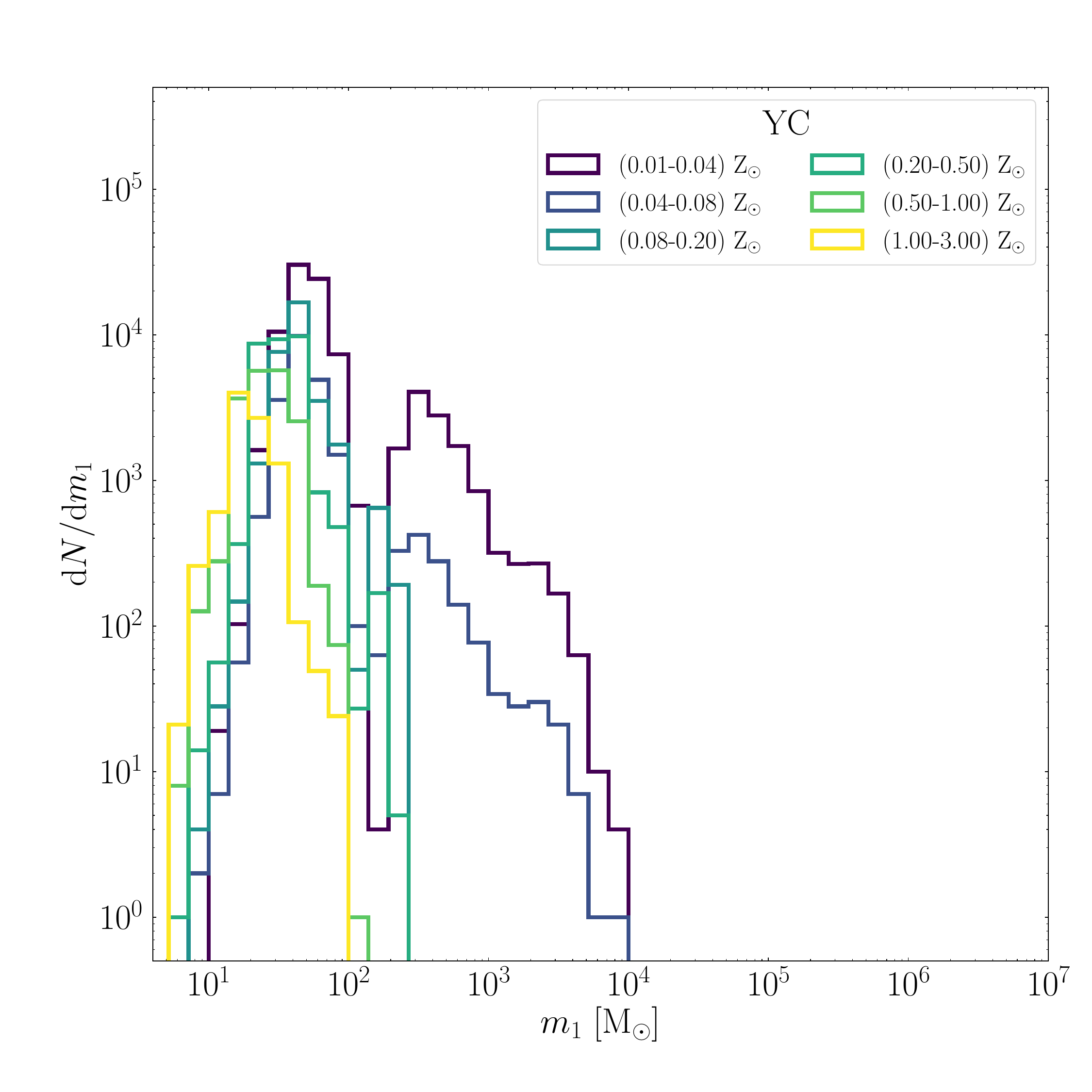}\\
    \includegraphics[width=0.95\columnwidth]{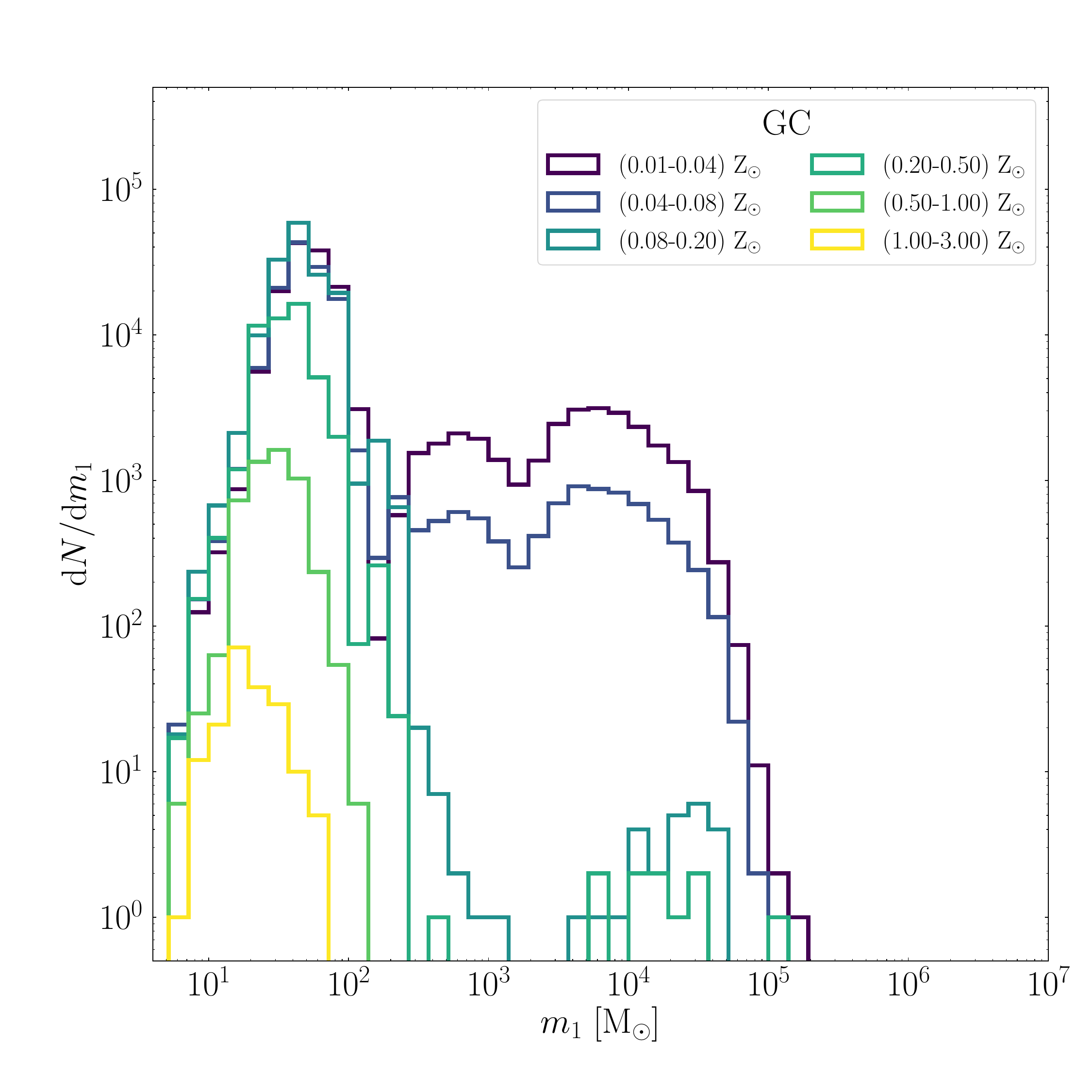}
    \includegraphics[width=0.95\columnwidth]{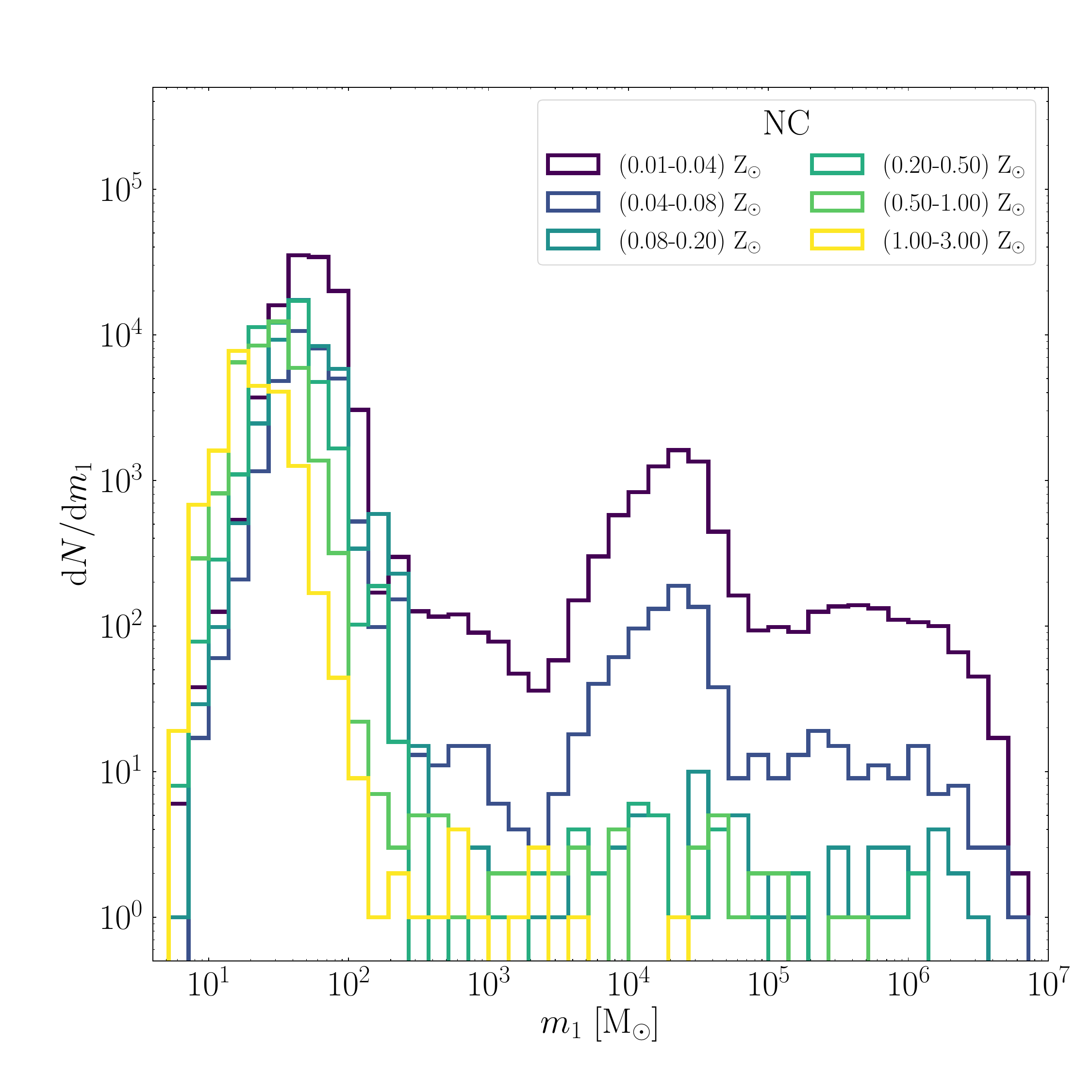}
    \caption{Primary mass distribution of merging BBHs in different environments. From left to right and from top to bottom: isolated binaries (IB), young (YC), globular (GC), and nuclear clusters (NC). Diffrent colors correspond to different metallicity bins.}
    \label{fig:gmass}
\end{figure*}

We first focus on the intrinsic properties of the simulated population of BBH mergers in our fiducial model. Figure~\ref{fig:gmass} shows the primary mass distribution for all merging BBHs dissected into different formation channels and metallicity bins. In the case of IBs, the distribution shows two different trends depending on the metallicity. Mergers originating in environments with metallicities below $0.04$ Z$_\odot$ show a nearly flat distribution extending up to $\sim 40 \, \Ms$, with two narrow peaks around $8$ and $35-40 \, \Ms$ and a broad peak covering the $15-25 \,\Ms$ mass range. More metal-rich environments, instead, produce BBH mergers with a primary mass distribution narrowly peaked around $8 \, \Ms$ and an extending tail up to $20-40 \, \Ms$. As expected, these features are fully consistent with previous results \citep{2023MNRAS.524..426I}.

The primary distribution of dynamical mergers in metal poor environments, $<0.1$ Z$_\odot$, exhibits a more complex morphology, with a clear peak around $40 \,\Ms$ common to all cluster types, and an extended tail peaking around $500 \,\Ms$ for YC, and covering a broad range, $\sim 10^3-10^7 \,\Ms$, in GCs and NCs. In these environments, stellar collisions and hierarchical mergers can efficiently form IMBHs and, in NCs, even objects in the supermassive BH mass range\footnote{In this paper, we label as IMBHs all objects with a mass $m_{\rm IMBH}>10^2\,\Ms$}.

We find that an IMBH is involved in at least one merger in $(6.7-8.6)\%$ of BBHs in YCs, $(8.4-10.9)\%$ of BBHs in GCs, and $(4.4-5.9)\%$ of mergers in NCs. The parameter that mostly affects these percentages, and brackets their boundaries, is the fraction of BHs forming in primordial binaries, $f_{\rm mix}$, with larger values of $f_{\rm mix}$ leading to a larger fraction of mergers involving an IMBH. 

Around $90\%$ of IMBHs with a mass $>10^3\,\Ms$ have an ancestor with an initial mass $m_{\rm pro} > 100\,\Ms$, mostly formed through stellar collisions. In the remaining $\sim 10\%$ of the cases, instead, IMBHs grow solely via hierarchical mergers from a stellar BH ancestor. 

BBH mergers from clusters with larger metallicities are characterized by a distribution with a single peak around $20-40\, \Ms$. Note that all the features described in this section are intrinsic of the population. To assess their impact on a realistic population of mergers, we must convolve different populations with their own merger efficiency, star formation history, and metallicity evolution.

\subsection{Building up a synthetic Universe: the cosmological merger rate density of stellar black hole binaries}
 
\begin{figure*}
    \centering
    \includegraphics[width=0.322\textwidth]{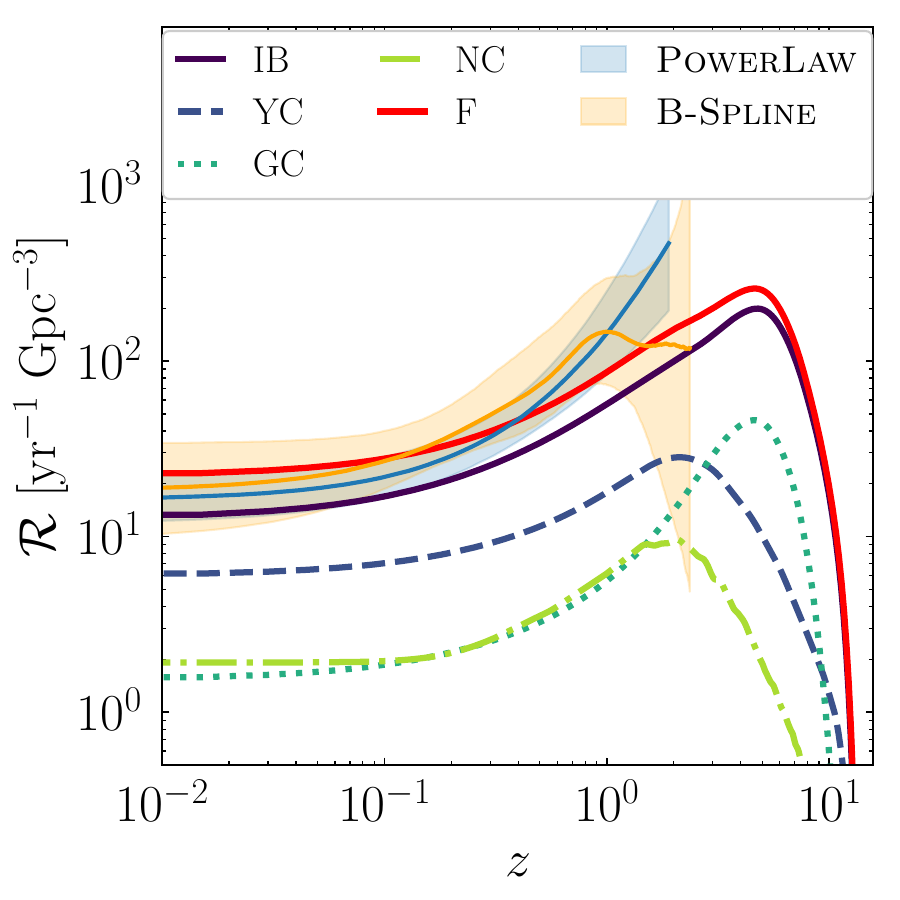}
    \includegraphics[width=0.322\textwidth]{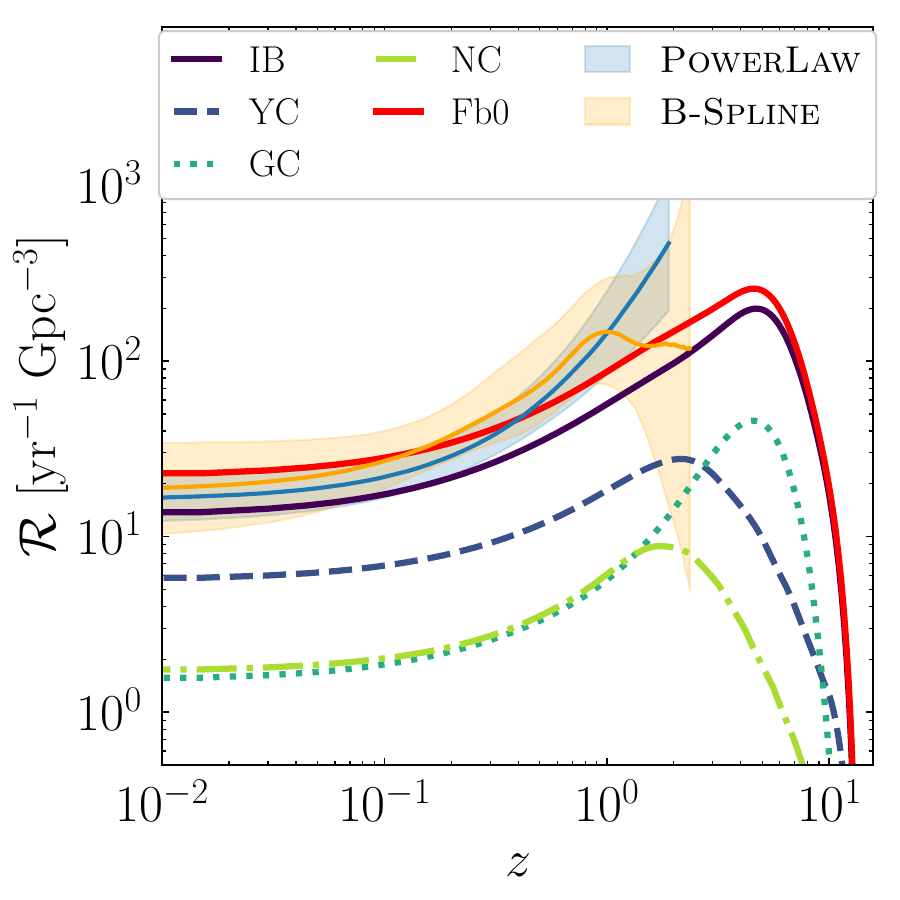}
    \includegraphics[width=0.322\textwidth]{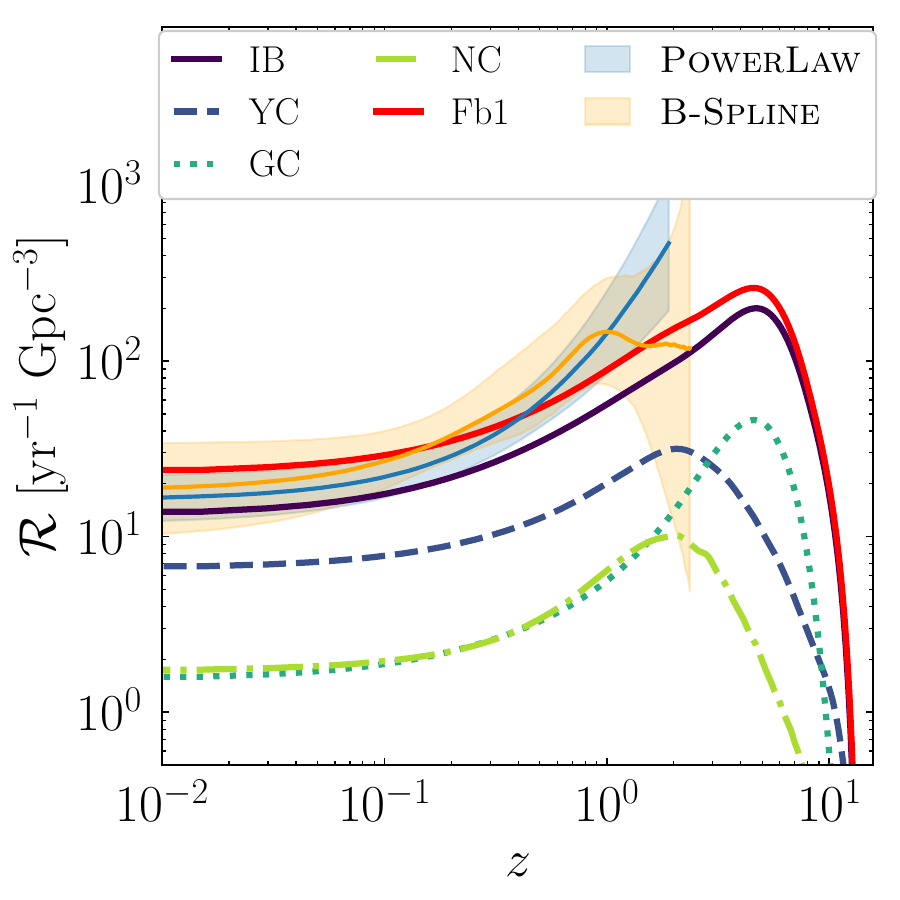} \\
    \includegraphics[width=0.322\textwidth]{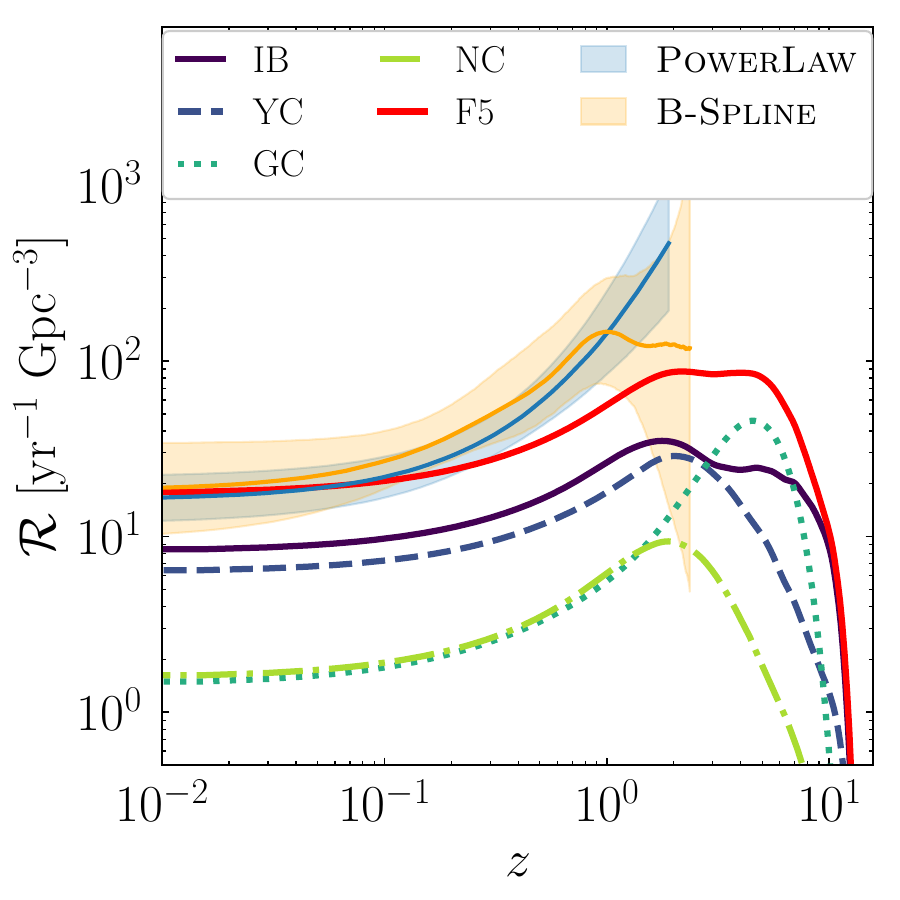} 
    \includegraphics[width=0.322\textwidth]{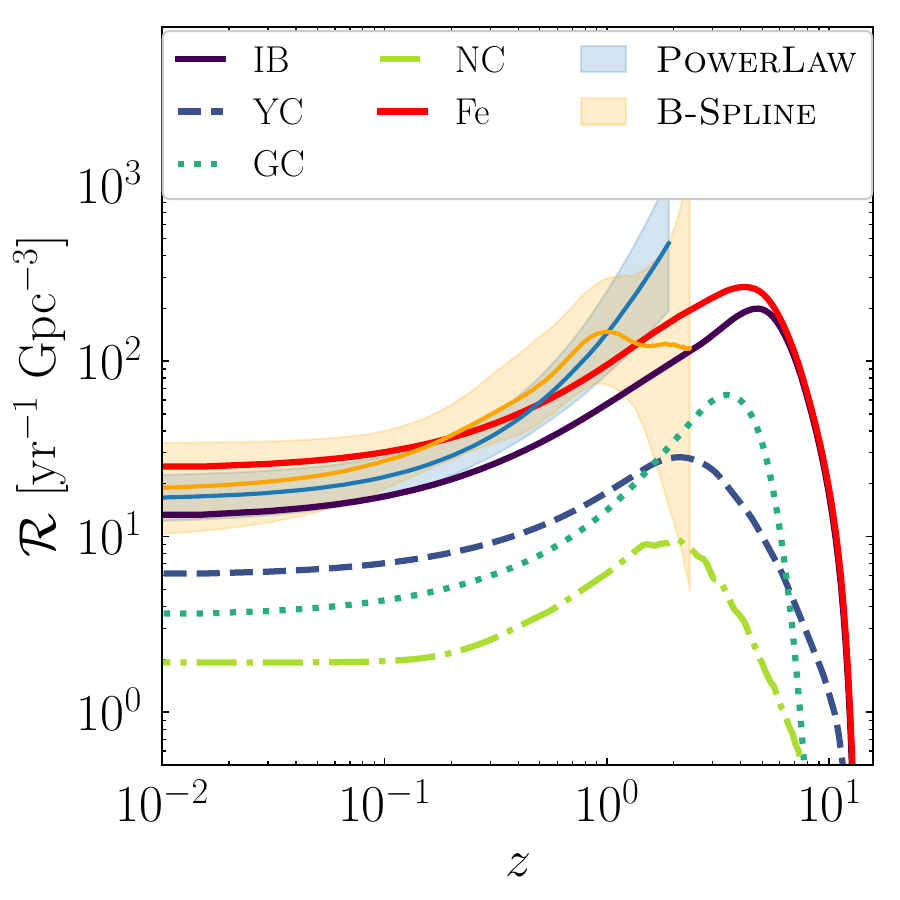} 
    \includegraphics[width=0.322\textwidth]{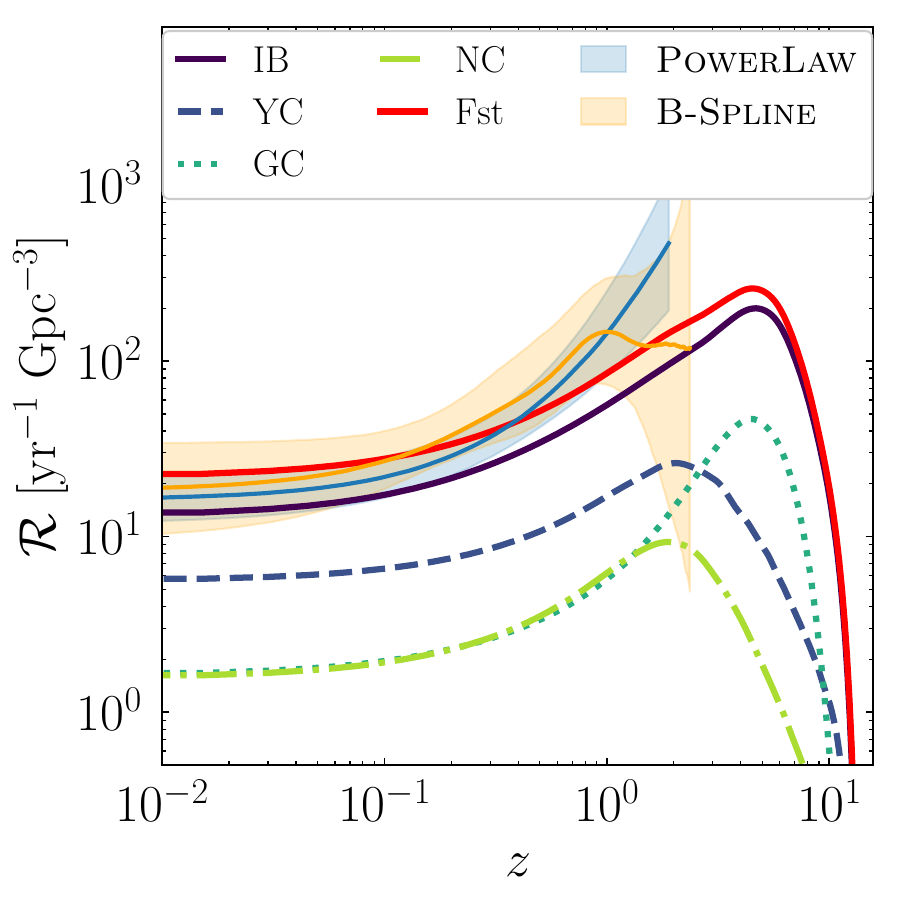} \\
    \caption{Merger rate density of all BBH mergers for different models. The straight blue and orange lines, as well as the shaded regions, correspond to the inferred rate from GWTC-4 according to a strongly modelled approach (\textsc{Power-law}, blue) and a weakly modelled approach (\textsc{B-SPLINE}, orange) \citep{2025arXiv250818083T}, and their $90\%$ confidence interval. In all panels, the red straight line identifies the total simulated merger rate density, while the other lines highlight the contributions from BBHs in isolated binaries (purple straight line, IB), young clusters (blue dashed line, YC), globular clusters (green dotted line, GC), and nuclear clusters (light green dash-dotted line, NC). From left to right, top row panels display the rate for the fiducial model (F) and its variations assuming a primordial binary fraction $f_{\rm mix} = 0$ (Fb0), and $1$ (Fb1). Similarly, bottom row panels display the rate for a model with $\alpha_{\rm CE}=5$ (F5), with a GC formation history as described in \citep{2020ApJ...898..152S} (Fe), and excluding dynamically formed upper-mass gap BHs and VMS remnants. }
    \label{fig:1}
\end{figure*}

Figure~\ref{fig:1} compares MRD profiles calculated for different models against the rate inferred from observed BBHs in GWTC-4, either assuming that the MRD follows a power-law (\textsc{Power Law} model), or adopting a weekly modelled approach to reconstruct the MRD (\textsc{B-SPLINE} model) \citep{2025arXiv250818082T,2025arXiv250818083T}. The last column in Table~\ref{tab:mrd} shows the local merger rate, $\mathcal{R}_{\rm loc}$, i.e. the MRD evaluated at redshift $z=0.2$ \citep{2023PhRvX..13a1048A}.  

\begin{table}[]
    \centering
    \caption{Simulated local merger rate. Column 1: model name. Column 2-4: local merger rate evaluated for first-generation mergers, high-generation mergers, and mergers involving an IMBH.}
\setlength{\tabcolsep}{3pt} 
    \renewcommand{\arraystretch}{1.3}
    \begin{tabular}{c|ccc}
\hline\hline
Model  & \multicolumn{3}{c}{$\mathcal{R}_{\rm loc}$ ($\yrgpc$)}\\[3pt]
  name  & 1g & $n$g & IMBH \\[3pt]
    \hline
F    &22.13 & 0.95 &0.0082\\[3pt]
Fb0  &22.0 & 1.01 &0.00089\\[3pt]
Fb1  &23.1 & 0.88 & 0.021\\[3pt]
F5   &17.5 & 0.06 &0.013\\[3pt] 
 Fe  &24.1 & 0.063 &0.016\\[3pt]
 Fb  &21.3 & 0.039 &0.0081\\[3pt]
Fmxl &22.4 & 0.005 &0.0055\\[3pt] 
Fgp  &21.9 & 0.043 &0.0073\\[3pt]
Fst  &21.9 & 0.045 &0.0081\\[3pt]
     \bottomrule             
   \end{tabular}
    \label{tab:mrd}
\end{table}

Overall, our models predict a local merger rate that ranges between $\Rate_{\rm loc}=(17.5-24.1)\yrgpc$, in broad agreement with LVK's latest predictions, i.e. $\Rate_{\rm LVK} = (14-26)\yrgpc$. Toward $z=0$, our simulated MRDs tend to decline more gently than the phenomenological functions adopted by LVK. 
{Within $z=1.5$, our models can be fit with a shallow power-law  with offset $\mathcal{R}_0\sim 22\yrgpc$ and slope $k_{\mathcal{R}} \sim 1.1$}. Moreover, our MRD profiles deviate from LVK results at redshift $z\gtrsim 1-1.5$, bending to follow the underlying SFR. On the one hand, the differences between \bpop models and the \textsc{Power-law} model arise from the fact that LVK analysis assumes an MRD that evolves with redshift following a power-law, whilst in our case it is the result of the combined effects of SFR and BBH formation and evolution processes. On the other hand, the \textsc{B-SPLINE} model peaks at lower redshift compared to \bpop. This may be due to a non-trivial combination of factors: the adopted SFR, the metallicity relation, and the physics of binary evolution. Indeed, the amplitude and location of the peak in our simulations slightly change across all models. In most models, the MRD maximizes at redsfhit $z\sim 5$, attaining maximum values of $R_{\rm peak}\sim 200\yrgpc$. The only exception is model F5, for which $\alpha_{\rm CE} = 5$, in which the peak is broadly distributed in the redshift range $z\simeq(2-6)$ and reaches $R_{\rm peak} \simeq (80-90)\yrgpc$. As evidenced by the plots, in most models the IB channel dominates the rate at all redshift, hence suggesting that -- in our models -- the MRD could be used to probe the IB formation scenario. Moreover, we see that each formation channel is characterized by a different evolution with redshift. This interesting feature suggests that the presence of multiple peaks in the observed MRD could carry information about the impact of different formation channels. In model F5, for instance, we see that the broad peak is due to the contribution, at lower redshifts, of IBs, NCs, and YCs and, at higher redshifts, of GCs.

By comparing the total number of mergers -- i.e. the integral of the MRD up to $z=15$ -- formed through different channels, we find that IBs constitute a fraction $f_{\rm GW,IB} = 0.69-0.73$ of the whole population of mergers, whilst dynamical mergers represent the remaining $f_{\rm GW,dyn} = 0.31-0.27$ in almost all models. 
The only exception is model F5, where the choice of $\alpha_{\rm CE}=5$ causes a significant decrease in the IB merger efficiency but has little effects on the dynamical mergers, leading to a population with $f_{\rm GW,IB} \sim 0.42$ members originating from IBs and $f_{\rm GW,dyn} = 0.68$ from dynamical environments. Therefore, the main effect of $\alpha_{\rm CE} = 5$ is to reduce the overall number of BBH mergers we obtain in our mock universe \citep[][Ugolini et al, in prep]{2023MNRAS.520.5724B}.

In the dynamical formation channel, binaries originating in YCs and GCs contribute comparably to the population, whereas binaries formed in NCs constitute roughly $10\%$ of the mergers produced through this channel. Table \ref{tab:break} summarizes the fractional contribution of each formation channel to the overall merger population.

\begin{table}[]
    \centering
    \caption{Fraction of mergers from different channels. Column 1: model name. Column 2-5: fraction of mergers from isolated binaries, young clusters, globular clusters, and nuclear clusters.}
\setlength{\tabcolsep}{3pt} 
    \renewcommand{\arraystretch}{1.3}
    \begin{tabular}{c|cccc}
\hline\hline
Model name & $p_{\rm IB}$ & $p_{\rm YC}$ & $p_{\rm GC}$ & $p_{\rm NC}$\\
\hline
F & $0.726$ & $0.109$ & $0.134$ & $0.032$\\ [3pt] 
Fb0 & $0.728$ & $0.107$ & $0.135$ & $0.030$\\ [3pt] 
Fb1 & $0.717$ & $0.118$ & $0.132$ & $0.033$\\ [3pt] 
F5 & $0.376$ & $0.248$ & $0.304$ & $0.072$\\ [3pt] 
Fe & $0.685$ & $0.103$ & $0.182$ & $0.030$\\ [3pt] 
Fb & $0.718$ & $0.108$ & $0.132$ & $0.042$\\ [3pt] 
Fmxl & $0.726$ & $0.111$ & $0.129$ & $0.034$\\ [3pt] 
Fgp & $0.730$ & $0.101$ & $0.138$ & $0.032$\\ [3pt] 
Fst & $0.732$ & $0.099$ & $0.137$ & $0.031$\\ [3pt] 
\bottomrule
\end{tabular}
    \label{tab:break}
\end{table}

The redshift-dependent metallicity distribution and the adopted star formation rate have a clear impact on the stellar population that mostly contribute to the cosmic population of BBHs. In the fiducial model, most mergers originate from stellar progenitors with metallicities in the range $Z = 10^{-3}-10^{-2}$. Around $\gtrsim 95\%$ of BBH mergers from NCs have progenitors with a metallicity in the range $Z = 0.004-0.02$, whilst for GCs this range reduces to $Z = 0.001-0.007$. Similarly, around $97\%$ of all mergers from the IB channel have progenitor metallicities in the range $Z = 0.0014-0.014$. 

\begin{figure}
    \centering
    \includegraphics[width=\columnwidth]{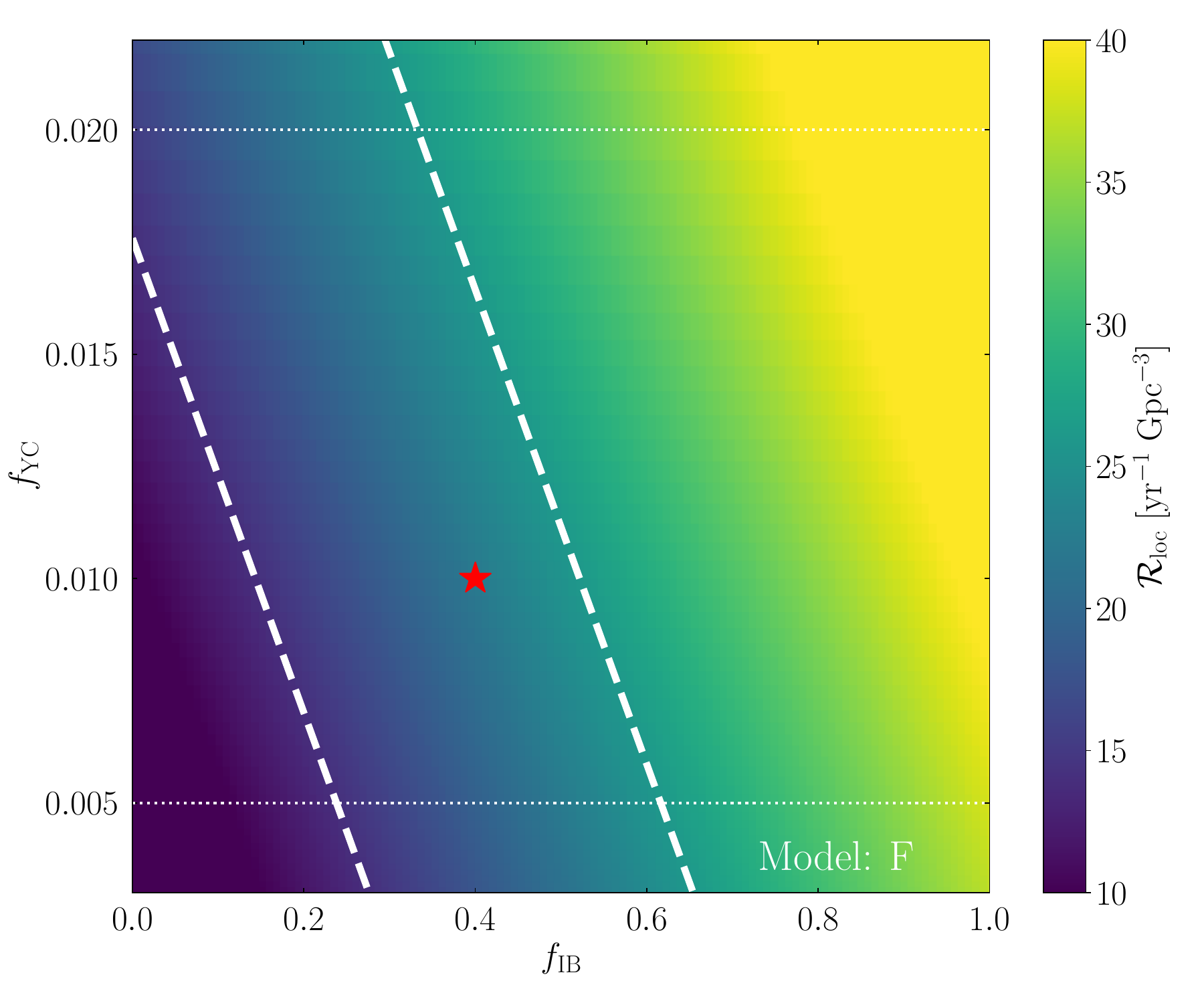}\\
    \includegraphics[width=\columnwidth]{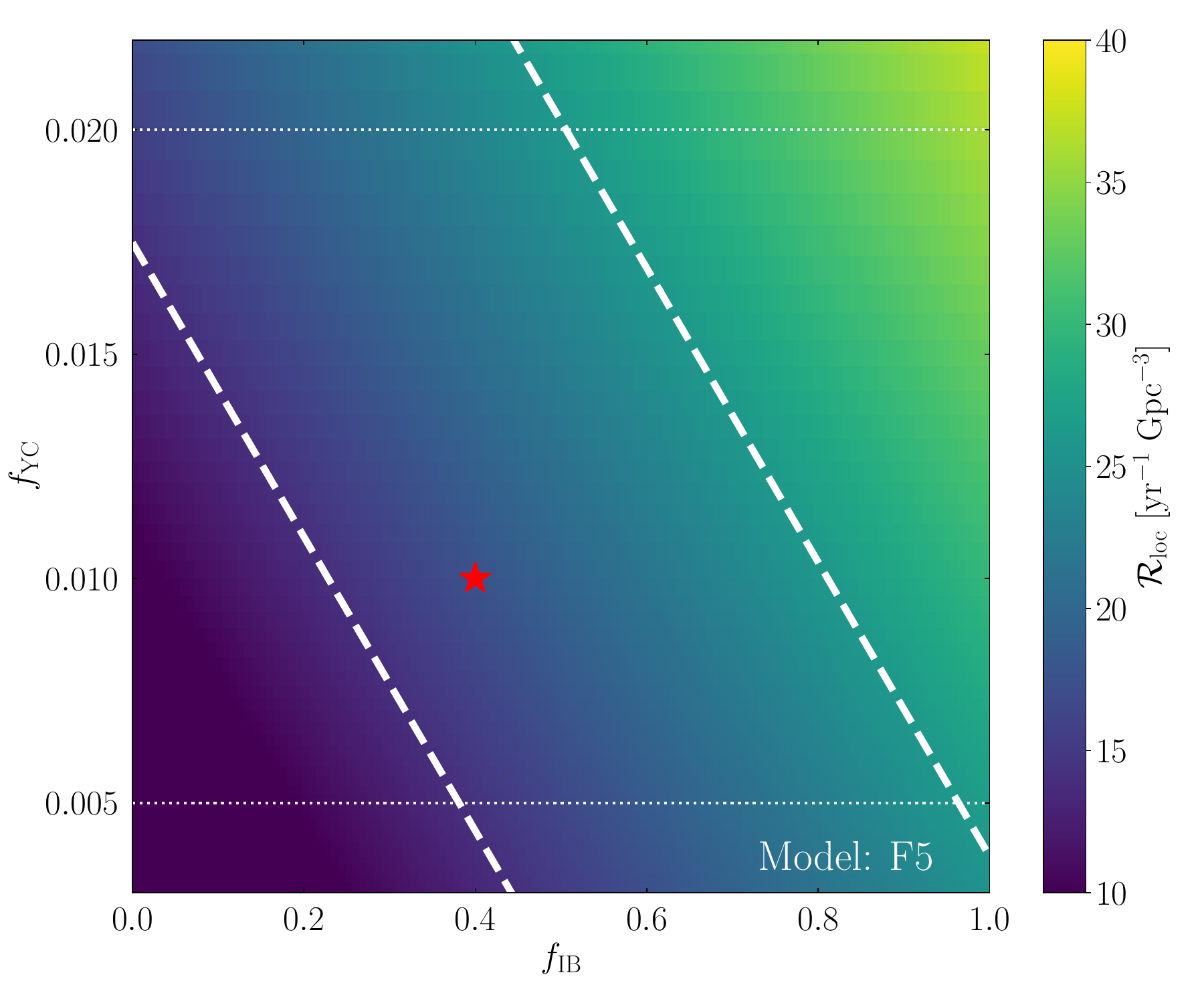}\\
    \caption{Merger rate density at redshift $z=0.2$ as a function of the fraction of isolated binaries ($f_{\rm IB}$) and the fraction of stars in YCs ($f_{\rm YC}$). The red star identifies the considered model. White dashed lines denote the loci of $f_{\rm IB}-f_{\rm YC}$ values required to reproduce the upper and lower bounds of the LVK-inferred rate. The horizontal, white, dotted lines mark the observational constraints on $f_{\rm YC}$.}
    \label{fig:2}
\end{figure}

The SFR functional form and normalization has a clear impact on the overall shape of the MRD, its local value, and the resulting mixing fractions among different formation channels. In these regards, Figure~\ref{fig:2} shows how the local MRD varies with the scaling factors adopted to quantify the fraction of binary stars in the field ($f_{\rm IB}$) or the amount of stellar mass bound in YCs ($f_{\rm YC}$), for the fiducial model and the model with $\alpha_{\rm CE}=5$. 

The plot identifies the range of $f_{\rm IB}-f_{\rm YC}$ values within which our model predicts a local merger rate compatible with GWTC-4 constraints. Moving inside this region has implications for the environments in which BBHs form. For example, observations of YCs suggest that they may constitute a fraction $f_{\rm YC} = 0.005-0.02$ of their host galaxy stellar mass \citep{2008MNRAS.390..759B}. Considering the fiducial model and the LVK merger rate estimate, we see that the extreme value $f_{\rm YC}=0.02$ would require a fraction of stars in stellar binaries $f_{\rm IB}\lesssim 0.3$, which however is in contrast with observations \citep[e.g.][]{1991A&A...248..485D,2010ApJS..190....1R}, especially among massive stars \citep{2012Sci...337..444S,2017ApJS..230...15M}. This comparison highlights how the merger rate of BBHs could be used to constrain the environments in which BBHs form and vice-versa, modulo the many uncertainties affecting theoretical models.
We also perform a similar exploration for GCs and NCs. To exceed the LVK upper boundary, the amount of stars forming in NCs or GCs should increase by a factor $\sim 4$ with respect to adopted values. However, this extreme choice would imply a local number density of NCs and GCs significantly larger than what observations suggest, up to a factor $2-5$. 

Models adopting a different star formation history for GCs and NCs (Fe and Fb in Table \ref{tab1}, respectively), are broadly consistent with the fiducial one, both in terms of local MRD and the distribution of population parameters, such as the primary mass. The little impact of the star cluster formation history on our results is due to the fact that IBs dominate the global population. 

Clearly, our results are bound to the properties of the synthetic Universe they refer to, and as such cannot predict the impact of additional formation channels or environments. However, the ``per-channel'' merger rate derived in this section represents an extensive quantity, and as such can be summed to the merger rate evaluated for any additional channel not considered here. This implies that, within observational constraints and the uncertainties associated with our models, additional processes should be characterized by an overall local merger rate $\mathcal{R}_{\rm others} < 3 \yrgpc$. This constraint could be combined with predictions from other channels, not explored here, such as the active galactic nuclei scenario \citep[e.g.][]{2020ApJ...898...25T,2022MNRAS.517.5827F,2023Univ....9..138A,2025MNRAS.544.4576R}.

\subsection{Main features of the fiducial population of mergers}

\begin{figure*}
    \includegraphics[width=0.325\textwidth]{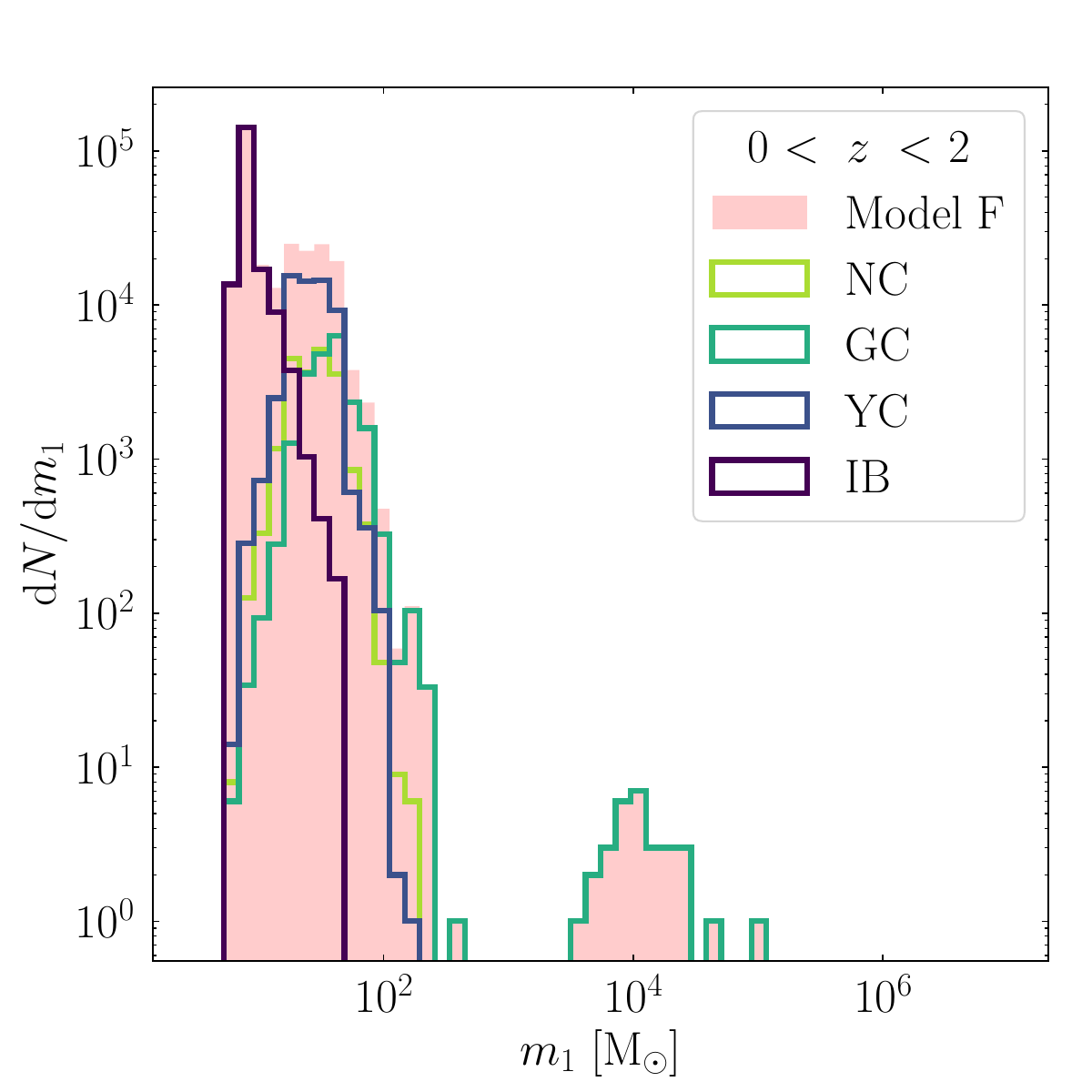}
    \includegraphics[width=0.325\textwidth]{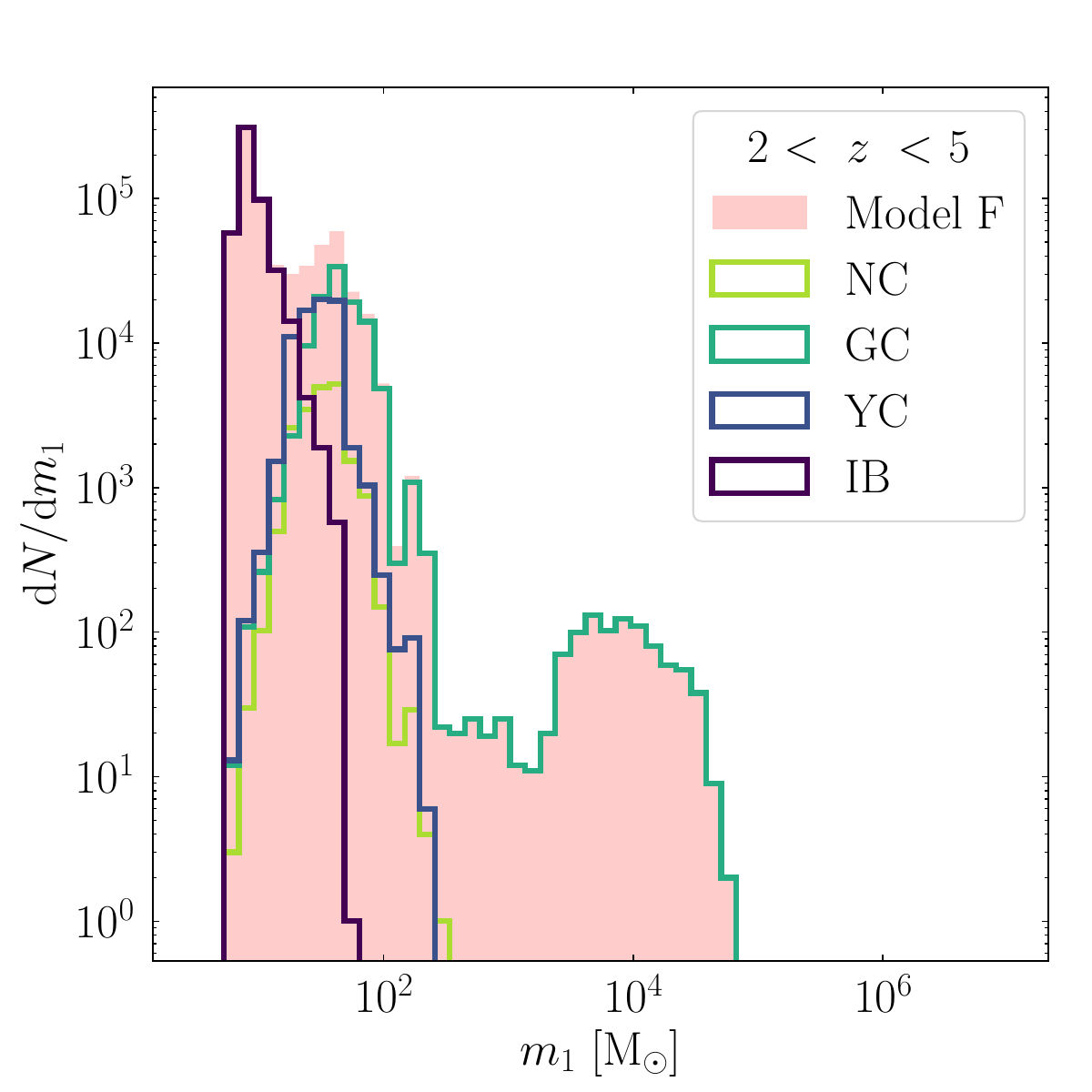}
    \includegraphics[width=0.325\textwidth]{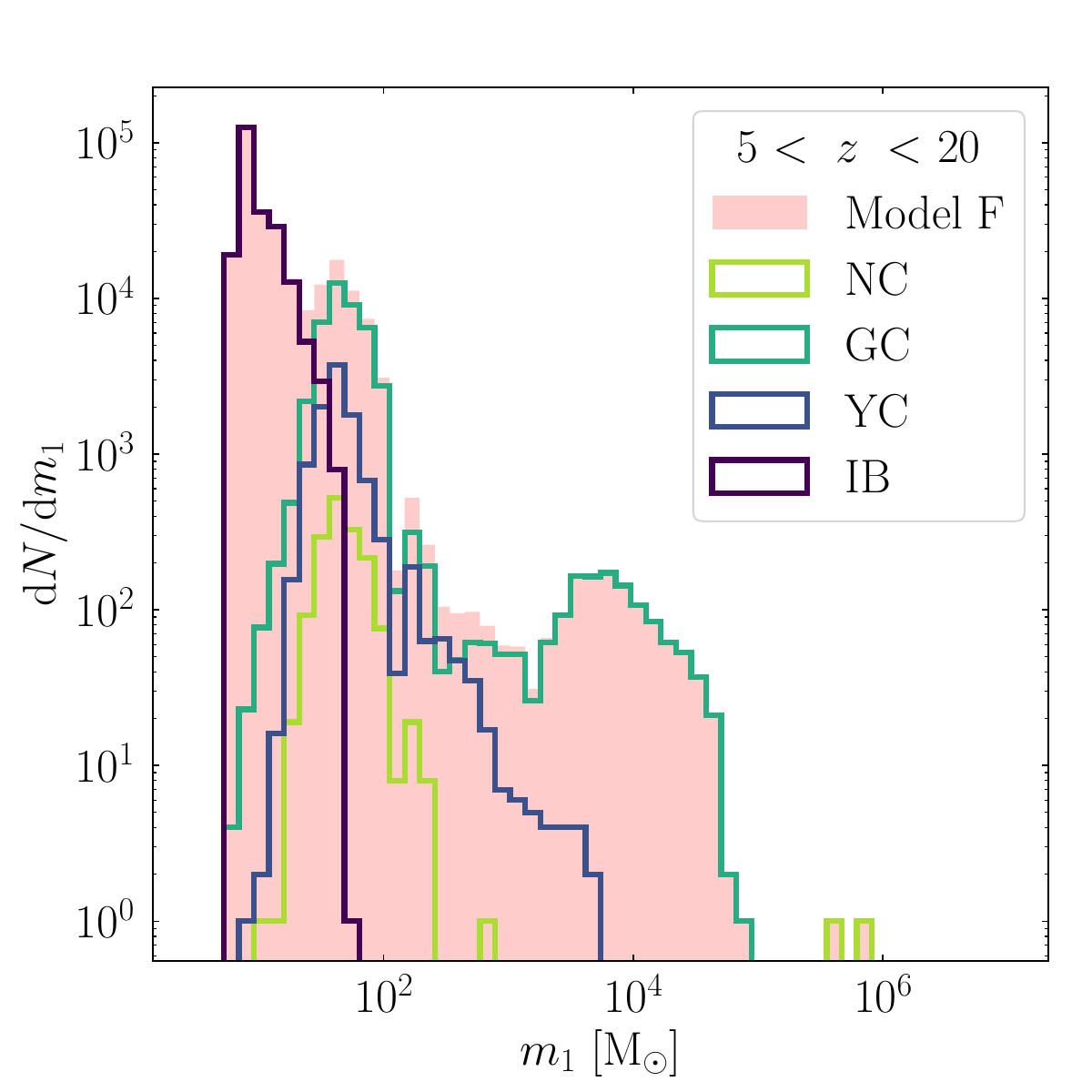}\\
    \includegraphics[width=0.325\textwidth]{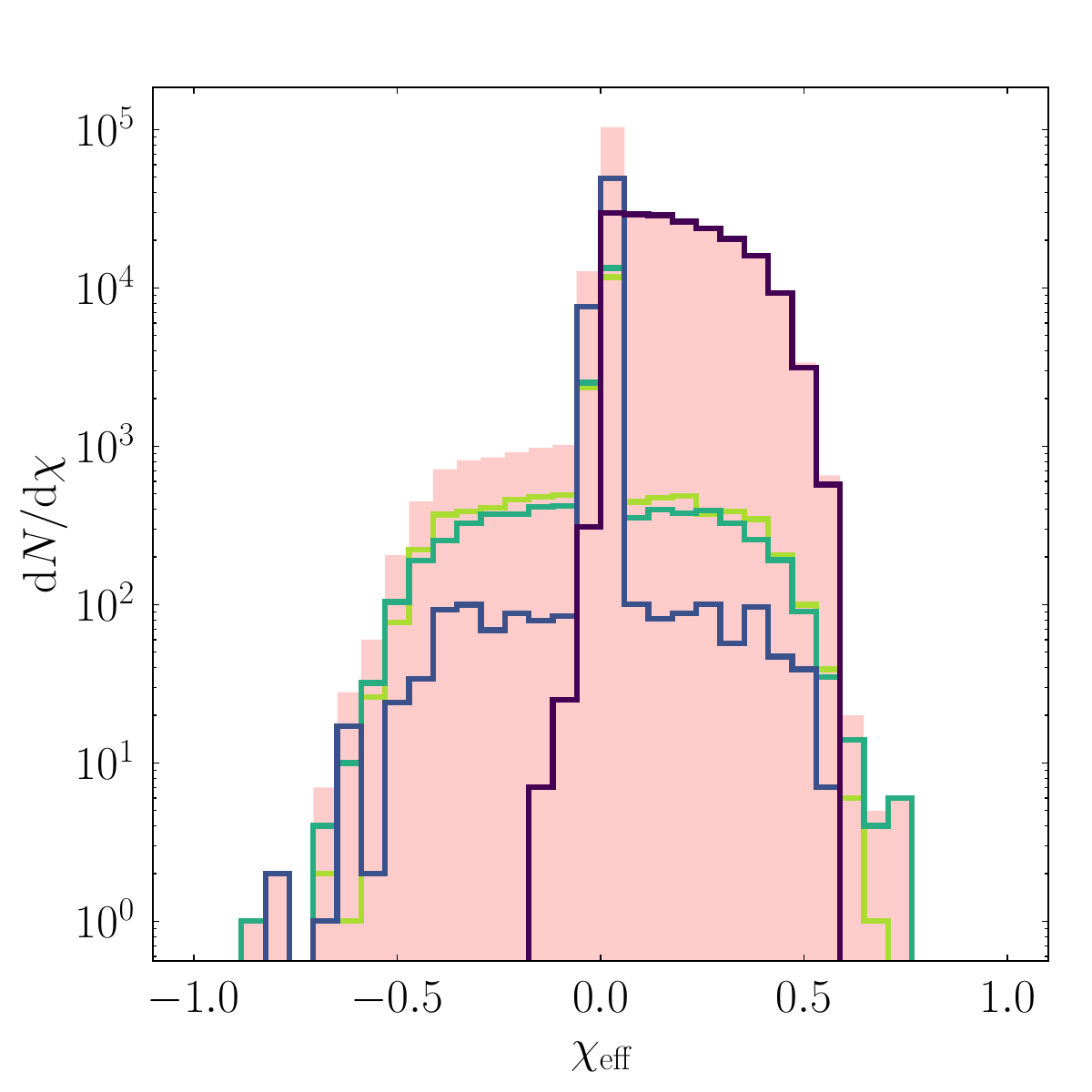}
    \includegraphics[width=0.325\textwidth]{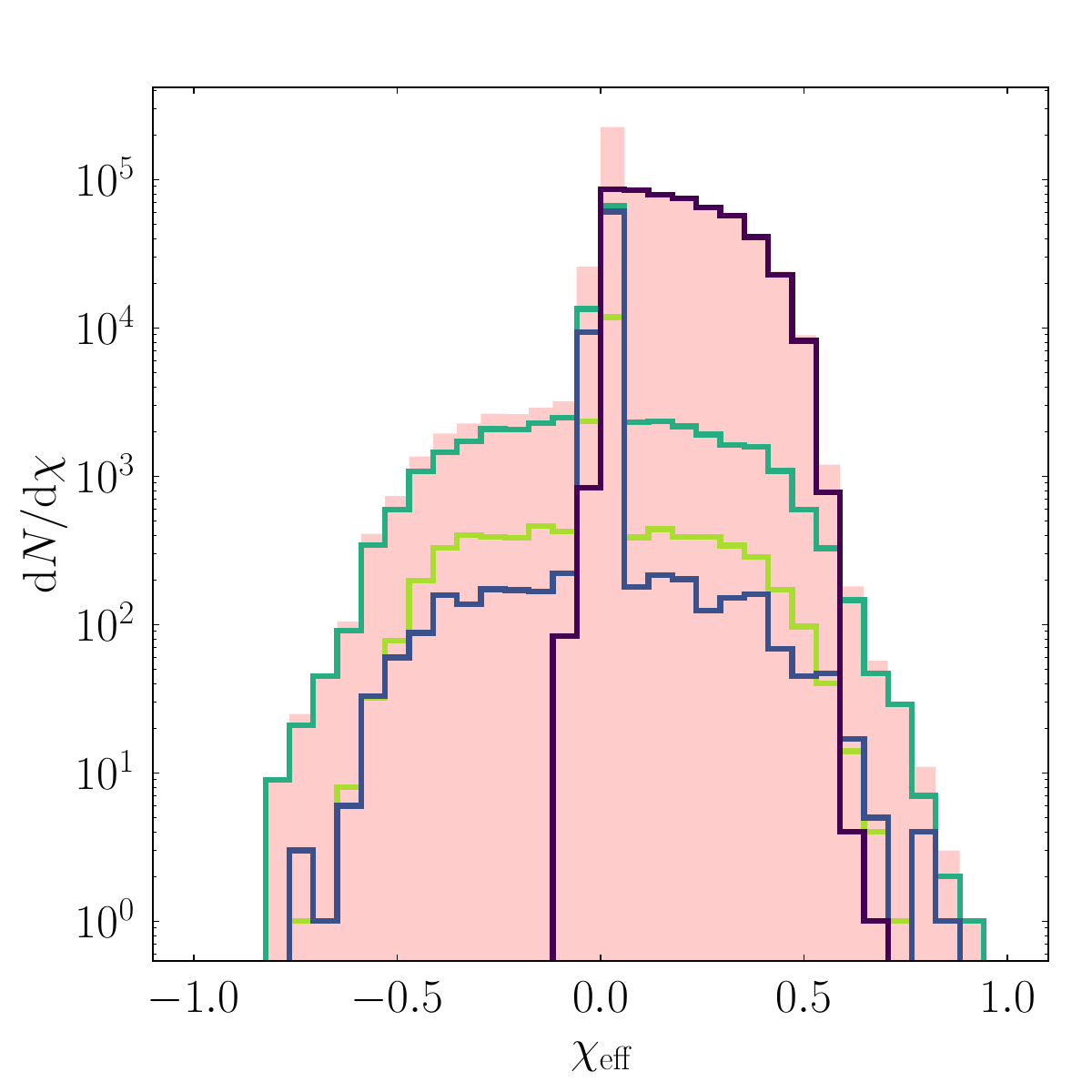}
    \includegraphics[width=0.325\textwidth]{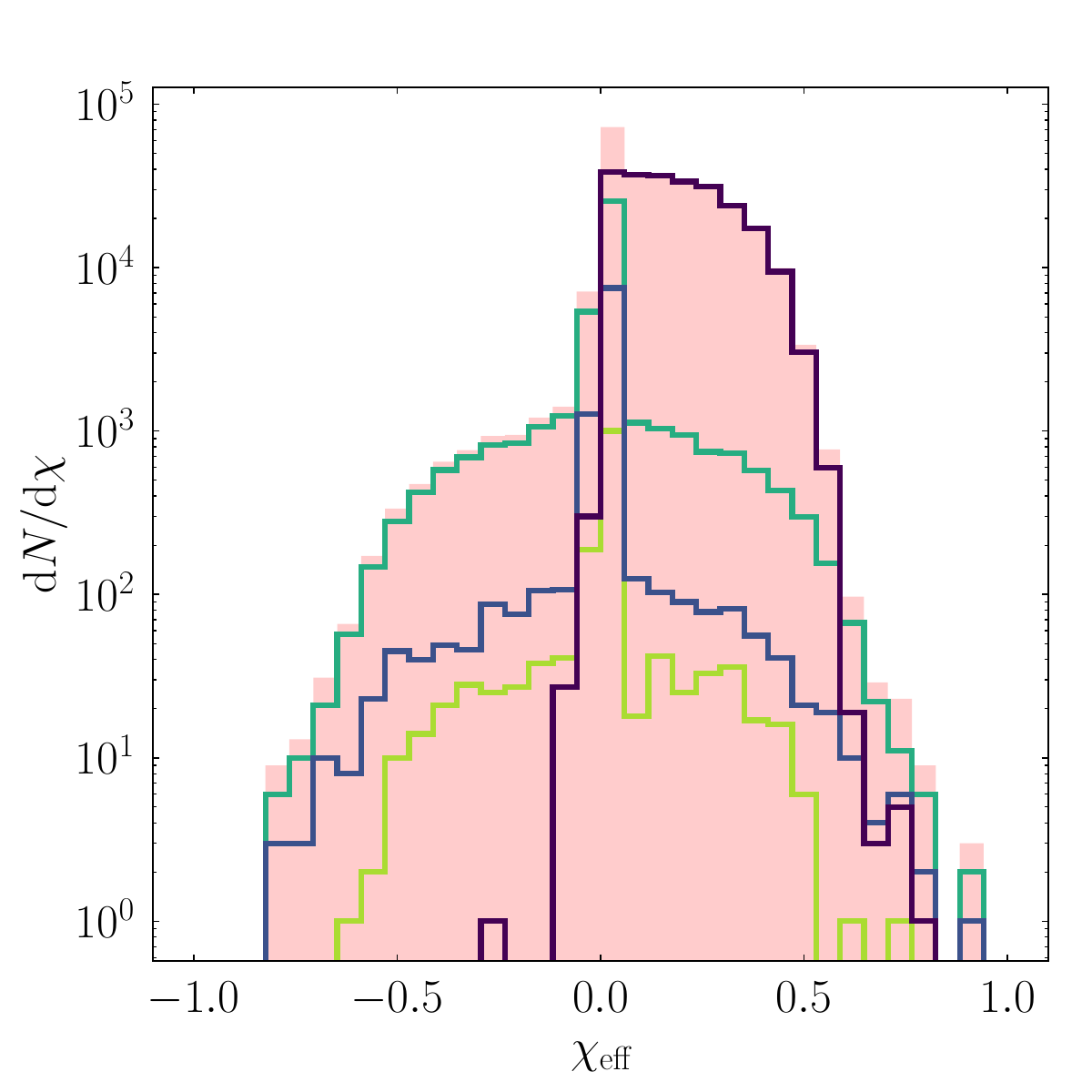}\\
    \includegraphics[width=0.325\textwidth]{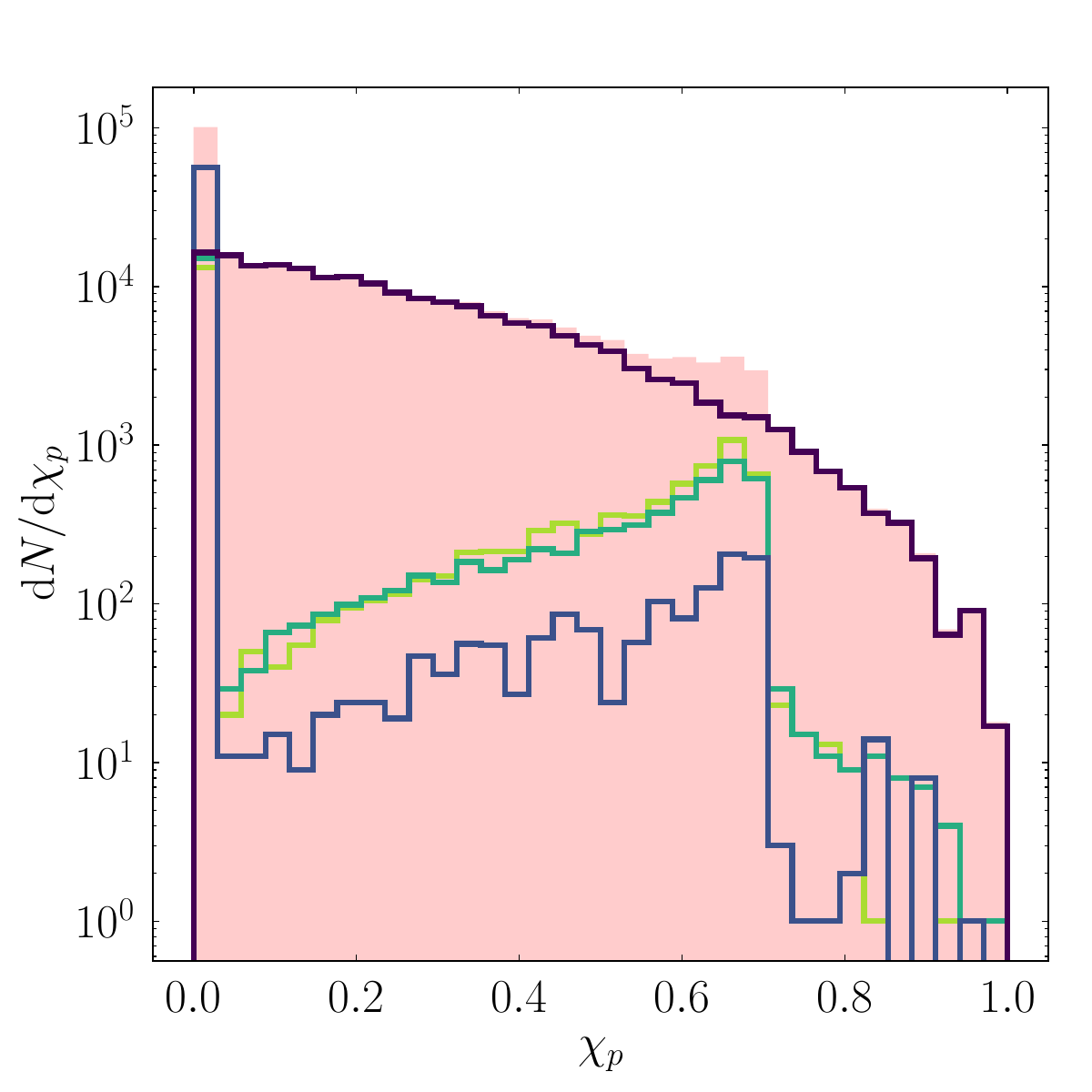}
    \includegraphics[width=0.325\textwidth]{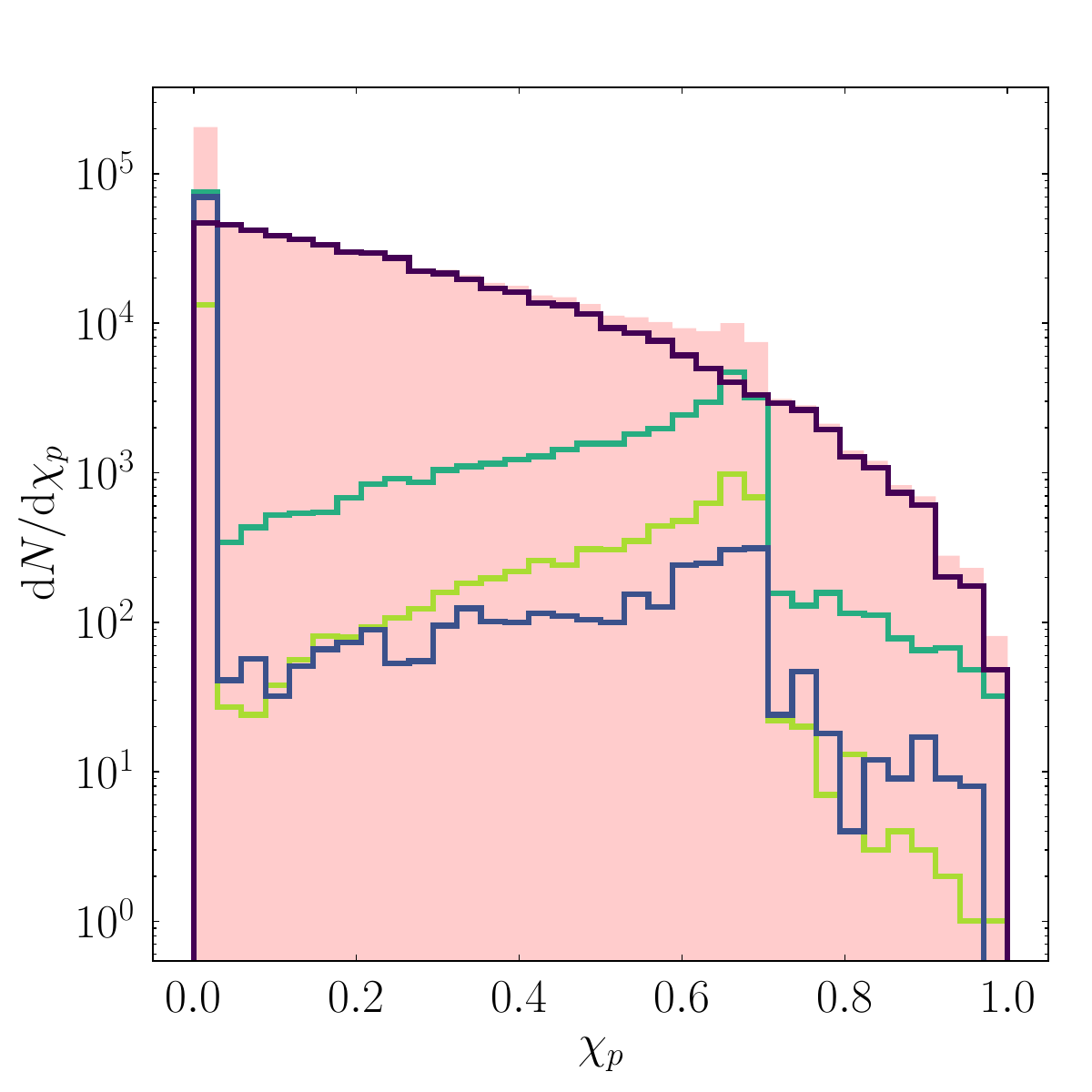}
    \includegraphics[width=0.325\textwidth]{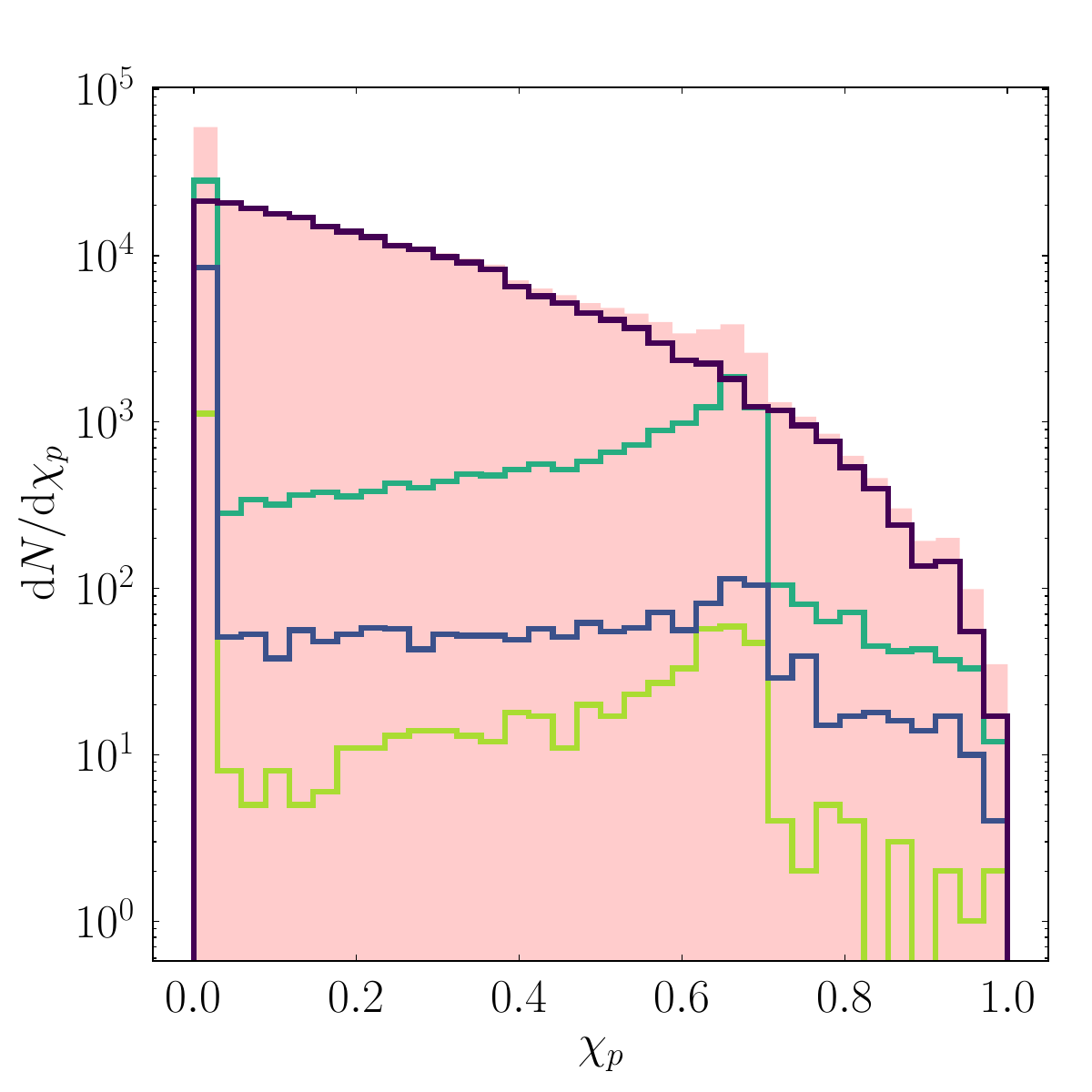}\\
    \caption{Top row panels: mass distribution of black hole binary mergers in different redshift bins extracted from the overall population (filled red steps) and mergers from isolated binaries (purple steps), young (blue steps), globular (green steps), and nuclear clusters (light green steps). Central row panels: same as above, but for the binary effective spin parameter. Bottom row panels: same as in top panels, but for the binary precession spin parameter.}
    \label{fig:3}
\end{figure*}

In this section we build up a mock catalogue of BBH mergers drawn accordingly to the cosmological MRDs discussed in the previous section. 
 
Figure~\ref{fig:3} shows, for the fiducial model and in different redshift bins, the distributions of primary mass, effective spin parameter, and precession spin parameter. In the following, we discuss the main properties of our mock sources. All results are based on 10 years of events, corresponding to $\sim 1.3\times 10^6$ BBH mergers. Hereafter, we refer to BHs that undergo a merger for the first time as $1$g BHs, and to higher-generation BHs as $n$g BHs. 

\subsubsection{Primary mass and mass-ratio distributions}

Overall, the primary mass distribution is characterized by a prominent peak at around $m_{\rm pk} = 8.6\,\Ms$ and a second broad peak covering the range $20-40\,\Ms$. These features persist across different redshift bins (top panels of Figure~\ref{fig:3}). Evident differences arise in the $m_1 > 100\,\Ms$ mass range, where high-redshift mergers ($z>2$) exhibit a clear sub-population of events involving an IMBH with mass $(10^3-10^6)\,\Ms$ that represents a fraction $\sim 0.002$ of the whole sample. At $z<2$, these massive mergers have primary masses clustering around $\sim 10^4\,\Ms$ and constitute a small fraction of the population, $\sim 10^{-4}$. The apparent dearth of massive BHs from NCs, where in principle the conditions for their formation is optimal, is due to the combination of SFR, metallicity distribution, and dynamics, as discussed in Appendix \ref{app:bigbh}. Dynamical and isolated binaries occupy clearly different regions of the $m_1$ distribution. Dynamical mergers dominate the mass distribution at $m_1 > 15\,\Ms$, where they constitute $\sim 90\%$ of all mergers. Isolated mergers, instead, dominate the primary mass distribution in the low-mass range, determining the peak at $8.6\,\Ms$.

\begin{figure}
    \centering
    \includegraphics[width=\columnwidth]{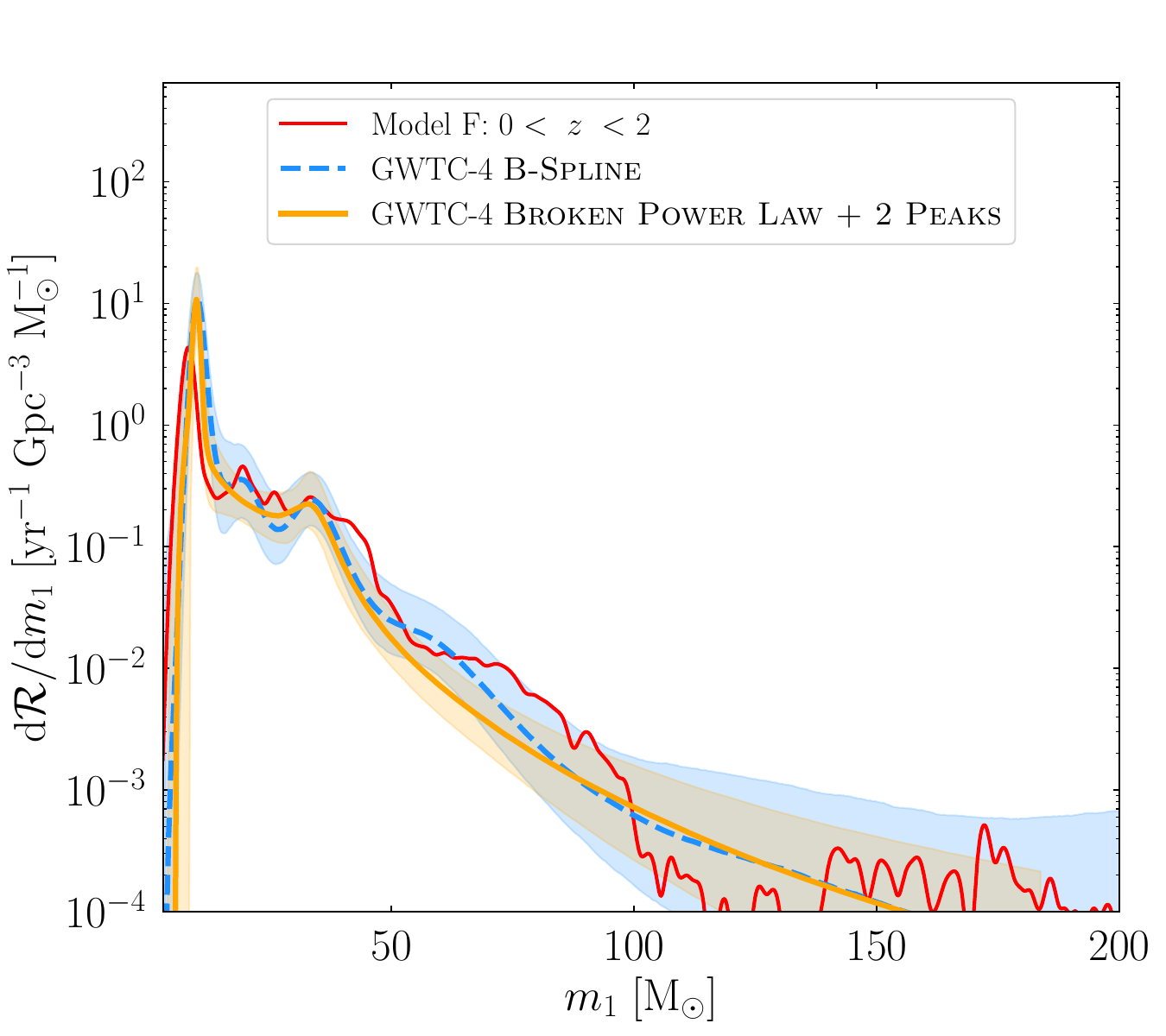}\\
    \includegraphics[width=\columnwidth]{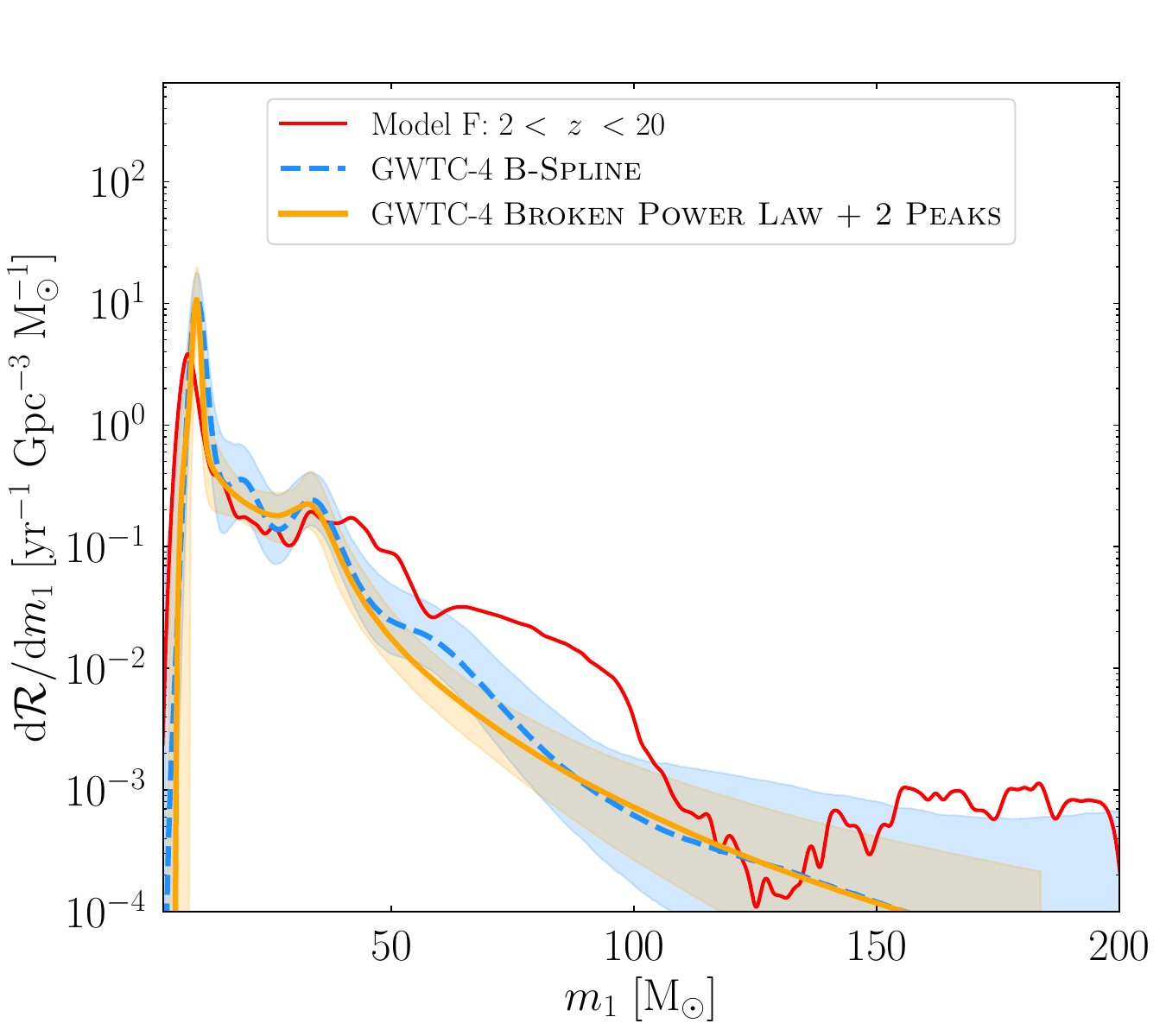}\\
    \caption{Differential merger rate density as a function of the primary mass for our fiducial model (red straight line), considering only mergers occurring at $z<2$ (top panel) or $z>2$ (bottom panel), compared against the rate inferred from GWTC-4 within the \textsc{B-Spline} (straight orange line) and the \textsc{Broken Power Law + 2 Peaks}  (dashed blue line) fit models \citep[dashed blue line][]{2025arXiv250818083T}.}
    \label{fig:5}
\end{figure}

To enable a more direct comparison with observational constraints, Figure~\ref{fig:5} compares the differential MRD, as a function of primary mass for our fiducial model and in different redshift bins\footnote{We adopted the Gaussian kernel density estimator \textsc{FFTKDE} implemented in the \textsc{KDEpy} package (see \url{https://kdepy.readthedocs.io/en/latest/_modules/KDEpy/FFTKDE.html}).}, to the inferred distribution from GWTC-4 \citep{2025arXiv250818083T}. 
At redshift $z<2$, this distribution clearly shows the primary peak around $8.6\,\Ms$ discussed above and a few secondary peaks at $20,~25,~35\,\Ms$, nicely reproducing the so-called ``bump'' at $\sim 35\Ms$ suggested by LVK results. In our model, this bump is determined by dynamical mergers, which constitute the $\sim 95\%$ of all mergers with a primary mass in the range $30-40\,\Ms$. Features in the primary mass distribution are also a natural outcome of the combined star formation history and metallicity-redshift dependence adopted. 

At $z<2$, the majority of simulated mergers have progenitors with a metallicity $Z > 0.003$ ($93.6\%$) and $0.001 < Z < 0.003$ ($6\%$). The population of BBHs at larger redshifts, instead, contains also a significant percentage -- $54\%$ -- of mergers with metal-poor progenitors ($Z < 0.003$). 

Next-generation GW detectors, like the Einstein Telescope (ET), will be able to observe up to $10^5$ BBH merger candidates per year \citep{2023JCAP...07..068B,2025arXiv250312263A}, possibly enabling a thorough study of these redshift-related signatures.

\begin{figure}
    \centering
    \includegraphics[width=\columnwidth]{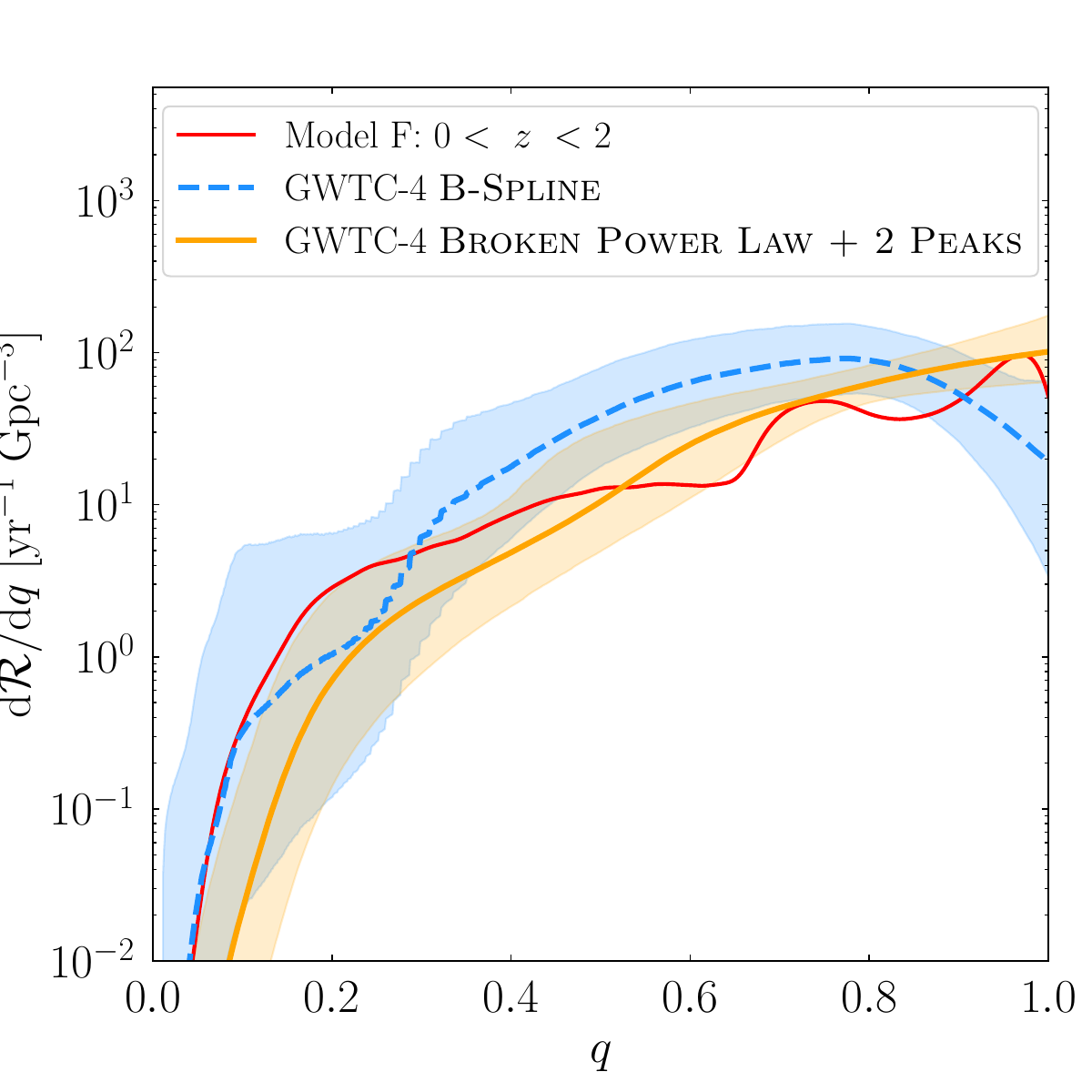}
    \caption{Differential merger rate as a function of the mass ratio for the reference model and for mergers occurring at a redshift $z<2$ (straight red line), compared to the rate inferred from GWTC-4 and described through the models \textsc{B-Spline} and \textsc{Broken Power Law + 2 Peaks} \citep{2025arXiv250818083T}.}
    \label{fig:ratio}
\end{figure}

Also the mass-ratio distribution, shown in Figure~\ref{fig:ratio}, resembles the observed distribution, although our model is characterized by multiple peaks around $q\simeq 0.6-0.9$, rather than a single peak around $0.8$ or a steady increase as suggested by LVK phenomenological models. The mass ratio distribution depends strongly on the BBH formation channels. In fact, mergers from the IB channel dominate the distribution at $q > 0.6$, where they represent over $83\%$ of the events, while dynamical mergers constitute $90\%$ of mergers with $q < 0.6$. 

\subsubsection{Spin distributions}
The effective spin distribution is broad and characterized by a bump around $\chi_{\rm eff} \sim 0$ (central panels of Figure \ref{fig:3}). The peculiar shape of the distribution results from three different contributions: a) a broad range of $\chi_{\rm eff}$ between 0 and 0.5 due to isolated binaries, reflecting the assumptions of mild spin alignment and low natal spin for first-born black holes; b) a clear peak around 0 due to dynamical binaries and driven by the assumption of low-natal spin for BHs formed from single stars; c) an extension down to $\chi_{\rm eff} < 0$ completely dominated by dynamical mergers. 

\begin{figure}
    \centering
    \includegraphics[width=\columnwidth]{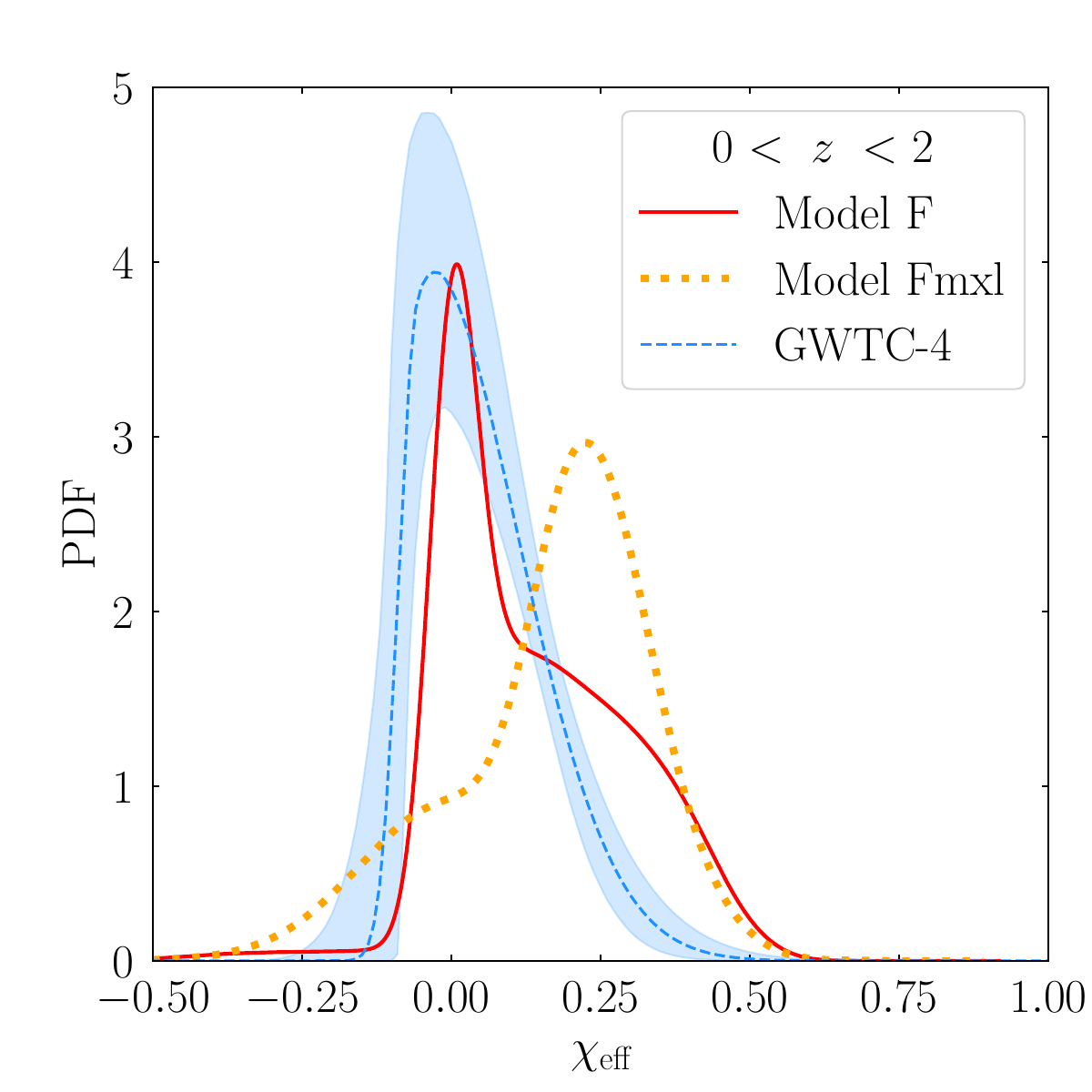}
    \caption{Effective spin parameter distribution inferred from GWTC-4 (dashed blue line), the fiducial model (F, red straight line) and the model in which BH natal spins follow a Maxwellian distribution (Fmxl, orange dotted line). We consider only mergers occurring at redshift $z<2$ and with a primary mass $<200\,\Ms$.}
    \label{fig:xfskew}
\end{figure}

In Figure~\ref{fig:xfskew}, we compare the skewed Gaussian distribution of $\chi_{\rm eff}$ inferred from GWTC-4 \citep{2025arXiv250818083T} and the results from the fiducial model and the Fmxl model, which relies on a Maxwellian distribution of BH natal spins (see Table~\ref{tab1}). The fiducial model leads to BBHs with a $\chi_{\rm eff}$ distribution that nicely resembles inferred data, mostly owing to the population of BHs with nearly zero spin. A Maxwellian distribution with $\sigma_\chi = 0.2$, instead, leads to a distribution narrowly peaked around $\chi_{\rm eff} \sim 0.25$, with a small population of objects, mostly formed dynamically, broadly distributed around $\chi_{\rm eff} \sim 0$.

\begin{figure}
    \centering
    \includegraphics[width=\columnwidth]{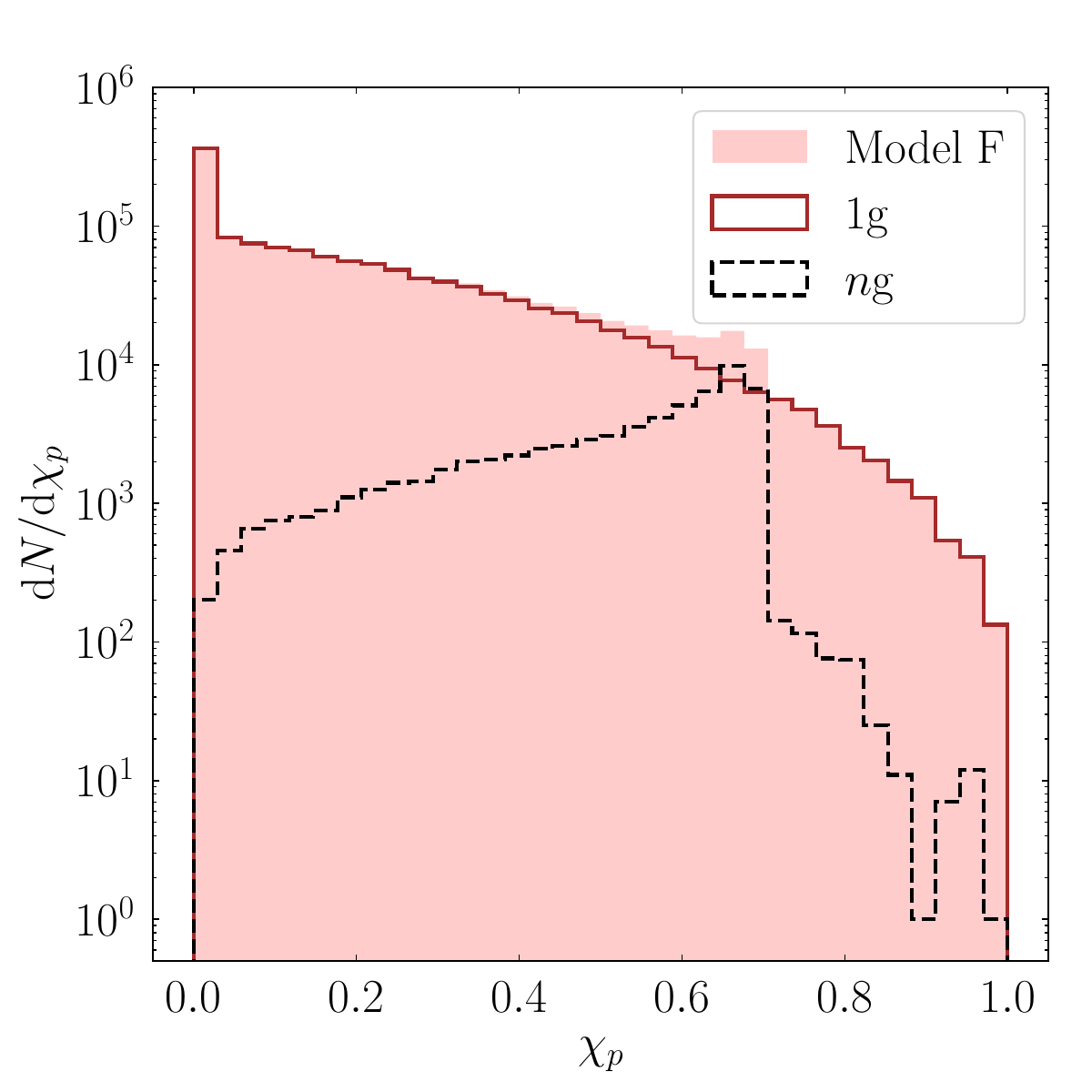}
    \caption{Precession spin parameter distribution for all BBH mergers in the fiducial model (red filled steps), and for mergers involving a first-generation primary (1g, open red steps), or a higher-generation primary ($n$g, dashed black steps).}
    \label{fig:chiprep}
\end{figure}

Another parameter that is generally identified as a possible indicator of the BBH formation channel is the precession spin parameter $\chi_p$. The distribution of $\chi_p$ in the fiducial model and for different redshift bins is shown in the bottom row panels of Figure~\ref{fig:3}. The distributions exhibit a clear peak around $\chi_p\sim0$, determined by the population of BHs born with nearly zero spin in our models, and a secondary, smaller peak around $\chi_p\sim 0.6-0.7$. The latter is due to the contribution of second-generation, and higher-generation, mergers, which are characterized by a post-merger spin narrowly peaked around $\sim 0.69$ \citep{2006PhRvD..73j4002B,2007PhRvD..76f4034B,2008PhRvD..77h1502H,2016ApJ...825L..19H}. This is further highlighted in Figure~\ref{fig:chiprep}, which shows the $\chi_p$ distribution for BBHs involving either a first- or higher-generation primary.

In the next section, we further focus on the sub-population of high-generation mergers and their features.

\subsubsection{Redshift evolution}

Several studies have used the observed population of GW sources to argue for redshift evolution in the distribution of BBH masses and spins \citep[see, e.g.,][and references therein]{2021ApJ...912...98F,2022ApJ...932L..19B,2026arXiv260103456F}. However, the limited reach of current detectors, combined with selection effects, make constraints on BBH parameter evolution highly uncertain. The GWTC-4 population analysis finds clear evidence for an evolving merger rate and a redshift-dependent broadening of the effective spin ($\chi_{\rm eff}$) distribution, but reports no statistically significant evidence for evolution in the BBH mass spectrum \citep{2025arXiv250818083T}.

From a theoretical standpoint, some degree of redshift evolution in BBH properties is generically expected, owing to the intrinsic coupling between binary formation channels, host environments, and cosmological star formation \citep[e.g.][]{2020MNRAS.492.2936A,2023MNRAS.520.5259A,2024A&A...688A.148T,2024ApJ...967...62Y}. Here, we investigate the redshift dependence of binary mass and effective spin across different evolutionary scenarios. 

\begin{figure}
    \centering
    \includegraphics[width=0.87\columnwidth]{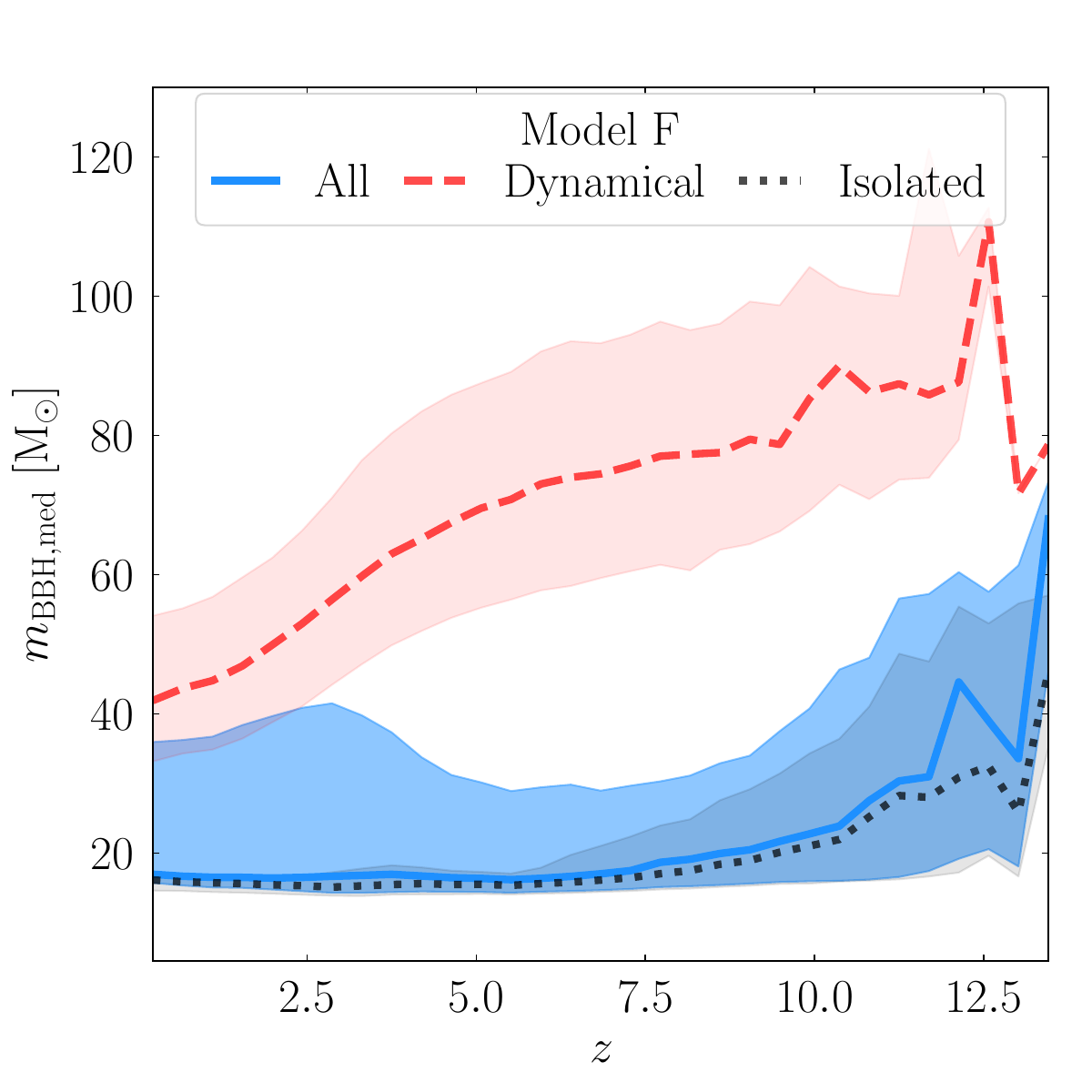}
    \includegraphics[width=0.87\columnwidth]{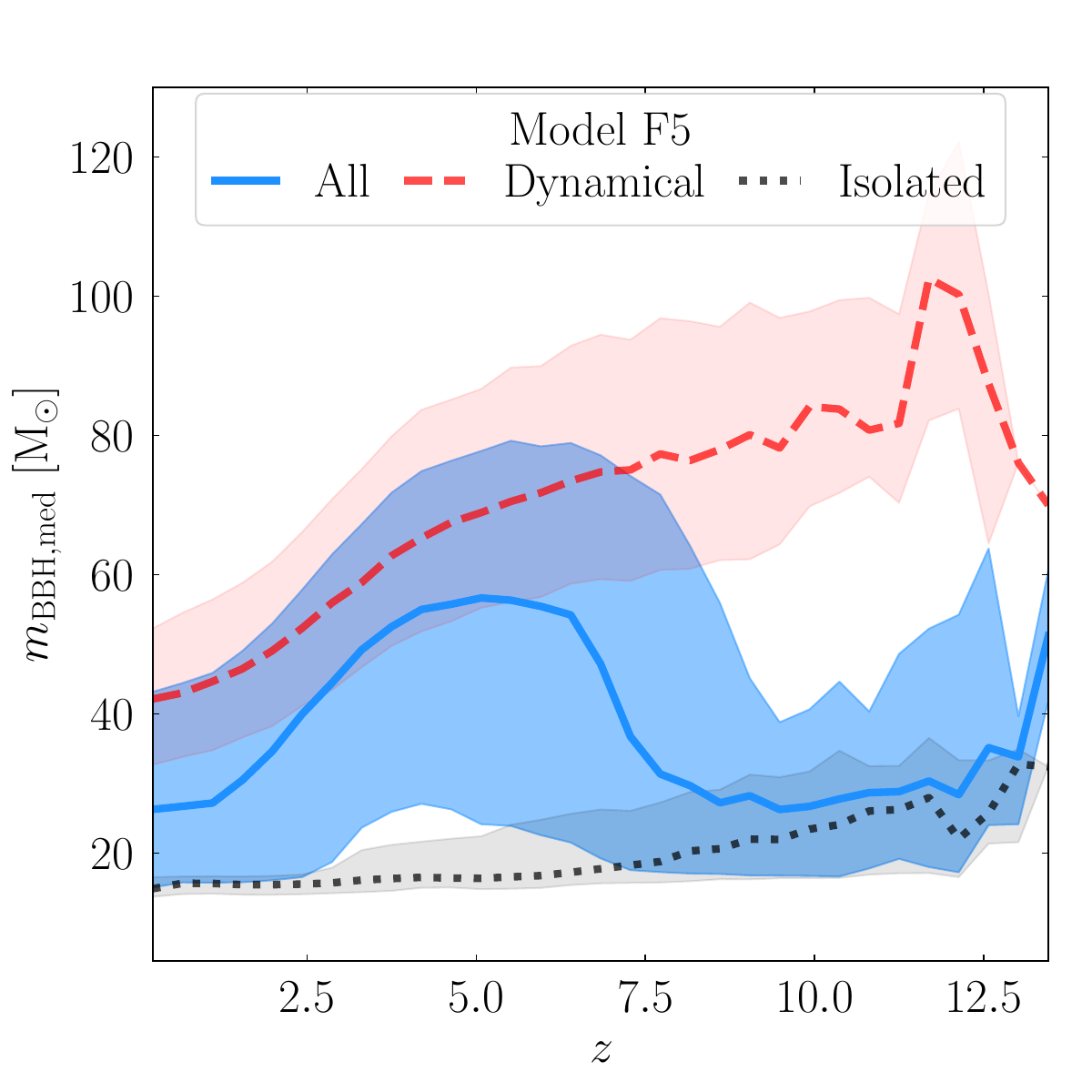}
    \includegraphics[width=0.87\columnwidth]{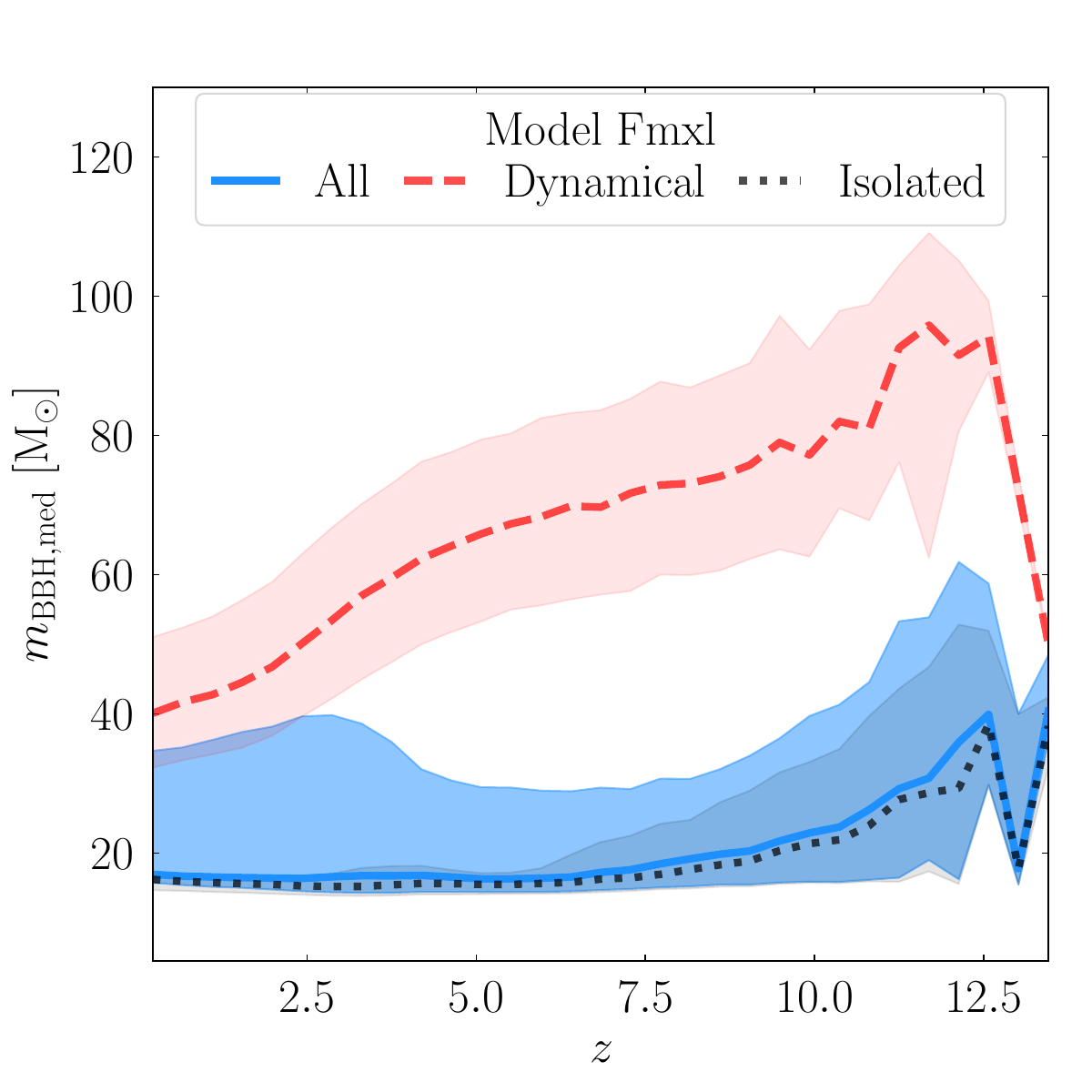}
    \caption{Median binary mass as a function of redshift for mergers in models F (top panel), F5 (central panel), and Fmxl (bottom panel). We show the median mass for the whole population (blue solid lines), as well as for dynamical (red dashed lines) and isolated mergers (black dotted lines). The shaded area encompasses the $50\%$ of the population.}
    \label{fig:placeholder}
\end{figure}

Figure~\ref{fig:placeholder} shows the redshift evolution of the median binary mass ($m_{\rm BBH, med}$) for the fiducial model (F), the model in which we assume $\alpha_{\rm CE}=5$ (F5), and the model adopting a Maxwellian natal spin distribution (Fmxl), separating the results for the isolated and dynamical formation channels.

In general, IBs exhibit a slight increase of the primary mass with redshift, with the median slowly shifting in the range $m_{\rm iso,med} \sim (20-40)\,\Ms$ over the redshift range $z<15$. Dynamical mergers, instead, exhibit a much steeper evolution, with the median monotonically increasing in the range $m_{\rm dyn,med} = 40-100\,\Ms$. Since in most models the majority of mergers formed in IBs, the global BBH mass evolution resembles that of isolated mergers. The only exception is model F5, in which dynamical mergers dominate the population. In this case, the redshift evolution of $m_{\rm BBH, med}$ is more complex showing a sudden increase from $m_{\rm BBH,med} = 25\,\Ms$ to $50\,\Ms$ at redshift $z\simeq 2-6$, a subsequent decrease lasting up to redshift $10$, and a shallower increase from $z=10-12$. This behavior is driven by the dominant impact of isolated(dynamical) mergers at $z<3$($z>3$). 

Interestingly, at a redshift $z<2$, i.e. within LVK detectors' reach, the median binary mass does not evolve appreciably, remaining constant around $m_{\rm BBH, med}\simeq 9.7\Ms$, in all models but in F5, which instead features a clear $m_{\rm BBH, med}$ increase already within this redshift range.  

It should be noted that a different stellar evolution paradigm can affect these results. Models in which the population of IB mergers is significantly contributed by binary stars undergoing Roche-lobe overflow could lead to a median mass that increases toward small redshift \cite[see e.g.][]{2022ApJ...931...17V}.

As shown in Figure~\ref{fig:3}, dynamically formed mergers dominate the population at primary masses $m_1 \gtrsim 20\,\Ms$. Consequently, the median mass of the high-mass sub-population is expected to reflect the characteristic mass growth associated with dynamical assembly.

Regarding the spins, instead, we explored the variation of  the median of $\chi_{\rm eff,med}$ for the same set of models. Overall, we do not find any significant evolution within $z<2$, a finding broadly consistent with GWTC-4 analysis \citep{2025arXiv250818083T}. The median effective spin attains a value $\chi_{\rm eff,med} \simeq 0.1$ in the fiducial model, and $\chi_{\rm eff,med} \simeq 0.2$ in the Fmxl model. These models favor positive $\chi_{\rm eff,med}$ values owing to isolated mergers, which dominate the distribution. In model F5, instead, the larger impact of dynamical mergers lead to $\chi_{\rm eff,med} \simeq 0$. 


\subsubsection{Repeated mergers}

In dynamical environments, BHs can undergo multiple merger events, provided that the recoil imparted onto the remnant of a merger is smaller than the environment's escape velocity. Multiple, or hierarchical, merger products are expected to feature peculiarities that can help distinguishing them from the overall population of mergers, such as a larger mass or a peculiar spin distribution. 
In the current version of \bpop, we assume that multiple mergers involve only the primary BH. This assumption relies on the fact that, after a few mergers, the primary BH rapidly becomes the object with the largest cross-section in the cluster, thus monopolizing the interactions and preventing the development of additional merger chains\footnote{A detailed study on the impact of high-generation secondaries is underway and will soon be published in a follow-up study (Ugolini et al., in prep.).}.

Globally, high-generation primary BHs represent a fraction $f_{\rm ng} = 0.008-0.112$ of the population, depending on the initial conditions. The largest value is obtained for model F5, owing to the larger fraction of dynamical mergers over the total, whilst the lowest is obtained for model Fmxl, owing to the fact that a Maxwellian distribution of spins favors larger post-merger kicks compared to other models. We find that mergers with primary masses $> 100\,\Ms$ constitute a fraction $f_{\rm IMRI}\simeq 0.001-0.0068$ of all BBHs. In the following, we refer to these systems as intermediate-mass ratio inspirals (IMRIs), though we note that the term ``IMRI'' is generally used to identify binaries with a mass ratio $q\sim 10^{-4}-10^{-2}$ \citep[see e.g.][]{2007CQGra..24R.113A}.

In terms of local merger rate, these numbers correspond to $\mathcal{R}_{\rm loc, ng} = (0.005-1.01)\yrgpc$ for high-generation mergers, and $\mathcal{R}_{\rm loc, 100} \sim (0.9-8.2)\times10^{-3}\yrgpc$ (see Table~\ref{tab:mrd}).
    
\begin{figure}
    \centering
    \includegraphics[width=\columnwidth]{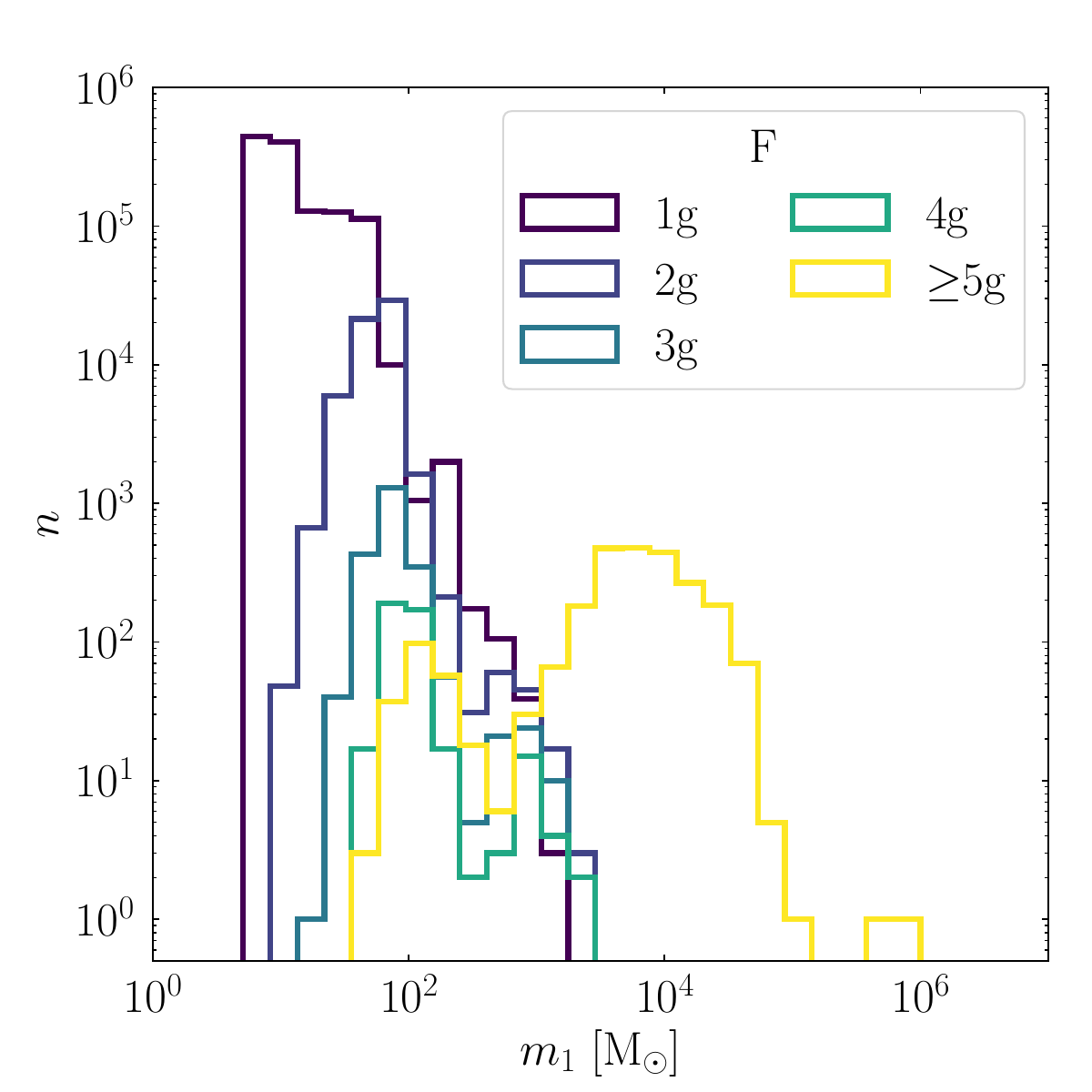}\\
    \includegraphics[width=\columnwidth]{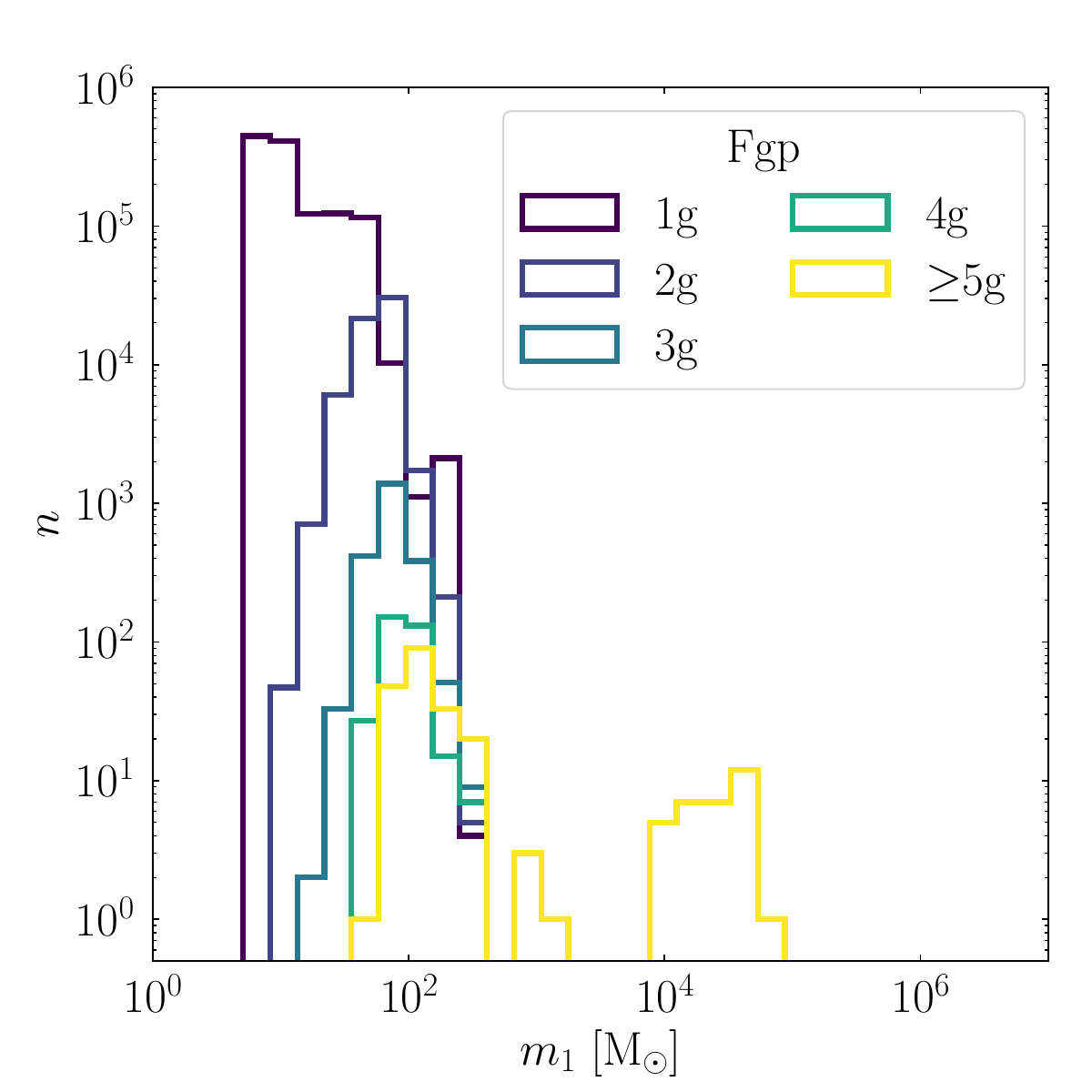}\\
    \caption{Primary mass distribution for all BBH mergers in model F (top panel) and Fgp (i.e. without including stellar collision products, bottom panel). Different colors correspond to different generations, with $1$g identifying all BHs that are undergoing their first merger, and $>2$g all BHs that already merged at least once.}
    \label{fig:repeated}
\end{figure}

To dissect the properties of multiple BH merger remnants in our models, we show in Figure~\ref{fig:repeated} the primary mass distribution of different generations of BHs for the fiducial model and for the model in which we neglect the impact of multiple stellar collisions (model Fgp, see Table~\ref{tab1}). 

In the fiducial model, the population of IMBHs with a mass in the range $m_{\rm IMBH} = (10^2-10^3)\,\Ms$, is dominated by first-generation (1g) IMBHs -- formed from stellar collisions -- and 2g IMBHs, which contribute for up to $25-50\%$. The population of higher-generation objects is negligible in this mass range. Interestingly, in the mass range $m_{\rm IMBH} = (100-250)\,\Ms$, almost $64\%$ of mergers involve a 1g IMBH regardless of the model. This owes to the impact of stellar mergers on the natal mass spectrum of BHs, as they can extend the maximum mass out to $\sim 250\,\Ms$, especially in low-metallicity environments. Hence, the inclusion of non-trivial stellar processes, like stellar mergers, may be essential to correctly interpret the origin of peculiar GW sources, like GW231123 \citep[see our recent work in][]{paiella_letter_2025}.  
The population of IMBHs with a mass $>10^3\,\Ms$, instead, is mostly comprised of objects that underwent more than six previous mergers ($>6$g), even in models in which the maximum natal mass allowed for BHs is $ < 100\,\Ms$, such as in model Fgp. More specifically, these IMBHs form through long chains counting $(10^2-10^3)$ mergers with secondary BHs with a median mass of $\sim 20-30\Ms$. 
When massive BH seeds are not included (models Fgp and Fst), the number of BBH mergers with $m_1>10^3\,\Ms$ is smaller by 1-2 orders of magnitude and the mass spectrum reaches a smaller maximum mass. Table~\ref{tab:ngen} summarizes the results on high-generation mergers for different models and mass ranges. 

\begin{table}[]
    \centering
    \caption{Percentage of high-generation mergers in different mass bins for the fiducial model}
    \begin{tabular}{c|c|c|c|c|c|c}
    \hline\hline
    $\delta m_1$ & \multicolumn{6}{c}{Generation (\%)}\\
    \hline
    M$_\odot$ &  $1$g & $2$g & $3$g & $4$g & $5$g & $\geq 6$g \\
    \hline\hline 
    \multicolumn{7}{c}{{\bf Fiducial}}\\ [3pt]
    \hline
    $0-50$ & $98.2$ & $1.78$ & $0.019$ & $0.001$ & $0$ & $0$ \\
    $50-100$ & $40.37$ & $56.61$ & $2.56$ & $0.38$ & $0.08$ & $0.01$ \\
    $100-250$ & $63.87$ & $21.42$ & $7.53$ & $3.68$ & $1.39$ & $2.11$ \\
    $250-10^3$ & $56.28$ & $23.19$ & $8.50$ & $3.72$ & $1.59$ & $6.73$ \\
    $10^3-10^4$ & $0.39$ & $1.76$ & $0.92$ & $0.46$ & $0.52$ & $95.95$ \\
    $10^4-10^8$ & $0$ & $0$ & $0$ & $0$ & $0$ & $100$ \\
    \hline
    \multicolumn{7}{c}{{\bf Model Fgp}}\\ [3pt]
    \hline
    $0-50$ & $98.184$ & $1.797$ & $0.019$ & $0$ & $0$ & $0$ \\
    $50-100$ & $40.51$ & $56.53$ & $2.583$ & $0.293$ & $0.075$ & $0.009$ \\
    $100-250$ & $64.44$ & $22.06$ & $7.59$ & $3.09$ & $1.22$ & $1.61$ \\
    $250-10^3$ & $10.20$ & $10.20$ & $18.37$ & $14.29$ & $8.16$ & $38.78$ \\
    $10^3-10^4$ & $0$ & $0$ & $0$ & $0$ & $0$ & $100$ \\
    $10^4-10^8$ & $0$ & $0$ & $0$ & $0$ & $0$ & $100$ \\
    \hline
    \end{tabular}
    \label{tab:ngen}
\end{table}

Repeated mergers are also expected to exhibit distinctive features in the spin distribution. In fact, the merger remnants of nearly equal mass, non-spinning, BHs are characterized by a quite precise final spin of $\chi_{\rm rem}\sim 0.69$ \citep{2006PhRvD..73j4002B,2007PhRvD..76f4034B,2008PhRvD..77h1502H,2016ApJ...825L..19H}, although this quantity can attain a more broad range of values for asymmetric binaries with highly spinning components \citep{2003ApJ...585L.101H}. 

Several works pointed out that mapping BH spins onto the primary mass and leveraging on the peculiar spin of $n$-th generation BHs ($n$g) can help identify a mass threshold above which hierarchical mergers dominate the primary mass distribution \citep[e.g.][]{2025arXiv250915646B,2025arXiv250904151T,2025arXiv250904637A,2026arXiv260103457V}. \cite{2025arXiv250904637A} found a threshold $\tilde{m}_1 \simeq 47\,\Ms$. Above $\tilde{m}_1$, the inferred GWTC-4 mass-ratio distribution is nearly flat, offering a further quantity to compare with dynamical BBH models. For example, Monte-Carlo simulations of GCs suggest that 1g+2g mergers should be characterized by a mass-ratio distribution narrowly peaking around $q\simeq 0.5$ \citep{2019PhRvD.100d3027R}, possibly suggesting that the tail of the mass distribution beyond $\simeq 40-50\,\Ms$ inferred from GWTC-4 is not necessarily dominated by $n$g merger products \citep[see e.g.][]{2025arXiv251018867R}.

\begin{figure}
    \centering
    \includegraphics[width=\columnwidth]{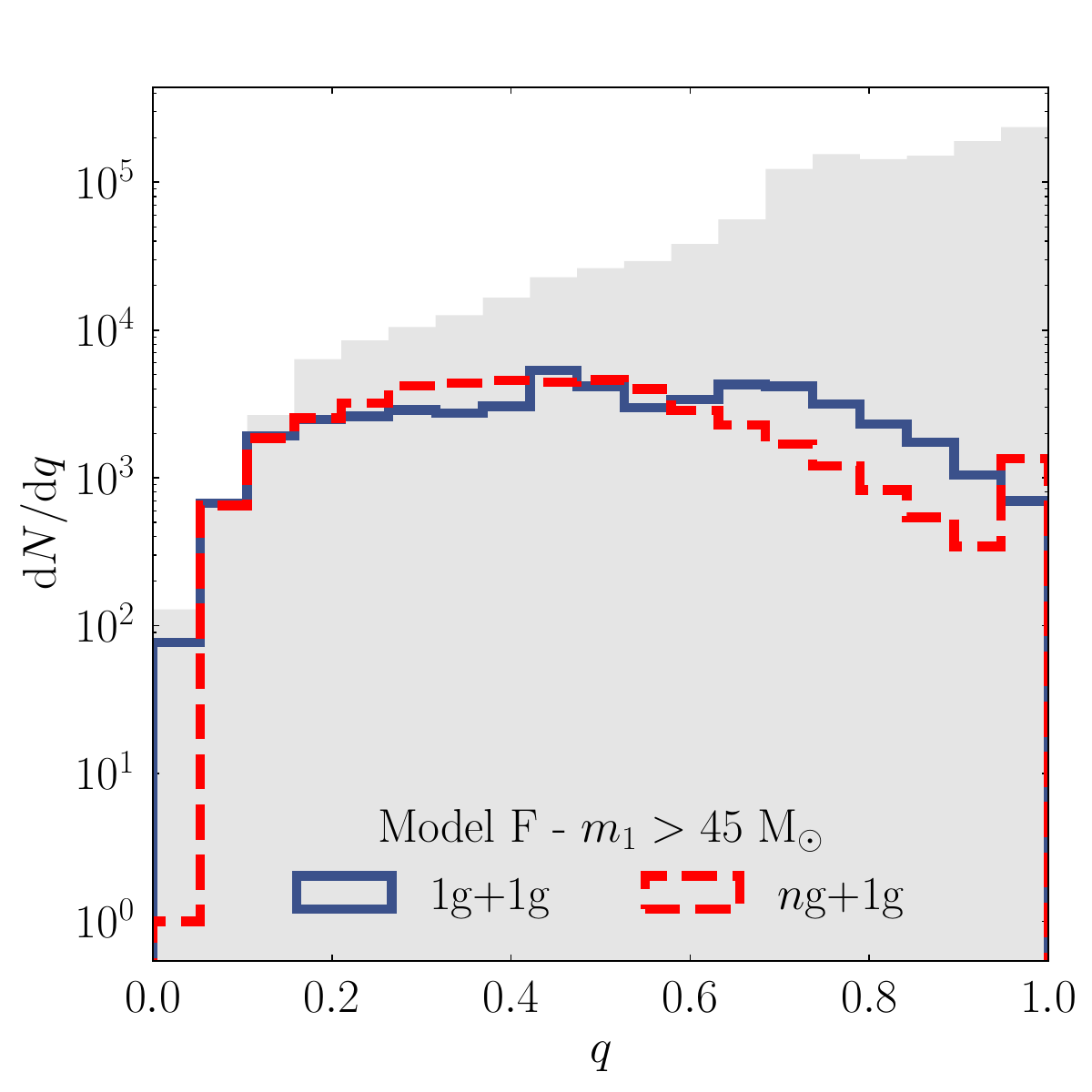}\\
    \includegraphics[width=\columnwidth]{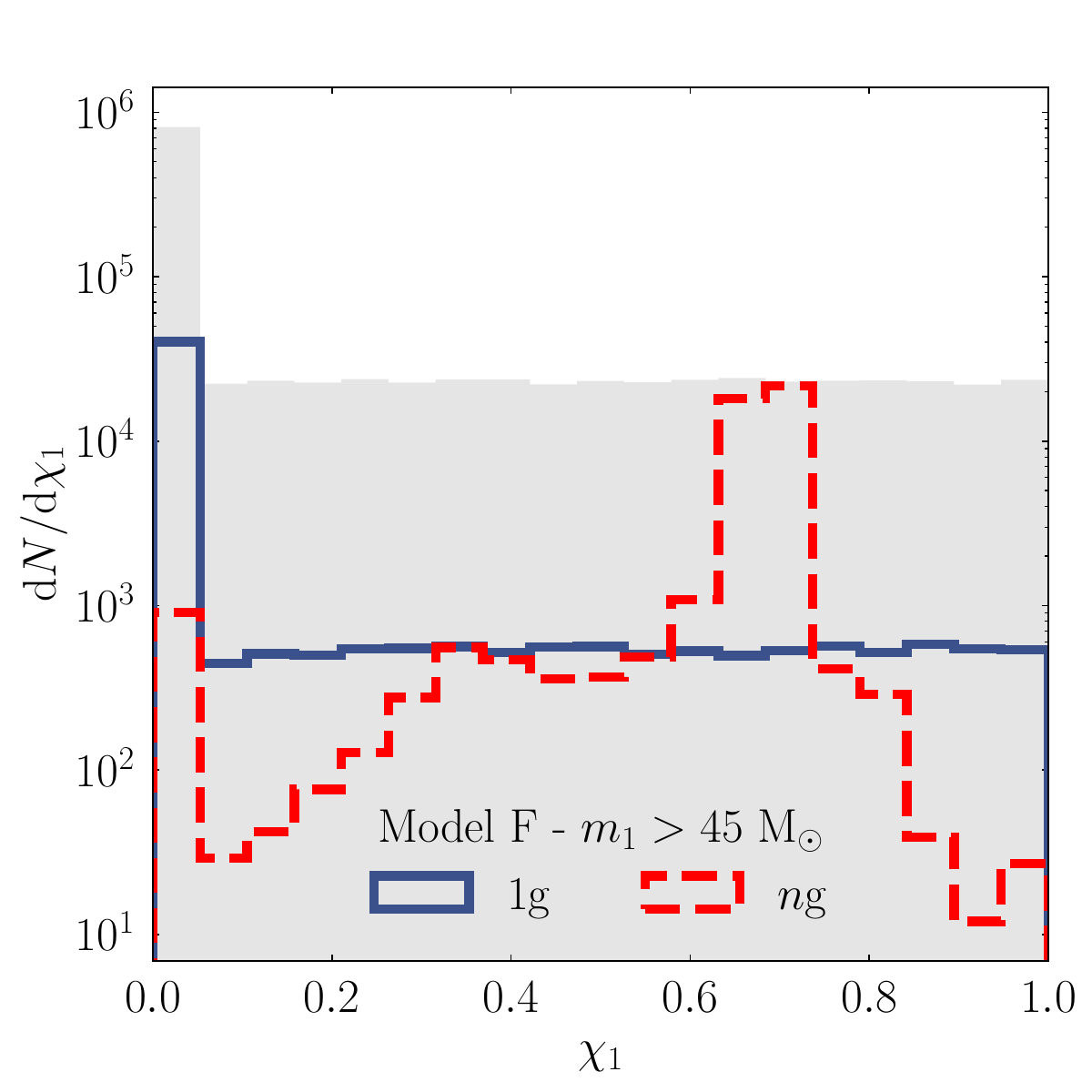}\\
    \caption{Top panel: Mass-ratio distribution for all first generation mergers (1g,  gray filled steps), and mergers with a primary mass $>45\,\Ms$, either 1g (blue straight line), or reaching a higher generation ($n$g, red dashed line). Bottom panel: same as above, but here we show the primary spin distribution.}
    \label{fig:rationext}
\end{figure}

To explore this feature in our models, in Figure~\ref{fig:rationext} we compare the $q$ distribution for all 1g+1g mergers to the one for 1g+1g and $n$g+1g mergers with a primary mass in the range $m_1=(45-500)\,\Ms$. Note that, in the fiducial model, more than $95\%$ of mergers in this mass range form in star clusters. Mergers with such a massive primary are characterized by a broad and nearly flat distribution, regardless of the generation of their progenitor. Our results are similar to those found with Model A in \cite{2022MNRAS.517.2953T}, though here we do not make any a priori assumptions on BH pairing. 

This feature persists in all our models, suggesting that a flat mass-ratio distribution at $m_1 > 45\,\Ms$ is a peculiarity of dynamical mergers involving both 1g and higher-generation primaries. 

\begin{figure}
    \centering
    \includegraphics[width=\columnwidth]{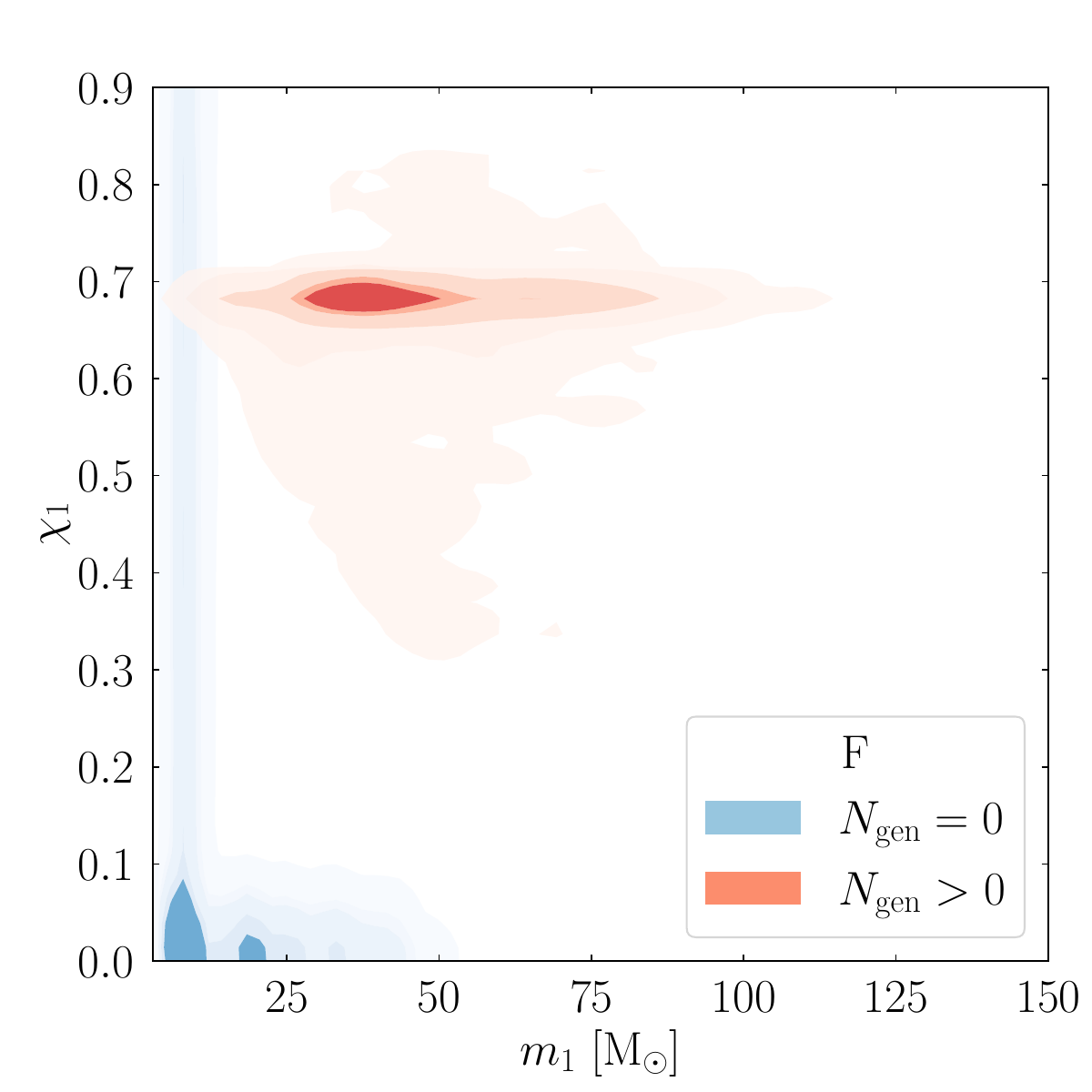}
    \includegraphics[width=\columnwidth]{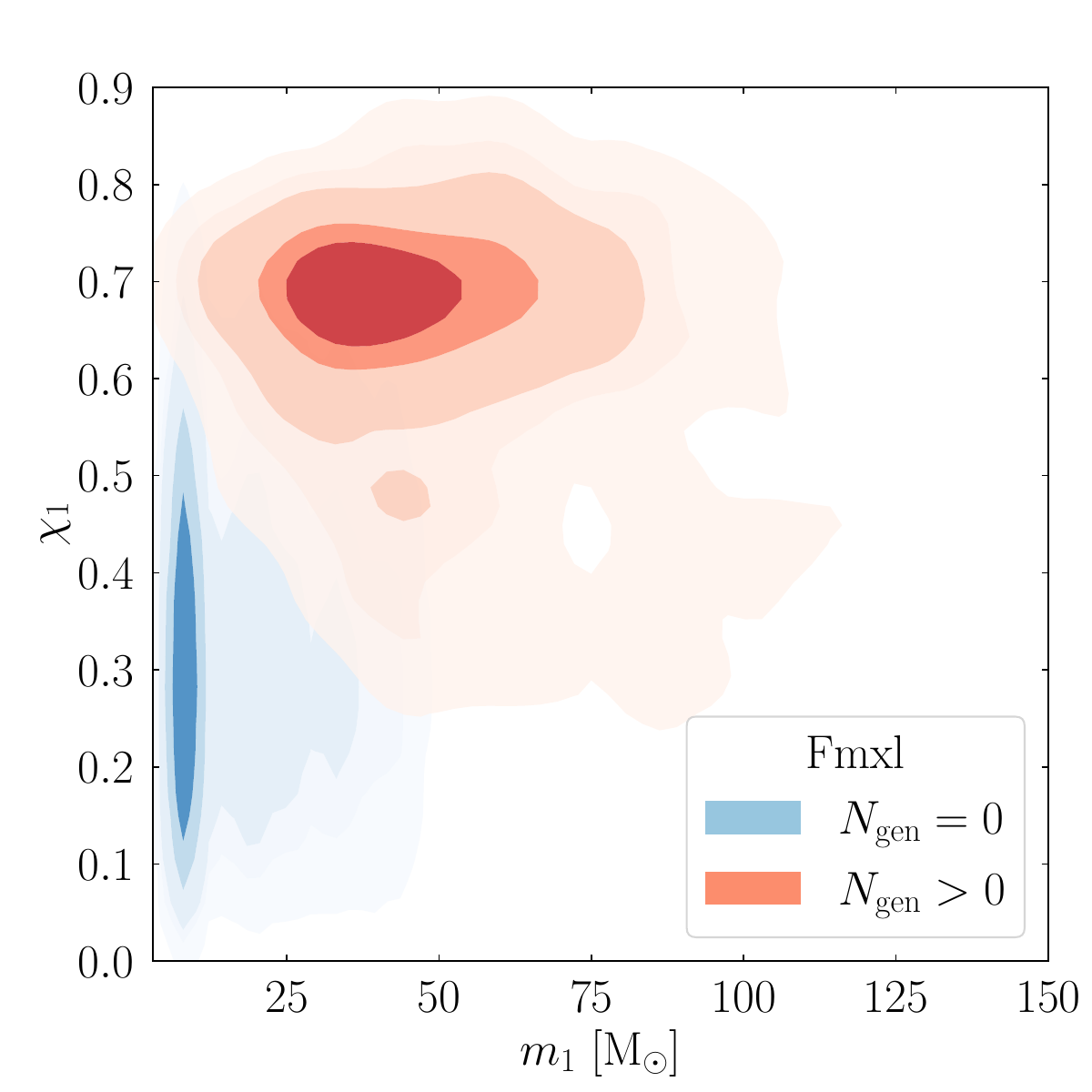}
    \caption{Combined distribution of primary mass and $\chi_{\rm eff}$ for all mergers in models F and Fmxl, dissected into 1g+1g (blue shaded areas) and 1g+$n$g (red shaded areas). From the outermost to the innermost contour, different levels encompass the 2nd, 10th, 20th, 50th, and 70th percentile, respectively. }
    \label{fig:high1}
\end{figure}

A further, clear peculiarity can be seen in the combined distribution of primary mass $m_1$ and spin $\chi_1$, which in Figure~\ref{fig:high1} is shown for the fiducial model and the alternative model (Fmxl) in which we assume a Maxwellian distribution with $\sigma_\chi=0.2$ for BH natal spins\footnote{Contours are realized through the \textsc{KDEPlot} utitlity of the Seaborn library (see \url{https://seaborn.pydata.org/generated/seaborn.kdeplot.html)}}. As expected from General Relativity, $n$g mergers tend to accumulate around a spin $0.4-0.8$, leading to two clearly distinguishable populations \citep[see also][]{2021NatAs...5..749G}, with $n$g mergers gathering around a typical primary mass of $\sim 40\,\Ms$.

\begin{figure*}
    \centering
    \includegraphics[width=\columnwidth]{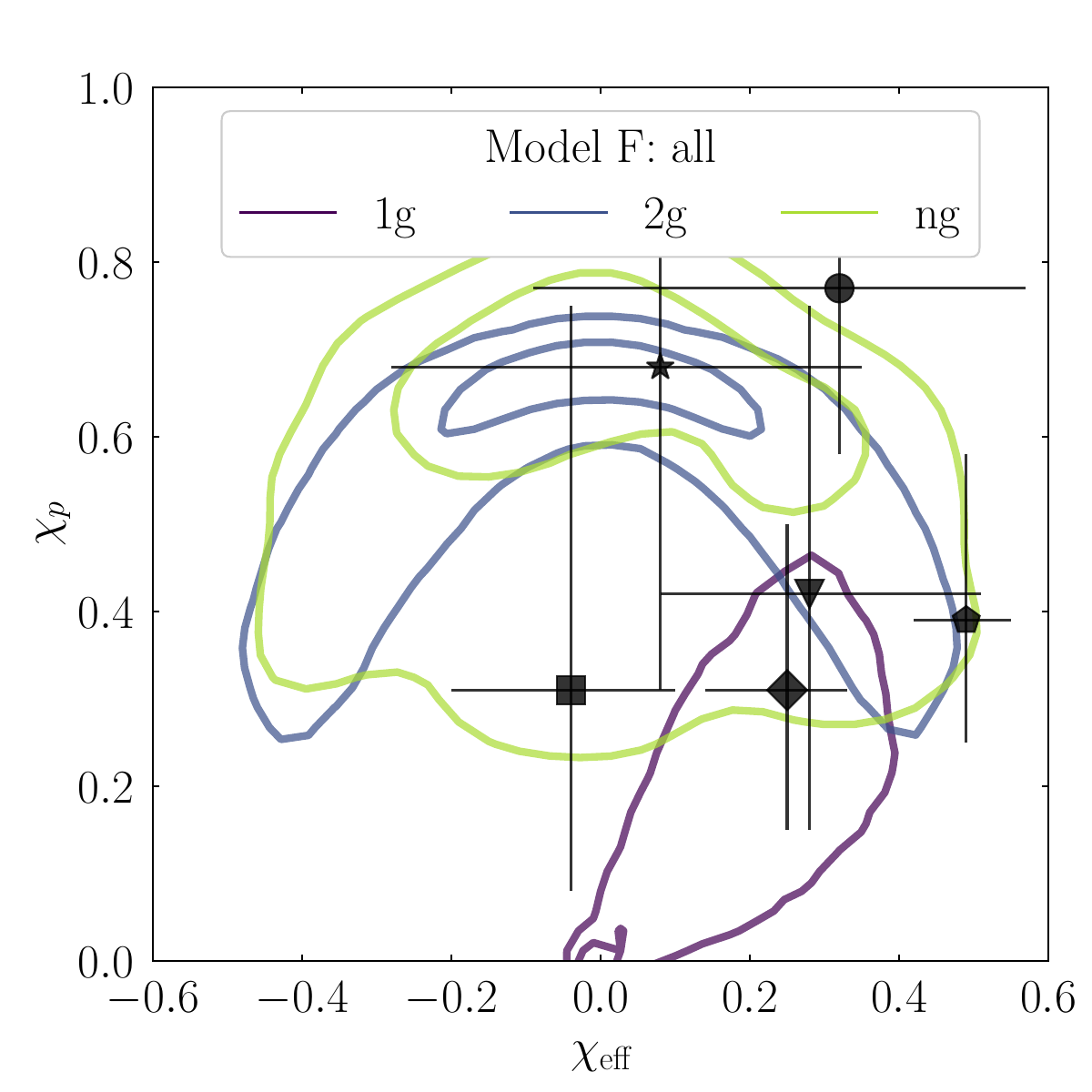}
    \includegraphics[width=\columnwidth]{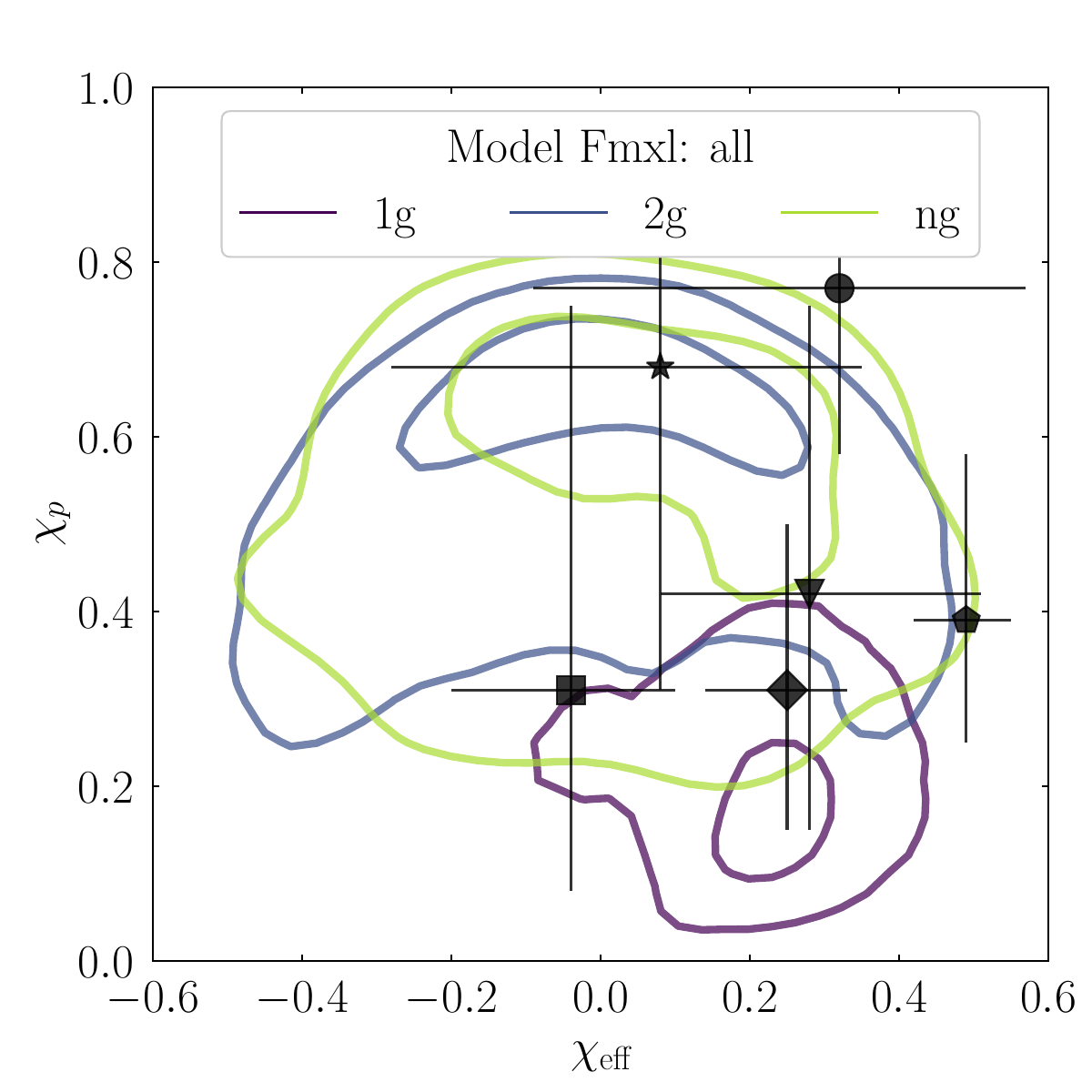}\\
    \caption{Combined distribution of $\chi_{\rm eff}$ and $\chi_{\rm eff}$ for mergers in GCs and IBs and with a primary mass $<200\Ms$, in models F (left panel) and Fmxl (right panel), dissected into 1g+1g (purple contours), 2g+1g (blue contours), and $n$g+1g (green contours) generation mergers. Contour lines represent the 25th and 75th percentiles of the distribution. Different markers identify different sources, namely GW231123 (dot), GW241011 (pentagon),  GW241110 (triangle), GW190521 (star), GW150914 (square), GW190412 (diamond).}
    \label{fig:high2}
\end{figure*}

Additional differences between 1g and $n$g mergers may be hidden in the distribution of $\chi_{\rm eff}$ and $\chi_p$, which directly depend on component spins and masses. Recently, \cite{2024ApJ...966L..16P} proposed that mergers involving high-generation components, for example, occupy a well-defined portion of the $\chi_{\rm eff}-\chi_p$ plane. Figure~\ref{fig:high2} shows the combined $\chi_{\rm eff}-\chi_p$ distribution for all BBH mergers for mergers with $m_1 < 200\,\Ms$.

High-generation mergers are distributed in a broad region in the semi-plane with $\chi_p > 0.2$. The overlap with first-generation mergers is minimal in the fiducial model, but becomes more evident in the Fmxl model, due to the different assumptions on BH natal spins. Nonetheless, we find that $\sim 25\%$ of high-generation mergers fall within a peanut-shaped region, which for 2g+1g mergers is enclosed roughly within $\chi_{\rm eff}=(-0.3,0.3)$ and $\chi_p = (0.6,0.7)$, and for higher-generation primaries is slightly larger, as shown in the Figure. Our results are consistent with the phenomenological model proposed by \cite{2024ApJ...966L..16P}. Given the negligible overlap between the distributions of dynamical 1g+1g and IB mergers, this region may represent an ideal spot to identify mergers involving a high-generation primary.

We overlay on the distribution a few GW sources for which a dynamical origin has been proposed, namely GW231123, GW241011, GW241110, GW190521, GW150914, and GW190412 \citep{2025arXiv250818083T}. Among all considered sources, only GW190521 falls in the peanut region identified above, and GW231123 seems an outlier of the ng+1g distribution, likely owing to its massive secondary. All other sources are located in regions with significant overlaps among different channels.

\subsection{\label{sec:ics} The impact of initial conditions}

In this section we explore how different sets of initial conditions affect our results and conclusions. 

\subsubsection{The impact of common envelope physics}

\begin{figure}
    \centering
    \includegraphics[width=\columnwidth]{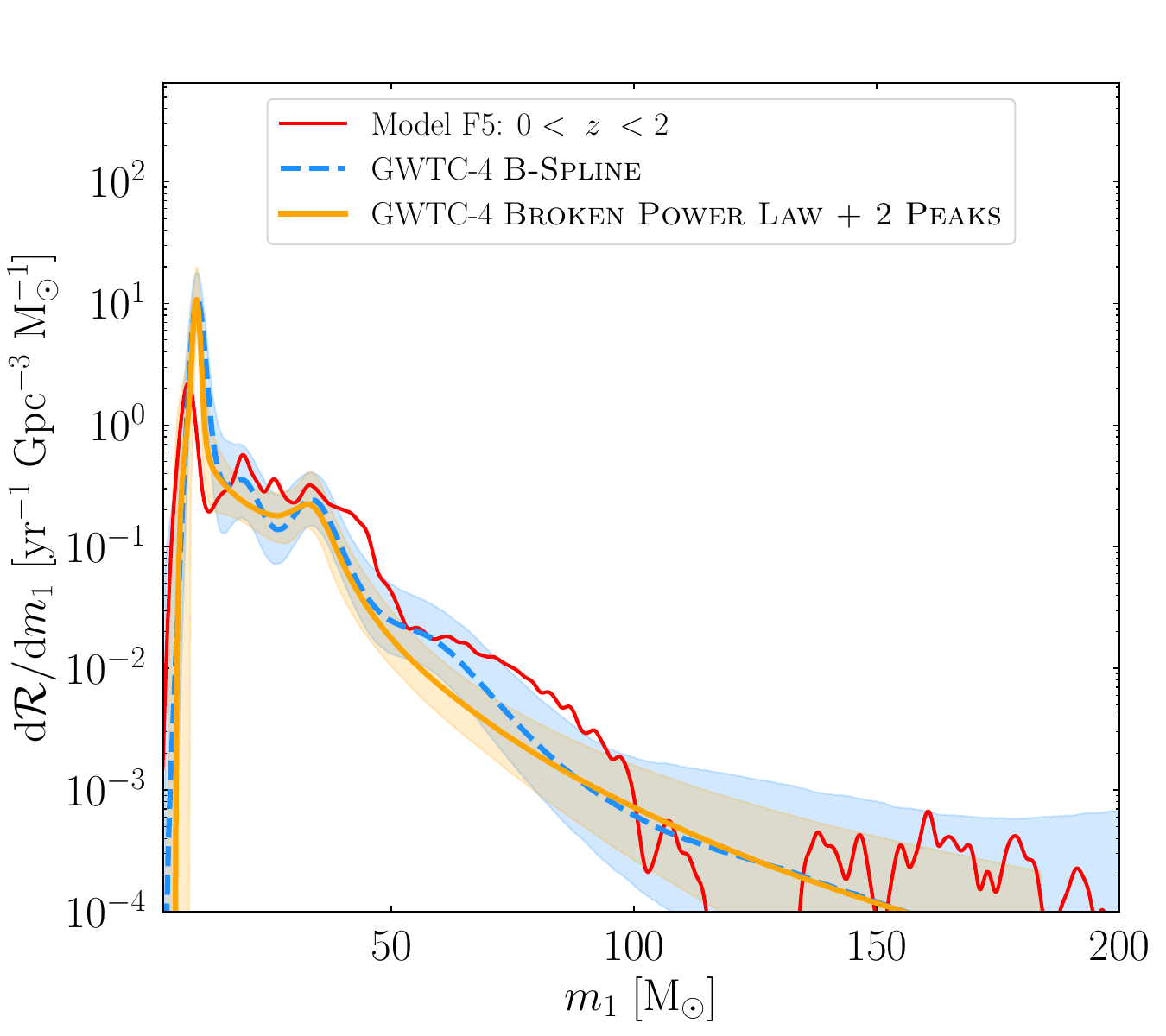}
    \caption{As in Figure \ref{fig:5}, but for model F5.}
    \label{fig:cenv}
\end{figure}

Increasing the value of $\alpha_{\rm CE}$ from 1 to 5 results in a significant overall decrease in the merger rate density of IB across all redshifts, leading to a difference in the local merger rate values between the two models of $\sim 25\%$. The resulting variation in the relative amount of BBHs from isolated and dynamical channels leads to the suppression of the tail at $m_1 < 40\,\Ms$, as shown in Figure \ref{fig:cenv}. This owes primarily to the significant reduction in the merger efficiency of isolated binaries at metallicities $Z > 0.001$ \citep{2023MNRAS.524..426I}. The resulting primary distribution is characterized by a smaller peak at $m_1\simeq 8-10\,\Ms$ compared to the fiducial model and LVK constraints, although maintaining an overall visual agreement with the inferred distribution at larger $m_1$ values.

\subsubsection{The impact of binary fraction in star clusters}

\begin{figure}
    \centering
    \includegraphics[width=\columnwidth]{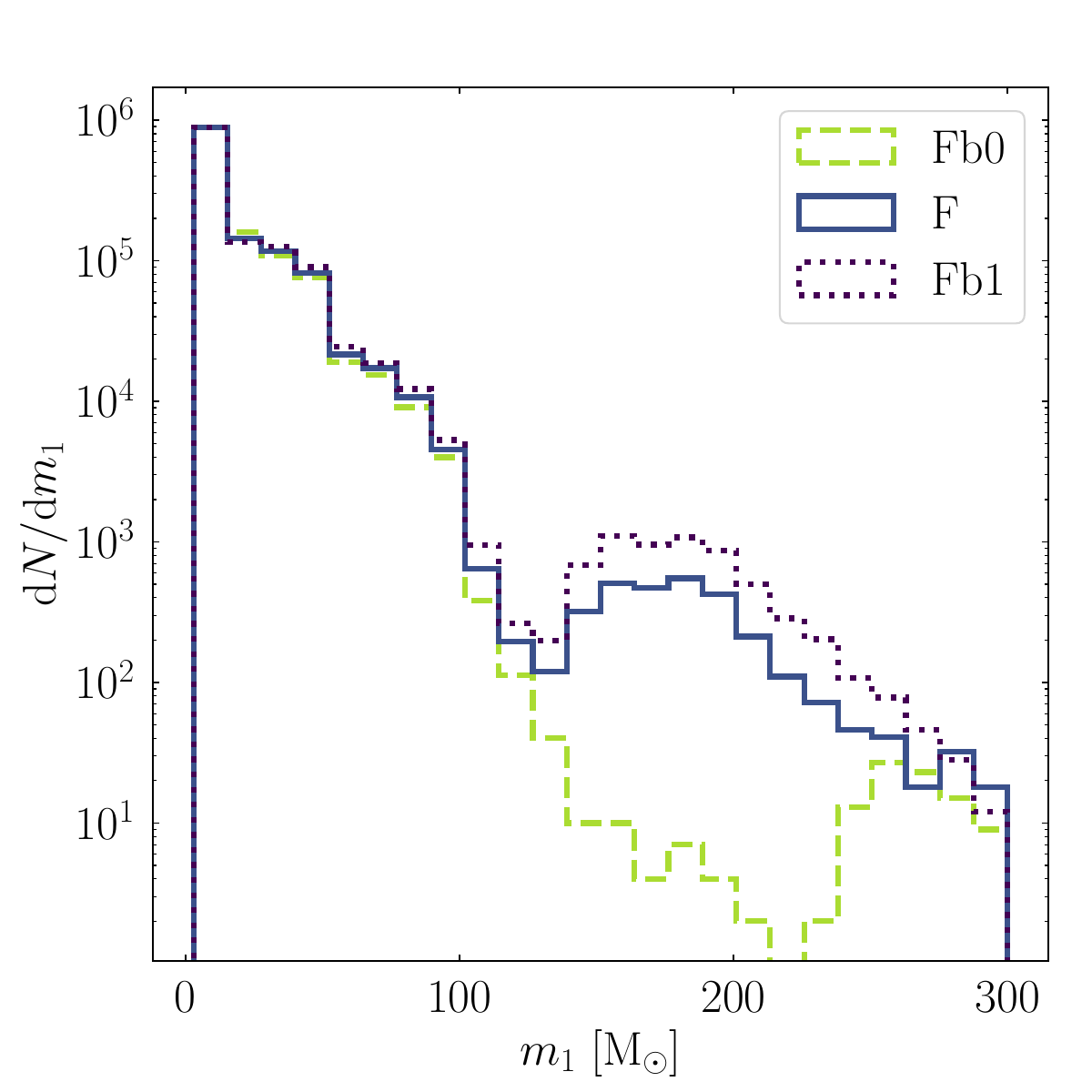}
    \caption{Primary mass distribution of all BBH mergers in models with $f_{\rm mix} = 0$ (Fb0, dashed light green line), $0.5$ (F, solid blue line), and $1$ (Fb1, dotted purple line).}
    \label{fig:mfmix}
\end{figure}

In our fiducial model, we assume that the BHs in dynamical mergers have a probability of $50\%$ to have formed from a single star or in a binary. This roughly corresponds to assuming that star clusters have, on average, an initial fraction of binaries $f_{\rm mix} = 0.5$. The mass distribution of BHs formed in binaries systematically extends toward larger values compared to that of BHs from single stars, regardless of the metallicity \citep[see e.g.][]{2023MNRAS.520.5259A}. Stellar collisions and matter accretion can indeed form BHs with masses in the range $50-200\,\Ms$, that will more likely interact with other BHs in star clusters, possibly promoting the formation of heavy mergers. Figure~\ref{fig:mfmix} shows the overall primary mass distribution of dynamical BBH mergers assuming $f_{\rm mix} = 0-0.5-1$, highlighting the effect of including BHs formed in binary stars. Compared to the case $f_{\rm mix} = 0$, the number of mergers with a primary more massive than $100\,\Ms$ increases by a factor 2 (3) when $f_{\rm mix} = 0.5$ ($1$), and up to a factor 4 (8) if the analysis is limited to mergers occurring at $z<2$. 

The resulting local merger rate for mergers with at least one component $m_1 > 100\,\Ms$ varies by a factor $\sim 10$, ranging between $\mathcal{R}_{\rm 100} \simeq (0.0009-0.021)\yrgpc$, where the lower bound corresponds to lower values of $f_{\rm mix}$. 

Such massive BHs lie at the edge, and even outside, the current detectors' sensitivity, making it hard to assess the properties of BBH mergers at the edge of the mass distribution. Next-generation observatories operating in different frequency bands, like ET \citep{2025arXiv250312263A}, the Laser Interferometer Space Antenna \citep[LISA][]{2024arXiv240207571C}, and possibly detectors in the deci-Hz frequency band like Lunar Gravitational Wave Antenna\citep[LGWA][]{2020CQGra..37u5011A,2025JCAP...01..108A}, will allow to synergetically probe the IMBH mass spectrum and unravel their properties.

\subsubsection{The impact of high-mass BHs}
Our fiducial model includes the possibility that sufficiently dense clusters undergo a series of stellar collisions that rapidly build an IMBH seed, as recently suggested by state-of-the-art numerical simulations \citep[e.g.][]{2024MNRAS.531.3770R, 2025MNRAS.543.2130R}. Additionally, we also include BHs forming from dynamical processes affecting primordial binaries, following a metallicity dependent approach that relies on numerical models of young massive clusters \citep{2020MNRAS.497.1043D,2023MNRAS.526..429A}. 

\begin{figure}
    \centering
    \includegraphics[width=\columnwidth]{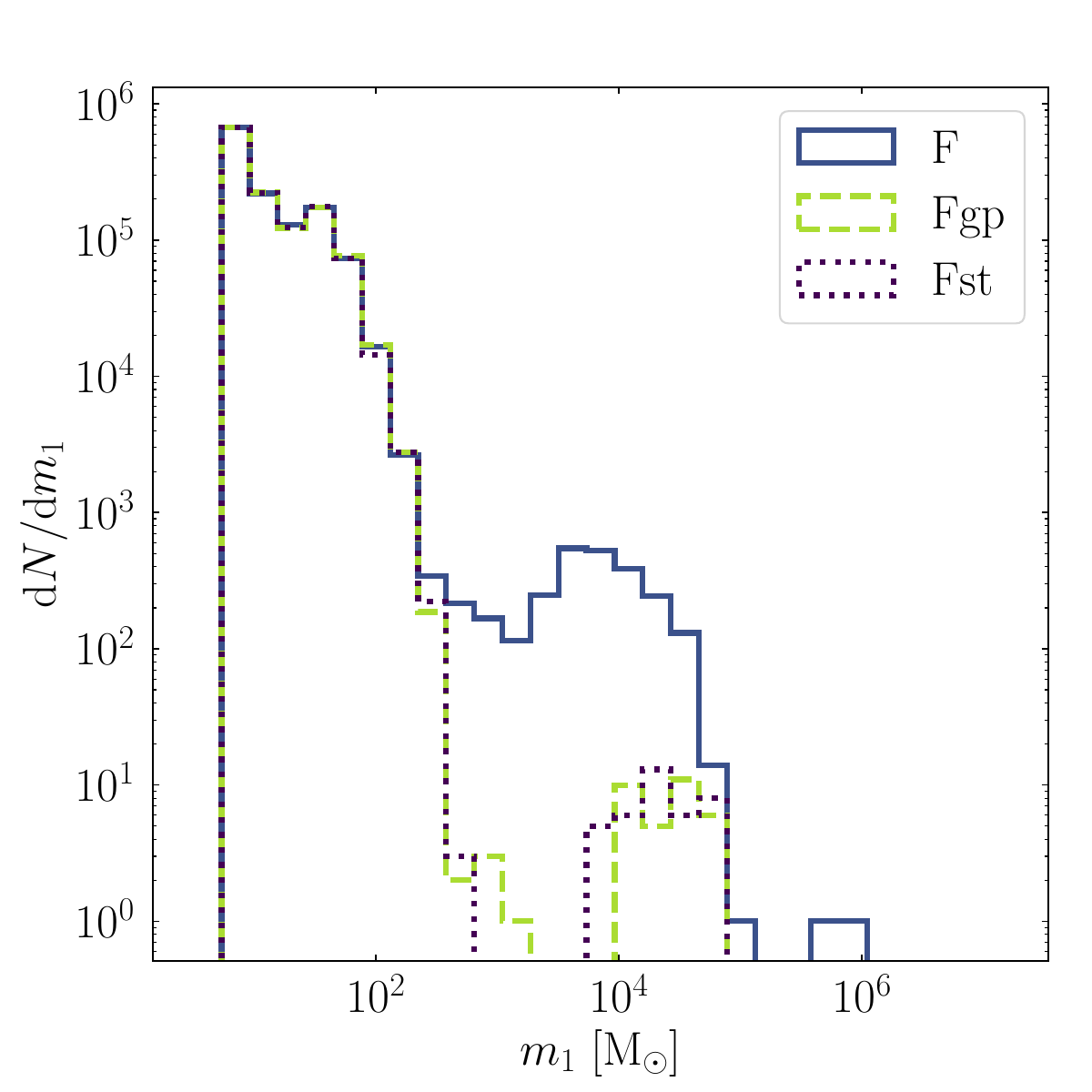}
    \caption{Primary mass distribution of all BBH mergers in the fiducial model (solid blue line), in the model where we do not include IMBH seeds from stellar collision products (Fgp, dashed light green line) and in the model where we exclude BHs with a natal mass in the upper mass gap (Fst, dotted purple line). }
    \label{fig:gap}
\end{figure}

We exclude stellar-collision products (model Fgp) and dynamically formed BHs in the upper mass gap (model Fst) to assess the impact of these features on the global BH mass spectrum. As shown in Figure~\ref{fig:gap}, this assumptions primarily affect only the high-mass tail at $m_1 > 300\,\Ms$. In the fiducial model, the fraction of such massive BHs is $f_{300} = 2.1\times10^{-3}$, while, in both models Fgp and Fst, this fraction is reduced to $f_{300} = 3.7\times10^{-5}$. Despite this substantial suppression, the effect on the local merger rate of binaries with $m_1 > 100\Ms$ remains negligible, with $\mathcal{R}_{\rm loc,100} = (0.0073-0.0081)\,\yrgpc$, since BHs with masses above $300\,\Ms$ contribute only a minor fraction of the overall population.

\subsubsection{The impact of spins}
\label{sec:spins}

Spin amplitude and alignment represent further parameters to assess BBH origins, especially in concert with the binary masses or mass ratios. Binaries with nearly equal masses and aligned spins -- i.e. $\chi_{\rm eff} > 0$ -- are generally associated with the IB scenario, although supernova explosions during the BH formation can tilt the spin vectors \citep[e.g.][]{2000ApJ...541..319K,2024arXiv241203461B}. However, similar features are also associated with formation in an AGN disk \citep{2020ApJ...899...26T,2024MNRAS.531.3479M}, unless BBHs retain a significant eccentricity \citep{2025arXiv251007952F}. BBHs formed through dynamical processes, instead, should feature a broad mass spectrum and spins isotropically oriented in space, leading to a $\chi_{\rm eff}$ distribution peaked around $0$. Additionally, BBHs involving high-generation BHs are characterized by spins peaking around $\sim 0.7$, and mass-ratios $q<1$. 

\begin{figure}
    \centering
    \includegraphics[width=0.85\columnwidth]{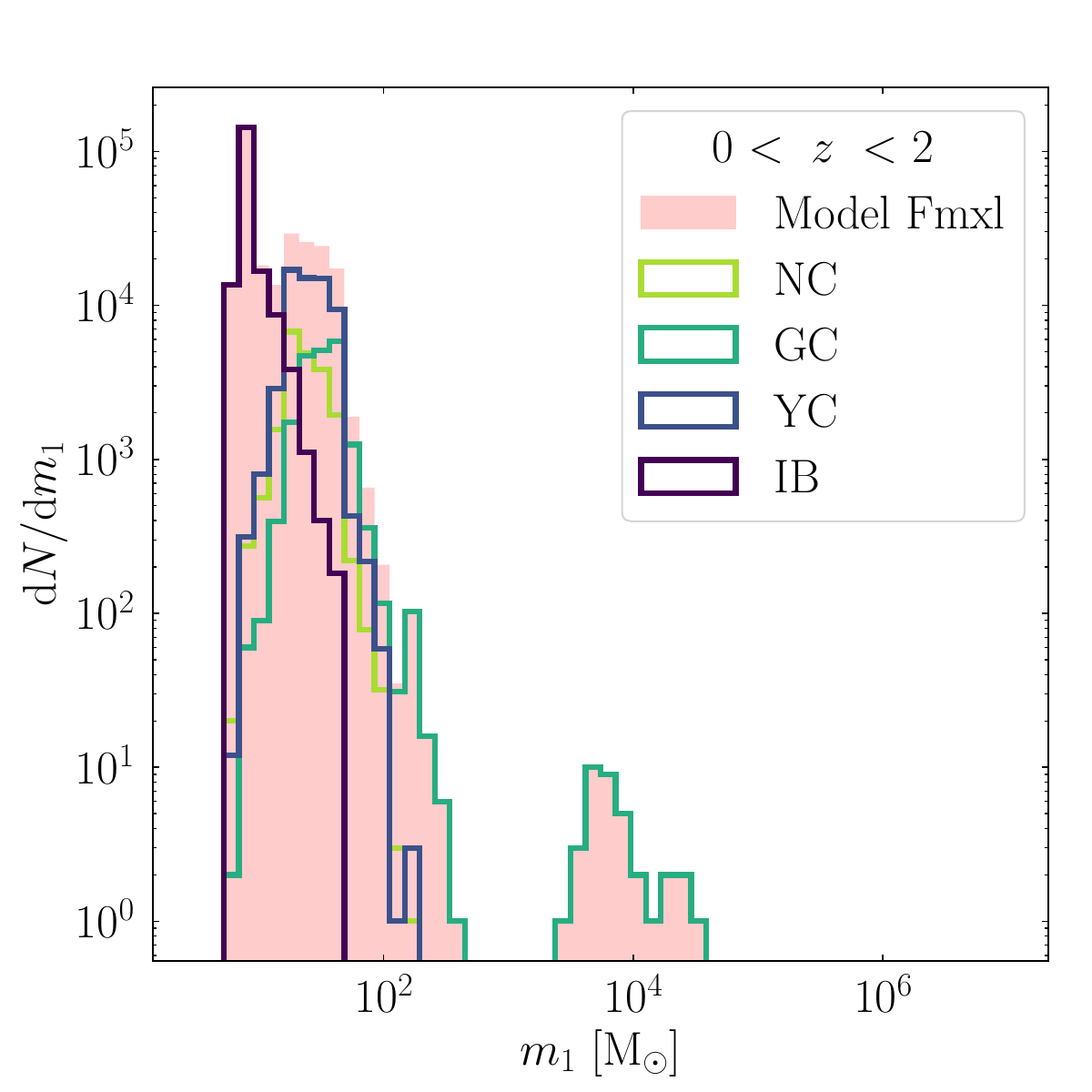}\\
    \includegraphics[width=0.85\columnwidth]{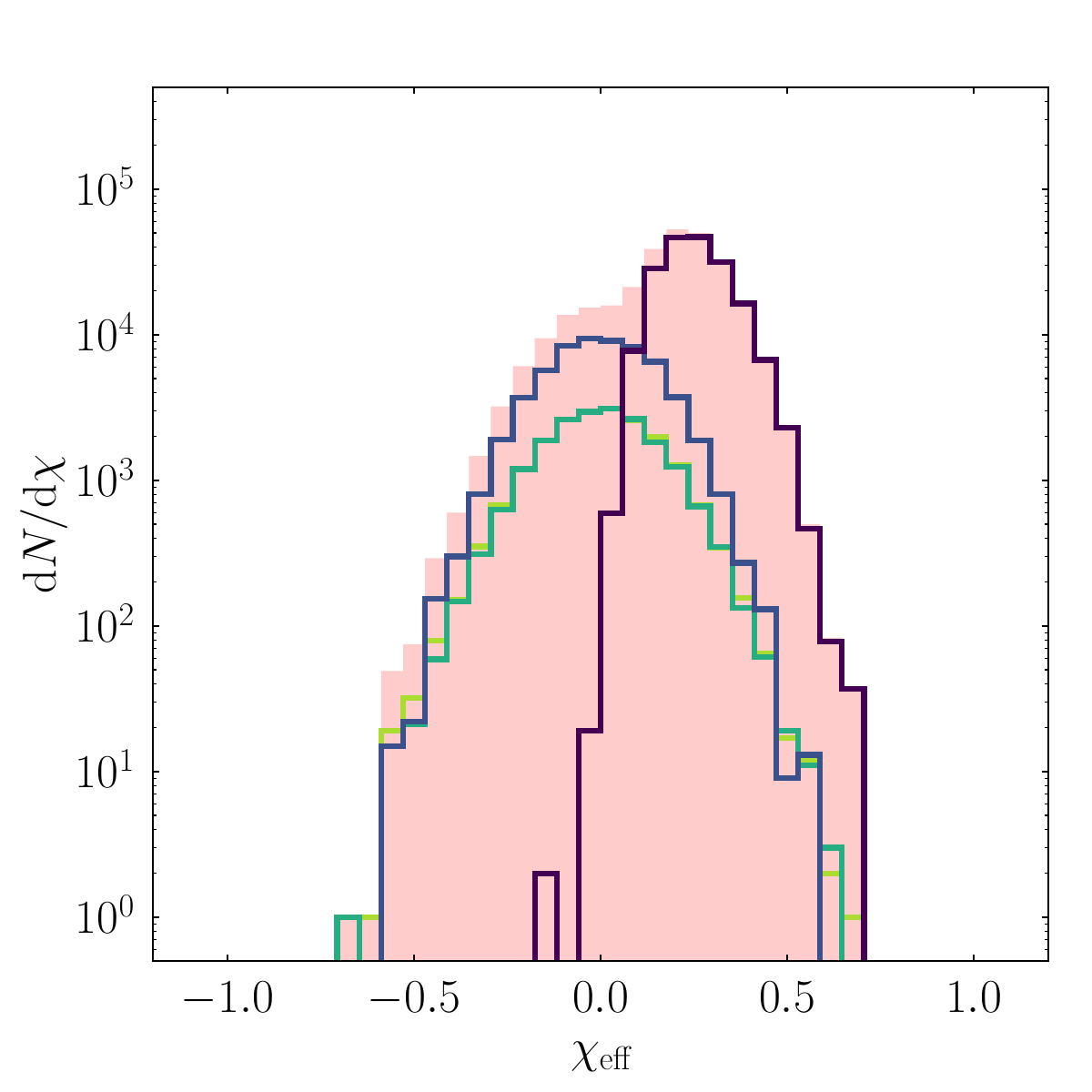}\\
    \includegraphics[width=0.85\columnwidth]{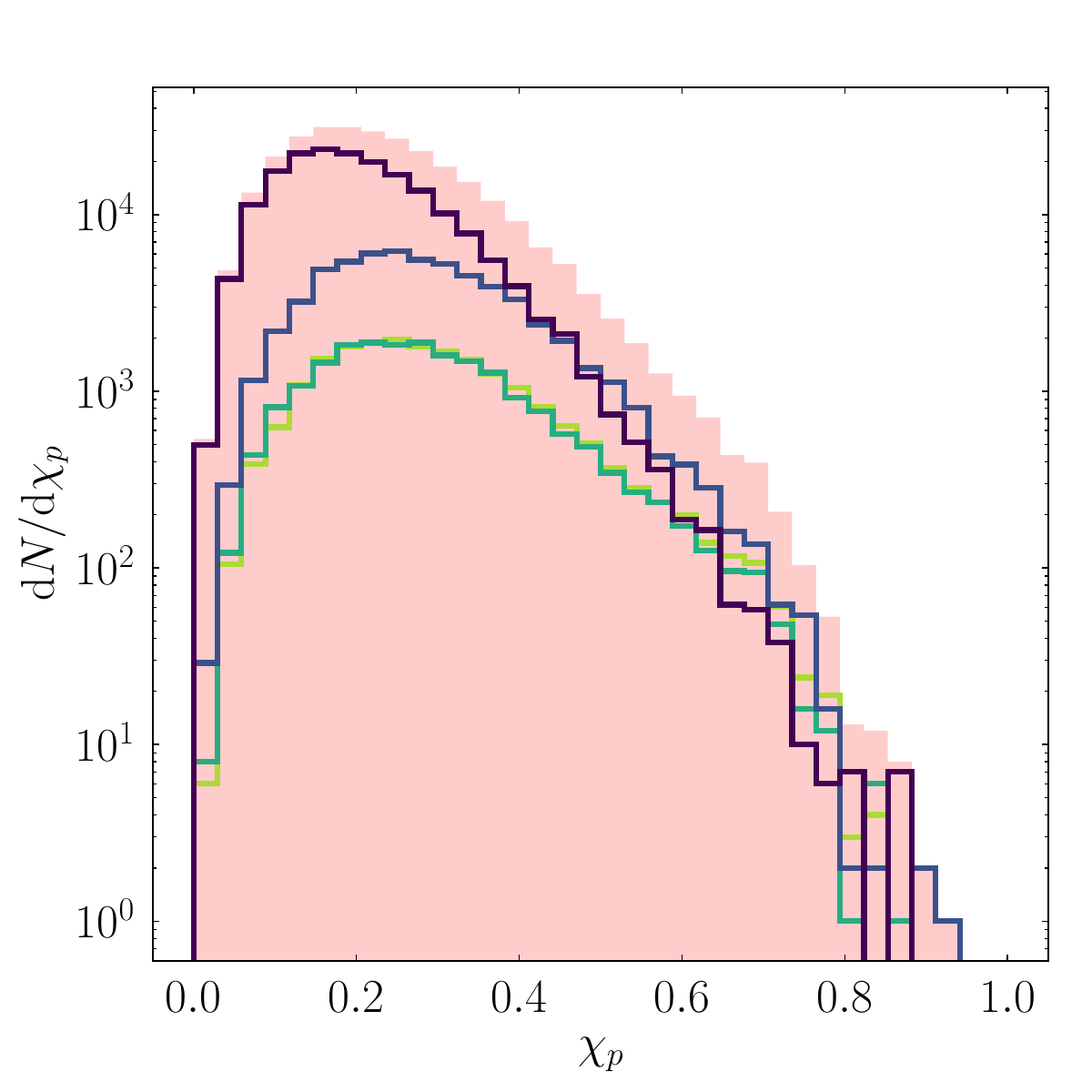}\\
    \caption{Distribution of primary masses (top panel), effective spin parameters (central panel) and precession spin parameters (bottom panel) for mergers at $z<2$ in model Fmxl, which assumes a Maxwellian distribution with $\sigma_{\chi} = 0.2$ for BH natal spins. The red shaded areas identify the overall distribution, while open steps split the population into different origins: nuclear clusters (light green), globular clusters (green), young clusters (blue), and isolated binaries (purple).}
    \label{fig:mxl}
\end{figure}

Hence, different formation channels leave fingerprints in the global spin parameters distributions that could help untangling them. For example, we have shown in Figure~\ref{fig:3} how dynamical mergers contribute to the broadening of the $\chi_{\rm eff}$ distribution toward negative values, and to the development of a small peak around $\chi_p\simeq 0.65-0.7$ due to high-generation mergers. Nevertheless, some of these features may be significantly influenced by BH natal spins. Figure~\ref{fig:mxl} shows the distribution of $\chi_{\rm eff}$ and $\chi_p$ for mergers occurring at $z<2$ in model Fmxl in which natal spins follow a Maxwellian distribution with $\sigma_\chi=0.2$. 

We see that a clear difference between dynamical and isolated mergers is still visible in the $\chi_{\rm eff}$ distribution. However, the adoption of a Maxwellian distribution for all BH natal spins reduces the probability for multiple mergers, as BH remnants receive larger kicks compared to the fiducial case (in which BHs can have zero-spins at birth depending on their formation history). This results in the absence of the bump at $\chi_p\sim 0.6-0.7$ which is instead present in the fiducial model. In these regards, it is interesting to note that GWTC-4 results suggest a skewed distribution for $\chi_{\rm eff}$ asymmetric around zero with support for positive values, and a $\chi_p$ distribution peaking around $\chi_p \simeq 0.2-0.4$, albeit the uncertainties related to the inference procedure \citep{2025arXiv250818083T}. 

Intrinsic correlations between spins and masses may translate into dependencies in the observable parameter space. For example, analysis of GWTC-4 data suggests a weak anti-correlation in the $q-\chi_{\rm eff}$ plane, although its statistical significance has reduced with respect to GWTC-3 \citep{2023PhRvX..13a1048A,2025arXiv250818083T}. 

\begin{figure}
    \centering
    \includegraphics[width=\columnwidth]{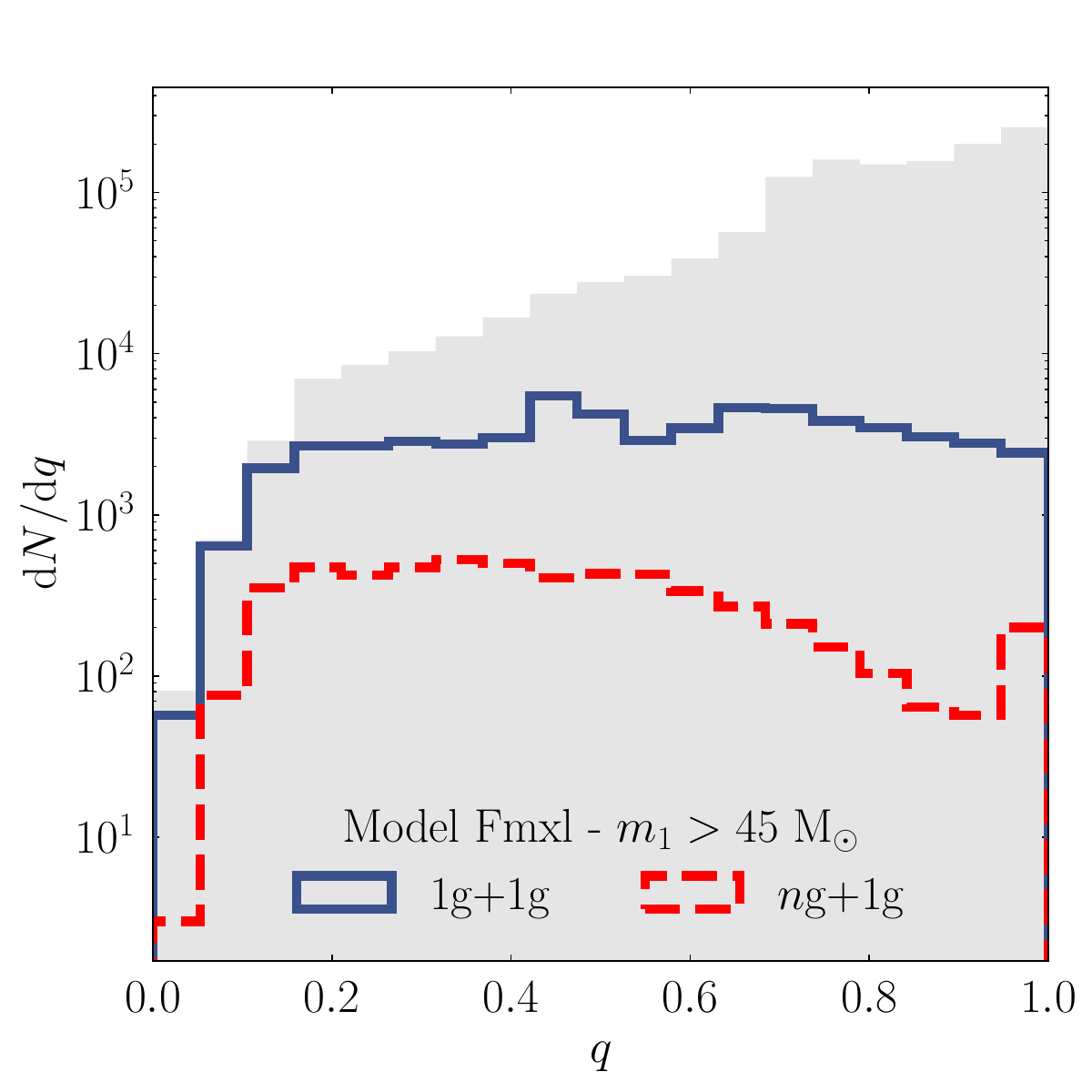}\\
    \includegraphics[width=\columnwidth]{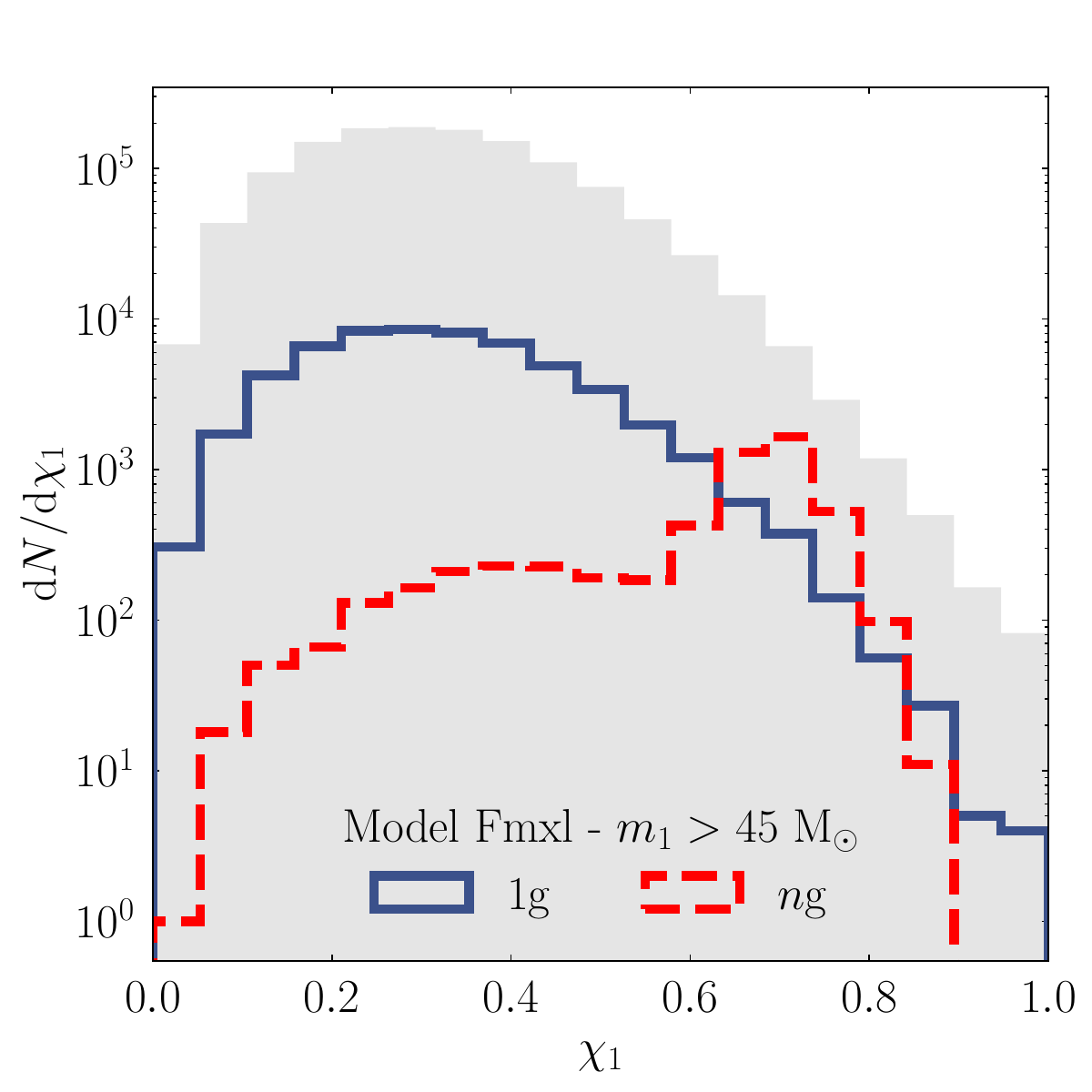}
    \caption{Same as Figure \ref{fig:rationext}, but for model Fmxl, which adopts a Maxwellian distribution for BH natal spins.}
    \label{fig:spI}
\end{figure}

In Figure~\ref{fig:rationext}, we have shown that BBHs with a primary mass $>45\,\Ms$, which according to our fiducial model exhibit a nearly flat mass ratio, have a distribution in $\chi_{\rm 1}$ that strongly depends on whether the primary underwent previous mergers or formed from stellar evolution processes. Figure~\ref{fig:spI} shows the same distributions for  BHs with mass $m_1>45\,\Ms$ in model Fmxl. Despite the lower number of high-generation BHs, even in this model we find a flat mass-ratio distribution. In contrast, the spin distribution differs substantially from that of the fiducial model, although it continues to show a peak around $\chi_{\rm 1}\simeq 0.6-0.8$ for high-generation mergers.

\begin{figure}
    \centering
    \includegraphics[width=\columnwidth]{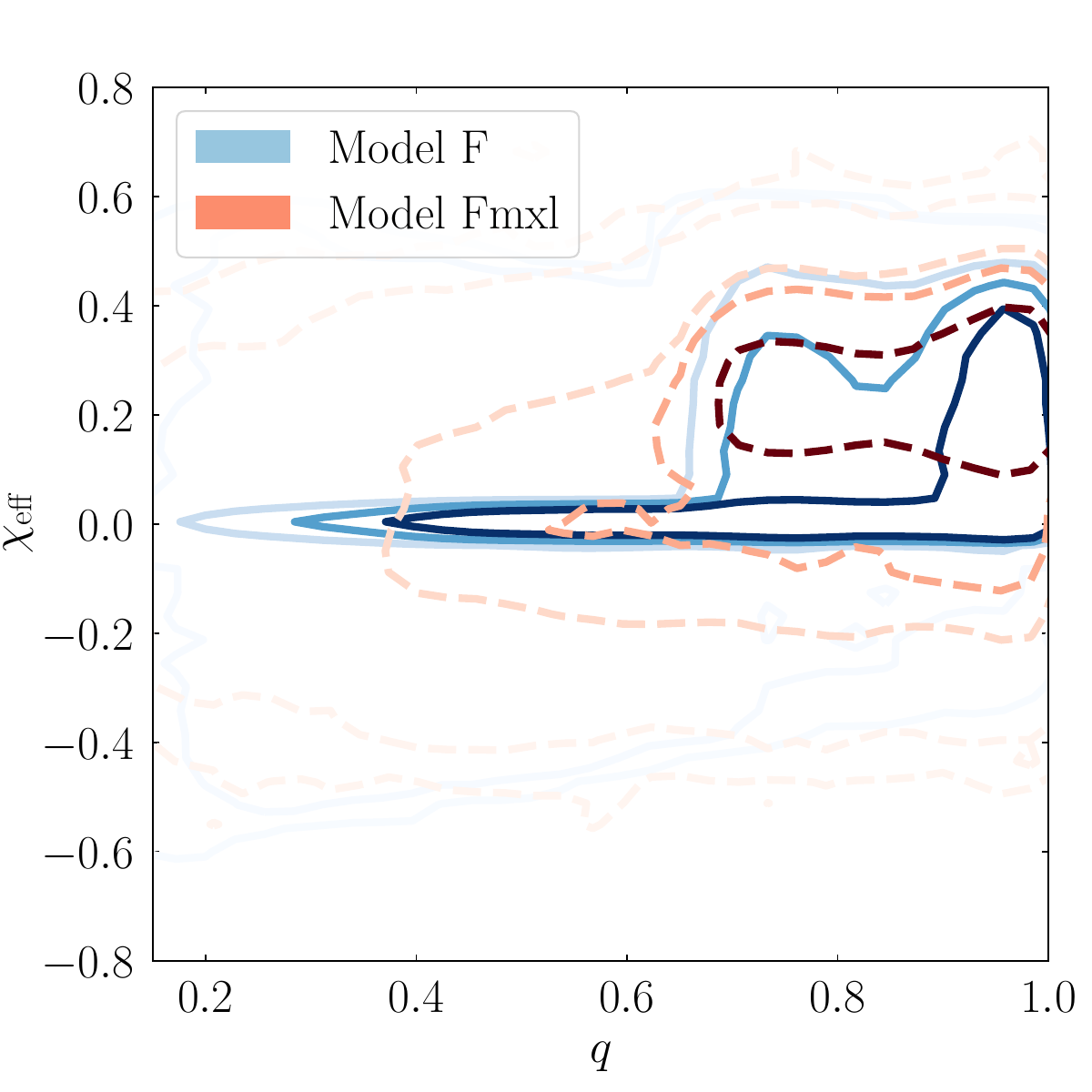}
    \caption{Mass ratio ($x$-axis) and effective spin parameter ($y$-axis) distribution for the fiducial model (blue solid contours) and for model Fmxl, which assumes a Maxwellian distribution for BH natal spins (red dashed contours). Contour plots identify, from the faintest to the more vivid lines, the 1st, 5th, 10th, 25th, and 50th percentiles of the distribution. }
    \label{fig:qxeff}
\end{figure}

We further investigate possible $q-\chi_{\rm eff}$ correlations between the two models in Figure~\ref{fig:qxeff}. Regardless of the adopted natal spin prescription, two populations arise in the $q-\chi_{\rm eff}$ plane due to their different formation channels. At mass ratios above $q > 0.6$, the bulk of the distribution lies in the range $\chi_{\rm eff} = (0-0.4)$, where mergers from the IB channel dominate. Dynamical mergers, instead, dominate the region at $q<0.6$, leading to a relatively narrow distribution of $\chi_{\rm eff}$, with the width of the distribution depending on the adopted natal spin prescriptions.

In Table~\ref{tab:2} we list the percentages of BBHs from the dynamical and isolated channels falling in different portions of the $q-\chi_{\rm eff}$ in the two models. We only consider mergers occurring at redshift $z<2$ and with a primary mass $m_1<200\,\Ms$. 

In the fiducial model, approximately $95.5\%$ of BBHs with $q > 0.6$ and $\chi_{\rm eff} > 0$ are formed through isolated binaries (IBs), whereas $99.8\%$ of BBHs with $q < 0.6$ and $\chi_{\rm eff} < 0$ have a dynamical origin. Although adopting an alternative natal spin distribution (model Fmxl) alters these fractions, the overall conclusion remains unchanged. This further highlights the potential of this combined analysis in disentangling the two formation channels for low-redshift stellar-mass BBH mergers.

\begingroup
\setlength{\tabcolsep}{10pt} 
\renewcommand{\arraystretch}{1.5} 
\begin{table}[]
    \centering
    \caption{Fraction of mergers dissected into formation channel, mass ratio, and effective spin parameter.}
    \renewcommand{\arraystretch}{1.3}
    \begin{tabular}{c|cc|cc}
\hline\hline
\multicolumn{5}{c}{{\bf Model F}}\\
\hline\hline
 & \multicolumn{2}{c|}{$q< 0.6$} & \multicolumn{2}{c}{$q>0.6$}\\[3pt]
 \hline
 & Dyn & Iso & Dyn & Iso\\
 \hline
    $\chi_{\rm eff} < 0$ &  0.998 & 0.002 & 0.962 & 0.038\\ [3pt]
    $\chi_{\rm eff} > 0$ &  0.757 & 0.243 & 0.045 & 0.955\\ [3pt]
\hline\hline
\multicolumn{5}{c}{{\bf Model Fmxl}}\\
\hline\hline
 & \multicolumn{2}{c|}{$q< 0.6$} & \multicolumn{2}{c}{$q>0.6$}\\[3pt]
 \hline
    $\chi_{\rm eff} < 0$ &  1 & 0 & 0.999 & 0.001 \\ [3pt]
    $ \chi_{\rm eff}> 0$ &  0.851 & 0.149 & 0.148 & 0.852 \\ [3pt]  
    \bottomrule             
    \end{tabular}
    \label{tab:2}
\end{table}
\endgroup

By restricting the analysis to mergers with a primary mass $m_1 > 20\,\Ms$, we find that dynamical BBHs completely dominate the population, representing the $82-100\%$ of all mergers. Still, relying on such evidences to assess the origin of observed sources is not trivial, owing to the large uncertainties associated with parameter estimation of observed sources. Among all considered GWTC-4 events, GW230919 is the only one for which $m_1>20\,\Ms$, $q > 0.6$, and $\chi_{\rm eff} > 0$ are confirmed at $90\%$ credible interval. In our models, $82.1\%$ (model F) and $93.5\%$ (model Fmxl) of the population in this portion of the parameter space is populated by dynamical mergers. Similarly, GW190403 and GW190412 have $m_1 > 20\,\Ms$, $q < 0.6$, and $\chi_{\rm eff} > 0$, hence lying in a region of the parameter space consisting of dynamical mergers only.

\subsection{\label{sec:selGW} Model selection with GW events}

In this section, we use GW events to compare different formation channels through the detectability-weighted Bayes factor \citep{2004AIPC..735..195L,mould2023}
\begin{equation}
    \mathcal{D} = \frac{p\left({\rm det}|{\rm dyn}\right)}{p\left({\rm det}|{\rm iso}\right)} \mathcal{B},
    \label{eq:dayes}
\end{equation} 
which conditions the Bayes factor, i.e. the ratio of probabilities to measure the source's set of parameters $\theta$ given a formation channel (assuming equal prior distributions), 
\begin{equation}
     \mathcal{B} = \frac{p\left(\theta|{\rm iso}\right)}{p\left(\theta|{\rm dyn}\right)},
     \label{eq:bayes}
\end{equation}
with the ratio of probabilities that a source is detectable given the dynamical, $p({\rm det}|{\rm dyn})$, or isolated channel, $p({\rm det}|{\rm iso})$.
The factor $\mathcal{D}$ accounts for the fact that detected GW signals belong to the population of detectable sources \citep{mould2023}. According to our definition, a value of $\mathcal{D} > 1$ ($<1$) implies that the isolated (dynamical) origin is more likely to produce the GW data, although strong (decisive) conclusion can be drawn only for $|\ln\mathcal{D}| > 2.3~(5)$ \citep{jeffreys1939}.

To evaluate the probability of having a detectable source from a given formation channel, $p({\rm det} | {\rm channel})$, we inject the population of simulated isolated and dynamical BBHs in \gwfish \citep{2023A&C....4200671D}, a code based on the Fisher-matrix approximation to perform signal-to-noise ratio (SNR) measurements and parameter estimation (PE) of GW signals, and calculate $p({\rm det} | {\rm channel})$ as the fraction of sources with an SNR$>8$, i.e. the threshold generally adopted to mark a source as detectable through matched-filtering searches \citep{1987thyg.book..330T,2012PhRvD..85l2006A}. 

In this analysis, we consider the set of observables defined by the primary mass, mass ratio, and effective spin parameters, i.e. ${\bf \theta} = [m_1, q, \chi_{\rm eff}]$. We apply the calculation to the 165 BBHs in GWTC-4 with a $p_{\rm astro} > 0.5$, and consider only mock BBHs with component masses $m_{1,2} = (3-500)\,\Ms$, merging at a redshift $z<2$. 

\begin{figure}
    \centering
    \includegraphics[width=\columnwidth]{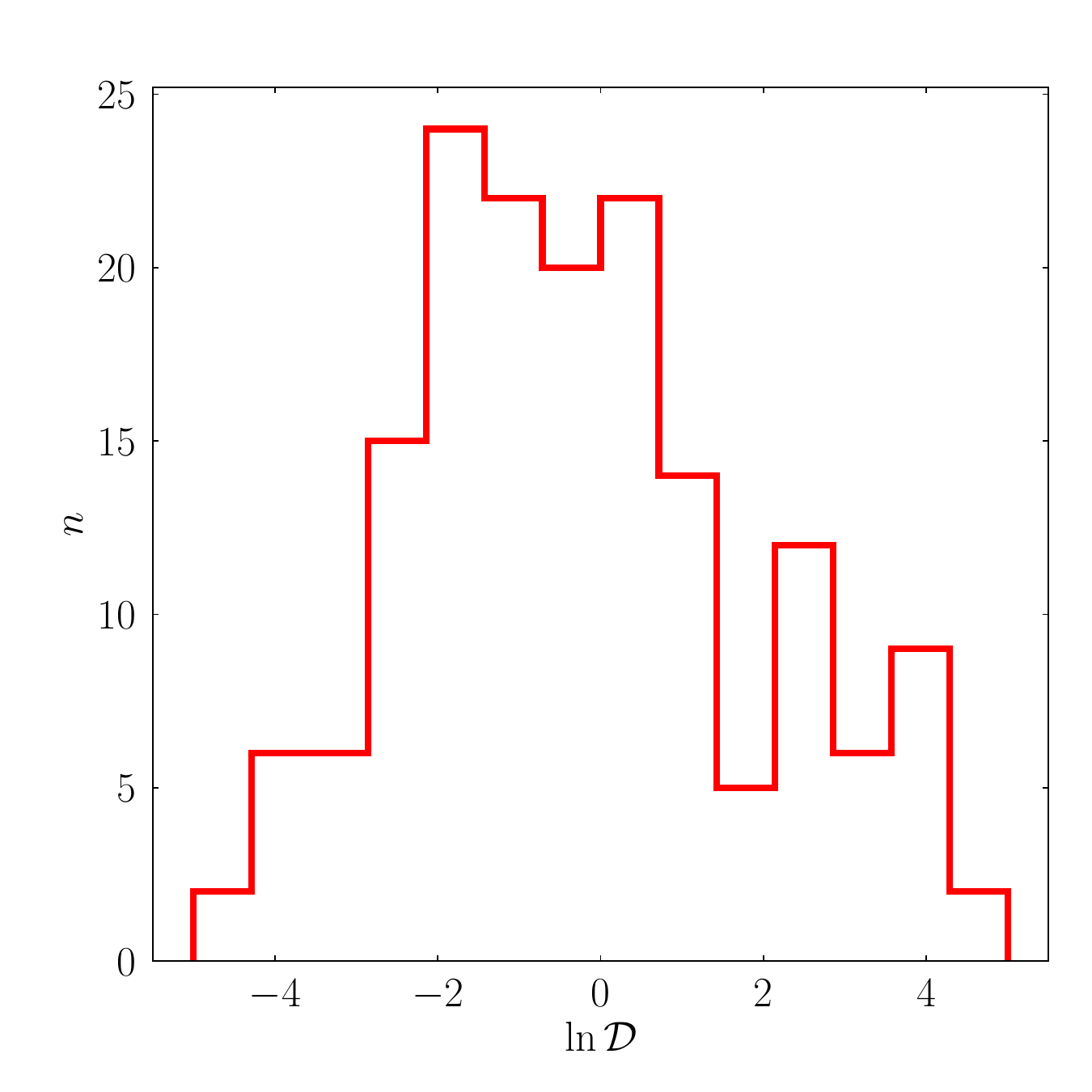}
    \caption{Distribution of the detectability-conditioned Bayes factor $\ln\mathcal{D}$ for all GW events with $m_{1,2} > 3\,\Ms$ and $p_{\rm astro} > 0.5$.}
    \label{fig:histD}
\end{figure}

The distribution of $\ln\mathcal{D}$ for all the GW events considered, shown in Figure~\ref{fig:histD}, reveals that almost all sources are characterized by $-3<\ln\mathcal{D}<3$. In fact, as summarized in Table~\ref{tab:Bayes}, we find 133 GW events with a $\ln\mathcal{D}$ in this range, and only one exceeding the threshold value for a significant assessment. This implies that, within the fiducial model, we are unable to draw decisive conclusions about most mergers. We perform the same analysis either adopting a larger SNR threshold, SNR$>12$, or calculating the unconditioned Bayes factor in Equation~\ref{eq:bayes}, thus neglecting the detectability-conditioned corrections, finding that the main conclusions of the analysis are not quantitatively affected. 

Rather than providing strong evidence in support of one formation channel or another, our analysis wants to highlight how the complexity behind astrophysical models can severely hinder the assessment of the origin of the sources. 

Among all mergers, GW231123 is the only one for which a dynamical origin is clearly favored over the isolated scenario, yielding $\ln\mathcal{D} = -7.6$. In the range $-5 < \ln \mathcal{D} < -3$ we find 13 sources, among which GW150914 ($\ln\mathcal{D} = -3.4$), GW190519 ($\ln\mathcal{D} = -4.2$), GW190521 ($\ln\mathcal{D} = -3.5$), GW230914 ($\ln\mathcal{D} = -3.9$), and GW231226 ($\ln\mathcal{D} = -4.6$). For sources with $\ln\mathcal{D} < -3$ we also calculate the Bayes factor to compare the sub-population of dynamical $n$g+1g and 1g+1g mergers. In this case, we adopt Equation~\ref{eq:bayes}, subtly assuming the same detection probability for mergers belonging to the same broad family (i.e. dynamical mergers). The largest values we found are for GW231123 ($\ln\mathcal{B} = 2.1$) and GW190519 ($\ln\mathcal{B} = 2.0$), mildly supporting a $n$g+1g origin.

In the isolated scenario, we find 18 sources with a $\ln \mathcal{D} > 3$, with the largest value found for GW191204 ($\ln \mathcal{B} \sim 5.1$). This sample includes GW230627 ($\ln\mathcal{D} = 4.8$) and GW231118 ($\ln\mathcal{D} = 4.7$).
We applied the same analysis to the recently discovered sources GW241011 and GW241110 \citep{Abac_2025}, whose highly spinning components seem to support a hierarchical origin. However, we find inconclusive values, namely $\ln \mathcal{D} = 0.6$ for GW241011 and $\ln \mathcal{D} = 0.4$ for GW241110.

\begin{table}[]
    \centering
    \setlength{\tabcolsep}{6pt} 
    \renewcommand{\arraystretch}{1.3}
    \caption{Number of GW events falling in different ranges of the detectability-conditioned Bayes factor.}   
    \begin{tabular}{c|c|c|c|c|c}
        \hline\hline
        &\multicolumn{5}{c}{$\ln\mathcal{D}$}\\
        \hline
         & $<-5$ & $(-5,-3)$ & $(-3,3)$ & $(3,5)$ & $>5$\\
        \hline
        $N_{\rm GW}$ & 1 & 13 & 133 & 18 & 0 \\
        \hline
    \end{tabular}
    \label{tab:Bayes}
\end{table}

\section{\label{sec:end} Conclusions}

In this work we used the \bpop code to examine how different formation channels, isolated evolution of stellar binaries and dynamical assembly in star clusters, shape the global properties of the population of BBH mergers.
We conduct a thorough investigation of the various factors shaping the final BBH merger population, including stellar metallicity evolution, cluster formation history, primordial binary fraction, common-envelope efficiency, and more.
In the following, we briefly summarize our main results.
\begin{itemize}
    \item The metallicity of stellar progenitors strongly affects the BBH primary mass spectrum both in isolated and dynamical mergers. In metal-poor environments, $Z \lesssim 0.1$\,Z$_\odot$, the primary mass of dynamical mergers can attain values as large as $m_{\rm1,~max} = 10^4,~10^5,~10^7\,\Ms$ in young, globular, and nuclear clusters, respectively, owing to the formation of massive stellar collision products and the onset of hierarchical BH mergers. Stellar winds and pair instability supernovae inhibit the formation of such massive BHs in environments more rich in metals [Figure \ref{fig:gmass}].
    \item The estimated local merger rate is in the range $\mathcal{R}_{\rm loc} = (17.5-24.1)\yrgpc$, which is consistent with LVK inferred limits. The largest variations are induced by the adopted common envelope $\alpha_{\rm CE}$ parameter, the fraction of binary stars initially present in star clusters, and the globular cluster formation history. The merger rate clearly increases up to redshift $z\simeq 3-6$ and subsequently declines, following the redshift evolution of the star formation rate adopted to model different environments [Figure \ref{fig:1}-\ref{fig:2}].
    \item The BBH merger fiducial population naturally reproduces the primary mass, mass ratio, and effective spin parameter distributions inferred from GWTC-4 data. At redshift $z<2$, the primary mass distribution in the fiducial model is characterized by a prominent peak at $m_{\rm pk}\simeq \,8.6\,\Ms$ and secondary peaks around $20, ~25, ~35\,\Ms$. The primary peak is due to isolated binaries, while the bump is determined by dynamical mergers, which constitute $95\%$ of all mergers with $m_1 = (30-40)\,\Ms$. 
    The mass ratio distribution is broadly consistent with LVK predictions, although we find several peaks around $q\sim 0.6-0.9$ rather than a smooth increasing trend. At $q > 0.6$, $86\%$ of mergers form from isolated binaries, whilst at lower values dynamical mergers represent the $90\%$ of the population, hence suggesting that the mass ratio could be a clear indicator of a BBH origin. In terms of spins, the effective spin parameter is well represented by a narrow distribution skewed toward positive values.
    This distribution owes to the mild alignment of spins adopted for isolated binaries (thus implying $\chi_{\rm eff} > 0$) and the fact that they  dominate the global population [Figure \ref{fig:3} and Figure \ref{fig:xfskew}]. 
    \item At larger redshifts, the features at $m_1 < 50\,\Ms$ remain, while the tail at high masses is populated by objects with masses up to $m_1 \lesssim 10^5\,\Ms$. The median binary mass is nearly constant out to redshift $z < 5$, $m_{\rm 1,med}\sim19\,\Ms$, consistent with the non-evolution of the binary mass inferred from GWTC-4 data. At $z>5$, the median mass increases owing to the larger fraction of dynamical mergers and the development of more massive BHs forming in metal-poor environments.  This evolution slightly changes if we assume a common-envelope parameter $\alpha_{\rm CE} = 5$. In this case, the fraction of dynamical mergers is much larger compared to other models, leading the median binary mass to initially increase, reaching a peak around $z\sim 5-6$, and rapidly drop beyond this redshift value. The median effective spin, $\chi_{\rm eff,med}$ does not evolve appreciably, attaining a stably positive value around $\chi_{\rm eff,med}\simeq 0-0.3$ out to redshift $z=15$.    
    \item A substantial fraction of mergers, $f_{n{\rm g}}\simeq 0.047-0.112$, involve high-generation BHs with masses up to $10^6\,\Ms$, depending on the environment and the initial conditions. These hierarchical merger products are characterized by a nearly flat mass spectrum and a peculiar spin distribution that exhibits a peak around $0.6-0.8$, as expected by numerical relativity. High-generation mergers locate in a peanut-shaped region of the plane defined by the effective spin parameter and the precession spin parameter. This region, roughly boxed between $\chi_{\rm eff} = [-0.2,0.2]$ and $\chi_p = [0.6,0.8]$ is poorly populated by first generation mergers, hence suggesting that sources with spin parameters falling in these ranges may have a dynamical origin. We compared the mock distribution with the inferred $\chi_{\rm eff},~\chi_p$ of some sources for which a dynamical origin has been invoked: GW231123, GW241011, GW241110, GW190521, GW150914, and GW190412. We find that only GW190521 has spins compatible with the ranges above, and GW231123 possibly belongs to the population of mergers with a primary of generation $>2$ [Figures \ref{fig:repeated}-\ref{fig:high2}].
    \item Within the explored parameter space, the features of our BBHs are robust. Common envelope physics has the largest impact on the population, affecting the height of the primary mass peak and the mutual fraction of dynamical and isolated mergers. A large primordial binary fraction in star clusters and the inclusion of stellar collision remnants affects mainly the population of mergers with a primary mass $m_1 > 100\,\Ms$, hence determining the fraction of mergers involving an IMBH and their mass distribution. Spins mostly affect the fraction of repeated mergers and the effective spin and precession spin distributions. In a physically motivated model, $\chi_{\rm eff}$ follows a skewed distribution with support for positive values, whilst the assumption of a common Maxwellian distribution for BH natal spins leads to a more balanced distribution peaking at positive values [Figures \ref{fig:cenv}-\ref{fig:spI}].
    \item Combining the information on mass, mass ratio, and spins, we find a region of the $q-\chi_{\rm eff}$ where one formation channel clearly dominates over the other. In general, binaries with $q<0.6$ ($>0.6$) and $\chi_{\rm eff}<0$ ($\chi_{\rm eff} > 0$) are more likely to form in a dynamical environment. More specifically, considering the semi-plane $q < 0.6$ and $\chi_{\rm eff}< 0$, $\sim 99.8-100\%$ of all mergers are dynamical, while in the case of $q > 0.6$ and $\chi_{\rm eff} > 0$, $\sim 85.2-95.5\%$ of all mergers are from isolated binaries [Table \ref{tab:2}, Figure \ref{fig:qxeff}]. 
    \item We use our mock catalog to determine the likelihood of different formation scenarios for the list of GW events with $p_{\rm astro}>0.5$ in GWTC-4, by computing the detectability-weighted Bayes factor, $\mathcal{D}$, between the isolated and dynamical channel. We find conclusive evidence ($|\ln \mathcal{D}| > 5$) only for one event, GW231123, for which our model supports a dynamical origin. A strong evidence ($2 < |\ln \mathcal{D}| < 5$) in support of a dynamical (isolated) origin is found for 13 (18) GW events, while for the remaining 133 sources the analysis led to inconclusive results. Rather than finding the origin of GW events, our analysis highlights how complexities in the underlying astrophysical processes can limit our ability to determine the origin of GW sources [Figure \ref{fig:histD} and Table \ref{tab:Bayes}]. 
\end{itemize}

\appendix

\section{\label{app:method}The \bpop code}
\label{appA}
The \bpop \citep[Binary merger POPulations][]{2019MNRAS.482.2991A, 2020ApJ...894..133A,2023MNRAS.520.5259A} code is a semi-analytical software to model BBH mergers forming from IBs and dynamical interactions in YCs, GCs, and NCs. The code implements several recipes to initialize the cosmic star formation, metallicity, star cluster masses and sizes and their long-term evolution, natal mass, spin, and kick of stellar BHs, properties of dynamically formed binaries, mass of BHs formed from stellar collisions and interactions. 
Moreover, \bpop handles multiple generation mergers in star clusters using numerical relativity fitting formulae to calculate the merger remnant mass and spin
\citep{2017PhRvD..95f4024J} and gravitational recoil \citep{2007PhRvL..98w1102C,2007PhRvL..98i1101G,2008PhRvD..77d4028L,2012PhRvD..85h4015L}. This enables the user to create a synthetic universe filled with an heterogeneous population of BBH mergers formed in different environments. 
We refer the interested reader to \citep{2020ApJ...894..133A, 2023MNRAS.520.5259A} for further details about the code features.

\begin{figure*}
    \centering
    \includegraphics[width=\textwidth]{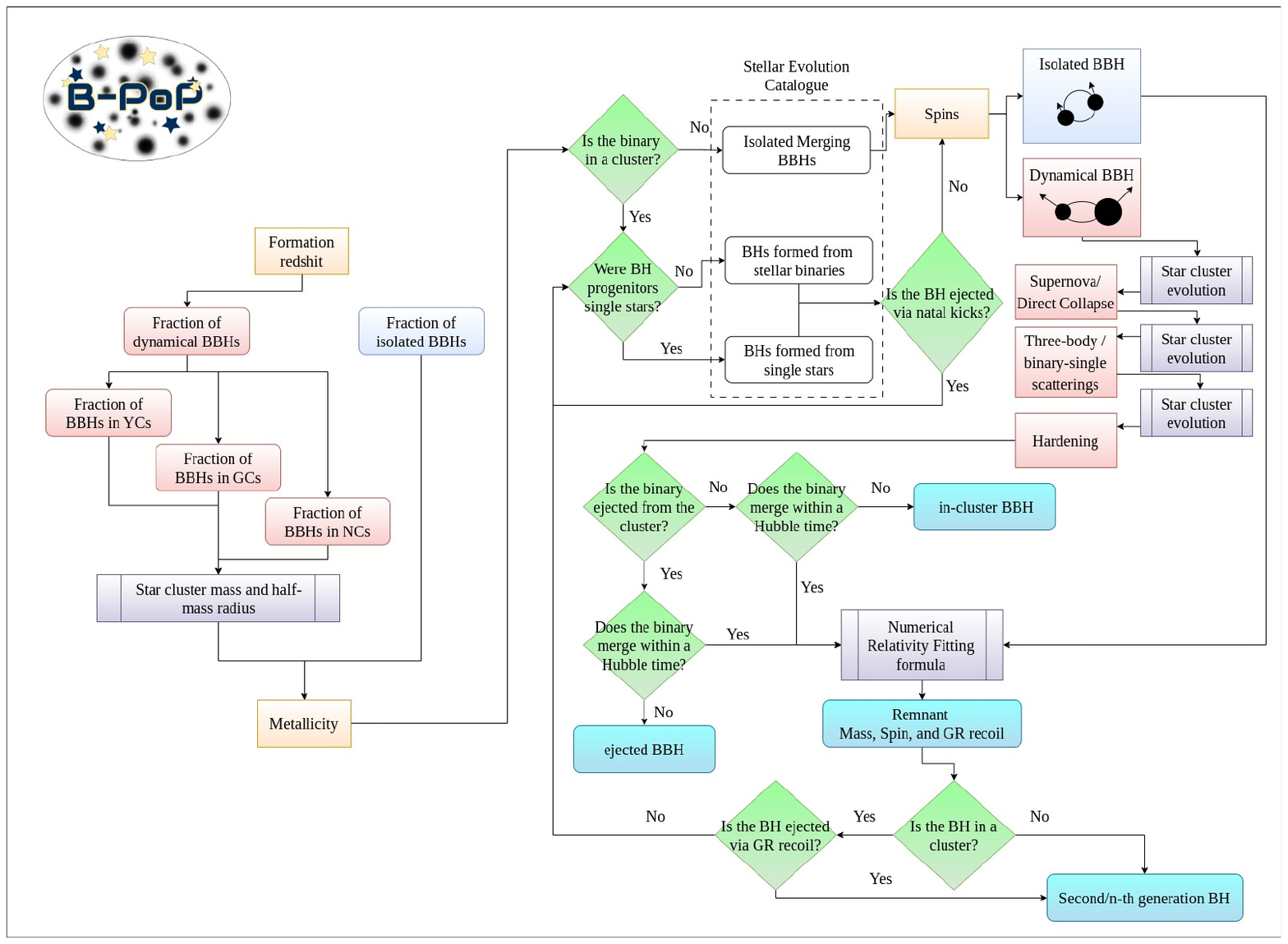}
    \caption{\bpop workflow.}
    \label{fig:workflow}
\end{figure*}

Recently, we upgraded \bpop to improve the treatment of the cosmic star formation history for different environments, the metallicity distribution and its dependence on the redshift, the dynamical evolution of star clusters and their initial properties, and to consider BHs forming in the IMBH mass range or inside the so-called upper-mass gap, i.e. the region of the mass spectrum within $60-300\,\Ms$ where BH progenitors are expected to undergo PISN or PPISN, hence leaving no remnant or collapsing to a final BH with mass $(40-60)\,\Ms$ \citep{2021ApJ...912L..31W,2016A&A...594A..97B}. 
In the following, we briefly describe how \bpop works and its new features. Figure \ref{fig:workflow} sketches \bpop workflow.

BH masses are drawn from pre-compiled catalogs of single BHs and BHs in binaries generated with a population synthesis tool. Previous works relied on catalogs generated with \mobse \citep{2018MNRAS.474.2959G,2018MNRAS.480.2011G,2020ApJ...891..141G}, while in the present work we used the \sevn code \citep{2015MNRAS.451.4086S,2017MNRAS.470.4739S,2023MNRAS.524..426I}, following initial conditions described in \citep{2023MNRAS.524..426I} and generating catalogs of BHs from single stars and binaries across 12 values of metallicities, between $Z = 0.0001-0.03$.

For each value of the metallicity, stellar evolution catalogs can contain several millions of objects, therefore the workflow of \bpop is devised to extract information from them in a computationally efficient way. 
\bpop first determines the number $n_i$ of BBHs that must be created for any formation channel, generating a sample of $N=n_i$ IDs, where $N$ represents the total number of sources input by the user. \bpop then assigns to each ID a formation time and redshift according to the star formation rate selected for each formation channel. At this point, \bpop assigns to every source a metallicity according to the adopted distribution, possibly redshift-dependent. IDs are sorted by metallicity, so that for every $Z$ value, the stellar evolution catalog is read and loaded into memory only once.

The population of isolated BBHs is generated first. \bpop select candidates from the catalog of BBH mergers generated by the stellar evolution tool. For each BH, \bpop assigns a natal spin and determines spin alignment according to the adopted distribution. 

The population of dynamical binaries is trickier. \bpop assigns to the binary a semi-major axis, eccentricity, and component masses, taking into account the possibility that a BH formed from a single star or in a binary system, and all the relevant dynamical processes that contribute to BHs pairing and coalescence, with the corresponding timescales. Even in this case, each BBH has component spins and spin alignment drawn from the adopted distribution. 

Alongside the formation process, \bpop takes also into account the evolution of the host cluster, which affects the time it takes for the BBH to form, shrink, and merge. If a dynamical BBH does not merge within a Hubble time, it is stored in a dedicated catalog. If it merges inside the cluster and the remnant's kick is smaller than the host cluster escape velocity, the BH can pair again and undergo multiple mergers. \bpop adopts numerical relativity fitting formulae to calculate remnant masses and spins \citep{2017PhRvD..95f4024J}, and post merger recoil kicks \citep{2007PhRvL..98w1102C,2008PhRvD..77d4028L,2012PhRvD..85h4015L}.

In the following subsection, we describe in more details the various modules of the code that have been upgraded with respect to the previous version \citep{2020ApJ...894..133A,2023MNRAS.520.5259A}.

\subsection{Stellar evolution of single and binary stars in \bpop}
\label{appA1}
In \bpop, the masses of BHs are drawn from pre-compiled catalogs containing remnant masses for single and binary stars. In this work, we rely on a catalog generated with the \sevn population synthesis code \citep{2015MNRAS.451.4086S,2019MNRAS.485..889S,2023MNRAS.524..426I}, assuming a \cite{2001MNRAS.322..231K} initial mass function for single stars and following the setup described in \cite{2023MNRAS.524..426I} to draw initial binary properties. The catalog consists of $10^7$ single stars and $10^7$ binary stars simulated for 12 different values of metallicity in the range $Z = 10^{-4}-0.03$, for a total of more than $2\times 10^8$ compact object progenitors. To explore the impact of common envelope on the overall BH distribution, we assume two values of the common envelope efficiency parameter, $\alpha = 1-5$, a rapid core-collapse supernova scheme \citep{2012ApJ...749...91F}, and the PISNe/PPISne described in \citep{2017MNRAS.470.4739S,2020ApJ...888...76M} based on models by \citep{2017ApJ...836..244W}.

BBH mergers forming from IBs are extracted directly from \sevn catalogs and characterized through the component masses and delay time, i.e. the time elapsed from the formation of the two stars to the final BBH merger. For dynamical mergers, instead, we consider only BHs with a natal kick smaller than the cluster escape velocity. For this sub-sample, we calculate the delay time as the time required for the two stellar progenitors to form, turn into BHs, find each other to form a tight pair, and harden up to the point at which a further interaction either kicks out the binary from the cluster or shrinks the binary sufficiently for GW emission to become the dominant mechanism driving the binary evolution. All the aforementioned phases are modeled through their associated timescales, and at each new timestep the timescales are used to update the star cluster mass and radius to ensure self-consistency between the dynamical evolution of the BBH and its parent cluster. All dynamical interactions affecting the formation and merger of compact binaries are assumed to occur in the cluster core, where interactions are harder and more frequent. At formation, the semi-major axis $a_{\rm BBH}$ of dynamical BBHs is drawn from a Gaussian distribution peaked around the hard-binary separation with dispersion $\sigma = 0.3$ and the further constrain $a_{\rm BBH}>0$, while the eccentricity is drawn from a thermal distribution, i.e. $p(e)\,{\rm d}e \propto e \,{\rm d}e$. 

Unlike the vast majority of other semi-analytical codes, \bpop enables the users to include masses, orbital parameters, formation time, and natal kicks of BHs formed not only from single stars, but also from binary stellar evolution processes. The probability that a BH formed from a single star or in a binary system is regulated through the $f_{\rm mix}$ parameter, such that $f_{\rm mix} = 0$ ($1$) implies that BHs are only produced by single stars (stellar binaries) \citep[see also][]{2020ApJ...894..133A,2023MNRAS.520.5259A}. In some sense, $f_{\rm mix}$ represents a proxy to the primordial binary fraction in star clusters. In this work, we adopt $f_{\rm mix} = 0-0.5-1$.

Even taking into account the evolution in binary stars, our catalogs cannot include BHs from progenitors that suffered from dynamical interactions before stellar evolution processes took place. This can include stellar merger products from prompt stellar collisions, evolution in stellar triples or close interactions, or matter accretion in dynamically formed binaries. The main effect of these processes is to produce BHs in the mass gap \citep[see e.g.][]{2020MNRAS.497.1043D,2024MNRAS.528.5140A}. To take into account this further feature, we combine results from state-of-the-art $N$-body simulations to derive a metallicity-dependent parameter that represents the probability that a BH has a mass in the range $60-100\,\Ms$, i.e.
\begin{equation}
p_{\rm gp} = 0.005 \left(1 + 44.4 \,\exp\left( \frac{-0.089\,Z}{0.0002} \right) \right).
\end{equation}
This relation reproduces the probability for a BBH to have at least one component with a mass in the upper-mass gap as found in \cite{2020MNRAS.497.1043D}. Additionally, we follow results from \citep{2024MNRAS.528.5140A}, and assume that the fraction of BBHs that merge and have at least one component, or both, in the mass gap is $p_{\rm GW,1gp} = 0.269$ and $p_{\rm GW,2gp} = 0.154$, respectively.
This choice leads to a probability for a star cluster with a metallicity $Z = 0.0002$ ($0.02$) of having a BBH merger with at least one component in the gap to be $\sim 8\%$ ($0.2\%$). This model is labeled as DC19 in \bpop. 
Alternatively, the PWL model enables the user to define $p_{\rm gp}$ and to assign BHs a mass drawn from a power-law with user-defined slope and boundaries (see Table~\ref{tab1}). 

The inclusion of BHs forming in the upper-mass gap through the interplay between stellar dynamics and evolution is still unable to capture additional pathways that can lead to the formation of IMBH {\it seeds}. For example, we expect that multiple stellar collisions in dense clusters can drive the formation of a VMS that eventually collapses to an IMBH with an initial mass $>100\,\Ms$, as observed in several state-of-the-art numerical simulations \citep{2021MNRAS.501.5257R, 2023MNRAS.526..429A, 2024MNRAS.531.3770R}. 

In our fiducial simulations, we assume that a VMS can form in $f_{\rm vms}=0.2$ of all clusters with an initial mass $>10^3\,\Ms$, a density $>10^5\Ms$ pc$^{-3}$, and a metallicity $Z < 0.001$\footnote{The latter condition is based on the fact that more metal-rich stars are expected to avoid PISN and collapse to BHs with a mass $\sim 20-30\,\Ms$ regardless of the mass of the progenitor \citep[][]{2017MNRAS.470.4739S}.}. In such a model, referred to as Bifrost and based on recent findings by \cite{2024MNRAS.531.3770R}, we assume that the VMS terminal mass is given by
\begin{equation}
    m_{\rm vms} = \min(2\times10^4\,\Ms, 0.02\,m_{\rm cl}), 
\end{equation}
and the resulting IMBH have a mass $m_{\rm imbh}\simeq\,0.9m_{\rm vms}$ \citep{2017MNRAS.470.4739S}.
Alternatively, \bpop implements two further options. In the PWL model, a user can define $f_{\rm vms}$ and assign the resulting IMBH a mass drawn from a power-law distribution with user-defined slope and boundaries. In the DENSITY model,the user has the same freedom as in PWL model, but VMS are assumed to form only in a fraction $f_{\rm vms}$ of clusters denser than $3\times10^5\Ms$ pc$^{-3}$ \citep{2024MNRAS.528.5119A}.

\subsection{Black hole natal spins}
\label{appA2}
The origin and properties of stellar BH natal spins are still unclear, but several recent works suggested that there may be a dependence on the last stages of stars' life, the possible evolution in a binary system, and the spin. 
\bpop implements several choices to distribute BH natal spins. Following \cite{2020AeA...635A..97B}, in the fiducial model presented here we assume that first-born BHs in stellar binaries and BHs from single stars have negligible spins \citep[see][]{2019ApJ...881L...1F}, while for second-born BHs we assume a spin randomly distributed between 0-1 \citep[see e.g.][]{2018A&A...616A..28Q}. The latter assumption is also used to assign spins to BHs from stellar collisions and those in the upper-mass gap, assuming that the complex physical processes leading to the formation of the BH progenitor can preserve a substantial amount of angular momentum that is not dissipated during BH formation \citep{2025arXiv250915619S}.
Alternatively, users can also select: a Maxwellian distribution with user-defined dispersion, a Gaussian distribution with user-defined mean and dispersion, a mass-dependent relation as defined in \cite{2016A&A...594A..97B} or \cite{2016MNRAS.458.3075A}, or zero-spin \citep{2019ApJ...881L...1F}.

Spin alignment is characterized through the polar angle $\theta_i$ between the $i$-th BH spin vector and the angular momentum. In \bpop, $\theta$ is assigned to each BH assuming either a uniform distribution in $\cos\theta$ or a cumulative distribution in the form \citep{2019MNRAS.482.2991A,2023MNRAS.520.5259A}
\begin{equation}
    P_{\theta} = \frac{1}{2}\left[\cos\theta + 1 \right]^{n_\theta+1},
    \label{ptheta}
\end{equation}
where the $n_\theta$ parameter determines the probability that the spin vectors are aligned within a certain discrepancy. 
In typical fiducial models, we adopt a flat distribution for dynamical BBHs, while for isolated binaries we adopt the $P_{\theta}$ distribution, assuming $n_\theta=8$, which implies a $20\%$($55\%$) probability to have $\theta_{1,2}$ differing by less than 5(20) per cent. The azimuthal angle $\Phi$ between the spin vectors, instead, is assumed to be isotropically distributed.

\subsection{Star cluster dynamics}
\label{appA3}
Following our previous papers \citep{2020ApJ...894..133A,2023MNRAS.520.5259A}, the initial mass of the cluster is assigned according to the mass distribution of observed GCs \citep{2010arXiv1012.3224H} and NCs \citep{2016MNRAS.457.2122G}, and further assuming that the YC mass distribution resembles that of GCs but is shifted to lower values by 2.5 dex. Similarly, the initial value of the half-mass radius is assigned according to observed clusters \cite[see e.g. Figure 4 in][]{2023MNRAS.520.5259A}. To take into account the fact that most observed clusters underwent slow expansion driven by internal (e.g. two-body relaxation) and external (e.g. tidal heating from the galactic field) processes, in the fiducial model we reduce the initial value of the half-mass radius by a factor extracted between 2 and 20 \citep[e.g.][]{2024MNRAS.528.5119A,2024MNRAS.531.3770R}. 

\bpop takes into account the dynamical evolution of star clusters by modeling the time variation of the bound mass, mass in the core, and half-mass and core radii. Given the initial cluster mass $m_{\rm cl,0}$, its time evolution is modeled as
\begin{equation}
    m_{\rm cl,bnd} = m_{\rm cl,0}\left(1+\frac{t}{t_{\rm st}}\right)^{k_1} \left(1-\frac{t}{f(m_{\rm cl,0})\,t_{\rm rlx}}\right),
\end{equation}
where $t_{\rm st} = 10\,\rm Myr$, $k_1 = -0.1$, and $t_{\rm rlx}$ is the relaxation time. The scaling parameter
\begin{equation}    
   f(m_{\rm cl,0}) = \alpha_{\rm bnd} \left(\frac{m_{\rm cl,0}}{10^6\,\Ms}\right)^{\beta_{\rm bnd}},
\end{equation}
with $\alpha_{\rm bnd}=100$ for YCs and $1000$ for GCs and NCs, and $\beta_{\rm bnd} = 0.6$, which determines how rapidly relaxation takes over in the cluster dispersal. The chosen parameters produce a mass and radius evolution in agreement with the \dragonii $N$-body simulations and consistent with the expected life cycle of clusters, as recently described in \cite{2023MNRAS.522.5340G}. 
Similarly, given an initial half-mass radius $r_{\rm hm,0}$, its evolution is modelled as
\begin{equation}
   r_{\rm hm} = r_{\rm hm,0} \left(1 + \frac{t}{t_0} \right)^\alpha_{\rm hm},
\end{equation}
where $t_0 = 5\,t_{\rm cc}\left[1+\left(r_{\rm hm,0}/1\,{\rm pc}\right)\right]$ and the slope is selected in the range $\alpha_{\rm hm} = 0.15-0.25$ at any timestep. This relation is tailored to represent the half-mass radius evolution observed in \dragonii simulations. 

Similarly, the time evolution of the core mass and radius is modeled as
\begin{equation}
    m_{\rm core} = \alpha_{\rm core}\,m_{\rm cl,0} \left(\frac{r_{\rm core}}{0.15}\right)^{\beta_{\rm core}},
\end{equation}
where $\alpha_{\rm core} = 0.03$, and $\beta_{\rm core} = 0.4$ if $t < 2\,t_{\rm cc}$ or $\beta_{\rm core}=1/1.3$ otherwise; and 
\begin{equation}
\frac{r_{\rm core}}{r_{\rm hm}} = 
\begin{cases}
\max\!\left(\tilde{r}_{\rm min},\,
0.2 \left(1-\frac{t}{2\,t_{\rm cc}}\right)^{0.53}\right)
& t \leq 2\,t_{\rm cc},\\[6pt]
\tilde{r}_{\rm min} + 0.2\left(\frac{t}{2\,t_{\rm cc}} - 1 \right)^{1/4}
& 2\,t_{\rm cc} < t \leq 5\, t_{\rm cc},\\[6pt]
\tilde{r}_{\rm min} - 0.8 + \left(\frac{t}{2\,t_{\rm cc}}-1\right)^{0.05}
& t > 5\,t_{\rm cc}.
\end{cases}
\end{equation}

\begin{figure}
    \centering
    \includegraphics[width=\columnwidth]{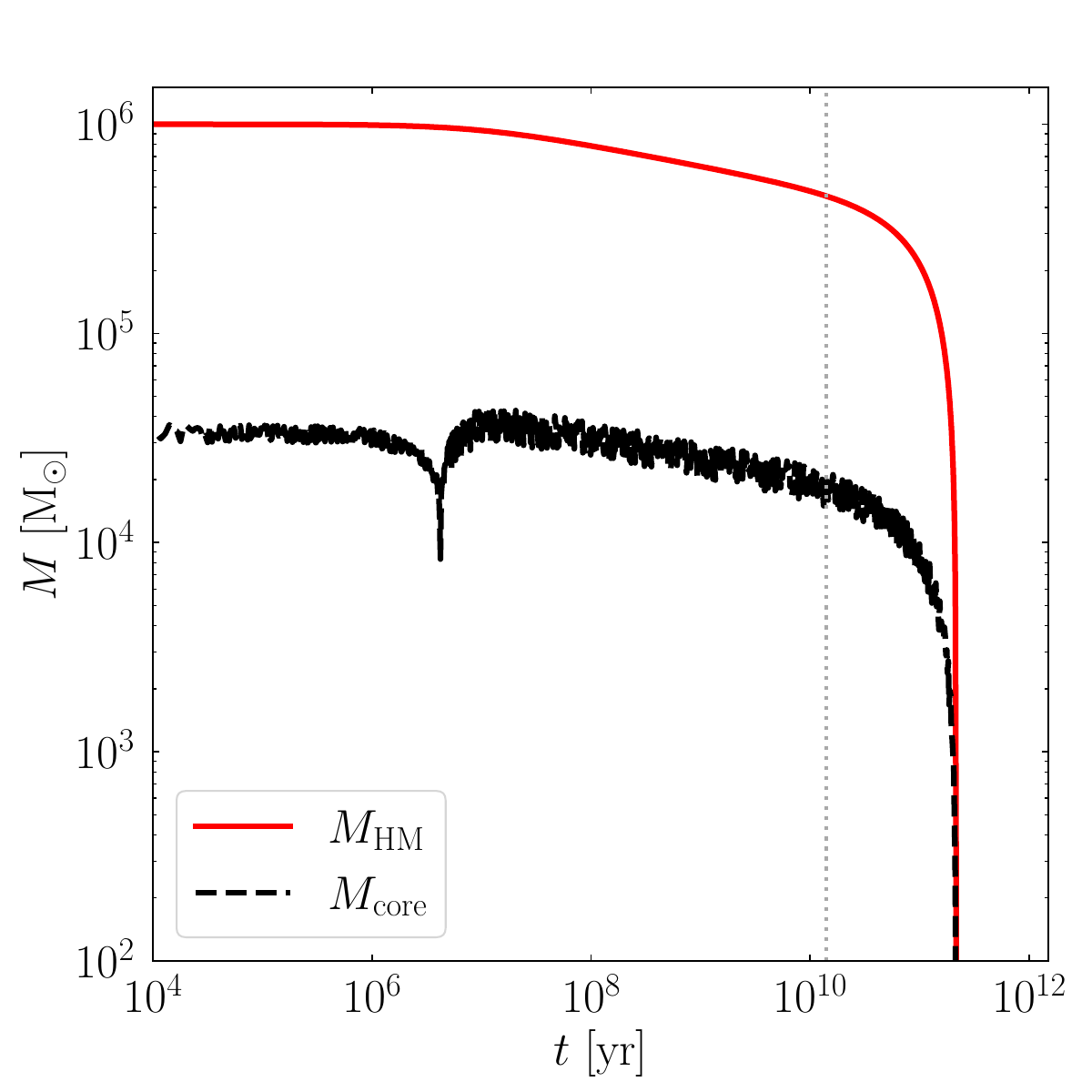}\\
    \includegraphics[width=\columnwidth]{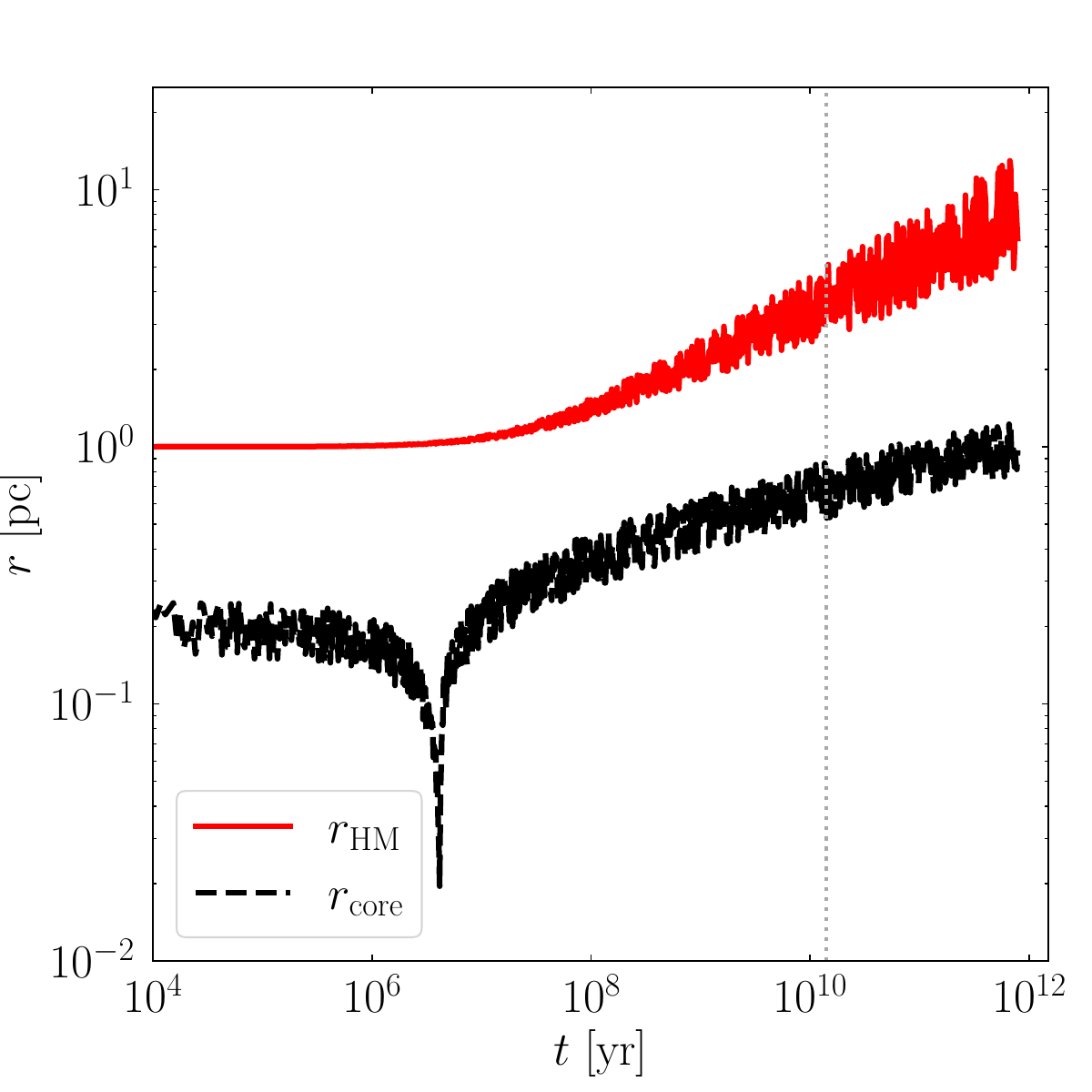}\\
    \caption{Time evolution of cluster properties in \bpop. Top panel: evolution of bound mass (red straight line) and core mass (black dashed line). Bottom panel: evolution of half-mass radius (red straight line) and core radius (black dashed line). The gray dotted line marks a Hubble time. The initial cluster is assumed to have an initial mass of $10^6\Ms$ and a half-mass radius of $1$ pc.}
    \label{fig:clevo}
\end{figure}

The maximum contraction $\tilde{r}_{\rm min}$ is determined by the formation of hard binaries that halt the core collapse and revert the process, driving the core rebound \citep{1975MNRAS.173..729H,1993ApJ...403..271G,2008gady.book.....B}. This generally happens when a handful of stars are in the core \citep{2008gady.book.....B}, but in our semi-analytical approach such number is ill-defined, as we cannot follow the cluster evolution on a star-by-star basis. We leverage on the concept of influence radius, which is the region of the cluster in which an object with mass $m_o$, or a collection of objects with total mass $m_o$, dominates dynamics \citep[e.g.][]{1972ApJ...178..371P,2013degn.book.....M}, assuming a velocity dispersion $\sigma^2$ 
\begin{equation}
    r_{\rm inf} = \frac{Gm_o}{\sigma^2}. 
\end{equation}
We also rely on the wandering radius, i.e. the region in which interactions with cluster members displace an object from its position \citep[e.g.][]{1976ApJ...209..214B,2011MNRAS.418.1308B}
\begin{equation}
    r_{\rm wan} = R_c\,\sqrt{\frac{\langle m_* \rangle}{m_o}},
\end{equation}
where $\langle m_*\rangle$ is the average stellar mass in the cluster.
The object will settle in the cluster center and dominate dynamics if $r_{\rm inf} / r_{\rm wan} < 1$, i.e. when 
\begin{equation}
    m_o \geq m_{o*} \equiv m_*^{1/3} M_c^{2/3}.
\end{equation}
Substituting $m_{o*}$ in the equation of the influence radius, and considering that $\sigma^2 \sim GM_c / R_c$, we obtain the minimum influence radius
\begin{equation}
    \frac{r_{\rm inf}}{R_c} = \left(\frac{m_*}{M_c}\right)^{1/3},
\end{equation}
and assume that this quantity represents a proxy to the minimum core radius, below which dynamical interactions halt and revert the collapse. The scheme above guarantees a smooth evolution of both global and central cluster quantities, as shown in Figure~\ref{fig:clevo} for two example clusters. 

\subsection{Black hole binary formation through dynamical interactions}
\label{appA4}

In \bpop, we assume that the BH reservoir is comparable to the number of BHs inside the cluster half-mass radius
\begin{equation}
    n_{\rm bhs} = f_{\rm bh} \,{} f_{\rm reten} \,{} f_{\rm encl}\,{} M_c,
\end{equation}
where $f_{\rm BH} = (8\pm0.8)\times 10^{-4}$ is the fraction of stars that evolve in a BH according to a Kroupa initial mass function \citep{2001MNRAS.322..231K}, $f_{\rm reten} = (0.5 \pm 0.3)$ is the retention fraction of BHs \citep{2015PhRvL.115e1101R, 2018MNRAS.479.4652A}, $f_{\rm encl}$ is the fraction of cluster mass enclosed within the cluster scale radius, to which we associate a $10\%$ uncertainty to mimic the combined effect of mass segregation and cluster contraction and expansion. 

Every time two BHs pair, $n_{\rm bhs}$ is decreased accordingly. In the case of repeated mergers, the merger chain is halted if the BBH or the merger remnant are ejected from the cluster, the new BBH has a merger time longer than the Hubble time, or the BH reservoir is emptied. 

Pairing and hardening processes in \bpop are treated through a series of characteristic timescales, which are updated accordingly to the evolution of cluster masses and radii. In the following, we briefly sketch how dynamics is treated in our code.

To calculate the delay time of dynamical mergers, for each BBH we firstly extract the cluster formation time, $t_{\rm for}$, from the adopted star formation rate. 

We then select the maximum lifetime of the progenitors, $t_{\rm birth}$, from the stellar evolution catalog. The two BHs drift to the cluster center via dynamical friction over a timescale \citep{1943ApJ....97..255C, 2008gady.book.....B}
\begin{equation}
t_{\rm df} = 0.42 \,{\rm Gyr}  \left( \frac{10\,m_*}{m_{\rm CO}} \right)  \left(\frac{r}{r_h}\right)^{1.76} \left(\frac{t_{\rm rel} }{4.2{\rm ~Gyr}}\right),
\end{equation}
which depends on the average stellar mass $m_*$, the mass of the drifting satellite $m_{\rm CO}$, its position inside the cluster $r$, and the relaxation time $t_{\rm rel}$ \citep{1987degc.book.....S, 2008gady.book.....B}:
\begin{equation}
t_{\rm rel} = 4.2 {\rm ~ Gyr}  \left(\frac{15}{\log \Lambda}\right) \left(\frac{r_h}{4\,{\rm pc}}\right)^{3/2} \sqrt{\frac{M_c}{10^7\,\Ms}},		
\end{equation}
where $\log\Lambda$ represents the Coulomb logarithm.
Although generally $t_{\rm df} > t_{\rm birth}$, it is also possible for BH progenitors to reach the cluster center before evolving into BHs. Therefore, we add to $t_{\rm for}$ the maximum between the stellar lifetime and the dynamical friction timescale. 

Once in the center, two BHs can pair up either via the interaction of three unbound bodies, which occurs over a time $t_{\rm 3bb}$ \citep{1993ApJ...418..147L} 
\begin{align}
t_{\rm 3bb} =\,& 4 {\rm ~Gyr} \left(\frac{10^6 \,{\rm \Ms~pc^{-3}}}{\rho_c}\right)^2 \left(\zeta^{-1}\frac{\sigma_c}{30{\rm ~km/s}}\right)^9 \times \nonumber \\
& \times \left(\frac{m_*}{m_{\rm CO}}\,10\right)^{9/2} \left(\frac{10}{m_{\rm CO}}\right)^{-5},
\label{t3bb}
\end{align}
or via interaction between a single object and a binary, characterized by a timescale 
$t_{\rm bs}$ \citep{2009ApJ...692..917M, 2016ApJ...831..187A}
\begin{align}
t_{\rm bs} =\,& 3{\rm ~Gyr} \left(\frac{0.01}{f_b}\right) \left(\frac{10^6{\rm~pc^{-3}}}{n_c}\right) \times \nonumber \\
& \times \left(\frac{\sigma_c}{30{\rm~km/s}}\right) \left(\frac{10\,\Ms}{a_h\,(M_1 + M_2 + m_p)}\right).
\end{align}

As shown in the equations above, the interaction rates and associated timescales crucially depend on the cluster number density ($n_c$) and matter density ($\rho_c$), and the cluster velocity dispersion ($\sigma_c$).

The binary-single interaction rate depends on the binary fraction $f_b$, the $\zeta\leq 1$ parameter, which is a proxy to the level of energy equipartition in the cluster, and the typical mass of the perturbers $m_p$. Interactions involve hard binaries, with a binding energy larger than the cluster average kinetic energy. This condition implies that hard binaries have a semi-major axis smaller than the threshold value
\begin{equation*}
a_h \simeq 59\,{\rm AU} \left(\frac{M_1+M_2}{30\,\Ms}\right)\left(\frac{30\,{\rm km/s}}{\sigma_c}\right)^2,
\end{equation*} 
usually referred to as hard-binary separation.
In \bpop, BBH semi-major axes are assigned either according to a Gaussian distribution centered around $a_h$ with a user defined dispersion, or a flat distribution between $0.1-0.3$ AU, following a setup similar to the one explored by \cite{2014ApJ...784...71S,2017ApJ...840L..14S}. The BBH eccentricity, instead, is drawn from a thermal distribution. 
Despite the limited choice available in the current version of the code, further recipes to initialize the BBH orbital parameters require minimal intervention on the code.

We take into account the uncertainties associated to the cluster mass and half-mass radius distributions, the cosmic star formation rate, and the small-scale physics regulating star cluster formation in galaxies by extracting each of the timescales listed above from a Gaussian distribution centered around the nominal value with a dispersion of $10\%$. 

The time needed for a hard binary to form is thus given by $t_{\rm hard} = t_{\rm for} + {\rm max}(t_{\rm birth}, t_{\rm df}) + {\rm min}(t_{\rm 3bb},t_{\rm bs}) + t_{\rm ex}$.

Hard binaries spend most of their lifetime involved in  binary-single interactions over a timescale 
\citep{2004ApJ...616..221G, 2016ApJ...831..187A}
\begin{align}
t_{1-2} =\,& \frac{0.02\,{\rm Gyr}}{\zeta}  \left(\frac{10^6 {\rm~pc^{-3}}}{n_c}\right) \left(\frac{\sigma_c}{30}\right) \times \nonumber \\
& \times \sqrt{\frac{10\,m_* }{ M_1+M_2}}  \left(\frac{0.05\,{\rm AU} }{ a_h}\right) \left(\frac{20}{M_1+M_2}\right).
\label{t12}
\end{align}

As the binary hardens, new interactions become rarer and more violent, slowing down the hardening rate and impinging a Newtonian recoil on the binary center of mass. If the recoil exceeds the cluster escape velocity $v_{\rm esc}$, the binary is ejected from the cluster. This can happen if the semi-major axis of the binary falls below a critical value $a_{\rm ej}$, defined as 
\begin{equation}
\label{eq:ejorin}
a_{\rm ej} = 0.07 \, {\rm AU} ~\frac{\mu\,{} m_p^2 }{ (M_1 + M_2 + m_p)(M_1+M_2)} \left(\frac{v_{\rm esc}}{50\,{\rm km/s}}\right)^{-2},
\end{equation}
where $\mu=M_1\,{}M_2/(M_1+M_2)$ is the reduced mass of the binary system.
Otherwise, the binary semi-major axis shrinks until it reaches a critical value below which GW emission becomes the dominant hardening mechanism, i.e.
\begin{align}
\label{eq:ejorin2}
a_{\rm gw} =\,& 0.05 {\rm AU} \left(\frac{M_1+M_2}{20\,\Ms}\right)^{3/5} \frac{ (M_2/M_1)^{1/5} }{ \left(1+M_2/M_1\right)^{2/5} }\times \nonumber \\
&\times \left(\frac{\sigma_c}{30\,{\rm km/s}}\right)^{1/5} 
\left(\frac{10^6\,{\rm \Ms \, pc^{-3}} }{\rho_c}\right)^{1/5},
\end{align}
The competing action of dynamical hardening, Newtonian recoil and GW hardening is determined by these two quantities. If $a_{\rm ej} > a_{\rm gw}$, the next interaction will likely eject the binary from the cluster. Otherwise, the binary will keep shrinking until it merges inside the cluster. 
The timescale for a hard binary to reach the critical semi-major axis $a_{\rm cri} = {\rm max}(a_{\rm ej}, a_{\rm gw})$ is \citep{2009A&A...494..539L, 2016ApJ...831..187A}
\begin{equation}
    t_{\rm cri} = 5\,t_{1-2} \frac{m_1 + m_2}{m_p}.
\end{equation}

If the binary is prone to ejection, $a_{\rm ej} > a_{\rm gw}$, in \bpop we recalculate the binary eccentricity and the GW-driven merger time, $t_{\rm gw}$, computed via the \cite{1964PhRv..136.1224P} formalism. If $t_{\rm gw} < t_{\rm cri}$ we label the binary as {\it in-cluster merger} and assume its merger time to be $t_{\rm mer} = t_{\rm gw}$, because the eccentricity change driven by interactions drives a prompt merger of the binary before it gets ejected from the cluster. Otherwise, we label the binary as {\it ejected} and its merger time is given by $t_{\rm mer} = t_{\rm cri}+t_{\rm gw}$.  
Finally, if $a_{\rm ej} \leq a_{\rm gw}$, we assume that the binary merges inside the cluster, assigning to the binary a merger time $t_{\rm mer} = t_{\rm cri}+t_{\rm gw}$.

\begin{figure}
    \centering
    \includegraphics[width=\columnwidth]{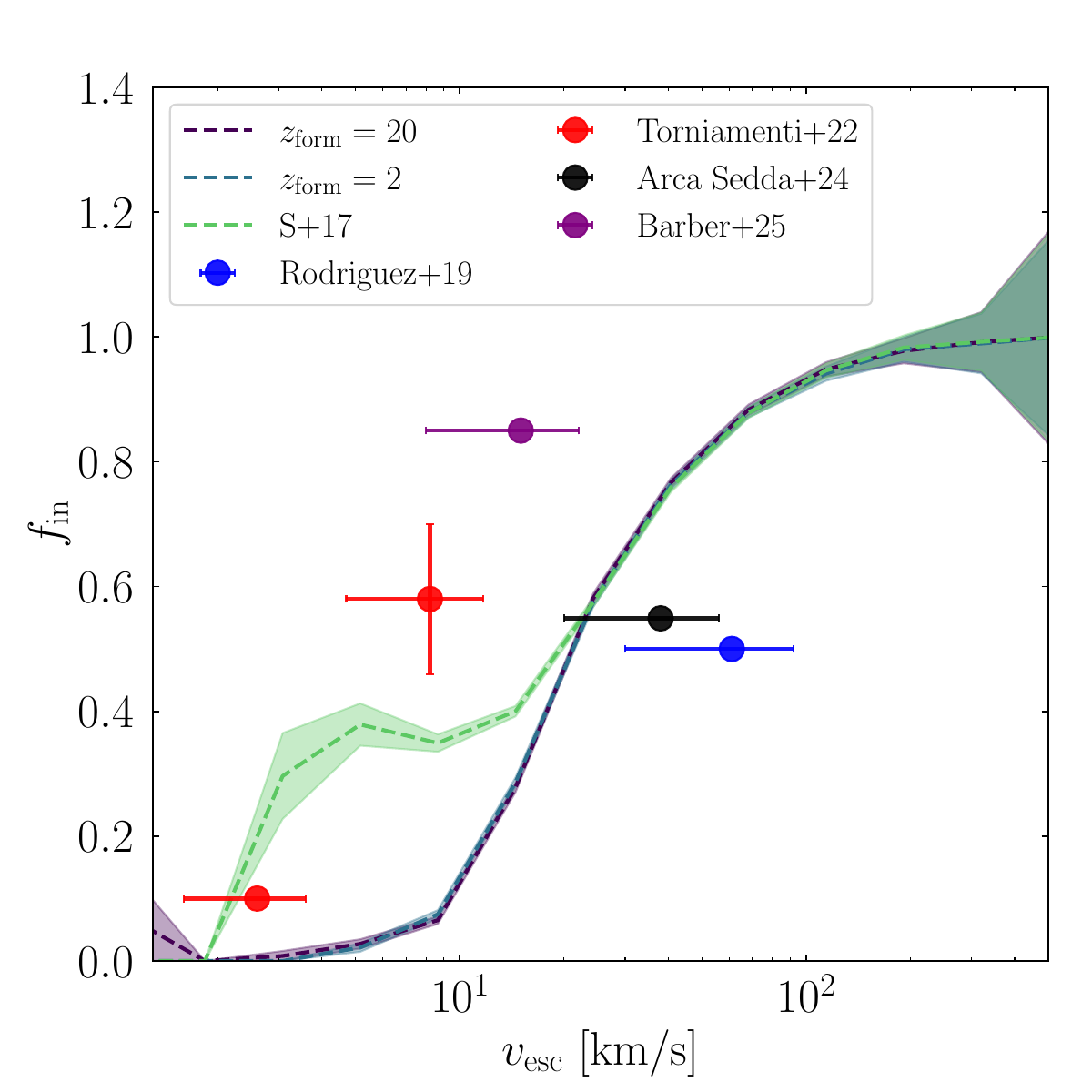}
    \caption{Fraction of mergers occurring inside the parent cluster as a function of the initial cluster escape velocity. \bpop models assume $f_{\rm mix} = 0.5$, a Gaussian distribution for the semi-major axis with dispersion $\sigma_a = 0.3$, and either a fixed formation time at $t_{\rm f} = 10$ Gyr (straight purple line), a fixed formation redshift at $z_{\rm f} = 4$ (dashed blue line), the MF17 star formation rate (dotted green line), even assuming an alternative, flat, distribution for BBH semi-major axis limited between $0.1-0.3$ AU (dot-dashed black line) inspired by \cite{2017ApJ...840L..14S}. Points represent $N$-body and Monte Carlo simulations by \cite{2019PhRvD.100d3027R}, \cite{2022MNRAS.517.2953T}, \cite{2024MNRAS.528.5140A}, and \cite{2025MNRAS.538..639B}. The error is calculated assuming that both the number of in-cluster merger and total number of mergers are characterized by Poissonian statistics.}
    \label{fig:incluster}
\end{figure}

Figure~\ref{fig:incluster} shows the fraction of mergers that occur inside the cluster, $f_{\rm in}$, for a population of GCs simulated with model AS20, which assumes a Gaussian distribution for semi-major axes with dispersion $\sigma_a = 0.3$ or a flat distribution in the range $0.1-0.3$ AU, and that all GCs have a metallicity $Z = 0.0002$ and formed either at a redshift $z\simeq 2$ or $z\sim 20$. We compare \bpop models to self-consistent $N$-body and Monte Carlo simulations representing low-mass young clusters \citep{2021MNRAS.500.3002B,2022MNRAS.517.2953T} and young massive and globular clusters \citep{2019PhRvD.100d3027R,2024MNRAS.528.5140A,2025MNRAS.538..639B}. The values of $f_{\rm in}$ calculated for $N$-body simulations take into account only BBH mergers triggered by dynamical processes. The clear difference among different $N$-body models likely depends on the masses and densities explored, the initial fraction of binaries, and the possible presence of an external tidal field. In the case of a Gaussian distribution for semi-major axes, \bpop models seem to balance the differences among $N$-body models, under-predicting the fraction of in-cluster mergers in low-mass or low-density clusters, with $v\sim$O$(10$ km$/$s$)$, and slightly over-predicting $f_{\rm in}$ for lighter or sparser clusters. The choice of  a flat semi-major axis in the range $0.1-0.3$ AU \cite[S+17,][]{2017ApJ...840L..14S} limits the BBH sampling to relatively tight systems that are mostly affected by dynamics but may represent a small population compared to the general population of BBHs in star clusters. We see that in this case $f_{\rm in}$ sensibly increases, but remains compatible with results from self-consistent simulations \citep{2022MNRAS.517.2953T,2025MNRAS.538..639B}.
Given the subtle assumptions on which both semi-analytic and $N$-body simulations rely on, it is not surprising that Figure~\ref{fig:incluster} does not show a perfect match among different results, but it serves to demonstrate that the simplistic approach used to model dynamics in \bpop leads to a reasonable ratio between in-cluster and ejected mergers.

According to the implemented procedure, the delay time of dynamical mergers is thus given by
$t_{\rm del} = t_{\rm hard} + t_{\rm mer}$.
If $t_{\rm del}$ is larger than a Hubble time ($t_{\rm H}$) we discard the BBH and proceed to the next one. 
If $t_{\rm del} < t_{\rm H}$, the BBH merges and the final mass, spin, and GW recoil kick are calculated. If the kick is larger than the cluster escape velocity \bpop halts the integration and proceeds to the next BBH. If $v_{\rm gw} < v_{\rm esc}$, the remnant is displaced from the center to a maximum distance $r_{\rm d} = r_h \sqrt{v_{\rm esc}^4 / (v_{\rm esc}^2-v_{\rm gw}^2)^2 - 1}$ \citep[see e.g.][]{2019MNRAS.486.5008A,2019MNRAS.488...47F}, and it is assumed to drift back to the cluster center over a dynamical friction timescale. 

At this point, we assign a companion to the remnant and recalculate all the dynamical timescales described above to check if the new BBH merges within a Hubble time. The procedure is iterated until either all BHs in the cluster are consumed in the same merger chain, or the BBH or its remnant are ejected, owing either to Newtonian ($a_{\rm ej} > a_{\rm gw}$) or relativistic kicks ($v_{\rm gw}>v_{\rm esc}$).

All equations above critically depend on the cluster structural properties, which change over time due to cluster relaxation and external processes. In \bpop, we model the cluster evolution according to the set of Equations described in the previous section, and adopt a recursive method to update the cluster mass, radius, density, and velocity dispersion at the end of every phase determining the dynamical evolution of BBHs. This inevitably affects the BBH delay time, possibly leading to the termination of a merger chain according to the criteria described in the previous paragraph. Additionally, the merger chain is halted if the cluster mass falls below $10\,{}\Ms$.

\subsection{Star Formation Cosmic History for different environments}
\label{appA5}

To determine the cosmic evolution of BBHs, we need to make assumptions on the SFR of the environments in which BBHs form. 

In the case of IBs, we adopt the broadly used SFR derived by \cite[][hereafter referred to as MF17 model]{2017ApJ...840...39M} 
\begin{equation}
\psi_{\rm MF17}(z) = \frac{0.01 \,(1+z)^{2.6} }{ 1 + \left[(1+z)/{3.2}\right]^{6.2}} ~\Ms\, {\rm yr}^{-1}\, {\rm Mpc}^{-3}.
\label{eq:sfr}
\end{equation} 
This relation is also adopted to model the cosmic evolution of BBHs forming in YCs, following the observational evidence that YCs are generally found in galactic discs and have metallicities that are compatible with the local metallicity gradient of the host galaxy \citep[e.g.][]{2016A&A...585A.150N}. 

Modeling the SFR of GCs and NCs is more complex, owing to the uncertain evolutionary pathways of these stellar systems. For GCs, we rely on the recent results of \cite[][hereafter referred to as EB19 model]{2019MNRAS.482.4528E}, according to which the formation of Galactic GCs can be compatible with a redshift-dependent Gaussian distribution \citep[see also][]{2022MNRAS.511.5797M}, i.e.:
\begin{equation}
    \psi_{\rm EB19}(z) = B_{\rm GC} \exp[-(z - z_{\rm GC})^2/(2\,\sigma_{\rm GC})^2].
    \label{eq:EB19}
\end{equation}

The scaling values $B_{\rm GC}, ~z_{\rm GC}, ~\sigma_{\rm GC}$ are rather uncertain \citep[see cfr.][]{2019MNRAS.482.4528E,2022MNRAS.511.5797M}, owing to the uncertainties in GC formation physics. Here, we either assume $B_{\rm GC} = 1.2\times10^{-4}\,\Ms~$yr$^{-1}$ Mpc$^{-3}$, $z_{\rm GC}=4.5$, $\sigma_{\rm GC}=2$ \citep{2019MNRAS.482.4528E}, or $B_{\rm GC} = 2\times10^{-4}\,\Ms~$yr$^{-1}$ Mpc$^{-3}$, $z_{\rm GC}=3.2$, $\sigma_{\rm GC}=1.5$ \citep{2022MNRAS.511.5797M}.

For NCs, instead, we assume either that they form in-situ, i.e. they grow in the center of the galaxy during its growth \citep[e.g.][]{2004ApJ...605L..13M,2007MNRAS.379...21N}, or through the collision and merger of massive clusters, the so-called dry-merger scenario \citep{1975ApJ...196..407T,1993ApJ...415..616C,2012ApJ...750..111A,2014MNRAS.444.3738A}. 
In the case of in-situ formation, we assume the same SFR in Equation~\ref{eq:sfr}, rescaled by a factor $f_{\rm NC} = 0.0005$. Note that this is a typical value for the observed NC-galaxy mass fraction in Milky Way-sized galaxies \citep{2020A&ARv..28....4N}, but it could be considerably larger in dwarf galaxies, up to $f_{\rm NC} = 0.01$. 
In the case of dry-merger formation, we assume a Gaussian SFR as in Equation~\ref{eq:EB19} \citep[see also][]{2022MNRAS.511.5797M}, assuming in this case a scaling factor $B_{\rm NC} = 5\times10^{-5}\,\Ms~$yr$^{-1}$ Mpc$^{-3}$, and $z_{\rm NC}=z_{\rm GC}$, $\sigma_{\rm NC}=\sigma_{\rm GC}$.

To distribute BBHs across different channels and calculate the corresponding merger rate density, we assume that the CSFR is well described by Equation~\ref{eq:sfr}, i.e. $\psi_{\rm tot} = \psi_{\rm MF17}$ \cite{2017ApJ...840...39M}, and that it is given by the sum of the SFR for single stars and IBs, and the SFR of different star cluster types, i.e. 
$\psi_{\rm tot} = \psi_{\rm s} + \psi_{\rm IB} + \psi_{\rm dyn}$, with  $\psi_{\rm dyn} = \psi_{\rm YC}+\psi_{\rm GC}+\psi_{\rm NC}$.

To ensure that the total CSFR is preserved, we weight the SFR of different channels as follows. We first consider the SFR for YCs, $\psi_{\rm YC}$, as
\begin{equation}
    \psi_{\rm YC} = {\rm max}\left[0,{\rm min}\left(f_{\rm YC}, 1-\frac{\psi_{\rm GC}+\psi_{\rm NC}}{\psi_{\rm tot}}\right)\right]\psi_{\rm tot},
\end{equation}
where $f_{\rm YC} = 0.01$ represents the fraction of star formation that goes in bound YCs \citep[e.g.][]{2009A&A...494..539L}.

Similarly, the SFR of IBs is modeled as
\begin{equation}
    \psi_{\rm IB} = {\rm max}\left[0,{\rm min}\left(f_{\rm IB}, 1-\frac{\psi_{\rm dyn}}{\psi_{\rm tot}}\right)\right]\psi_{\rm tot},
\end{equation}
where $f_{\rm IB}$ roughly represents the fraction of field stars born in binaries.

In the extreme case that all field stars form in binaries, $f_{\rm IB} = 1$, the difference $\psi_{\rm tot} - (\psi_{\rm IB}+\psi_{\rm YC}+\psi_{\rm GC}+\psi_{\rm NC})=0$. If, instead, $f_{\rm IB} < 1$, this difference returns roughly the fraction of single stars in the field, owing to the little impact of star cluster formation on the total CSFR.

Given the adopted relation between the CSFR and the characteristic rate of different channels, the parameters $f_{\rm IB}$ and $f_{\rm YC}$ can be used to re-weight the mutual contribution of different channels to the CSFR, and consequently to the BBH merger rate, without the need to run additional simulations.

\subsection{Metallicity distribution}
\label{appA6}
In \bpop, the metal enrichment in galaxies as a function of the redshift is modeled via the functional form \citep{2017ApJ...840...39M,2020AeA...635A..97B}
\begin{equation}
     {\rm Log} \left\langle Z/{\rm Z}_\odot \right\rangle = \alpha + \beta z^\gamma,
\end{equation}
with coefficients based on previous works by \cite{2012ApJ...755...89R,2017ApJ...840...39M,2018MNRAS.480.2011G,2018ApJ...860..100D,2020ApJ...891..141G,2020AeA...635A..97B}. Note that the choice $\alpha = 0.153$, $\beta = -0.074$, and $\gamma = 1.34$ corresponds to the \cite{2017ApJ...840...39M} model. 
The dispersion around the mean value is modeled through a log-normal distribution \citep[see][]{2020AeA...635A..97B,2022MNRAS.511.5797M}:
\begin{align}
    p(z, Z) =& \frac{1}{\sqrt{2\pi \sigma_Z^2}} \times \nonumber \\
    &\times \exp \left\{ -\frac{\left[{\rm Log} (Z/{\rm Z}_\odot) - \langle {\rm Log} (Z/{\rm Z}_\odot
    )\rangle\right]^2}{2\,\sigma_Z^2} \right\},
    \label{eq:logZ}
\end{align}
where 
\begin{equation}
    \displaystyle \left\langle {\rm Log} \left( \frac{Z}{{\rm Z}_\odot} \right) \right\rangle = {\rm Log}\left\langle\frac{Z}{{\rm Z}_\odot}\right\rangle + A(Z),
\end{equation}
and either $A(Z) = 0$ following \cite{2019ApJ...886...25B} or:
\begin{equation}
     A(Z) = \frac{\ln(10)\,\sigma_Z^2}{2},
\end{equation}
following \cite{2022MNRAS.511.5797M}.

Additionally, in the case of GCs, \bpop implements the metallicity-redshift relation described in \cite{2019MNRAS.482.4528E} and \cite{2016MNRAS.456.2140M}, according to which the metallicity depends on the redshift and the available stellar mass as 
\begin{equation}
    {\rm Log}\left(\frac{Z}{{\rm Z}_\odot}\right) = \alpha\left[{\rm Log}\left(\frac{M}{\Ms}\right) - 10\right] + \beta \exp\left(-\gamma z\right) - \delta.
\end{equation}
The values of parameters $(\alpha, \beta, \gamma, \delta)$ depend on the considered galaxy components, being equal to $(0.35,~ 0.93,~ 0.43,~ 1.05)$ for the metallicity of the galaxy gaseous component, or $(0.4,~ 0.67,~ 0.5,~ 1.04)$ for the stellar component. Here we have adopted the parameters characterizing the stellar component. 
 
The selection of the BBH component masses relies on a limited number of catalogs, each linked to a specific metallicity value. In \bpop, the metallicity obtained from the chosen distribution is matched to the nearest available value in the catalogs for each BBH. If the selected metallicity falls outside the range available in the catalogs, \bpop assigns to the source the corresponding limiting value. In current runs, our metallicity ranges from $Z = 0.0001$ to $0.03$.

\section{\label{app:bigbh} The dearth of massive IMBH-BH mergers from galactic nuclei}

\begin{figure}
    \centering
    \includegraphics[width=\columnwidth]{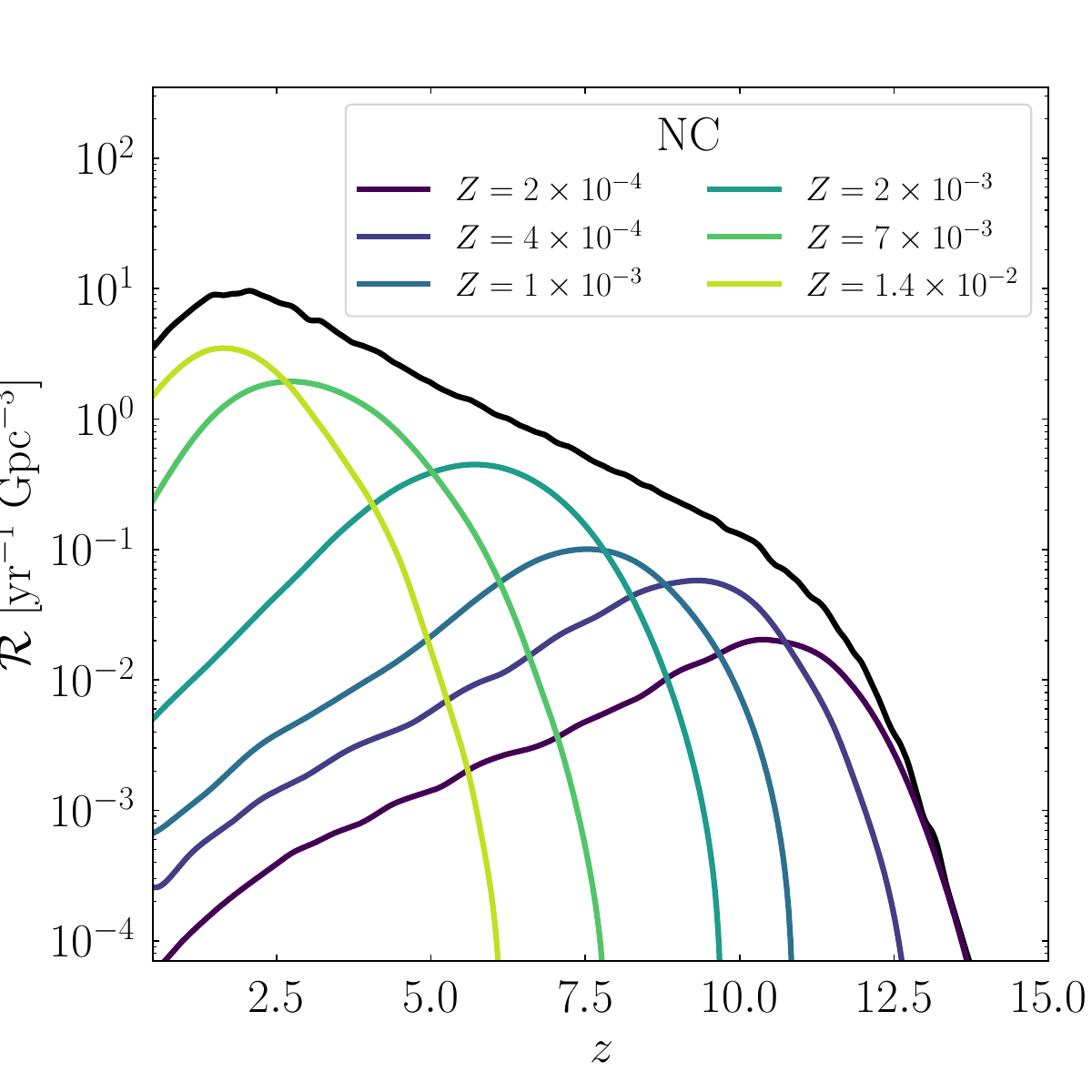}
    \caption{Merger rate density profile for NCs in the fiducial model for the overall population (black line), and divided by metallicity (colored lines), as a function of the redshift. For the sake of visibility, we only show 6 values of the metallicity out of the total 15 adopted values.}
    \label{fig:mrdNC}
\end{figure}

The apparent, and counterintuitive, lack of massive IMBHs from NCs in Figure~\ref{fig:3} owes to the procedures implemented in our code to model IMBH seeding, calculate the MRD and extract the catalogs. First of all, we find that IMBHs from NCs with a terminal mass $>10^5\,\Ms$ form only in environments with a metallicity $Z < 10^{-3}$, primarily because we assume no IMBH seeds can form from the collapse of a VMS if $Z$ exceeds this threshold. Second, the fraction of mergers from NCs in the fiducial model is expected to be $f_{\rm gw,NC} \sim 0.038$. Third, hierarchical mergers in these particular environments proceed in a catastrophic way for the entire lifetime of the cluster, such that in almost all models the last merger occurs at $z<0.05$. Fourth, as shown in Figure~\ref{fig:mrdNC}, the contribution to the local MRD from metal-poor NCs is $10^{-5}$ times smaller compared to that of more metal-rich NCs. The combination of all the aforementioned features makes it quite unlikely for BHs heavier than $10^5\,\Ms$ forming in NC to be included in our mock catalogs. 

\acknowledgments{The authors thank Maya Fishbach, Mario Spera, Ulyana Dupletsa, Matthew Mould, Jan Harms, and the anonymous referee for insightful comments and suggestions. MAS acknowledges funding from the European Union’s Horizon 2020 research and innovation programme under the Marie Skłodowska-Curie grant agreement No.~101025436 (project GRACE-BH) and from the MERAC Foundation through the 2023 MERAC prize. MAS and MB acknowledge the ACME project which has received funding from the European Union's Horizon Europe Research and Innovation programme under Grant Agreement No.~101131928. This material is based upon work supported by NSF's LIGO Laboratory which is a major facility fully funded by the National Science Foundation. The authors are grateful for computational resources provided by the LIGO Lab (CIT, LHO, LLO) and supported by the National Science Foundation Grants PHY-0757058 and PHY-0823459. 
}


\bibliographystyle{apsrev4-2}
\makeatletter
\def\bib@maxauthors{3} 
\def\bib@minauthors{3} 
\makeatother

\bibliography{example}

 \newcommand{\noop}[1]{}
\begin{thebibliography}{195}%
\makeatletter
\providecommand \@ifxundefined [1]{%
 \@ifx{#1\undefined}
}%
\providecommand \@ifnum [1]{%
 \ifnum #1\expandafter \@firstoftwo
 \else \expandafter \@secondoftwo
 \fi
}%
\providecommand \@ifx [1]{%
 \ifx #1\expandafter \@firstoftwo
 \else \expandafter \@secondoftwo
 \fi
}%
\providecommand \natexlab [1]{#1}%
\providecommand \enquote  [1]{``#1''}%
\providecommand \bibnamefont  [1]{#1}%
\providecommand \bibfnamefont [1]{#1}%
\providecommand \citenamefont [1]{#1}%
\providecommand \href@noop [0]{\@secondoftwo}%
\providecommand \href [0]{\begingroup \@sanitize@url \@href}%
\providecommand \@href[1]{\@@startlink{#1}\@@href}%
\providecommand \@@href[1]{\endgroup#1\@@endlink}%
\providecommand \@sanitize@url [0]{\catcode `\\12\catcode `\$12\catcode
  `\&12\catcode `\#12\catcode `\^12\catcode `\_12\catcode `\%12\relax}%
\providecommand \@@startlink[1]{}%
\providecommand \@@endlink[0]{}%
\providecommand \url  [0]{\begingroup\@sanitize@url \@url }%
\providecommand \@url [1]{\endgroup\@href {#1}{\urlprefix }}%
\providecommand \urlprefix  [0]{URL }%
\providecommand \Eprint [0]{\href }%
\providecommand \doibase [0]{https://doi.org/}%
\providecommand \selectlanguage [0]{\@gobble}%
\providecommand \bibinfo  [0]{\@secondoftwo}%
\providecommand \bibfield  [0]{\@secondoftwo}%
\providecommand \translation [1]{[#1]}%
\providecommand \BibitemOpen [0]{}%
\providecommand \bibitemStop [0]{}%
\providecommand \bibitemNoStop [0]{.\EOS\space}%
\providecommand \EOS [0]{\spacefactor3000\relax}%
\providecommand \BibitemShut  [1]{\csname bibitem#1\endcsname}%
\let\auto@bib@innerbib\@empty
\bibitem [{\citenamefont {{Harry}}\ and\ \citenamefont {{LIGO Scientific
  Collaboration}}(2010)}]{2010CQGra..27h4006H}%
  \BibitemOpen
  \bibfield  {author} {\bibinfo {author} {\bibfnamefont {G.~M.}\ \bibnamefont
  {{Harry}}}\ and\ \bibinfo {author} {\bibnamefont {{LIGO Scientific
  Collaboration}}},\ }\href {https://doi.org/10.1088/0264-9381/27/8/084006}
  {\bibfield  {journal} {\bibinfo  {journal} {Classical and Quantum Gravity}\
  }\textbf {\bibinfo {volume} {27}},\ \bibinfo {eid} {084006} (\bibinfo {year}
  {2010})}\BibitemShut {NoStop}%
\bibitem [{\citenamefont {{Aso}}\ \emph {et~al.}(2013)\citenamefont {{Aso}},
  \citenamefont {{Michimura}}, \citenamefont {{Somiya}}, \citenamefont
  {{Ando}}, \citenamefont {{Miyakawa}}, \citenamefont {{Sekiguchi}},
  \citenamefont {{Tatsumi}},\ and\ \citenamefont
  {{Yamamoto}}}]{2013PhRvD..88d3007A}%
  \BibitemOpen
  \bibfield  {author} {\bibinfo {author} {\bibfnamefont {Y.}~\bibnamefont
  {{Aso}}}, \bibinfo {author} {\bibfnamefont {Y.}~\bibnamefont {{Michimura}}},
  \bibinfo {author} {\bibfnamefont {K.}~\bibnamefont {{Somiya}}}, \bibinfo
  {author} {\bibfnamefont {M.}~\bibnamefont {{Ando}}}, \bibinfo {author}
  {\bibfnamefont {O.}~\bibnamefont {{Miyakawa}}}, \bibinfo {author}
  {\bibfnamefont {T.}~\bibnamefont {{Sekiguchi}}}, \bibinfo {author}
  {\bibfnamefont {D.}~\bibnamefont {{Tatsumi}}},\ and\ \bibinfo {author}
  {\bibfnamefont {H.}~\bibnamefont {{Yamamoto}}},\ }\href
  {https://doi.org/10.1103/PhysRevD.88.043007} {\bibfield  {journal} {\bibinfo
  {journal} {\prd}\ }\textbf {\bibinfo {volume} {88}},\ \bibinfo {eid} {043007}
  (\bibinfo {year} {2013})},\ \Eprint {https://arxiv.org/abs/1306.6747}
  {arXiv:1306.6747 [gr-qc]} \BibitemShut {NoStop}%
\bibitem [{\citenamefont {{LIGO Scientific Collaboration}}\ \emph
  {et~al.}(2015)\citenamefont {{LIGO Scientific Collaboration}}, \citenamefont
  {{Aasi}}, \citenamefont {{Abbott}}, \citenamefont {{Abbott}}, \citenamefont
  {{Abbott}}, \citenamefont {{Abernathy}}, \citenamefont {{Ackley}},
  \citenamefont {{Adams}}, \citenamefont {{Adams}}, \citenamefont {{Addesso}}
  \emph {et~al.}}]{2015CQGra..32g4001L}%
  \BibitemOpen
  \bibfield  {author} {\bibinfo {author} {\bibnamefont {{LIGO Scientific
  Collaboration}}}, \bibinfo {author} {\bibfnamefont {J.}~\bibnamefont
  {{Aasi}}}, \bibinfo {author} {\bibfnamefont {B.~P.}\ \bibnamefont
  {{Abbott}}}, \bibinfo {author} {\bibfnamefont {R.}~\bibnamefont {{Abbott}}},
  \bibinfo {author} {\bibfnamefont {T.}~\bibnamefont {{Abbott}}}, \bibinfo
  {author} {\bibfnamefont {M.~R.}\ \bibnamefont {{Abernathy}}}, \bibinfo
  {author} {\bibfnamefont {K.}~\bibnamefont {{Ackley}}}, \bibinfo {author}
  {\bibfnamefont {C.}~\bibnamefont {{Adams}}}, \bibinfo {author} {\bibfnamefont
  {T.}~\bibnamefont {{Adams}}}, \bibinfo {author} {\bibfnamefont
  {P.}~\bibnamefont {{Addesso}}}, \emph {et~al.},\ }\href
  {https://doi.org/10.1088/0264-9381/32/7/074001} {\bibfield  {journal}
  {\bibinfo  {journal} {Classical and Quantum Gravity}\ }\textbf {\bibinfo
  {volume} {32}},\ \bibinfo {eid} {074001} (\bibinfo {year} {2015})},\ \Eprint
  {https://arxiv.org/abs/1411.4547} {arXiv:1411.4547 [gr-qc]} \BibitemShut
  {NoStop}%
\bibitem [{\citenamefont {{Acernese}}\ \emph {et~al.}(2015)\citenamefont
  {{Acernese}}, \citenamefont {{Agathos}}, \citenamefont {{Agatsuma}},
  \citenamefont {{Aisa}}, \citenamefont {{Allemandou}}, \citenamefont
  {{Allocca}}, \citenamefont {{Amarni}}, \citenamefont {{Astone}},
  \citenamefont {{Balestri}}, \citenamefont {{Ballardin}}, \citenamefont
  {{Barone}}, \citenamefont {{Baronick}} \emph {et~al.}}]{2015CQGra..32b4001A}%
  \BibitemOpen
  \bibfield  {author} {\bibinfo {author} {\bibfnamefont {F.}~\bibnamefont
  {{Acernese}}}, \bibinfo {author} {\bibfnamefont {M.}~\bibnamefont
  {{Agathos}}}, \bibinfo {author} {\bibfnamefont {K.}~\bibnamefont
  {{Agatsuma}}}, \bibinfo {author} {\bibfnamefont {D.}~\bibnamefont {{Aisa}}},
  \bibinfo {author} {\bibfnamefont {N.}~\bibnamefont {{Allemandou}}}, \bibinfo
  {author} {\bibfnamefont {A.}~\bibnamefont {{Allocca}}}, \bibinfo {author}
  {\bibfnamefont {J.}~\bibnamefont {{Amarni}}}, \bibinfo {author}
  {\bibfnamefont {P.}~\bibnamefont {{Astone}}}, \bibinfo {author}
  {\bibfnamefont {G.}~\bibnamefont {{Balestri}}}, \bibinfo {author}
  {\bibfnamefont {G.}~\bibnamefont {{Ballardin}}}, \bibinfo {author}
  {\bibfnamefont {F.}~\bibnamefont {{Barone}}}, \bibinfo {author}
  {\bibfnamefont {J.-P.}\ \bibnamefont {{Baronick}}}, \emph {et~al.},\ }\href
  {https://doi.org/10.1088/0264-9381/32/2/024001} {\bibfield  {journal}
  {\bibinfo  {journal} {Classical and Quantum Gravity}\ }\textbf {\bibinfo
  {volume} {32}},\ \bibinfo {eid} {024001} (\bibinfo {year} {2015})},\ \Eprint
  {https://arxiv.org/abs/1408.3978} {arXiv:1408.3978 [gr-qc]} \BibitemShut
  {NoStop}%
\bibitem [{\citenamefont {{Abbott}}\ \emph {et~al.}(2018)\citenamefont
  {{Abbott}}, \citenamefont {{Abbott}}, \citenamefont {{Abbott}}, \citenamefont
  {{Abernathy}}, \citenamefont {{Acernese}}, \citenamefont {{Ackley}},
  \citenamefont {{Adams}}, \citenamefont {{Adams}}, \citenamefont {{Addesso}},
  \citenamefont {{Adhikari}}, \citenamefont {{Adya}}, \citenamefont
  {{Affeldt}}, \citenamefont {{Agathos}}, \citenamefont {{Agatsuma}},
  \citenamefont {{Kagra Collaboration}},\ and\ \citenamefont {{VIRGO
  Collaboration}}}]{2018LRR....21....3A}%
  \BibitemOpen
  \bibfield  {author} {\bibinfo {author} {\bibfnamefont {B.~P.}\ \bibnamefont
  {{Abbott}}}, \bibinfo {author} {\bibfnamefont {R.}~\bibnamefont {{Abbott}}},
  \bibinfo {author} {\bibfnamefont {T.~D.}\ \bibnamefont {{Abbott}}}, \bibinfo
  {author} {\bibfnamefont {M.~R.}\ \bibnamefont {{Abernathy}}}, \bibinfo
  {author} {\bibfnamefont {F.}~\bibnamefont {{Acernese}}}, \bibinfo {author}
  {\bibfnamefont {K.}~\bibnamefont {{Ackley}}}, \bibinfo {author}
  {\bibfnamefont {C.}~\bibnamefont {{Adams}}}, \bibinfo {author} {\bibfnamefont
  {T.}~\bibnamefont {{Adams}}}, \bibinfo {author} {\bibfnamefont
  {P.}~\bibnamefont {{Addesso}}}, \bibinfo {author} {\bibfnamefont {R.~X.}\
  \bibnamefont {{Adhikari}}}, \bibinfo {author} {\bibfnamefont {V.~B.}\
  \bibnamefont {{Adya}}}, \bibinfo {author} {\bibfnamefont {C.}~\bibnamefont
  {{Affeldt}}}, \bibinfo {author} {\bibfnamefont {M.}~\bibnamefont
  {{Agathos}}}, \bibinfo {author} {\bibfnamefont {K.}~\bibnamefont
  {{Agatsuma}}}, \bibinfo {author} {\bibfnamefont {L.~S.~C.}\ \bibnamefont
  {{Kagra Collaboration}}},\ and\ \bibinfo {author} {\bibnamefont {{VIRGO
  Collaboration}}},\ }\href {https://doi.org/10.1007/s41114-018-0012-9}
  {\bibfield  {journal} {\bibinfo  {journal} {Living Reviews in Relativity}\
  }\textbf {\bibinfo {volume} {21}},\ \bibinfo {eid} {3} (\bibinfo {year}
  {2018})},\ \Eprint {https://arxiv.org/abs/1304.0670} {arXiv:1304.0670
  [gr-qc]} \BibitemShut {NoStop}%
\bibitem [{\citenamefont {{Abbott}}\ \emph
  {et~al.}(2020{\natexlab{a}})\citenamefont {{Abbott}}, \citenamefont
  {{Abbott}}, \citenamefont {{Abbott}}, \citenamefont {{Abraham}},
  \citenamefont {{Acernese}} \emph {et~al.}}]{2020LRR....23....3A}%
  \BibitemOpen
  \bibfield  {author} {\bibinfo {author} {\bibfnamefont {B.~P.}\ \bibnamefont
  {{Abbott}}}, \bibinfo {author} {\bibfnamefont {R.}~\bibnamefont {{Abbott}}},
  \bibinfo {author} {\bibfnamefont {T.~D.}\ \bibnamefont {{Abbott}}}, \bibinfo
  {author} {\bibfnamefont {S.}~\bibnamefont {{Abraham}}}, \bibinfo {author}
  {\bibfnamefont {F.}~\bibnamefont {{Acernese}}}, \emph {et~al.},\ }\href
  {https://doi.org/10.1007/s41114-020-00026-9} {\bibfield  {journal} {\bibinfo
  {journal} {Living Reviews in Relativity}\ }\textbf {\bibinfo {volume} {23}},\
  \bibinfo {eid} {3} (\bibinfo {year} {2020}{\natexlab{a}})}\BibitemShut
  {NoStop}%
\bibitem [{\citenamefont {{Akutsu}}\ \emph {et~al.}(2021)\citenamefont
  {{Akutsu}}, \citenamefont {{Ando}}, \citenamefont {{Arai}}, \citenamefont
  {{Arai}}, \citenamefont {{Araki}}, \citenamefont {{Araya}}, \citenamefont
  {{Aritomi}}, \citenamefont {{Aso}}, \citenamefont {{Bae}} \emph
  {et~al.}}]{2021PTEP.2021eA101A}%
  \BibitemOpen
  \bibfield  {author} {\bibinfo {author} {\bibfnamefont {T.}~\bibnamefont
  {{Akutsu}}}, \bibinfo {author} {\bibfnamefont {M.}~\bibnamefont {{Ando}}},
  \bibinfo {author} {\bibfnamefont {K.}~\bibnamefont {{Arai}}}, \bibinfo
  {author} {\bibfnamefont {Y.}~\bibnamefont {{Arai}}}, \bibinfo {author}
  {\bibfnamefont {S.}~\bibnamefont {{Araki}}}, \bibinfo {author} {\bibfnamefont
  {A.}~\bibnamefont {{Araya}}}, \bibinfo {author} {\bibfnamefont
  {N.}~\bibnamefont {{Aritomi}}}, \bibinfo {author} {\bibfnamefont
  {Y.}~\bibnamefont {{Aso}}}, \bibinfo {author} {\bibfnamefont
  {S.}~\bibnamefont {{Bae}}}, \emph {et~al.},\ }\href
  {https://doi.org/10.1093/ptep/ptaa125} {\bibfield  {journal} {\bibinfo
  {journal} {Progress of Theoretical and Experimental Physics}\ }\textbf
  {\bibinfo {volume} {2021}},\ \bibinfo {eid} {05A101} (\bibinfo {year}
  {2021})},\ \Eprint {https://arxiv.org/abs/2005.05574} {arXiv:2005.05574
  [physics.ins-det]} \BibitemShut {NoStop}%
\bibitem [{\citenamefont {{The LIGO Scientific Collaboration}}\ \emph
  {et~al.}(2025{\natexlab{a}})\citenamefont {{The LIGO Scientific
  Collaboration}}, \citenamefont {{the Virgo Collaboration}}, \citenamefont
  {{the KAGRA Collaboration}}, \citenamefont {{Abac}}, \citenamefont
  {{Abouelfettouh}}, \citenamefont {{Acernese}}, \citenamefont {{Ackley}},
  \citenamefont {{Adamcewicz}}, \citenamefont {{Adhicary}} \emph
  {et~al.}}]{2025arXiv250818082T}%
  \BibitemOpen
  \bibfield  {author} {\bibinfo {author} {\bibnamefont {{The LIGO Scientific
  Collaboration}}}, \bibinfo {author} {\bibnamefont {{the Virgo
  Collaboration}}}, \bibinfo {author} {\bibnamefont {{the KAGRA
  Collaboration}}}, \bibinfo {author} {\bibfnamefont {A.~G.}\ \bibnamefont
  {{Abac}}}, \bibinfo {author} {\bibfnamefont {I.}~\bibnamefont
  {{Abouelfettouh}}}, \bibinfo {author} {\bibfnamefont {F.}~\bibnamefont
  {{Acernese}}}, \bibinfo {author} {\bibfnamefont {K.}~\bibnamefont
  {{Ackley}}}, \bibinfo {author} {\bibfnamefont {C.}~\bibnamefont
  {{Adamcewicz}}}, \bibinfo {author} {\bibfnamefont {S.}~\bibnamefont
  {{Adhicary}}}, \emph {et~al.},\ }\href
  {https://doi.org/10.48550/arXiv.2508.18082} {\bibfield  {journal} {\bibinfo
  {journal} {arXiv e-prints}\ ,\ \bibinfo {eid} {arXiv:2508.18082}} (\bibinfo
  {year} {2025}{\natexlab{a}})},\ \Eprint {https://arxiv.org/abs/2508.18082}
  {arXiv:2508.18082 [gr-qc]} \BibitemShut {NoStop}%
\bibitem [{\citenamefont {{Mandel}}\ and\ \citenamefont
  {{Farmer}}(2022)}]{2022PhR...955....1M}%
  \BibitemOpen
  \bibfield  {author} {\bibinfo {author} {\bibfnamefont {I.}~\bibnamefont
  {{Mandel}}}\ and\ \bibinfo {author} {\bibfnamefont {A.}~\bibnamefont
  {{Farmer}}},\ }\href {https://doi.org/10.1016/j.physrep.2022.01.003}
  {\bibfield  {journal} {\bibinfo  {journal} {\physrep}\ }\textbf {\bibinfo
  {volume} {955}},\ \bibinfo {pages} {1} (\bibinfo {year} {2022})},\ \Eprint
  {https://arxiv.org/abs/1806.05820} {arXiv:1806.05820 [astro-ph.HE]}
  \BibitemShut {NoStop}%
\bibitem [{\citenamefont {{Spera}}\ \emph {et~al.}(2022)\citenamefont
  {{Spera}}, \citenamefont {{Trani}},\ and\ \citenamefont
  {{Mencagli}}}]{2022Galax..10...76S}%
  \BibitemOpen
  \bibfield  {author} {\bibinfo {author} {\bibfnamefont {M.}~\bibnamefont
  {{Spera}}}, \bibinfo {author} {\bibfnamefont {A.~A.}\ \bibnamefont
  {{Trani}}},\ and\ \bibinfo {author} {\bibfnamefont {M.}~\bibnamefont
  {{Mencagli}}},\ }\href {https://doi.org/10.3390/galaxies10040076} {\bibfield
  {journal} {\bibinfo  {journal} {Galaxies}\ }\textbf {\bibinfo {volume}
  {10}},\ \bibinfo {eid} {76} (\bibinfo {year} {2022})},\ \Eprint
  {https://arxiv.org/abs/2206.15392} {arXiv:2206.15392 [astro-ph.HE]}
  \BibitemShut {NoStop}%
\bibitem [{\citenamefont {{Arca Sedda}}\ and\ \citenamefont
  {{Benacquista}}(2019)}]{2019MNRAS.482.2991A}%
  \BibitemOpen
  \bibfield  {author} {\bibinfo {author} {\bibfnamefont {M.}~\bibnamefont
  {{Arca Sedda}}}\ and\ \bibinfo {author} {\bibfnamefont {M.}~\bibnamefont
  {{Benacquista}}},\ }\href {https://doi.org/10.1093/mnras/sty2764} {\bibfield
  {journal} {\bibinfo  {journal} {\mnras}\ }\textbf {\bibinfo {volume} {482}},\
  \bibinfo {pages} {2991} (\bibinfo {year} {2019})},\ \Eprint
  {https://arxiv.org/abs/1806.01285} {arXiv:1806.01285 [astro-ph.GA]}
  \BibitemShut {NoStop}%
\bibitem [{\citenamefont {{Arca Sedda}}\ \emph
  {et~al.}(2020{\natexlab{a}})\citenamefont {{Arca Sedda}}, \citenamefont
  {{Mapelli}}, \citenamefont {{Spera}}, \citenamefont {{Benacquista}},\ and\
  \citenamefont {{Giacobbo}}}]{2020ApJ...894..133A}%
  \BibitemOpen
  \bibfield  {author} {\bibinfo {author} {\bibfnamefont {M.}~\bibnamefont
  {{Arca Sedda}}}, \bibinfo {author} {\bibfnamefont {M.}~\bibnamefont
  {{Mapelli}}}, \bibinfo {author} {\bibfnamefont {M.}~\bibnamefont {{Spera}}},
  \bibinfo {author} {\bibfnamefont {M.}~\bibnamefont {{Benacquista}}},\ and\
  \bibinfo {author} {\bibfnamefont {N.}~\bibnamefont {{Giacobbo}}},\ }\href
  {https://doi.org/10.3847/1538-4357/ab88b2} {\bibfield  {journal} {\bibinfo
  {journal} {\apj}\ }\textbf {\bibinfo {volume} {894}},\ \bibinfo {eid} {133}
  (\bibinfo {year} {2020}{\natexlab{a}})},\ \Eprint
  {https://arxiv.org/abs/2003.07409} {arXiv:2003.07409 [astro-ph.GA]}
  \BibitemShut {NoStop}%
\bibitem [{\citenamefont {{Bavera}}\ \emph {et~al.}(2020)\citenamefont
  {{Bavera}}, \citenamefont {{Fragos}}, \citenamefont {{Qin}}, \citenamefont
  {{Zapartas}}, \citenamefont {{Neijssel}}, \citenamefont {{Mandel}},
  \citenamefont {{Batta}}, \citenamefont {{Gaebel}}, \citenamefont
  {{Kimball}},\ and\ \citenamefont {{Stevenson}}}]{2020AeA...635A..97B}%
  \BibitemOpen
  \bibfield  {author} {\bibinfo {author} {\bibfnamefont {S.~S.}\ \bibnamefont
  {{Bavera}}}, \bibinfo {author} {\bibfnamefont {T.}~\bibnamefont {{Fragos}}},
  \bibinfo {author} {\bibfnamefont {Y.}~\bibnamefont {{Qin}}}, \bibinfo
  {author} {\bibfnamefont {E.}~\bibnamefont {{Zapartas}}}, \bibinfo {author}
  {\bibfnamefont {C.~J.}\ \bibnamefont {{Neijssel}}}, \bibinfo {author}
  {\bibfnamefont {I.}~\bibnamefont {{Mandel}}}, \bibinfo {author}
  {\bibfnamefont {A.}~\bibnamefont {{Batta}}}, \bibinfo {author} {\bibfnamefont
  {S.~M.}\ \bibnamefont {{Gaebel}}}, \bibinfo {author} {\bibfnamefont
  {C.}~\bibnamefont {{Kimball}}},\ and\ \bibinfo {author} {\bibfnamefont
  {S.}~\bibnamefont {{Stevenson}}},\ }\href
  {https://doi.org/10.1051/0004-6361/201936204} {\bibfield  {journal} {\bibinfo
   {journal} {\aap}\ }\textbf {\bibinfo {volume} {635}},\ \bibinfo {eid} {A97}
  (\bibinfo {year} {2020})},\ \Eprint {https://arxiv.org/abs/1906.12257}
  {arXiv:1906.12257 [astro-ph.HE]} \BibitemShut {NoStop}%
\bibitem [{\citenamefont {{Zevin}}\ \emph {et~al.}(2021)\citenamefont
  {{Zevin}}, \citenamefont {{Bavera}}, \citenamefont {{Berry}}, \citenamefont
  {{Kalogera}}, \citenamefont {{Fragos}}, \citenamefont {{Marchant}},
  \citenamefont {{Rodriguez}}, \citenamefont {{Antonini}}, \citenamefont
  {{Holz}},\ and\ \citenamefont {{Pankow}}}]{2021ApJ...910..152Z}%
  \BibitemOpen
  \bibfield  {author} {\bibinfo {author} {\bibfnamefont {M.}~\bibnamefont
  {{Zevin}}}, \bibinfo {author} {\bibfnamefont {S.~S.}\ \bibnamefont
  {{Bavera}}}, \bibinfo {author} {\bibfnamefont {C.~P.~L.}\ \bibnamefont
  {{Berry}}}, \bibinfo {author} {\bibfnamefont {V.}~\bibnamefont {{Kalogera}}},
  \bibinfo {author} {\bibfnamefont {T.}~\bibnamefont {{Fragos}}}, \bibinfo
  {author} {\bibfnamefont {P.}~\bibnamefont {{Marchant}}}, \bibinfo {author}
  {\bibfnamefont {C.~L.}\ \bibnamefont {{Rodriguez}}}, \bibinfo {author}
  {\bibfnamefont {F.}~\bibnamefont {{Antonini}}}, \bibinfo {author}
  {\bibfnamefont {D.~E.}\ \bibnamefont {{Holz}}},\ and\ \bibinfo {author}
  {\bibfnamefont {C.}~\bibnamefont {{Pankow}}},\ }\href
  {https://doi.org/10.3847/1538-4357/abe40e} {\bibfield  {journal} {\bibinfo
  {journal} {\apj}\ }\textbf {\bibinfo {volume} {910}},\ \bibinfo {eid} {152}
  (\bibinfo {year} {2021})},\ \Eprint {https://arxiv.org/abs/2011.10057}
  {arXiv:2011.10057 [astro-ph.HE]} \BibitemShut {NoStop}%
\bibitem [{\citenamefont {{Mapelli}}\ \emph {et~al.}(2022)\citenamefont
  {{Mapelli}}, \citenamefont {{Bouffanais}}, \citenamefont {{Santoliquido}},
  \citenamefont {{Arca Sedda}},\ and\ \citenamefont
  {{Artale}}}]{2022MNRAS.511.5797M}%
  \BibitemOpen
  \bibfield  {author} {\bibinfo {author} {\bibfnamefont {M.}~\bibnamefont
  {{Mapelli}}}, \bibinfo {author} {\bibfnamefont {Y.}~\bibnamefont
  {{Bouffanais}}}, \bibinfo {author} {\bibfnamefont {F.}~\bibnamefont
  {{Santoliquido}}}, \bibinfo {author} {\bibfnamefont {M.}~\bibnamefont {{Arca
  Sedda}}},\ and\ \bibinfo {author} {\bibfnamefont {M.~C.}\ \bibnamefont
  {{Artale}}},\ }\href {https://doi.org/10.1093/mnras/stac422} {\bibfield
  {journal} {\bibinfo  {journal} {\mnras}\ }\textbf {\bibinfo {volume} {511}},\
  \bibinfo {pages} {5797} (\bibinfo {year} {2022})},\ \Eprint
  {https://arxiv.org/abs/2109.06222} {arXiv:2109.06222 [astro-ph.HE]}
  \BibitemShut {NoStop}%
\bibitem [{\citenamefont {{Arca Sedda}}\ \emph
  {et~al.}(2023{\natexlab{a}})\citenamefont {{Arca Sedda}}, \citenamefont
  {{Mapelli}}, \citenamefont {{Benacquista}},\ and\ \citenamefont
  {{Spera}}}]{2023MNRAS.520.5259A}%
  \BibitemOpen
  \bibfield  {author} {\bibinfo {author} {\bibfnamefont {M.}~\bibnamefont
  {{Arca Sedda}}}, \bibinfo {author} {\bibfnamefont {M.}~\bibnamefont
  {{Mapelli}}}, \bibinfo {author} {\bibfnamefont {M.}~\bibnamefont
  {{Benacquista}}},\ and\ \bibinfo {author} {\bibfnamefont {M.}~\bibnamefont
  {{Spera}}},\ }\href {https://doi.org/10.1093/mnras/stad331} {\bibfield
  {journal} {\bibinfo  {journal} {\mnras}\ }\textbf {\bibinfo {volume} {520}},\
  \bibinfo {pages} {5259} (\bibinfo {year} {2023}{\natexlab{a}})},\ \Eprint
  {https://arxiv.org/abs/2109.12119} {arXiv:2109.12119 [astro-ph.GA]}
  \BibitemShut {NoStop}%
\bibitem [{\citenamefont {{Dominik}}\ \emph {et~al.}(2012)\citenamefont
  {{Dominik}}, \citenamefont {{Belczynski}}, \citenamefont {{Fryer}},
  \citenamefont {{Holz}}, \citenamefont {{Berti}}, \citenamefont {{Bulik}},
  \citenamefont {{Mandel}},\ and\ \citenamefont
  {{O'Shaughnessy}}}]{2012ApJ...759...52D}%
  \BibitemOpen
  \bibfield  {author} {\bibinfo {author} {\bibfnamefont {M.}~\bibnamefont
  {{Dominik}}}, \bibinfo {author} {\bibfnamefont {K.}~\bibnamefont
  {{Belczynski}}}, \bibinfo {author} {\bibfnamefont {C.}~\bibnamefont
  {{Fryer}}}, \bibinfo {author} {\bibfnamefont {D.~E.}\ \bibnamefont {{Holz}}},
  \bibinfo {author} {\bibfnamefont {E.}~\bibnamefont {{Berti}}}, \bibinfo
  {author} {\bibfnamefont {T.}~\bibnamefont {{Bulik}}}, \bibinfo {author}
  {\bibfnamefont {I.}~\bibnamefont {{Mandel}}},\ and\ \bibinfo {author}
  {\bibfnamefont {R.}~\bibnamefont {{O'Shaughnessy}}},\ }\href
  {https://doi.org/10.1088/0004-637X/759/1/52} {\bibfield  {journal} {\bibinfo
  {journal} {\apj}\ }\textbf {\bibinfo {volume} {759}},\ \bibinfo {eid} {52}
  (\bibinfo {year} {2012})},\ \Eprint {https://arxiv.org/abs/1202.4901}
  {arXiv:1202.4901 [astro-ph.HE]} \BibitemShut {NoStop}%
\bibitem [{\citenamefont {{Belczynski}}(2020)}]{2020ApJ...905L..15B}%
  \BibitemOpen
  \bibfield  {author} {\bibinfo {author} {\bibfnamefont {K.}~\bibnamefont
  {{Belczynski}}},\ }\href {https://doi.org/10.3847/2041-8213/abcbf1}
  {\bibfield  {journal} {\bibinfo  {journal} {\apjl}\ }\textbf {\bibinfo
  {volume} {905}},\ \bibinfo {eid} {L15} (\bibinfo {year} {2020})},\ \Eprint
  {https://arxiv.org/abs/2009.13526} {arXiv:2009.13526 [astro-ph.HE]}
  \BibitemShut {NoStop}%
\bibitem [{\citenamefont {{Belczynski}}\ \emph {et~al.}(2002)\citenamefont
  {{Belczynski}}, \citenamefont {{Kalogera}},\ and\ \citenamefont
  {{Bulik}}}]{2002ApJ...572..407B}%
  \BibitemOpen
  \bibfield  {author} {\bibinfo {author} {\bibfnamefont {K.}~\bibnamefont
  {{Belczynski}}}, \bibinfo {author} {\bibfnamefont {V.}~\bibnamefont
  {{Kalogera}}},\ and\ \bibinfo {author} {\bibfnamefont {T.}~\bibnamefont
  {{Bulik}}},\ }\href {https://doi.org/10.1086/340304} {\bibfield  {journal}
  {\bibinfo  {journal} {\apj}\ }\textbf {\bibinfo {volume} {572}},\ \bibinfo
  {pages} {407} (\bibinfo {year} {2002})},\ \Eprint
  {https://arxiv.org/abs/astro-ph/0111452} {arXiv:astro-ph/0111452 [astro-ph]}
  \BibitemShut {NoStop}%
\bibitem [{\citenamefont {{Iorio}}\ \emph {et~al.}(2023)\citenamefont
  {{Iorio}}, \citenamefont {{Mapelli}}, \citenamefont {{Costa}}, \citenamefont
  {{Spera}}, \citenamefont {{Escobar}}, \citenamefont {{Sgalletta}},
  \citenamefont {{Trani}}, \citenamefont {{Korb}}, \citenamefont
  {{Santoliquido}}, \citenamefont {{Dall'Amico}}, \citenamefont {{Gaspari}},\
  and\ \citenamefont {{Bressan}}}]{2023MNRAS.524..426I}%
  \BibitemOpen
  \bibfield  {author} {\bibinfo {author} {\bibfnamefont {G.}~\bibnamefont
  {{Iorio}}}, \bibinfo {author} {\bibfnamefont {M.}~\bibnamefont {{Mapelli}}},
  \bibinfo {author} {\bibfnamefont {G.}~\bibnamefont {{Costa}}}, \bibinfo
  {author} {\bibfnamefont {M.}~\bibnamefont {{Spera}}}, \bibinfo {author}
  {\bibfnamefont {G.~J.}\ \bibnamefont {{Escobar}}}, \bibinfo {author}
  {\bibfnamefont {C.}~\bibnamefont {{Sgalletta}}}, \bibinfo {author}
  {\bibfnamefont {A.~A.}\ \bibnamefont {{Trani}}}, \bibinfo {author}
  {\bibfnamefont {E.}~\bibnamefont {{Korb}}}, \bibinfo {author} {\bibfnamefont
  {F.}~\bibnamefont {{Santoliquido}}}, \bibinfo {author} {\bibfnamefont
  {M.}~\bibnamefont {{Dall'Amico}}}, \bibinfo {author} {\bibfnamefont
  {N.}~\bibnamefont {{Gaspari}}},\ and\ \bibinfo {author} {\bibfnamefont
  {A.}~\bibnamefont {{Bressan}}},\ }\href
  {https://doi.org/10.1093/mnras/stad1630} {\bibfield  {journal} {\bibinfo
  {journal} {\mnras}\ }\textbf {\bibinfo {volume} {524}},\ \bibinfo {pages}
  {426} (\bibinfo {year} {2023})},\ \Eprint {https://arxiv.org/abs/2211.11774}
  {arXiv:2211.11774 [astro-ph.HE]} \BibitemShut {NoStop}%
\bibitem [{\citenamefont {{Spera}}\ \emph {et~al.}(2019)\citenamefont
  {{Spera}}, \citenamefont {{Mapelli}}, \citenamefont {{Giacobbo}},
  \citenamefont {{Trani}}, \citenamefont {{Bressan}},\ and\ \citenamefont
  {{Costa}}}]{2019MNRAS.485..889S}%
  \BibitemOpen
  \bibfield  {author} {\bibinfo {author} {\bibfnamefont {M.}~\bibnamefont
  {{Spera}}}, \bibinfo {author} {\bibfnamefont {M.}~\bibnamefont {{Mapelli}}},
  \bibinfo {author} {\bibfnamefont {N.}~\bibnamefont {{Giacobbo}}}, \bibinfo
  {author} {\bibfnamefont {A.~A.}\ \bibnamefont {{Trani}}}, \bibinfo {author}
  {\bibfnamefont {A.}~\bibnamefont {{Bressan}}},\ and\ \bibinfo {author}
  {\bibfnamefont {G.}~\bibnamefont {{Costa}}},\ }\href
  {https://doi.org/10.1093/mnras/stz359} {\bibfield  {journal} {\bibinfo
  {journal} {\mnras}\ }\textbf {\bibinfo {volume} {485}},\ \bibinfo {pages}
  {889} (\bibinfo {year} {2019})},\ \Eprint {https://arxiv.org/abs/1809.04605}
  {arXiv:1809.04605 [astro-ph.HE]} \BibitemShut {NoStop}%
\bibitem [{\citenamefont {{Belczynski}}\ \emph {et~al.}(2016)\citenamefont
  {{Belczynski}}, \citenamefont {{Heger}}, \citenamefont {{Gladysz}},
  \citenamefont {{Ruiter}}, \citenamefont {{Woosley}}, \citenamefont
  {{Wiktorowicz}}, \citenamefont {{Chen}}, \citenamefont {{Bulik}},
  \citenamefont {{O'Shaughnessy}}, \citenamefont {{Holz}}, \citenamefont
  {{Fryer}},\ and\ \citenamefont {{Berti}}}]{2016A&A...594A..97B}%
  \BibitemOpen
  \bibfield  {author} {\bibinfo {author} {\bibfnamefont {K.}~\bibnamefont
  {{Belczynski}}}, \bibinfo {author} {\bibfnamefont {A.}~\bibnamefont
  {{Heger}}}, \bibinfo {author} {\bibfnamefont {W.}~\bibnamefont {{Gladysz}}},
  \bibinfo {author} {\bibfnamefont {A.~J.}\ \bibnamefont {{Ruiter}}}, \bibinfo
  {author} {\bibfnamefont {S.}~\bibnamefont {{Woosley}}}, \bibinfo {author}
  {\bibfnamefont {G.}~\bibnamefont {{Wiktorowicz}}}, \bibinfo {author}
  {\bibfnamefont {H.~Y.}\ \bibnamefont {{Chen}}}, \bibinfo {author}
  {\bibfnamefont {T.}~\bibnamefont {{Bulik}}}, \bibinfo {author} {\bibfnamefont
  {R.}~\bibnamefont {{O'Shaughnessy}}}, \bibinfo {author} {\bibfnamefont
  {D.~E.}\ \bibnamefont {{Holz}}}, \bibinfo {author} {\bibfnamefont {C.~L.}\
  \bibnamefont {{Fryer}}},\ and\ \bibinfo {author} {\bibfnamefont
  {E.}~\bibnamefont {{Berti}}},\ }\href
  {https://doi.org/10.1051/0004-6361/201628980} {\bibfield  {journal} {\bibinfo
   {journal} {\aap}\ }\textbf {\bibinfo {volume} {594}},\ \bibinfo {eid} {A97}
  (\bibinfo {year} {2016})},\ \Eprint {https://arxiv.org/abs/1607.03116}
  {arXiv:1607.03116 [astro-ph.HE]} \BibitemShut {NoStop}%
\bibitem [{\citenamefont {{Stevenson}}\ \emph {et~al.}(2019)\citenamefont
  {{Stevenson}}, \citenamefont {{Sampson}}, \citenamefont {{Powell}},
  \citenamefont {{Vigna-G{\'o}mez}}, \citenamefont {{Neijssel}}, \citenamefont
  {{Sz{\'e}csi}},\ and\ \citenamefont {{Mandel}}}]{2019ApJ...882..121S}%
  \BibitemOpen
  \bibfield  {author} {\bibinfo {author} {\bibfnamefont {S.}~\bibnamefont
  {{Stevenson}}}, \bibinfo {author} {\bibfnamefont {M.}~\bibnamefont
  {{Sampson}}}, \bibinfo {author} {\bibfnamefont {J.}~\bibnamefont {{Powell}}},
  \bibinfo {author} {\bibfnamefont {A.}~\bibnamefont {{Vigna-G{\'o}mez}}},
  \bibinfo {author} {\bibfnamefont {C.~J.}\ \bibnamefont {{Neijssel}}},
  \bibinfo {author} {\bibfnamefont {D.}~\bibnamefont {{Sz{\'e}csi}}},\ and\
  \bibinfo {author} {\bibfnamefont {I.}~\bibnamefont {{Mandel}}},\ }\href
  {https://doi.org/10.3847/1538-4357/ab3981} {\bibfield  {journal} {\bibinfo
  {journal} {\apj}\ }\textbf {\bibinfo {volume} {882}},\ \bibinfo {eid} {121}
  (\bibinfo {year} {2019})},\ \Eprint {https://arxiv.org/abs/1904.02821}
  {arXiv:1904.02821 [astro-ph.HE]} \BibitemShut {NoStop}%
\bibitem [{\citenamefont {{Farmer}}\ \emph {et~al.}(2019)\citenamefont
  {{Farmer}}, \citenamefont {{Renzo}}, \citenamefont {{de Mink}}, \citenamefont
  {{Marchant}},\ and\ \citenamefont {{Justham}}}]{2019ApJ...887...53F}%
  \BibitemOpen
  \bibfield  {author} {\bibinfo {author} {\bibfnamefont {R.}~\bibnamefont
  {{Farmer}}}, \bibinfo {author} {\bibfnamefont {M.}~\bibnamefont {{Renzo}}},
  \bibinfo {author} {\bibfnamefont {S.~E.}\ \bibnamefont {{de Mink}}}, \bibinfo
  {author} {\bibfnamefont {P.}~\bibnamefont {{Marchant}}},\ and\ \bibinfo
  {author} {\bibfnamefont {S.}~\bibnamefont {{Justham}}},\ }\href
  {https://doi.org/10.3847/1538-4357/ab518b} {\bibfield  {journal} {\bibinfo
  {journal} {\apj}\ }\textbf {\bibinfo {volume} {887}},\ \bibinfo {eid} {53}
  (\bibinfo {year} {2019})},\ \Eprint {https://arxiv.org/abs/1910.12874}
  {arXiv:1910.12874 [astro-ph.SR]} \BibitemShut {NoStop}%
\bibitem [{\citenamefont {{Godfrey}}\ \emph {et~al.}(2023)\citenamefont
  {{Godfrey}}, \citenamefont {{Edelman}},\ and\ \citenamefont
  {{Farr}}}]{2023arXiv230401288G}%
  \BibitemOpen
  \bibfield  {author} {\bibinfo {author} {\bibfnamefont {J.}~\bibnamefont
  {{Godfrey}}}, \bibinfo {author} {\bibfnamefont {B.}~\bibnamefont
  {{Edelman}}},\ and\ \bibinfo {author} {\bibfnamefont {B.}~\bibnamefont
  {{Farr}}},\ }\href {https://doi.org/10.48550/arXiv.2304.01288} {\bibfield
  {journal} {\bibinfo  {journal} {arXiv e-prints}\ ,\ \bibinfo {eid}
  {arXiv:2304.01288}} (\bibinfo {year} {2023})},\ \Eprint
  {https://arxiv.org/abs/2304.01288} {arXiv:2304.01288 [astro-ph.HE]}
  \BibitemShut {NoStop}%
\bibitem [{\citenamefont {{Lee}}(1993)}]{1993ApJ...418..147L}%
  \BibitemOpen
  \bibfield  {author} {\bibinfo {author} {\bibfnamefont {M.~H.}\ \bibnamefont
  {{Lee}}},\ }\href {https://doi.org/10.1086/173378} {\bibfield  {journal}
  {\bibinfo  {journal} {\apj}\ }\textbf {\bibinfo {volume} {418}},\ \bibinfo
  {pages} {147} (\bibinfo {year} {1993})}\BibitemShut {NoStop}%
\bibitem [{\citenamefont {{Downing}}\ \emph {et~al.}(2010)\citenamefont
  {{Downing}}, \citenamefont {{Benacquista}}, \citenamefont {{Giersz}},\ and\
  \citenamefont {{Spurzem}}}]{2010MNRAS.407.1946D}%
  \BibitemOpen
  \bibfield  {author} {\bibinfo {author} {\bibfnamefont {J.~M.~B.}\
  \bibnamefont {{Downing}}}, \bibinfo {author} {\bibfnamefont {M.~J.}\
  \bibnamefont {{Benacquista}}}, \bibinfo {author} {\bibfnamefont
  {M.}~\bibnamefont {{Giersz}}},\ and\ \bibinfo {author} {\bibfnamefont
  {R.}~\bibnamefont {{Spurzem}}},\ }\href
  {https://doi.org/10.1111/j.1365-2966.2010.17040.x} {\bibfield  {journal}
  {\bibinfo  {journal} {\mnras}\ }\textbf {\bibinfo {volume} {407}},\ \bibinfo
  {pages} {1946} (\bibinfo {year} {2010})},\ \Eprint
  {https://arxiv.org/abs/0910.0546} {arXiv:0910.0546 [astro-ph.SR]}
  \BibitemShut {NoStop}%
\bibitem [{\citenamefont {{Banerjee}}\ \emph {et~al.}(2010)\citenamefont
  {{Banerjee}}, \citenamefont {{Baumgardt}},\ and\ \citenamefont
  {{Kroupa}}}]{2010MNRAS.402..371B}%
  \BibitemOpen
  \bibfield  {author} {\bibinfo {author} {\bibfnamefont {S.}~\bibnamefont
  {{Banerjee}}}, \bibinfo {author} {\bibfnamefont {H.}~\bibnamefont
  {{Baumgardt}}},\ and\ \bibinfo {author} {\bibfnamefont {P.}~\bibnamefont
  {{Kroupa}}},\ }\href {https://doi.org/10.1111/j.1365-2966.2009.15880.x}
  {\bibfield  {journal} {\bibinfo  {journal} {\mnras}\ }\textbf {\bibinfo
  {volume} {402}},\ \bibinfo {pages} {371} (\bibinfo {year} {2010})},\ \Eprint
  {https://arxiv.org/abs/0910.3954} {arXiv:0910.3954 [astro-ph.SR]}
  \BibitemShut {NoStop}%
\bibitem [{\citenamefont {{Rodriguez}}\ \emph {et~al.}(2015)\citenamefont
  {{Rodriguez}}, \citenamefont {{Morscher}}, \citenamefont {{Pattabiraman}},
  \citenamefont {{Chatterjee}}, \citenamefont {{Haster}},\ and\ \citenamefont
  {{Rasio}}}]{2015PhRvL.115e1101R}%
  \BibitemOpen
  \bibfield  {author} {\bibinfo {author} {\bibfnamefont {C.~L.}\ \bibnamefont
  {{Rodriguez}}}, \bibinfo {author} {\bibfnamefont {M.}~\bibnamefont
  {{Morscher}}}, \bibinfo {author} {\bibfnamefont {B.}~\bibnamefont
  {{Pattabiraman}}}, \bibinfo {author} {\bibfnamefont {S.}~\bibnamefont
  {{Chatterjee}}}, \bibinfo {author} {\bibfnamefont {C.-J.}\ \bibnamefont
  {{Haster}}},\ and\ \bibinfo {author} {\bibfnamefont {F.~A.}\ \bibnamefont
  {{Rasio}}},\ }\href {https://doi.org/10.1103/PhysRevLett.115.051101}
  {\bibfield  {journal} {\bibinfo  {journal} {\prl}\ }\textbf {\bibinfo
  {volume} {115}},\ \bibinfo {eid} {051101} (\bibinfo {year} {2015})},\ \Eprint
  {https://arxiv.org/abs/1505.00792} {arXiv:1505.00792 [astro-ph.HE]}
  \BibitemShut {NoStop}%
\bibitem [{\citenamefont {{Antonini}}\ and\ \citenamefont
  {{Rasio}}(2016)}]{2016ApJ...831..187A}%
  \BibitemOpen
  \bibfield  {author} {\bibinfo {author} {\bibfnamefont {F.}~\bibnamefont
  {{Antonini}}}\ and\ \bibinfo {author} {\bibfnamefont {F.~A.}\ \bibnamefont
  {{Rasio}}},\ }\href {https://doi.org/10.3847/0004-637X/831/2/187} {\bibfield
  {journal} {\bibinfo  {journal} {\apj}\ }\textbf {\bibinfo {volume} {831}},\
  \bibinfo {eid} {187} (\bibinfo {year} {2016})},\ \Eprint
  {https://arxiv.org/abs/1606.04889} {arXiv:1606.04889 [astro-ph.HE]}
  \BibitemShut {NoStop}%
\bibitem [{\citenamefont {{Askar}}\ \emph {et~al.}(2017)\citenamefont
  {{Askar}}, \citenamefont {{Szkudlarek}}, \citenamefont
  {{Gondek-Rosi{\'n}ska}}, \citenamefont {{Giersz}},\ and\ \citenamefont
  {{Bulik}}}]{2017MNRAS.464L..36A}%
  \BibitemOpen
  \bibfield  {author} {\bibinfo {author} {\bibfnamefont {A.}~\bibnamefont
  {{Askar}}}, \bibinfo {author} {\bibfnamefont {M.}~\bibnamefont
  {{Szkudlarek}}}, \bibinfo {author} {\bibfnamefont {D.}~\bibnamefont
  {{Gondek-Rosi{\'n}ska}}}, \bibinfo {author} {\bibfnamefont {M.}~\bibnamefont
  {{Giersz}}},\ and\ \bibinfo {author} {\bibfnamefont {T.}~\bibnamefont
  {{Bulik}}},\ }\href {https://doi.org/10.1093/mnrasl/slw177} {\bibfield
  {journal} {\bibinfo  {journal} {\mnras}\ }\textbf {\bibinfo {volume} {464}},\
  \bibinfo {pages} {L36} (\bibinfo {year} {2017})},\ \Eprint
  {https://arxiv.org/abs/1608.02520} {arXiv:1608.02520 [astro-ph.HE]}
  \BibitemShut {NoStop}%
\bibitem [{\citenamefont {{Banerjee}}(2018)}]{2018MNRAS.473..909B}%
  \BibitemOpen
  \bibfield  {author} {\bibinfo {author} {\bibfnamefont {S.}~\bibnamefont
  {{Banerjee}}},\ }\href {https://doi.org/10.1093/mnras/stx2347} {\bibfield
  {journal} {\bibinfo  {journal} {\mnras}\ }\textbf {\bibinfo {volume} {473}},\
  \bibinfo {pages} {909} (\bibinfo {year} {2018})},\ \Eprint
  {https://arxiv.org/abs/1707.00922} {arXiv:1707.00922 [astro-ph.HE]}
  \BibitemShut {NoStop}%
\bibitem [{\citenamefont {{Di Carlo}}\ \emph {et~al.}(2019)\citenamefont {{Di
  Carlo}}, \citenamefont {{Giacobbo}}, \citenamefont {{Mapelli}}, \citenamefont
  {{Pasquato}}, \citenamefont {{Spera}}, \citenamefont {{Wang}},\ and\
  \citenamefont {{Haardt}}}]{2019MNRAS.487.2947D}%
  \BibitemOpen
  \bibfield  {author} {\bibinfo {author} {\bibfnamefont {U.~N.}\ \bibnamefont
  {{Di Carlo}}}, \bibinfo {author} {\bibfnamefont {N.}~\bibnamefont
  {{Giacobbo}}}, \bibinfo {author} {\bibfnamefont {M.}~\bibnamefont
  {{Mapelli}}}, \bibinfo {author} {\bibfnamefont {M.}~\bibnamefont
  {{Pasquato}}}, \bibinfo {author} {\bibfnamefont {M.}~\bibnamefont {{Spera}}},
  \bibinfo {author} {\bibfnamefont {L.}~\bibnamefont {{Wang}}},\ and\ \bibinfo
  {author} {\bibfnamefont {F.}~\bibnamefont {{Haardt}}},\ }\href
  {https://doi.org/10.1093/mnras/stz1453} {\bibfield  {journal} {\bibinfo
  {journal} {\mnras}\ }\textbf {\bibinfo {volume} {487}},\ \bibinfo {pages}
  {2947} (\bibinfo {year} {2019})},\ \Eprint {https://arxiv.org/abs/1901.00863}
  {arXiv:1901.00863 [astro-ph.HE]} \BibitemShut {NoStop}%
\bibitem [{\citenamefont {{Rodriguez}}\ \emph {et~al.}(2019)\citenamefont
  {{Rodriguez}}, \citenamefont {{Zevin}}, \citenamefont {{Amaro-Seoane}},
  \citenamefont {{Chatterjee}}, \citenamefont {{Kremer}}, \citenamefont
  {{Rasio}},\ and\ \citenamefont {{Ye}}}]{2019PhRvD.100d3027R}%
  \BibitemOpen
  \bibfield  {author} {\bibinfo {author} {\bibfnamefont {C.~L.}\ \bibnamefont
  {{Rodriguez}}}, \bibinfo {author} {\bibfnamefont {M.}~\bibnamefont
  {{Zevin}}}, \bibinfo {author} {\bibfnamefont {P.}~\bibnamefont
  {{Amaro-Seoane}}}, \bibinfo {author} {\bibfnamefont {S.}~\bibnamefont
  {{Chatterjee}}}, \bibinfo {author} {\bibfnamefont {K.}~\bibnamefont
  {{Kremer}}}, \bibinfo {author} {\bibfnamefont {F.~A.}\ \bibnamefont
  {{Rasio}}},\ and\ \bibinfo {author} {\bibfnamefont {C.~S.}\ \bibnamefont
  {{Ye}}},\ }\href {https://doi.org/10.1103/PhysRevD.100.043027} {\bibfield
  {journal} {\bibinfo  {journal} {\prd}\ }\textbf {\bibinfo {volume} {100}},\
  \bibinfo {eid} {043027} (\bibinfo {year} {2019})},\ \Eprint
  {https://arxiv.org/abs/1906.10260} {arXiv:1906.10260 [astro-ph.HE]}
  \BibitemShut {NoStop}%
\bibitem [{\citenamefont {{Tagawa}}\ \emph {et~al.}(2021)\citenamefont
  {{Tagawa}}, \citenamefont {{Kocsis}}, \citenamefont {{Haiman}}, \citenamefont
  {{Bartos}}, \citenamefont {{Omukai}},\ and\ \citenamefont
  {{Samsing}}}]{2021ApJ...908..194T}%
  \BibitemOpen
  \bibfield  {author} {\bibinfo {author} {\bibfnamefont {H.}~\bibnamefont
  {{Tagawa}}}, \bibinfo {author} {\bibfnamefont {B.}~\bibnamefont {{Kocsis}}},
  \bibinfo {author} {\bibfnamefont {Z.}~\bibnamefont {{Haiman}}}, \bibinfo
  {author} {\bibfnamefont {I.}~\bibnamefont {{Bartos}}}, \bibinfo {author}
  {\bibfnamefont {K.}~\bibnamefont {{Omukai}}},\ and\ \bibinfo {author}
  {\bibfnamefont {J.}~\bibnamefont {{Samsing}}},\ }\href
  {https://doi.org/10.3847/1538-4357/abd555} {\bibfield  {journal} {\bibinfo
  {journal} {\apj}\ }\textbf {\bibinfo {volume} {908}},\ \bibinfo {eid} {194}
  (\bibinfo {year} {2021})},\ \Eprint {https://arxiv.org/abs/2012.00011}
  {arXiv:2012.00011 [astro-ph.HE]} \BibitemShut {NoStop}%
\bibitem [{\citenamefont {{Banerjee}}(2022)}]{2022A&A...665A..20B}%
  \BibitemOpen
  \bibfield  {author} {\bibinfo {author} {\bibfnamefont {S.}~\bibnamefont
  {{Banerjee}}},\ }\href {https://doi.org/10.1051/0004-6361/202142331}
  {\bibfield  {journal} {\bibinfo  {journal} {\aap}\ }\textbf {\bibinfo
  {volume} {665}},\ \bibinfo {eid} {A20} (\bibinfo {year} {2022})},\ \Eprint
  {https://arxiv.org/abs/2109.14612} {arXiv:2109.14612 [astro-ph.HE]}
  \BibitemShut {NoStop}%
\bibitem [{\citenamefont {{Chattopadhyay}}\ \emph {et~al.}(2022)\citenamefont
  {{Chattopadhyay}}, \citenamefont {{Hurley}}, \citenamefont {{Stevenson}},\
  and\ \citenamefont {{Raidani}}}]{2022MNRAS.513.4527C}%
  \BibitemOpen
  \bibfield  {author} {\bibinfo {author} {\bibfnamefont {D.}~\bibnamefont
  {{Chattopadhyay}}}, \bibinfo {author} {\bibfnamefont {J.}~\bibnamefont
  {{Hurley}}}, \bibinfo {author} {\bibfnamefont {S.}~\bibnamefont
  {{Stevenson}}},\ and\ \bibinfo {author} {\bibfnamefont {A.}~\bibnamefont
  {{Raidani}}},\ }\href {https://doi.org/10.1093/mnras/stac1163} {\bibfield
  {journal} {\bibinfo  {journal} {\mnras}\ }\textbf {\bibinfo {volume} {513}},\
  \bibinfo {pages} {4527} (\bibinfo {year} {2022})},\ \Eprint
  {https://arxiv.org/abs/2202.08924} {arXiv:2202.08924 [astro-ph.GA]}
  \BibitemShut {NoStop}%
\bibitem [{\citenamefont {{Torniamenti}}\ \emph {et~al.}(2022)\citenamefont
  {{Torniamenti}}, \citenamefont {{Rastello}}, \citenamefont {{Mapelli}},
  \citenamefont {{Di Carlo}}, \citenamefont {{Ballone}},\ and\ \citenamefont
  {{Pasquato}}}]{2022MNRAS.517.2953T}%
  \BibitemOpen
  \bibfield  {author} {\bibinfo {author} {\bibfnamefont {S.}~\bibnamefont
  {{Torniamenti}}}, \bibinfo {author} {\bibfnamefont {S.}~\bibnamefont
  {{Rastello}}}, \bibinfo {author} {\bibfnamefont {M.}~\bibnamefont
  {{Mapelli}}}, \bibinfo {author} {\bibfnamefont {U.~N.}\ \bibnamefont {{Di
  Carlo}}}, \bibinfo {author} {\bibfnamefont {A.}~\bibnamefont {{Ballone}}},\
  and\ \bibinfo {author} {\bibfnamefont {M.}~\bibnamefont {{Pasquato}}},\
  }\href {https://doi.org/10.1093/mnras/stac2841} {\bibfield  {journal}
  {\bibinfo  {journal} {\mnras}\ }\textbf {\bibinfo {volume} {517}},\ \bibinfo
  {pages} {2953} (\bibinfo {year} {2022})},\ \Eprint
  {https://arxiv.org/abs/2203.08163} {arXiv:2203.08163 [astro-ph.GA]}
  \BibitemShut {NoStop}%
\bibitem [{\citenamefont {{Arca sedda}}\ \emph {et~al.}(2024)\citenamefont
  {{Arca sedda}}, \citenamefont {{Kamlah}}, \citenamefont {{Spurzem}},
  \citenamefont {{Rizzuto}}, \citenamefont {{Giersz}}, \citenamefont {{Naab}},\
  and\ \citenamefont {{Berczik}}}]{2024MNRAS.528.5140A}%
  \BibitemOpen
  \bibfield  {author} {\bibinfo {author} {\bibfnamefont {M.}~\bibnamefont
  {{Arca sedda}}}, \bibinfo {author} {\bibfnamefont {A.~W.~H.}\ \bibnamefont
  {{Kamlah}}}, \bibinfo {author} {\bibfnamefont {R.}~\bibnamefont {{Spurzem}}},
  \bibinfo {author} {\bibfnamefont {F.~P.}\ \bibnamefont {{Rizzuto}}}, \bibinfo
  {author} {\bibfnamefont {M.}~\bibnamefont {{Giersz}}}, \bibinfo {author}
  {\bibfnamefont {T.}~\bibnamefont {{Naab}}},\ and\ \bibinfo {author}
  {\bibfnamefont {P.}~\bibnamefont {{Berczik}}},\ }\href
  {https://doi.org/10.1093/mnras/stad3951} {\bibfield  {journal} {\bibinfo
  {journal} {\mnras}\ }\textbf {\bibinfo {volume} {528}},\ \bibinfo {pages}
  {5140} (\bibinfo {year} {2024})},\ \Eprint {https://arxiv.org/abs/2307.04807}
  {arXiv:2307.04807 [astro-ph.HE]} \BibitemShut {NoStop}%
\bibitem [{\citenamefont {{Barber}}\ and\ \citenamefont
  {{Antonini}}(2025)}]{2025MNRAS.538..639B}%
  \BibitemOpen
  \bibfield  {author} {\bibinfo {author} {\bibfnamefont {J.}~\bibnamefont
  {{Barber}}}\ and\ \bibinfo {author} {\bibfnamefont {F.}~\bibnamefont
  {{Antonini}}},\ }\href {https://doi.org/10.1093/mnras/staf279} {\bibfield
  {journal} {\bibinfo  {journal} {\mnras}\ }\textbf {\bibinfo {volume} {538}},\
  \bibinfo {pages} {639} (\bibinfo {year} {2025})},\ \Eprint
  {https://arxiv.org/abs/2410.03832} {arXiv:2410.03832 [astro-ph.GA]}
  \BibitemShut {NoStop}%
\bibitem [{\citenamefont {{Abbott}}\ \emph
  {et~al.}(2020{\natexlab{b}})\citenamefont {{Abbott}}, \citenamefont
  {{Abbott}}, \citenamefont {{Abraham}}, \citenamefont {{LIGO Scientific
  Collaboration}},\ and\ \citenamefont {{Virgo
  Collaboration}}}]{2020PhRvL.125j1102A}%
  \BibitemOpen
  \bibfield  {author} {\bibinfo {author} {\bibfnamefont {R.}~\bibnamefont
  {{Abbott}}}, \bibinfo {author} {\bibfnamefont {T.~D.}\ \bibnamefont
  {{Abbott}}}, \bibinfo {author} {\bibfnamefont {S.}~\bibnamefont {{Abraham}}},
  \bibinfo {author} {\bibnamefont {{LIGO Scientific Collaboration}}},\ and\
  \bibinfo {author} {\bibnamefont {{Virgo Collaboration}}},\ }\href
  {https://doi.org/10.1103/PhysRevLett.125.101102} {\bibfield  {journal}
  {\bibinfo  {journal} {\prl}\ }\textbf {\bibinfo {volume} {125}},\ \bibinfo
  {eid} {101102} (\bibinfo {year} {2020}{\natexlab{b}})},\ \Eprint
  {https://arxiv.org/abs/2009.01075} {arXiv:2009.01075 [gr-qc]} \BibitemShut
  {NoStop}%
\bibitem [{\citenamefont {{The LIGO Scientific Collaboration}}\ \emph
  {et~al.}(2025{\natexlab{b}})\citenamefont {{The LIGO Scientific
  Collaboration}}, \citenamefont {{the Virgo Collaboration}}, \citenamefont
  {{the KAGRA Collaboration}}, \citenamefont {{Abac}}, \citenamefont
  {{Abouelfettouh}}, \citenamefont {{Acernese}}, \citenamefont {{Ackley}},
  \citenamefont {{Adamcewicz}}, \citenamefont {{Adhicary}}, \citenamefont
  {{Adhikari}}, \citenamefont {{Adhikari}}, \citenamefont {{Adhikari}},
  \citenamefont {{Adkins}}, \citenamefont {{Afroz}}, \citenamefont {{Agapito}},
  \citenamefont {{Agarwal}} \emph {et~al.}}]{2025arXiv250708219T}%
  \BibitemOpen
  \bibfield  {author} {\bibinfo {author} {\bibnamefont {{The LIGO Scientific
  Collaboration}}}, \bibinfo {author} {\bibnamefont {{the Virgo
  Collaboration}}}, \bibinfo {author} {\bibnamefont {{the KAGRA
  Collaboration}}}, \bibinfo {author} {\bibfnamefont {A.~G.}\ \bibnamefont
  {{Abac}}}, \bibinfo {author} {\bibfnamefont {I.}~\bibnamefont
  {{Abouelfettouh}}}, \bibinfo {author} {\bibfnamefont {F.}~\bibnamefont
  {{Acernese}}}, \bibinfo {author} {\bibfnamefont {K.}~\bibnamefont
  {{Ackley}}}, \bibinfo {author} {\bibfnamefont {C.}~\bibnamefont
  {{Adamcewicz}}}, \bibinfo {author} {\bibfnamefont {S.}~\bibnamefont
  {{Adhicary}}}, \bibinfo {author} {\bibfnamefont {D.}~\bibnamefont
  {{Adhikari}}}, \bibinfo {author} {\bibfnamefont {N.}~\bibnamefont
  {{Adhikari}}}, \bibinfo {author} {\bibfnamefont {R.~X.}\ \bibnamefont
  {{Adhikari}}}, \bibinfo {author} {\bibfnamefont {V.~K.}\ \bibnamefont
  {{Adkins}}}, \bibinfo {author} {\bibfnamefont {S.}~\bibnamefont {{Afroz}}},
  \bibinfo {author} {\bibfnamefont {A.}~\bibnamefont {{Agapito}}}, \bibinfo
  {author} {\bibfnamefont {D.}~\bibnamefont {{Agarwal}}}, \emph {et~al.},\
  }\href {https://doi.org/10.48550/arXiv.2507.08219} {\bibfield  {journal}
  {\bibinfo  {journal} {arXiv e-prints}\ ,\ \bibinfo {eid} {arXiv:2507.08219}}
  (\bibinfo {year} {2025}{\natexlab{b}})},\ \Eprint
  {https://arxiv.org/abs/2507.08219} {arXiv:2507.08219 [astro-ph.HE]}
  \BibitemShut {NoStop}%
\bibitem [{\citenamefont {{Popa}}\ and\ \citenamefont {{de
  Mink}}(2025)}]{2025ApJ...995L..76P}%
  \BibitemOpen
  \bibfield  {author} {\bibinfo {author} {\bibfnamefont {S.~A.}\ \bibnamefont
  {{Popa}}}\ and\ \bibinfo {author} {\bibfnamefont {S.~E.}\ \bibnamefont {{de
  Mink}}},\ }\href {https://doi.org/10.3847/2041-8213/ae20f1} {\bibfield
  {journal} {\bibinfo  {journal} {\apjl}\ }\textbf {\bibinfo {volume} {995}},\
  \bibinfo {eid} {L76} (\bibinfo {year} {2025})},\ \Eprint
  {https://arxiv.org/abs/2509.00154} {arXiv:2509.00154 [astro-ph.HE]}
  \BibitemShut {NoStop}%
\bibitem [{\citenamefont {{Tanikawa}}\ \emph {et~al.}(2025)\citenamefont
  {{Tanikawa}}, \citenamefont {{Liu}}, \citenamefont {{Wu}}, \citenamefont
  {{Fujii}},\ and\ \citenamefont {{Wang}}}]{2025arXiv250801135T}%
  \BibitemOpen
  \bibfield  {author} {\bibinfo {author} {\bibfnamefont {A.}~\bibnamefont
  {{Tanikawa}}}, \bibinfo {author} {\bibfnamefont {S.}~\bibnamefont {{Liu}}},
  \bibinfo {author} {\bibfnamefont {W.}~\bibnamefont {{Wu}}}, \bibinfo {author}
  {\bibfnamefont {M.~S.}\ \bibnamefont {{Fujii}}},\ and\ \bibinfo {author}
  {\bibfnamefont {L.}~\bibnamefont {{Wang}}},\ }\href
  {https://doi.org/10.48550/arXiv.2508.01135} {\bibfield  {journal} {\bibinfo
  {journal} {arXiv e-prints}\ ,\ \bibinfo {eid} {arXiv:2508.01135}} (\bibinfo
  {year} {2025})},\ \Eprint {https://arxiv.org/abs/2508.01135}
  {arXiv:2508.01135 [astro-ph.SR]} \BibitemShut {NoStop}%
\bibitem [{\citenamefont {{K{\i}ro{\u{g}}lu}}\ \emph
  {et~al.}(2025)\citenamefont {{K{\i}ro{\u{g}}lu}}, \citenamefont {{Kremer}},\
  and\ \citenamefont {{Rasio}}}]{2025ApJ...994L..37K}%
  \BibitemOpen
  \bibfield  {author} {\bibinfo {author} {\bibfnamefont {F.}~\bibnamefont
  {{K{\i}ro{\u{g}}lu}}}, \bibinfo {author} {\bibfnamefont {K.}~\bibnamefont
  {{Kremer}}},\ and\ \bibinfo {author} {\bibfnamefont {F.~A.}\ \bibnamefont
  {{Rasio}}},\ }\href {https://doi.org/10.3847/2041-8213/ae1eeb} {\bibfield
  {journal} {\bibinfo  {journal} {\apjl}\ }\textbf {\bibinfo {volume} {994}},\
  \bibinfo {eid} {L37} (\bibinfo {year} {2025})},\ \Eprint
  {https://arxiv.org/abs/2509.05415} {arXiv:2509.05415 [astro-ph.HE]}
  \BibitemShut {NoStop}%
\bibitem [{\citenamefont {{Bartos}}\ and\ \citenamefont
  {{Haiman}}(2025)}]{2025arXiv250808558B}%
  \BibitemOpen
  \bibfield  {author} {\bibinfo {author} {\bibfnamefont {I.}~\bibnamefont
  {{Bartos}}}\ and\ \bibinfo {author} {\bibfnamefont {Z.}~\bibnamefont
  {{Haiman}}},\ }\href {https://doi.org/10.48550/arXiv.2508.08558} {\bibfield
  {journal} {\bibinfo  {journal} {arXiv e-prints}\ ,\ \bibinfo {eid}
  {arXiv:2508.08558}} (\bibinfo {year} {2025})},\ \Eprint
  {https://arxiv.org/abs/2508.08558} {arXiv:2508.08558 [astro-ph.HE]}
  \BibitemShut {NoStop}%
\bibitem [{\citenamefont {{Li}}\ and\ \citenamefont
  {{Fan}}(2025)}]{2025arXiv250908298L}%
  \BibitemOpen
  \bibfield  {author} {\bibinfo {author} {\bibfnamefont {G.-P.}\ \bibnamefont
  {{Li}}}\ and\ \bibinfo {author} {\bibfnamefont {X.-L.}\ \bibnamefont
  {{Fan}}},\ }\href {https://doi.org/10.48550/arXiv.2509.08298} {\bibfield
  {journal} {\bibinfo  {journal} {arXiv e-prints}\ ,\ \bibinfo {eid}
  {arXiv:2509.08298}} (\bibinfo {year} {2025})},\ \Eprint
  {https://arxiv.org/abs/2509.08298} {arXiv:2509.08298 [astro-ph.HE]}
  \BibitemShut {NoStop}%
\bibitem [{\citenamefont {{Paiella}}\ \emph {et~al.}(2025)\citenamefont
  {{Paiella}}, \citenamefont {{Ugolini}}, \citenamefont {{Spera}},
  \citenamefont {{Branchesi}},\ and\ \citenamefont {{Arca
  Sedda}}}]{paiella_letter_2025}%
  \BibitemOpen
  \bibfield  {author} {\bibinfo {author} {\bibfnamefont {L.}~\bibnamefont
  {{Paiella}}}, \bibinfo {author} {\bibfnamefont {C.}~\bibnamefont
  {{Ugolini}}}, \bibinfo {author} {\bibfnamefont {M.}~\bibnamefont {{Spera}}},
  \bibinfo {author} {\bibfnamefont {M.}~\bibnamefont {{Branchesi}}},\ and\
  \bibinfo {author} {\bibfnamefont {M.}~\bibnamefont {{Arca Sedda}}},\ }\href
  {https://doi.org/10.48550/arXiv.2509.10609} {\bibfield  {journal} {\bibinfo
  {journal} {arXiv e-prints}\ ,\ \bibinfo {eid} {arXiv:2509.10609}} (\bibinfo
  {year} {2025})},\ \Eprint {https://arxiv.org/abs/2509.10609}
  {arXiv:2509.10609 [astro-ph.GA]} \BibitemShut {NoStop}%
\bibitem [{\citenamefont {{Yuan}}\ \emph {et~al.}(2025)\citenamefont {{Yuan}},
  \citenamefont {{Chen}},\ and\ \citenamefont {{Liu}}}]{2025PhRvD.112h1306Y}%
  \BibitemOpen
  \bibfield  {author} {\bibinfo {author} {\bibfnamefont {C.}~\bibnamefont
  {{Yuan}}}, \bibinfo {author} {\bibfnamefont {Z.-C.}\ \bibnamefont {{Chen}}},\
  and\ \bibinfo {author} {\bibfnamefont {L.}~\bibnamefont {{Liu}}},\ }\href
  {https://doi.org/10.1103/2vfn-48kh} {\bibfield  {journal} {\bibinfo
  {journal} {\prd}\ }\textbf {\bibinfo {volume} {112}},\ \bibinfo {eid}
  {L081306} (\bibinfo {year} {2025})},\ \Eprint
  {https://arxiv.org/abs/2507.15701} {arXiv:2507.15701 [astro-ph.CO]}
  \BibitemShut {NoStop}%
\bibitem [{\citenamefont {{De Luca}}\ \emph {et~al.}(2025)\citenamefont {{De
  Luca}}, \citenamefont {{Franciolini}},\ and\ \citenamefont
  {{Riotto}}}]{2025arXiv250809965D}%
  \BibitemOpen
  \bibfield  {author} {\bibinfo {author} {\bibfnamefont {V.}~\bibnamefont {{De
  Luca}}}, \bibinfo {author} {\bibfnamefont {G.}~\bibnamefont
  {{Franciolini}}},\ and\ \bibinfo {author} {\bibfnamefont {A.}~\bibnamefont
  {{Riotto}}},\ }\href {https://doi.org/10.48550/arXiv.2508.09965} {\bibfield
  {journal} {\bibinfo  {journal} {arXiv e-prints}\ ,\ \bibinfo {eid}
  {arXiv:2508.09965}} (\bibinfo {year} {2025})},\ \Eprint
  {https://arxiv.org/abs/2508.09965} {arXiv:2508.09965 [astro-ph.CO]}
  \BibitemShut {NoStop}%
\bibitem [{\citenamefont {{Fishbach}}\ and\ \citenamefont
  {{Holz}}(2017)}]{2017ApJ...851L..25F}%
  \BibitemOpen
  \bibfield  {author} {\bibinfo {author} {\bibfnamefont {M.}~\bibnamefont
  {{Fishbach}}}\ and\ \bibinfo {author} {\bibfnamefont {D.~E.}\ \bibnamefont
  {{Holz}}},\ }\href {https://doi.org/10.3847/2041-8213/aa9bf6} {\bibfield
  {journal} {\bibinfo  {journal} {\apjl}\ }\textbf {\bibinfo {volume} {851}},\
  \bibinfo {eid} {L25} (\bibinfo {year} {2017})},\ \Eprint
  {https://arxiv.org/abs/1709.08584} {arXiv:1709.08584 [astro-ph.HE]}
  \BibitemShut {NoStop}%
\bibitem [{\citenamefont {{Talbot}}\ and\ \citenamefont
  {{Thrane}}(2018)}]{2018ApJ...856..173T}%
  \BibitemOpen
  \bibfield  {author} {\bibinfo {author} {\bibfnamefont {C.}~\bibnamefont
  {{Talbot}}}\ and\ \bibinfo {author} {\bibfnamefont {E.}~\bibnamefont
  {{Thrane}}},\ }\href {https://doi.org/10.3847/1538-4357/aab34c} {\bibfield
  {journal} {\bibinfo  {journal} {\apj}\ }\textbf {\bibinfo {volume} {856}},\
  \bibinfo {eid} {173} (\bibinfo {year} {2018})},\ \Eprint
  {https://arxiv.org/abs/1801.02699} {arXiv:1801.02699 [astro-ph.HE]}
  \BibitemShut {NoStop}%
\bibitem [{\citenamefont {{Roulet}}\ and\ \citenamefont
  {{Zaldarriaga}}(2019)}]{2019MNRAS.484.4216R}%
  \BibitemOpen
  \bibfield  {author} {\bibinfo {author} {\bibfnamefont {J.}~\bibnamefont
  {{Roulet}}}\ and\ \bibinfo {author} {\bibfnamefont {M.}~\bibnamefont
  {{Zaldarriaga}}},\ }\href {https://doi.org/10.1093/mnras/stz226} {\bibfield
  {journal} {\bibinfo  {journal} {\mnras}\ }\textbf {\bibinfo {volume} {484}},\
  \bibinfo {pages} {4216} (\bibinfo {year} {2019})},\ \Eprint
  {https://arxiv.org/abs/1806.10610} {arXiv:1806.10610 [astro-ph.HE]}
  \BibitemShut {NoStop}%
\bibitem [{\citenamefont {{Tiwari}}\ and\ \citenamefont
  {{Fairhurst}}(2021)}]{2021ApJ...913L..19T}%
  \BibitemOpen
  \bibfield  {author} {\bibinfo {author} {\bibfnamefont {V.}~\bibnamefont
  {{Tiwari}}}\ and\ \bibinfo {author} {\bibfnamefont {S.}~\bibnamefont
  {{Fairhurst}}},\ }\href {https://doi.org/10.3847/2041-8213/abfbe7} {\bibfield
   {journal} {\bibinfo  {journal} {\apjl}\ }\textbf {\bibinfo {volume} {913}},\
  \bibinfo {eid} {L19} (\bibinfo {year} {2021})},\ \Eprint
  {https://arxiv.org/abs/2011.04502} {arXiv:2011.04502 [astro-ph.HE]}
  \BibitemShut {NoStop}%
\bibitem [{\citenamefont {{Talbot}}\ and\ \citenamefont
  {{Thrane}}(2017)}]{2017PhRvD..96b3012T}%
  \BibitemOpen
  \bibfield  {author} {\bibinfo {author} {\bibfnamefont {C.}~\bibnamefont
  {{Talbot}}}\ and\ \bibinfo {author} {\bibfnamefont {E.}~\bibnamefont
  {{Thrane}}},\ }\href {https://doi.org/10.1103/PhysRevD.96.023012} {\bibfield
  {journal} {\bibinfo  {journal} {\prd}\ }\textbf {\bibinfo {volume} {96}},\
  \bibinfo {eid} {023012} (\bibinfo {year} {2017})},\ \Eprint
  {https://arxiv.org/abs/1704.08370} {arXiv:1704.08370 [astro-ph.HE]}
  \BibitemShut {NoStop}%
\bibitem [{\citenamefont {{Wysocki}}\ \emph {et~al.}(2019)\citenamefont
  {{Wysocki}}, \citenamefont {{Lange}},\ and\ \citenamefont
  {{O'Shaughnessy}}}]{2019PhRvD.100d3012W}%
  \BibitemOpen
  \bibfield  {author} {\bibinfo {author} {\bibfnamefont {D.}~\bibnamefont
  {{Wysocki}}}, \bibinfo {author} {\bibfnamefont {J.}~\bibnamefont {{Lange}}},\
  and\ \bibinfo {author} {\bibfnamefont {R.}~\bibnamefont {{O'Shaughnessy}}},\
  }\href {https://doi.org/10.1103/PhysRevD.100.043012} {\bibfield  {journal}
  {\bibinfo  {journal} {\prd}\ }\textbf {\bibinfo {volume} {100}},\ \bibinfo
  {eid} {043012} (\bibinfo {year} {2019})},\ \Eprint
  {https://arxiv.org/abs/1805.06442} {arXiv:1805.06442 [gr-qc]} \BibitemShut
  {NoStop}%
\bibitem [{\citenamefont {{Miller}}\ \emph {et~al.}(2020)\citenamefont
  {{Miller}}, \citenamefont {{Callister}},\ and\ \citenamefont
  {{Farr}}}]{2020ApJ...895..128M}%
  \BibitemOpen
  \bibfield  {author} {\bibinfo {author} {\bibfnamefont {S.}~\bibnamefont
  {{Miller}}}, \bibinfo {author} {\bibfnamefont {T.~A.}\ \bibnamefont
  {{Callister}}},\ and\ \bibinfo {author} {\bibfnamefont {W.~M.}\ \bibnamefont
  {{Farr}}},\ }\href {https://doi.org/10.3847/1538-4357/ab80c0} {\bibfield
  {journal} {\bibinfo  {journal} {\apj}\ }\textbf {\bibinfo {volume} {895}},\
  \bibinfo {eid} {128} (\bibinfo {year} {2020})},\ \Eprint
  {https://arxiv.org/abs/2001.06051} {arXiv:2001.06051 [astro-ph.HE]}
  \BibitemShut {NoStop}%
\bibitem [{\citenamefont {{Roulet}}\ \emph {et~al.}(2021)\citenamefont
  {{Roulet}}, \citenamefont {{Chia}}, \citenamefont {{Olsen}}, \citenamefont
  {{Dai}}, \citenamefont {{Venumadhav}}, \citenamefont {{Zackay}},\ and\
  \citenamefont {{Zaldarriaga}}}]{2021PhRvD.104h3010R}%
  \BibitemOpen
  \bibfield  {author} {\bibinfo {author} {\bibfnamefont {J.}~\bibnamefont
  {{Roulet}}}, \bibinfo {author} {\bibfnamefont {H.~S.}\ \bibnamefont
  {{Chia}}}, \bibinfo {author} {\bibfnamefont {S.}~\bibnamefont {{Olsen}}},
  \bibinfo {author} {\bibfnamefont {L.}~\bibnamefont {{Dai}}}, \bibinfo
  {author} {\bibfnamefont {T.}~\bibnamefont {{Venumadhav}}}, \bibinfo {author}
  {\bibfnamefont {B.}~\bibnamefont {{Zackay}}},\ and\ \bibinfo {author}
  {\bibfnamefont {M.}~\bibnamefont {{Zaldarriaga}}},\ }\href
  {https://doi.org/10.1103/PhysRevD.104.083010} {\bibfield  {journal} {\bibinfo
   {journal} {\prd}\ }\textbf {\bibinfo {volume} {104}},\ \bibinfo {eid}
  {083010} (\bibinfo {year} {2021})},\ \Eprint
  {https://arxiv.org/abs/2105.10580} {arXiv:2105.10580 [astro-ph.HE]}
  \BibitemShut {NoStop}%
\bibitem [{\citenamefont {{Golomb}}\ and\ \citenamefont
  {{Talbot}}(2023)}]{2023PhRvD.108j3009G}%
  \BibitemOpen
  \bibfield  {author} {\bibinfo {author} {\bibfnamefont {J.}~\bibnamefont
  {{Golomb}}}\ and\ \bibinfo {author} {\bibfnamefont {C.}~\bibnamefont
  {{Talbot}}},\ }\href {https://doi.org/10.1103/PhysRevD.108.103009} {\bibfield
   {journal} {\bibinfo  {journal} {\prd}\ }\textbf {\bibinfo {volume} {108}},\
  \bibinfo {eid} {103009} (\bibinfo {year} {2023})},\ \Eprint
  {https://arxiv.org/abs/2210.12287} {arXiv:2210.12287 [astro-ph.HE]}
  \BibitemShut {NoStop}%
\bibitem [{\citenamefont {{Roulet}}\ \emph {et~al.}(2020)\citenamefont
  {{Roulet}}, \citenamefont {{Venumadhav}}, \citenamefont {{Zackay}},
  \citenamefont {{Dai}},\ and\ \citenamefont
  {{Zaldarriaga}}}]{2020PhRvD.102l3022R}%
  \BibitemOpen
  \bibfield  {author} {\bibinfo {author} {\bibfnamefont {J.}~\bibnamefont
  {{Roulet}}}, \bibinfo {author} {\bibfnamefont {T.}~\bibnamefont
  {{Venumadhav}}}, \bibinfo {author} {\bibfnamefont {B.}~\bibnamefont
  {{Zackay}}}, \bibinfo {author} {\bibfnamefont {L.}~\bibnamefont {{Dai}}},\
  and\ \bibinfo {author} {\bibfnamefont {M.}~\bibnamefont {{Zaldarriaga}}},\
  }\href {https://doi.org/10.1103/PhysRevD.102.123022} {\bibfield  {journal}
  {\bibinfo  {journal} {\prd}\ }\textbf {\bibinfo {volume} {102}},\ \bibinfo
  {eid} {123022} (\bibinfo {year} {2020})},\ \Eprint
  {https://arxiv.org/abs/2008.07014} {arXiv:2008.07014 [astro-ph.HE]}
  \BibitemShut {NoStop}%
\bibitem [{\citenamefont {{Edelman}}\ \emph {et~al.}(2023)\citenamefont
  {{Edelman}}, \citenamefont {{Farr}},\ and\ \citenamefont
  {{Doctor}}}]{2023ApJ...946...16E}%
  \BibitemOpen
  \bibfield  {author} {\bibinfo {author} {\bibfnamefont {B.}~\bibnamefont
  {{Edelman}}}, \bibinfo {author} {\bibfnamefont {B.}~\bibnamefont {{Farr}}},\
  and\ \bibinfo {author} {\bibfnamefont {Z.}~\bibnamefont {{Doctor}}},\ }\href
  {https://doi.org/10.3847/1538-4357/acb5ed} {\bibfield  {journal} {\bibinfo
  {journal} {\apj}\ }\textbf {\bibinfo {volume} {946}},\ \bibinfo {eid} {16}
  (\bibinfo {year} {2023})},\ \Eprint {https://arxiv.org/abs/2210.12834}
  {arXiv:2210.12834 [astro-ph.HE]} \BibitemShut {NoStop}%
\bibitem [{\citenamefont {{Callister}}\ and\ \citenamefont
  {{Farr}}(2024)}]{2024PhRvX..14b1005C}%
  \BibitemOpen
  \bibfield  {author} {\bibinfo {author} {\bibfnamefont {T.~A.}\ \bibnamefont
  {{Callister}}}\ and\ \bibinfo {author} {\bibfnamefont {W.~M.}\ \bibnamefont
  {{Farr}}},\ }\href {https://doi.org/10.1103/PhysRevX.14.021005} {\bibfield
  {journal} {\bibinfo  {journal} {Physical Review X}\ }\textbf {\bibinfo
  {volume} {14}},\ \bibinfo {eid} {021005} (\bibinfo {year} {2024})},\ \Eprint
  {https://arxiv.org/abs/2302.07289} {arXiv:2302.07289 [astro-ph.HE]}
  \BibitemShut {NoStop}%
\bibitem [{\citenamefont {{Afroz}}\ and\ \citenamefont
  {{Mukherjee}}(2025{\natexlab{a}})}]{2025PhRvD.112b3531A}%
  \BibitemOpen
  \bibfield  {author} {\bibinfo {author} {\bibfnamefont {S.}~\bibnamefont
  {{Afroz}}}\ and\ \bibinfo {author} {\bibfnamefont {S.}~\bibnamefont
  {{Mukherjee}}},\ }\href {https://doi.org/10.1103/7zc2-g9vq} {\bibfield
  {journal} {\bibinfo  {journal} {\prd}\ }\textbf {\bibinfo {volume} {112}},\
  \bibinfo {eid} {023531} (\bibinfo {year} {2025}{\natexlab{a}})},\ \Eprint
  {https://arxiv.org/abs/2411.07304} {arXiv:2411.07304 [astro-ph.HE]}
  \BibitemShut {NoStop}%
\bibitem [{\citenamefont {{Afroz}}\ and\ \citenamefont
  {{Mukherjee}}(2025{\natexlab{b}})}]{2025arXiv250909123A}%
  \BibitemOpen
  \bibfield  {author} {\bibinfo {author} {\bibfnamefont {S.}~\bibnamefont
  {{Afroz}}}\ and\ \bibinfo {author} {\bibfnamefont {S.}~\bibnamefont
  {{Mukherjee}}},\ }\href {https://doi.org/10.48550/arXiv.2509.09123}
  {\bibfield  {journal} {\bibinfo  {journal} {arXiv e-prints}\ ,\ \bibinfo
  {eid} {arXiv:2509.09123}} (\bibinfo {year} {2025}{\natexlab{b}})},\ \Eprint
  {https://arxiv.org/abs/2509.09123} {arXiv:2509.09123 [astro-ph.HE]}
  \BibitemShut {NoStop}%
\bibitem [{\citenamefont {{Ray}}\ \emph {et~al.}(2025)\citenamefont {{Ray}},
  \citenamefont {{Hernandez}}, \citenamefont {{Breivik}},\ and\ \citenamefont
  {{Creighton}}}]{2025ApJ...991...17R}%
  \BibitemOpen
  \bibfield  {author} {\bibinfo {author} {\bibfnamefont {A.}~\bibnamefont
  {{Ray}}}, \bibinfo {author} {\bibfnamefont {I.~M.}\ \bibnamefont
  {{Hernandez}}}, \bibinfo {author} {\bibfnamefont {K.}~\bibnamefont
  {{Breivik}}},\ and\ \bibinfo {author} {\bibfnamefont {J.}~\bibnamefont
  {{Creighton}}},\ }\href {https://doi.org/10.3847/1538-4357/adf22a} {\bibfield
   {journal} {\bibinfo  {journal} {\apj}\ }\textbf {\bibinfo {volume} {991}},\
  \bibinfo {eid} {17} (\bibinfo {year} {2025})},\ \Eprint
  {https://arxiv.org/abs/2404.03166} {arXiv:2404.03166 [astro-ph.HE]}
  \BibitemShut {NoStop}%
\bibitem [{\citenamefont {{Kishore Roy}}\ \emph {et~al.}(2025)\citenamefont
  {{Kishore Roy}}, \citenamefont {{van Son}},\ and\ \citenamefont
  {{Farr}}}]{2025CQGra..42v5008K}%
  \BibitemOpen
  \bibfield  {author} {\bibinfo {author} {\bibfnamefont {S.}~\bibnamefont
  {{Kishore Roy}}}, \bibinfo {author} {\bibfnamefont {L.~A.~C.}\ \bibnamefont
  {{van Son}}},\ and\ \bibinfo {author} {\bibfnamefont {W.~M.}\ \bibnamefont
  {{Farr}}},\ }\href {https://doi.org/10.1088/1361-6382/ae1921} {\bibfield
  {journal} {\bibinfo  {journal} {Classical and Quantum Gravity}\ }\textbf
  {\bibinfo {volume} {42}},\ \bibinfo {eid} {225008} (\bibinfo {year}
  {2025})},\ \Eprint {https://arxiv.org/abs/2507.01086} {arXiv:2507.01086
  [astro-ph.HE]} \BibitemShut {NoStop}%
\bibitem [{\citenamefont {{Farah}}\ \emph {et~al.}(2024)\citenamefont
  {{Farah}}, \citenamefont {{Fishbach}},\ and\ \citenamefont
  {{Holz}}}]{2024ApJ...962...69F}%
  \BibitemOpen
  \bibfield  {author} {\bibinfo {author} {\bibfnamefont {A.~M.}\ \bibnamefont
  {{Farah}}}, \bibinfo {author} {\bibfnamefont {M.}~\bibnamefont
  {{Fishbach}}},\ and\ \bibinfo {author} {\bibfnamefont {D.~E.}\ \bibnamefont
  {{Holz}}},\ }\href {https://doi.org/10.3847/1538-4357/ad0558} {\bibfield
  {journal} {\bibinfo  {journal} {\apj}\ }\textbf {\bibinfo {volume} {962}},\
  \bibinfo {eid} {69} (\bibinfo {year} {2024})},\ \Eprint
  {https://arxiv.org/abs/2308.05102} {arXiv:2308.05102 [astro-ph.HE]}
  \BibitemShut {NoStop}%
\bibitem [{\citenamefont {{Belczynski}}\ \emph {et~al.}(2008)\citenamefont
  {{Belczynski}}, \citenamefont {{Kalogera}}, \citenamefont {{Rasio}},
  \citenamefont {{Taam}}, \citenamefont {{Zezas}}, \citenamefont {{Bulik}},
  \citenamefont {{Maccarone}},\ and\ \citenamefont
  {{Ivanova}}}]{2008ApJS..174..223B}%
  \BibitemOpen
  \bibfield  {author} {\bibinfo {author} {\bibfnamefont {K.}~\bibnamefont
  {{Belczynski}}}, \bibinfo {author} {\bibfnamefont {V.}~\bibnamefont
  {{Kalogera}}}, \bibinfo {author} {\bibfnamefont {F.~A.}\ \bibnamefont
  {{Rasio}}}, \bibinfo {author} {\bibfnamefont {R.~E.}\ \bibnamefont {{Taam}}},
  \bibinfo {author} {\bibfnamefont {A.}~\bibnamefont {{Zezas}}}, \bibinfo
  {author} {\bibfnamefont {T.}~\bibnamefont {{Bulik}}}, \bibinfo {author}
  {\bibfnamefont {T.~J.}\ \bibnamefont {{Maccarone}}},\ and\ \bibinfo {author}
  {\bibfnamefont {N.}~\bibnamefont {{Ivanova}}},\ }\href
  {https://doi.org/10.1086/521026} {\bibfield  {journal} {\bibinfo  {journal}
  {\apjs}\ }\textbf {\bibinfo {volume} {174}},\ \bibinfo {pages} {223}
  (\bibinfo {year} {2008})},\ \Eprint {https://arxiv.org/abs/astro-ph/0511811}
  {arXiv:astro-ph/0511811 [astro-ph]} \BibitemShut {NoStop}%
\bibitem [{\citenamefont {{Hurley}}\ \emph {et~al.}(2002)\citenamefont
  {{Hurley}}, \citenamefont {{Tout}},\ and\ \citenamefont
  {{Pols}}}]{2002MNRAS.329..897H}%
  \BibitemOpen
  \bibfield  {author} {\bibinfo {author} {\bibfnamefont {J.~R.}\ \bibnamefont
  {{Hurley}}}, \bibinfo {author} {\bibfnamefont {C.~A.}\ \bibnamefont
  {{Tout}}},\ and\ \bibinfo {author} {\bibfnamefont {O.~R.}\ \bibnamefont
  {{Pols}}},\ }\href {https://doi.org/10.1046/j.1365-8711.2002.05038.x}
  {\bibfield  {journal} {\bibinfo  {journal} {\mnras}\ }\textbf {\bibinfo
  {volume} {329}},\ \bibinfo {pages} {897} (\bibinfo {year} {2002})},\ \Eprint
  {https://arxiv.org/abs/astro-ph/0201220} {arXiv:astro-ph/0201220 [astro-ph]}
  \BibitemShut {NoStop}%
\bibitem [{\citenamefont {{Antonini}}\ and\ \citenamefont
  {{Gieles}}(2020)}]{2020MNRAS.492.2936A}%
  \BibitemOpen
  \bibfield  {author} {\bibinfo {author} {\bibfnamefont {F.}~\bibnamefont
  {{Antonini}}}\ and\ \bibinfo {author} {\bibfnamefont {M.}~\bibnamefont
  {{Gieles}}},\ }\href {https://doi.org/10.1093/mnras/stz3584} {\bibfield
  {journal} {\bibinfo  {journal} {\mnras}\ }\textbf {\bibinfo {volume} {492}},\
  \bibinfo {pages} {2936} (\bibinfo {year} {2020})},\ \Eprint
  {https://arxiv.org/abs/1906.11855} {arXiv:1906.11855 [astro-ph.HE]}
  \BibitemShut {NoStop}%
\bibitem [{\citenamefont {{Kritos}}\ \emph {et~al.}(2024)\citenamefont
  {{Kritos}}, \citenamefont {{Strokov}}, \citenamefont {{Baibhav}},\ and\
  \citenamefont {{Berti}}}]{2024PhRvD.110d3023K}%
  \BibitemOpen
  \bibfield  {author} {\bibinfo {author} {\bibfnamefont {K.}~\bibnamefont
  {{Kritos}}}, \bibinfo {author} {\bibfnamefont {V.}~\bibnamefont {{Strokov}}},
  \bibinfo {author} {\bibfnamefont {V.}~\bibnamefont {{Baibhav}}},\ and\
  \bibinfo {author} {\bibfnamefont {E.}~\bibnamefont {{Berti}}},\ }\href
  {https://doi.org/10.1103/PhysRevD.110.043023} {\bibfield  {journal} {\bibinfo
   {journal} {\prd}\ }\textbf {\bibinfo {volume} {110}},\ \bibinfo {eid}
  {043023} (\bibinfo {year} {2024})},\ \Eprint
  {https://arxiv.org/abs/2210.10055} {arXiv:2210.10055 [astro-ph.HE]}
  \BibitemShut {NoStop}%
\bibitem [{\citenamefont {{Mapelli}}\ \emph {et~al.}(2021)\citenamefont
  {{Mapelli}}, \citenamefont {{Santoliquido}}, \citenamefont {{Bouffanais}},
  \citenamefont {{Arca Sedda}}, \citenamefont {{Artale}},\ and\ \citenamefont
  {{Ballone}}}]{2021Symm...13.1678M}%
  \BibitemOpen
  \bibfield  {author} {\bibinfo {author} {\bibfnamefont {M.}~\bibnamefont
  {{Mapelli}}}, \bibinfo {author} {\bibfnamefont {F.}~\bibnamefont
  {{Santoliquido}}}, \bibinfo {author} {\bibfnamefont {Y.}~\bibnamefont
  {{Bouffanais}}}, \bibinfo {author} {\bibfnamefont {M.~A.}\ \bibnamefont
  {{Arca Sedda}}}, \bibinfo {author} {\bibfnamefont {M.~C.}\ \bibnamefont
  {{Artale}}},\ and\ \bibinfo {author} {\bibfnamefont {A.}~\bibnamefont
  {{Ballone}}},\ }\href {https://doi.org/10.3390/sym13091678} {\bibfield
  {journal} {\bibinfo  {journal} {Symmetry}\ }\textbf {\bibinfo {volume}
  {13}},\ \bibinfo {pages} {1678} (\bibinfo {year} {2021})},\ \Eprint
  {https://arxiv.org/abs/2007.15022} {arXiv:2007.15022 [astro-ph.HE]}
  \BibitemShut {NoStop}%
\bibitem [{\citenamefont {{Antonini}}\ \emph {et~al.}(2019)\citenamefont
  {{Antonini}}, \citenamefont {{Gieles}},\ and\ \citenamefont
  {{Gualandris}}}]{2019MNRAS.486.5008A}%
  \BibitemOpen
  \bibfield  {author} {\bibinfo {author} {\bibfnamefont {F.}~\bibnamefont
  {{Antonini}}}, \bibinfo {author} {\bibfnamefont {M.}~\bibnamefont
  {{Gieles}}},\ and\ \bibinfo {author} {\bibfnamefont {A.}~\bibnamefont
  {{Gualandris}}},\ }\href {https://doi.org/10.1093/mnras/stz1149} {\bibfield
  {journal} {\bibinfo  {journal} {\mnras}\ }\textbf {\bibinfo {volume} {486}},\
  \bibinfo {pages} {5008} (\bibinfo {year} {2019})},\ \Eprint
  {https://arxiv.org/abs/1811.03640} {arXiv:1811.03640 [astro-ph.HE]}
  \BibitemShut {NoStop}%
\bibitem [{\citenamefont {{Kritos}}\ \emph {et~al.}(2022)\citenamefont
  {{Kritos}}, \citenamefont {{Strokov}}, \citenamefont {{Baibhav}},\ and\
  \citenamefont {{Berti}}}]{2022arXiv221010055K}%
  \BibitemOpen
  \bibfield  {author} {\bibinfo {author} {\bibfnamefont {K.}~\bibnamefont
  {{Kritos}}}, \bibinfo {author} {\bibfnamefont {V.}~\bibnamefont {{Strokov}}},
  \bibinfo {author} {\bibfnamefont {V.}~\bibnamefont {{Baibhav}}},\ and\
  \bibinfo {author} {\bibfnamefont {E.}~\bibnamefont {{Berti}}},\ }\href@noop
  {} {\bibfield  {journal} {\bibinfo  {journal} {arXiv e-prints}\ ,\ \bibinfo
  {eid} {arXiv:2210.10055}} (\bibinfo {year} {2022})},\ \Eprint
  {https://arxiv.org/abs/2210.10055} {arXiv:2210.10055 [astro-ph.HE]}
  \BibitemShut {NoStop}%
\bibitem [{\citenamefont {{Jim{\'e}nez-Forteza}}\ \emph
  {et~al.}(2017)\citenamefont {{Jim{\'e}nez-Forteza}}, \citenamefont
  {{Keitel}}, \citenamefont {{Husa}}, \citenamefont {{Hannam}}, \citenamefont
  {{Khan}},\ and\ \citenamefont {{P{\"u}rrer}}}]{2017PhRvD..95f4024J}%
  \BibitemOpen
  \bibfield  {author} {\bibinfo {author} {\bibfnamefont {X.}~\bibnamefont
  {{Jim{\'e}nez-Forteza}}}, \bibinfo {author} {\bibfnamefont {D.}~\bibnamefont
  {{Keitel}}}, \bibinfo {author} {\bibfnamefont {S.}~\bibnamefont {{Husa}}},
  \bibinfo {author} {\bibfnamefont {M.}~\bibnamefont {{Hannam}}}, \bibinfo
  {author} {\bibfnamefont {S.}~\bibnamefont {{Khan}}},\ and\ \bibinfo {author}
  {\bibfnamefont {M.}~\bibnamefont {{P{\"u}rrer}}},\ }\href
  {https://doi.org/10.1103/PhysRevD.95.064024} {\bibfield  {journal} {\bibinfo
  {journal} {\prd}\ }\textbf {\bibinfo {volume} {95}},\ \bibinfo {eid} {064024}
  (\bibinfo {year} {2017})},\ \Eprint {https://arxiv.org/abs/1611.00332}
  {arXiv:1611.00332 [gr-qc]} \BibitemShut {NoStop}%
\bibitem [{\citenamefont {{Campanelli}}\ \emph {et~al.}(2007)\citenamefont
  {{Campanelli}}, \citenamefont {{Lousto}}, \citenamefont {{Zlochower}},\ and\
  \citenamefont {{Merritt}}}]{2007PhRvL..98w1102C}%
  \BibitemOpen
  \bibfield  {author} {\bibinfo {author} {\bibfnamefont {M.}~\bibnamefont
  {{Campanelli}}}, \bibinfo {author} {\bibfnamefont {C.~O.}\ \bibnamefont
  {{Lousto}}}, \bibinfo {author} {\bibfnamefont {Y.}~\bibnamefont
  {{Zlochower}}},\ and\ \bibinfo {author} {\bibfnamefont {D.}~\bibnamefont
  {{Merritt}}},\ }\href {https://doi.org/10.1103/PhysRevLett.98.231102}
  {\bibfield  {journal} {\bibinfo  {journal} {\prl}\ }\textbf {\bibinfo
  {volume} {98}},\ \bibinfo {eid} {231102} (\bibinfo {year} {2007})},\ \Eprint
  {https://arxiv.org/abs/gr-qc/0702133} {arXiv:gr-qc/0702133 [gr-qc]}
  \BibitemShut {NoStop}%
\bibitem [{\citenamefont {{Gonz{\'a}lez}}\ \emph {et~al.}(2007)\citenamefont
  {{Gonz{\'a}lez}}, \citenamefont {{Sperhake}}, \citenamefont {{Br{\"u}gmann}},
  \citenamefont {{Hannam}},\ and\ \citenamefont
  {{Husa}}}]{2007PhRvL..98i1101G}%
  \BibitemOpen
  \bibfield  {author} {\bibinfo {author} {\bibfnamefont {J.~A.}\ \bibnamefont
  {{Gonz{\'a}lez}}}, \bibinfo {author} {\bibfnamefont {U.}~\bibnamefont
  {{Sperhake}}}, \bibinfo {author} {\bibfnamefont {B.}~\bibnamefont
  {{Br{\"u}gmann}}}, \bibinfo {author} {\bibfnamefont {M.}~\bibnamefont
  {{Hannam}}},\ and\ \bibinfo {author} {\bibfnamefont {S.}~\bibnamefont
  {{Husa}}},\ }\href {https://doi.org/10.1103/PhysRevLett.98.091101} {\bibfield
   {journal} {\bibinfo  {journal} {\prl}\ }\textbf {\bibinfo {volume} {98}},\
  \bibinfo {eid} {091101} (\bibinfo {year} {2007})},\ \Eprint
  {https://arxiv.org/abs/gr-qc/0610154} {arXiv:gr-qc/0610154 [gr-qc]}
  \BibitemShut {NoStop}%
\bibitem [{\citenamefont {{Lousto}}\ and\ \citenamefont
  {{Zlochower}}(2008)}]{2008PhRvD..77d4028L}%
  \BibitemOpen
  \bibfield  {author} {\bibinfo {author} {\bibfnamefont {C.~O.}\ \bibnamefont
  {{Lousto}}}\ and\ \bibinfo {author} {\bibfnamefont {Y.}~\bibnamefont
  {{Zlochower}}},\ }\href {https://doi.org/10.1103/PhysRevD.77.044028}
  {\bibfield  {journal} {\bibinfo  {journal} {\prd}\ }\textbf {\bibinfo
  {volume} {77}},\ \bibinfo {eid} {044028} (\bibinfo {year} {2008})},\ \Eprint
  {https://arxiv.org/abs/0708.4048} {arXiv:0708.4048 [gr-qc]} \BibitemShut
  {NoStop}%
\bibitem [{\citenamefont {{Lousto}}\ \emph {et~al.}(2012)\citenamefont
  {{Lousto}}, \citenamefont {{Zlochower}}, \citenamefont {{Dotti}},\ and\
  \citenamefont {{Volonteri}}}]{2012PhRvD..85h4015L}%
  \BibitemOpen
  \bibfield  {author} {\bibinfo {author} {\bibfnamefont {C.~O.}\ \bibnamefont
  {{Lousto}}}, \bibinfo {author} {\bibfnamefont {Y.}~\bibnamefont
  {{Zlochower}}}, \bibinfo {author} {\bibfnamefont {M.}~\bibnamefont
  {{Dotti}}},\ and\ \bibinfo {author} {\bibfnamefont {M.}~\bibnamefont
  {{Volonteri}}},\ }\href {https://doi.org/10.1103/PhysRevD.85.084015}
  {\bibfield  {journal} {\bibinfo  {journal} {\prd}\ }\textbf {\bibinfo
  {volume} {85}},\ \bibinfo {eid} {084015} (\bibinfo {year} {2012})},\ \Eprint
  {https://arxiv.org/abs/1201.1923} {arXiv:1201.1923 [gr-qc]} \BibitemShut
  {NoStop}%
\bibitem [{\citenamefont {{Spera}}\ \emph {et~al.}(2015)\citenamefont
  {{Spera}}, \citenamefont {{Mapelli}},\ and\ \citenamefont
  {{Bressan}}}]{2015MNRAS.451.4086S}%
  \BibitemOpen
  \bibfield  {author} {\bibinfo {author} {\bibfnamefont {M.}~\bibnamefont
  {{Spera}}}, \bibinfo {author} {\bibfnamefont {M.}~\bibnamefont {{Mapelli}}},\
  and\ \bibinfo {author} {\bibfnamefont {A.}~\bibnamefont {{Bressan}}},\ }\href
  {https://doi.org/10.1093/mnras/stv1161} {\bibfield  {journal} {\bibinfo
  {journal} {\mnras}\ }\textbf {\bibinfo {volume} {451}},\ \bibinfo {pages}
  {4086} (\bibinfo {year} {2015})},\ \Eprint {https://arxiv.org/abs/1505.05201}
  {arXiv:1505.05201 [astro-ph.SR]} \BibitemShut {NoStop}%
\bibitem [{\citenamefont {{Woosley}}\ and\ \citenamefont
  {{Heger}}(2021)}]{2021ApJ...912L..31W}%
  \BibitemOpen
  \bibfield  {author} {\bibinfo {author} {\bibfnamefont {S.~E.}\ \bibnamefont
  {{Woosley}}}\ and\ \bibinfo {author} {\bibfnamefont {A.}~\bibnamefont
  {{Heger}}},\ }\href {https://doi.org/10.3847/2041-8213/abf2c4} {\bibfield
  {journal} {\bibinfo  {journal} {\apjl}\ }\textbf {\bibinfo {volume} {912}},\
  \bibinfo {eid} {L31} (\bibinfo {year} {2021})},\ \Eprint
  {https://arxiv.org/abs/2103.07933} {arXiv:2103.07933 [astro-ph.SR]}
  \BibitemShut {NoStop}%
\bibitem [{\citenamefont {{Di Carlo}}\ \emph {et~al.}(2020)\citenamefont {{Di
  Carlo}}, \citenamefont {{Mapelli}}, \citenamefont {{Bouffanais}},
  \citenamefont {{Giacobbo}}, \citenamefont {{Santoliquido}}, \citenamefont
  {{Bressan}}, \citenamefont {{Spera}},\ and\ \citenamefont
  {{Haardt}}}]{2020MNRAS.497.1043D}%
  \BibitemOpen
  \bibfield  {author} {\bibinfo {author} {\bibfnamefont {U.~N.}\ \bibnamefont
  {{Di Carlo}}}, \bibinfo {author} {\bibfnamefont {M.}~\bibnamefont
  {{Mapelli}}}, \bibinfo {author} {\bibfnamefont {Y.}~\bibnamefont
  {{Bouffanais}}}, \bibinfo {author} {\bibfnamefont {N.}~\bibnamefont
  {{Giacobbo}}}, \bibinfo {author} {\bibfnamefont {F.}~\bibnamefont
  {{Santoliquido}}}, \bibinfo {author} {\bibfnamefont {A.}~\bibnamefont
  {{Bressan}}}, \bibinfo {author} {\bibfnamefont {M.}~\bibnamefont {{Spera}}},\
  and\ \bibinfo {author} {\bibfnamefont {F.}~\bibnamefont {{Haardt}}},\ }\href
  {https://doi.org/10.1093/mnras/staa1997} {\bibfield  {journal} {\bibinfo
  {journal} {\mnras}\ }\textbf {\bibinfo {volume} {497}},\ \bibinfo {pages}
  {1043} (\bibinfo {year} {2020})},\ \Eprint {https://arxiv.org/abs/1911.01434}
  {arXiv:1911.01434 [astro-ph.HE]} \BibitemShut {NoStop}%
\bibitem [{\citenamefont {{Kremer}}\ \emph {et~al.}(2020)\citenamefont
  {{Kremer}}, \citenamefont {{Spera}}, \citenamefont {{Becker}}, \citenamefont
  {{Chatterjee}}, \citenamefont {{Di Carlo}}, \citenamefont {{Fragione}},
  \citenamefont {{Rodriguez}}, \citenamefont {{Ye}},\ and\ \citenamefont
  {{Rasio}}}]{2020ApJ...903...45K}%
  \BibitemOpen
  \bibfield  {author} {\bibinfo {author} {\bibfnamefont {K.}~\bibnamefont
  {{Kremer}}}, \bibinfo {author} {\bibfnamefont {M.}~\bibnamefont {{Spera}}},
  \bibinfo {author} {\bibfnamefont {D.}~\bibnamefont {{Becker}}}, \bibinfo
  {author} {\bibfnamefont {S.}~\bibnamefont {{Chatterjee}}}, \bibinfo {author}
  {\bibfnamefont {U.~N.}\ \bibnamefont {{Di Carlo}}}, \bibinfo {author}
  {\bibfnamefont {G.}~\bibnamefont {{Fragione}}}, \bibinfo {author}
  {\bibfnamefont {C.~L.}\ \bibnamefont {{Rodriguez}}}, \bibinfo {author}
  {\bibfnamefont {C.~S.}\ \bibnamefont {{Ye}}},\ and\ \bibinfo {author}
  {\bibfnamefont {F.~A.}\ \bibnamefont {{Rasio}}},\ }\href
  {https://doi.org/10.3847/1538-4357/abb945} {\bibfield  {journal} {\bibinfo
  {journal} {\apj}\ }\textbf {\bibinfo {volume} {903}},\ \bibinfo {eid} {45}
  (\bibinfo {year} {2020})},\ \Eprint {https://arxiv.org/abs/2006.10771}
  {arXiv:2006.10771 [astro-ph.HE]} \BibitemShut {NoStop}%
\bibitem [{\citenamefont {{Kroupa}}(2001)}]{2001MNRAS.322..231K}%
  \BibitemOpen
  \bibfield  {author} {\bibinfo {author} {\bibfnamefont {P.}~\bibnamefont
  {{Kroupa}}},\ }\href {https://doi.org/10.1046/j.1365-8711.2001.04022.x}
  {\bibfield  {journal} {\bibinfo  {journal} {\mnras}\ }\textbf {\bibinfo
  {volume} {322}},\ \bibinfo {pages} {231} (\bibinfo {year} {2001})},\ \Eprint
  {https://arxiv.org/abs/astro-ph/0009005} {arXiv:astro-ph/0009005 [astro-ph]}
  \BibitemShut {NoStop}%
\bibitem [{\citenamefont {{Gonz{\'a}lez}}\ \emph {et~al.}(2021)\citenamefont
  {{Gonz{\'a}lez}}, \citenamefont {{Kremer}}, \citenamefont {{Chatterjee}},
  \citenamefont {{Fragione}}, \citenamefont {{Rodriguez}}, \citenamefont
  {{Weatherford}}, \citenamefont {{Ye}},\ and\ \citenamefont
  {{Rasio}}}]{2021ApJ...908L..29G}%
  \BibitemOpen
  \bibfield  {author} {\bibinfo {author} {\bibfnamefont {E.}~\bibnamefont
  {{Gonz{\'a}lez}}}, \bibinfo {author} {\bibfnamefont {K.}~\bibnamefont
  {{Kremer}}}, \bibinfo {author} {\bibfnamefont {S.}~\bibnamefont
  {{Chatterjee}}}, \bibinfo {author} {\bibfnamefont {G.}~\bibnamefont
  {{Fragione}}}, \bibinfo {author} {\bibfnamefont {C.~L.}\ \bibnamefont
  {{Rodriguez}}}, \bibinfo {author} {\bibfnamefont {N.~C.}\ \bibnamefont
  {{Weatherford}}}, \bibinfo {author} {\bibfnamefont {C.~S.}\ \bibnamefont
  {{Ye}}},\ and\ \bibinfo {author} {\bibfnamefont {F.~A.}\ \bibnamefont
  {{Rasio}}},\ }\href {https://doi.org/10.3847/2041-8213/abdf5b} {\bibfield
  {journal} {\bibinfo  {journal} {\apjl}\ }\textbf {\bibinfo {volume} {908}},\
  \bibinfo {eid} {L29} (\bibinfo {year} {2021})},\ \Eprint
  {https://arxiv.org/abs/2012.10497} {arXiv:2012.10497 [astro-ph.HE]}
  \BibitemShut {NoStop}%
\bibitem [{\citenamefont {{Rizzuto}}\ \emph {et~al.}(2021)\citenamefont
  {{Rizzuto}}, \citenamefont {{Naab}}, \citenamefont {{Spurzem}}, \citenamefont
  {{Giersz}}, \citenamefont {{Ostriker}}, \citenamefont {{Stone}},
  \citenamefont {{Wang}}, \citenamefont {{Berczik}},\ and\ \citenamefont
  {{Rampp}}}]{2021MNRAS.501.5257R}%
  \BibitemOpen
  \bibfield  {author} {\bibinfo {author} {\bibfnamefont {F.~P.}\ \bibnamefont
  {{Rizzuto}}}, \bibinfo {author} {\bibfnamefont {T.}~\bibnamefont {{Naab}}},
  \bibinfo {author} {\bibfnamefont {R.}~\bibnamefont {{Spurzem}}}, \bibinfo
  {author} {\bibfnamefont {M.}~\bibnamefont {{Giersz}}}, \bibinfo {author}
  {\bibfnamefont {J.~P.}\ \bibnamefont {{Ostriker}}}, \bibinfo {author}
  {\bibfnamefont {N.~C.}\ \bibnamefont {{Stone}}}, \bibinfo {author}
  {\bibfnamefont {L.}~\bibnamefont {{Wang}}}, \bibinfo {author} {\bibfnamefont
  {P.}~\bibnamefont {{Berczik}}},\ and\ \bibinfo {author} {\bibfnamefont
  {M.}~\bibnamefont {{Rampp}}},\ }\href
  {https://doi.org/10.1093/mnras/staa3634} {\bibfield  {journal} {\bibinfo
  {journal} {\mnras}\ }\textbf {\bibinfo {volume} {501}},\ \bibinfo {pages}
  {5257} (\bibinfo {year} {2021})},\ \Eprint {https://arxiv.org/abs/2008.09571}
  {arXiv:2008.09571 [astro-ph.GA]} \BibitemShut {NoStop}%
\bibitem [{\citenamefont {{Arca Sedda}}\ \emph
  {et~al.}(2023{\natexlab{b}})\citenamefont {{Arca Sedda}}, \citenamefont
  {{Kamlah}}, \citenamefont {{Spurzem}}, \citenamefont {{Rizzuto}},
  \citenamefont {{Naab}}, \citenamefont {{Giersz}},\ and\ \citenamefont
  {{Berczik}}}]{2023MNRAS.526..429A}%
  \BibitemOpen
  \bibfield  {author} {\bibinfo {author} {\bibfnamefont {M.}~\bibnamefont
  {{Arca Sedda}}}, \bibinfo {author} {\bibfnamefont {A.~W.~H.}\ \bibnamefont
  {{Kamlah}}}, \bibinfo {author} {\bibfnamefont {R.}~\bibnamefont {{Spurzem}}},
  \bibinfo {author} {\bibfnamefont {F.~P.}\ \bibnamefont {{Rizzuto}}}, \bibinfo
  {author} {\bibfnamefont {T.}~\bibnamefont {{Naab}}}, \bibinfo {author}
  {\bibfnamefont {M.}~\bibnamefont {{Giersz}}},\ and\ \bibinfo {author}
  {\bibfnamefont {P.}~\bibnamefont {{Berczik}}},\ }\href
  {https://doi.org/10.1093/mnras/stad2292} {\bibfield  {journal} {\bibinfo
  {journal} {\mnras}\ }\textbf {\bibinfo {volume} {526}},\ \bibinfo {pages}
  {429} (\bibinfo {year} {2023}{\natexlab{b}})},\ \Eprint
  {https://arxiv.org/abs/2307.04806} {arXiv:2307.04806 [astro-ph.GA]}
  \BibitemShut {NoStop}%
\bibitem [{\citenamefont {{Rantala}}\ \emph {et~al.}(2024)\citenamefont
  {{Rantala}}, \citenamefont {{Naab}},\ and\ \citenamefont
  {{Lah{\'e}n}}}]{2024MNRAS.531.3770R}%
  \BibitemOpen
  \bibfield  {author} {\bibinfo {author} {\bibfnamefont {A.}~\bibnamefont
  {{Rantala}}}, \bibinfo {author} {\bibfnamefont {T.}~\bibnamefont {{Naab}}},\
  and\ \bibinfo {author} {\bibfnamefont {N.}~\bibnamefont {{Lah{\'e}n}}},\
  }\href {https://doi.org/10.1093/mnras/stae1413} {\bibfield  {journal}
  {\bibinfo  {journal} {\mnras}\ }\textbf {\bibinfo {volume} {531}},\ \bibinfo
  {pages} {3770} (\bibinfo {year} {2024})},\ \Eprint
  {https://arxiv.org/abs/2403.10602} {arXiv:2403.10602 [astro-ph.GA]}
  \BibitemShut {NoStop}%
\bibitem [{\citenamefont {{Rantala}}\ \emph {et~al.}(2026)\citenamefont
  {{Rantala}}, \citenamefont {{Naab}}, \citenamefont {{Lah{\'e}n}},
  \citenamefont {{Reuter}}, \citenamefont {{Rampp}}, \citenamefont
  {{Chru{\'s}li{\'n}ska}},\ and\ \citenamefont
  {{Reinoso}}}]{2026arXiv260107917R}%
  \BibitemOpen
  \bibfield  {author} {\bibinfo {author} {\bibfnamefont {A.}~\bibnamefont
  {{Rantala}}}, \bibinfo {author} {\bibfnamefont {T.}~\bibnamefont {{Naab}}},
  \bibinfo {author} {\bibfnamefont {N.}~\bibnamefont {{Lah{\'e}n}}}, \bibinfo
  {author} {\bibfnamefont {K.}~\bibnamefont {{Reuter}}}, \bibinfo {author}
  {\bibfnamefont {M.}~\bibnamefont {{Rampp}}}, \bibinfo {author} {\bibfnamefont
  {M.}~\bibnamefont {{Chru{\'s}li{\'n}ska}}},\ and\ \bibinfo {author}
  {\bibfnamefont {B.}~\bibnamefont {{Reinoso}}},\ }\href
  {https://doi.org/10.48550/arXiv.2601.07917} {\bibfield  {journal} {\bibinfo
  {journal} {arXiv e-prints}\ ,\ \bibinfo {eid} {arXiv:2601.07917}} (\bibinfo
  {year} {2026})},\ \Eprint {https://arxiv.org/abs/2601.07917}
  {arXiv:2601.07917 [astro-ph.GA]} \BibitemShut {NoStop}%
\bibitem [{\citenamefont {{Spera}}\ and\ \citenamefont
  {{Mapelli}}(2017)}]{2017MNRAS.470.4739S}%
  \BibitemOpen
  \bibfield  {author} {\bibinfo {author} {\bibfnamefont {M.}~\bibnamefont
  {{Spera}}}\ and\ \bibinfo {author} {\bibfnamefont {M.}~\bibnamefont
  {{Mapelli}}},\ }\href {https://doi.org/10.1093/mnras/stx1576} {\bibfield
  {journal} {\bibinfo  {journal} {\mnras}\ }\textbf {\bibinfo {volume} {470}},\
  \bibinfo {pages} {4739} (\bibinfo {year} {2017})},\ \Eprint
  {https://arxiv.org/abs/1706.06109} {arXiv:1706.06109 [astro-ph.SR]}
  \BibitemShut {NoStop}%
\bibitem [{\citenamefont {{Fuller}}\ and\ \citenamefont
  {{Ma}}(2019)}]{2019ApJ...881L...1F}%
  \BibitemOpen
  \bibfield  {author} {\bibinfo {author} {\bibfnamefont {J.}~\bibnamefont
  {{Fuller}}}\ and\ \bibinfo {author} {\bibfnamefont {L.}~\bibnamefont
  {{Ma}}},\ }\href {https://doi.org/10.3847/2041-8213/ab339b} {\bibfield
  {journal} {\bibinfo  {journal} {\apjl}\ }\textbf {\bibinfo {volume} {881}},\
  \bibinfo {eid} {L1} (\bibinfo {year} {2019})},\ \Eprint
  {https://arxiv.org/abs/1907.03714} {arXiv:1907.03714 [astro-ph.SR]}
  \BibitemShut {NoStop}%
\bibitem [{\citenamefont {{Abbott}}\ \emph {et~al.}(2023)\citenamefont
  {{Abbott}}, \citenamefont {{Abbott}}, \citenamefont {{Acernese}},
  \citenamefont {{Ackley}}, \citenamefont {{Adams}}, \citenamefont
  {{Adhikari}}, \citenamefont {{Adhikari}}, \citenamefont {{Adya}},
  \citenamefont {{Affeldt}}, \citenamefont {{Agarwal}}, \citenamefont
  {{Agathos}}, \citenamefont {{Agatsuma}}, \citenamefont {{Aggarwal}},
  \citenamefont {{Aguiar}}, \citenamefont {{Aiello}}, \citenamefont {{Ain}},
  \citenamefont {{Ajith}} \emph {et~al.}}]{2023PhRvX..13a1048A}%
  \BibitemOpen
  \bibfield  {author} {\bibinfo {author} {\bibfnamefont {R.}~\bibnamefont
  {{Abbott}}}, \bibinfo {author} {\bibfnamefont {T.~D.}\ \bibnamefont
  {{Abbott}}}, \bibinfo {author} {\bibfnamefont {F.}~\bibnamefont
  {{Acernese}}}, \bibinfo {author} {\bibfnamefont {K.}~\bibnamefont
  {{Ackley}}}, \bibinfo {author} {\bibfnamefont {C.}~\bibnamefont {{Adams}}},
  \bibinfo {author} {\bibfnamefont {N.}~\bibnamefont {{Adhikari}}}, \bibinfo
  {author} {\bibfnamefont {R.~X.}\ \bibnamefont {{Adhikari}}}, \bibinfo
  {author} {\bibfnamefont {V.~B.}\ \bibnamefont {{Adya}}}, \bibinfo {author}
  {\bibfnamefont {C.}~\bibnamefont {{Affeldt}}}, \bibinfo {author}
  {\bibfnamefont {D.}~\bibnamefont {{Agarwal}}}, \bibinfo {author}
  {\bibfnamefont {M.}~\bibnamefont {{Agathos}}}, \bibinfo {author}
  {\bibfnamefont {K.}~\bibnamefont {{Agatsuma}}}, \bibinfo {author}
  {\bibfnamefont {N.}~\bibnamefont {{Aggarwal}}}, \bibinfo {author}
  {\bibfnamefont {O.~D.}\ \bibnamefont {{Aguiar}}}, \bibinfo {author}
  {\bibfnamefont {L.}~\bibnamefont {{Aiello}}}, \bibinfo {author}
  {\bibfnamefont {A.}~\bibnamefont {{Ain}}}, \bibinfo {author} {\bibfnamefont
  {P.}~\bibnamefont {{Ajith}}}, \emph {et~al.},\ }\href
  {https://doi.org/10.1103/PhysRevX.13.011048} {\bibfield  {journal} {\bibinfo
  {journal} {Physical Review X}\ }\textbf {\bibinfo {volume} {13}},\ \bibinfo
  {eid} {011048} (\bibinfo {year} {2023})},\ \Eprint
  {https://arxiv.org/abs/2111.03634} {arXiv:2111.03634 [astro-ph.HE]}
  \BibitemShut {NoStop}%
\bibitem [{\citenamefont {{The LIGO Scientific Collaboration}}\ \emph
  {et~al.}(2025{\natexlab{c}})\citenamefont {{The LIGO Scientific
  Collaboration}}, \citenamefont {{the Virgo Collaboration}}, \citenamefont
  {{the KAGRA Collaboration}}, \citenamefont {{Abac}}, \citenamefont
  {{Abouelfettouh}}, \citenamefont {{Acernese}}, \citenamefont {{Ackley}},\
  and\ \citenamefont {{Adamcewicz}}}]{2025arXiv250818083T}%
  \BibitemOpen
  \bibfield  {author} {\bibinfo {author} {\bibnamefont {{The LIGO Scientific
  Collaboration}}}, \bibinfo {author} {\bibnamefont {{the Virgo
  Collaboration}}}, \bibinfo {author} {\bibnamefont {{the KAGRA
  Collaboration}}}, \bibinfo {author} {\bibfnamefont {A.~G.}\ \bibnamefont
  {{Abac}}}, \bibinfo {author} {\bibfnamefont {I.}~\bibnamefont
  {{Abouelfettouh}}}, \bibinfo {author} {\bibfnamefont {F.}~\bibnamefont
  {{Acernese}}}, \bibinfo {author} {\bibfnamefont {K.}~\bibnamefont
  {{Ackley}}},\ and\ \bibinfo {author} {\bibfnamefont {C.~e.~a.}\ \bibnamefont
  {{Adamcewicz}}},\ }\href {https://doi.org/10.48550/arXiv.2508.18083}
  {\bibfield  {journal} {\bibinfo  {journal} {arXiv e-prints}\ ,\ \bibinfo
  {eid} {arXiv:2508.18083}} (\bibinfo {year} {2025}{\natexlab{c}})},\ \Eprint
  {https://arxiv.org/abs/2508.18083} {arXiv:2508.18083 [astro-ph.HE]}
  \BibitemShut {NoStop}%
\bibitem [{\citenamefont {{Racine}}(2008)}]{2008PhRvD..78d4021R}%
  \BibitemOpen
  \bibfield  {author} {\bibinfo {author} {\bibfnamefont {{\'E}.}~\bibnamefont
  {{Racine}}},\ }\href {https://doi.org/10.1103/PhysRevD.78.044021} {\bibfield
  {journal} {\bibinfo  {journal} {\prd}\ }\textbf {\bibinfo {volume} {78}},\
  \bibinfo {eid} {044021} (\bibinfo {year} {2008})},\ \Eprint
  {https://arxiv.org/abs/0803.1820} {arXiv:0803.1820 [gr-qc]} \BibitemShut
  {NoStop}%
\bibitem [{\citenamefont {{Santamar{\'\i}a}}\ \emph {et~al.}(2010)\citenamefont
  {{Santamar{\'\i}a}}, \citenamefont {{Ohme}}, \citenamefont {{Ajith}},
  \citenamefont {{Br{\"u}gmann}}, \citenamefont {{Dorband}}, \citenamefont
  {{Hannam}}, \citenamefont {{Husa}}, \citenamefont {{M{\"o}sta}},
  \citenamefont {{Pollney}}, \citenamefont {{Reisswig}}, \citenamefont
  {{Robinson}}, \citenamefont {{Seiler}},\ and\ \citenamefont
  {{Krishnan}}}]{2010PhRvD..82f4016S}%
  \BibitemOpen
  \bibfield  {author} {\bibinfo {author} {\bibfnamefont {L.}~\bibnamefont
  {{Santamar{\'\i}a}}}, \bibinfo {author} {\bibfnamefont {F.}~\bibnamefont
  {{Ohme}}}, \bibinfo {author} {\bibfnamefont {P.}~\bibnamefont {{Ajith}}},
  \bibinfo {author} {\bibfnamefont {B.}~\bibnamefont {{Br{\"u}gmann}}},
  \bibinfo {author} {\bibfnamefont {N.}~\bibnamefont {{Dorband}}}, \bibinfo
  {author} {\bibfnamefont {M.}~\bibnamefont {{Hannam}}}, \bibinfo {author}
  {\bibfnamefont {S.}~\bibnamefont {{Husa}}}, \bibinfo {author} {\bibfnamefont
  {P.}~\bibnamefont {{M{\"o}sta}}}, \bibinfo {author} {\bibfnamefont
  {D.}~\bibnamefont {{Pollney}}}, \bibinfo {author} {\bibfnamefont
  {C.}~\bibnamefont {{Reisswig}}}, \bibinfo {author} {\bibfnamefont {E.~L.}\
  \bibnamefont {{Robinson}}}, \bibinfo {author} {\bibfnamefont
  {J.}~\bibnamefont {{Seiler}}},\ and\ \bibinfo {author} {\bibfnamefont
  {B.}~\bibnamefont {{Krishnan}}},\ }\href
  {https://doi.org/10.1103/PhysRevD.82.064016} {\bibfield  {journal} {\bibinfo
  {journal} {\prd}\ }\textbf {\bibinfo {volume} {82}},\ \bibinfo {eid} {064016}
  (\bibinfo {year} {2010})},\ \Eprint {https://arxiv.org/abs/1005.3306}
  {arXiv:1005.3306 [gr-qc]} \BibitemShut {NoStop}%
\bibitem [{\citenamefont {{Ajith}}\ \emph {et~al.}(2011)\citenamefont
  {{Ajith}}, \citenamefont {{Hannam}}, \citenamefont {{Husa}}, \citenamefont
  {{Chen}}, \citenamefont {{Br{\"u}gmann}}, \citenamefont {{Dorband}},
  \citenamefont {{M{\"u}ller}}, \citenamefont {{Ohme}}, \citenamefont
  {{Pollney}}, \citenamefont {{Reisswig}}, \citenamefont {{Santamar{\'\i}a}},\
  and\ \citenamefont {{Seiler}}}]{2011PhRvL.106x1101A}%
  \BibitemOpen
  \bibfield  {author} {\bibinfo {author} {\bibfnamefont {P.}~\bibnamefont
  {{Ajith}}}, \bibinfo {author} {\bibfnamefont {M.}~\bibnamefont {{Hannam}}},
  \bibinfo {author} {\bibfnamefont {S.}~\bibnamefont {{Husa}}}, \bibinfo
  {author} {\bibfnamefont {Y.}~\bibnamefont {{Chen}}}, \bibinfo {author}
  {\bibfnamefont {B.}~\bibnamefont {{Br{\"u}gmann}}}, \bibinfo {author}
  {\bibfnamefont {N.}~\bibnamefont {{Dorband}}}, \bibinfo {author}
  {\bibfnamefont {D.}~\bibnamefont {{M{\"u}ller}}}, \bibinfo {author}
  {\bibfnamefont {F.}~\bibnamefont {{Ohme}}}, \bibinfo {author} {\bibfnamefont
  {D.}~\bibnamefont {{Pollney}}}, \bibinfo {author} {\bibfnamefont
  {C.}~\bibnamefont {{Reisswig}}}, \bibinfo {author} {\bibfnamefont
  {L.}~\bibnamefont {{Santamar{\'\i}a}}},\ and\ \bibinfo {author}
  {\bibfnamefont {J.}~\bibnamefont {{Seiler}}},\ }\href
  {https://doi.org/10.1103/PhysRevLett.106.241101} {\bibfield  {journal}
  {\bibinfo  {journal} {\prl}\ }\textbf {\bibinfo {volume} {106}},\ \bibinfo
  {eid} {241101} (\bibinfo {year} {2011})},\ \Eprint
  {https://arxiv.org/abs/0909.2867} {arXiv:0909.2867 [gr-qc]} \BibitemShut
  {NoStop}%
\bibitem [{\citenamefont {{Schmidt}}\ \emph {et~al.}(2015)\citenamefont
  {{Schmidt}}, \citenamefont {{Ohme}},\ and\ \citenamefont
  {{Hannam}}}]{2015PhRvD..91b4043S}%
  \BibitemOpen
  \bibfield  {author} {\bibinfo {author} {\bibfnamefont {P.}~\bibnamefont
  {{Schmidt}}}, \bibinfo {author} {\bibfnamefont {F.}~\bibnamefont {{Ohme}}},\
  and\ \bibinfo {author} {\bibfnamefont {M.}~\bibnamefont {{Hannam}}},\ }\href
  {https://doi.org/10.1103/PhysRevD.91.024043} {\bibfield  {journal} {\bibinfo
  {journal} {\prd}\ }\textbf {\bibinfo {volume} {91}},\ \bibinfo {eid} {024043}
  (\bibinfo {year} {2015})},\ \Eprint {https://arxiv.org/abs/1408.1810}
  {arXiv:1408.1810 [gr-qc]} \BibitemShut {NoStop}%
\bibitem [{\citenamefont {{Gerosa}}\ \emph {et~al.}(2021)\citenamefont
  {{Gerosa}}, \citenamefont {{Mould}}, \citenamefont {{Gangardt}},
  \citenamefont {{Schmidt}}, \citenamefont {{Pratten}},\ and\ \citenamefont
  {{Thomas}}}]{2021PhRvD.103f4067G}%
  \BibitemOpen
  \bibfield  {author} {\bibinfo {author} {\bibfnamefont {D.}~\bibnamefont
  {{Gerosa}}}, \bibinfo {author} {\bibfnamefont {M.}~\bibnamefont {{Mould}}},
  \bibinfo {author} {\bibfnamefont {D.}~\bibnamefont {{Gangardt}}}, \bibinfo
  {author} {\bibfnamefont {P.}~\bibnamefont {{Schmidt}}}, \bibinfo {author}
  {\bibfnamefont {G.}~\bibnamefont {{Pratten}}},\ and\ \bibinfo {author}
  {\bibfnamefont {L.~M.}\ \bibnamefont {{Thomas}}},\ }\href
  {https://doi.org/10.1103/PhysRevD.103.064067} {\bibfield  {journal} {\bibinfo
   {journal} {\prd}\ }\textbf {\bibinfo {volume} {103}},\ \bibinfo {eid}
  {064067} (\bibinfo {year} {2021})},\ \Eprint
  {https://arxiv.org/abs/2011.11948} {arXiv:2011.11948 [gr-qc]} \BibitemShut
  {NoStop}%
\bibitem [{\citenamefont {{Arca Sedda}}\ \emph {et~al.}(2024)\citenamefont
  {{Arca Sedda}}, \citenamefont {{Kamlah}}, \citenamefont {{Spurzem}},
  \citenamefont {{Giersz}}, \citenamefont {{Berczik}}, \citenamefont
  {{Rastello}}, \citenamefont {{Iorio}}, \citenamefont {{Mapelli}},
  \citenamefont {{Gatto}},\ and\ \citenamefont
  {{Grebel}}}]{2024MNRAS.528.5119A}%
  \BibitemOpen
  \bibfield  {author} {\bibinfo {author} {\bibfnamefont {M.}~\bibnamefont
  {{Arca Sedda}}}, \bibinfo {author} {\bibfnamefont {A.~W.~H.}\ \bibnamefont
  {{Kamlah}}}, \bibinfo {author} {\bibfnamefont {R.}~\bibnamefont {{Spurzem}}},
  \bibinfo {author} {\bibfnamefont {M.}~\bibnamefont {{Giersz}}}, \bibinfo
  {author} {\bibfnamefont {P.}~\bibnamefont {{Berczik}}}, \bibinfo {author}
  {\bibfnamefont {S.}~\bibnamefont {{Rastello}}}, \bibinfo {author}
  {\bibfnamefont {G.}~\bibnamefont {{Iorio}}}, \bibinfo {author} {\bibfnamefont
  {M.}~\bibnamefont {{Mapelli}}}, \bibinfo {author} {\bibfnamefont
  {M.}~\bibnamefont {{Gatto}}},\ and\ \bibinfo {author} {\bibfnamefont {E.~K.}\
  \bibnamefont {{Grebel}}},\ }\href {https://doi.org/10.1093/mnras/stad3952}
  {\bibfield  {journal} {\bibinfo  {journal} {\mnras}\ }\textbf {\bibinfo
  {volume} {528}},\ \bibinfo {pages} {5119} (\bibinfo {year} {2024})},\ \Eprint
  {https://arxiv.org/abs/2307.04805} {arXiv:2307.04805 [astro-ph.GA]}
  \BibitemShut {NoStop}%
\bibitem [{\citenamefont {{Gieles}}\ and\ \citenamefont
  {{Gnedin}}(2023)}]{2023MNRAS.522.5340G}%
  \BibitemOpen
  \bibfield  {author} {\bibinfo {author} {\bibfnamefont {M.}~\bibnamefont
  {{Gieles}}}\ and\ \bibinfo {author} {\bibfnamefont {O.~Y.}\ \bibnamefont
  {{Gnedin}}},\ }\href {https://doi.org/10.1093/mnras/stad1287} {\bibfield
  {journal} {\bibinfo  {journal} {\mnras}\ }\textbf {\bibinfo {volume} {522}},\
  \bibinfo {pages} {5340} (\bibinfo {year} {2023})},\ \Eprint
  {https://arxiv.org/abs/2303.03791} {arXiv:2303.03791 [astro-ph.GA]}
  \BibitemShut {NoStop}%
\bibitem [{\citenamefont {{Madau}}\ and\ \citenamefont
  {{Fragos}}(2017)}]{2017ApJ...840...39M}%
  \BibitemOpen
  \bibfield  {author} {\bibinfo {author} {\bibfnamefont {P.}~\bibnamefont
  {{Madau}}}\ and\ \bibinfo {author} {\bibfnamefont {T.}~\bibnamefont
  {{Fragos}}},\ }\href {https://doi.org/10.3847/1538-4357/aa6af9} {\bibfield
  {journal} {\bibinfo  {journal} {\apj}\ }\textbf {\bibinfo {volume} {840}},\
  \bibinfo {eid} {39} (\bibinfo {year} {2017})},\ \Eprint
  {https://arxiv.org/abs/1606.07887} {arXiv:1606.07887 [astro-ph.GA]}
  \BibitemShut {NoStop}%
\bibitem [{\citenamefont {{Bastian}}(2008)}]{2008MNRAS.390..759B}%
  \BibitemOpen
  \bibfield  {author} {\bibinfo {author} {\bibfnamefont {N.}~\bibnamefont
  {{Bastian}}},\ }\href {https://doi.org/10.1111/j.1365-2966.2008.13775.x}
  {\bibfield  {journal} {\bibinfo  {journal} {\mnras}\ }\textbf {\bibinfo
  {volume} {390}},\ \bibinfo {pages} {759} (\bibinfo {year} {2008})},\ \Eprint
  {https://arxiv.org/abs/0807.4687} {arXiv:0807.4687 [astro-ph]} \BibitemShut
  {NoStop}%
\bibitem [{\citenamefont {{Neumayer}}\ \emph {et~al.}(2020)\citenamefont
  {{Neumayer}}, \citenamefont {{Seth}},\ and\ \citenamefont
  {{B{\"o}ker}}}]{2020A&ARv..28....4N}%
  \BibitemOpen
  \bibfield  {author} {\bibinfo {author} {\bibfnamefont {N.}~\bibnamefont
  {{Neumayer}}}, \bibinfo {author} {\bibfnamefont {A.}~\bibnamefont {{Seth}}},\
  and\ \bibinfo {author} {\bibfnamefont {T.}~\bibnamefont {{B{\"o}ker}}},\
  }\href {https://doi.org/10.1007/s00159-020-00125-0} {\bibfield  {journal}
  {\bibinfo  {journal} {\aapr}\ }\textbf {\bibinfo {volume} {28}},\ \bibinfo
  {eid} {4} (\bibinfo {year} {2020})},\ \Eprint
  {https://arxiv.org/abs/2001.03626} {arXiv:2001.03626 [astro-ph.GA]}
  \BibitemShut {NoStop}%
\bibitem [{\citenamefont {{El-Badry}}\ \emph {et~al.}(2019)\citenamefont
  {{El-Badry}}, \citenamefont {{Quataert}}, \citenamefont {{Weisz}},
  \citenamefont {{Choksi}},\ and\ \citenamefont
  {{Boylan-Kolchin}}}]{2019MNRAS.482.4528E}%
  \BibitemOpen
  \bibfield  {author} {\bibinfo {author} {\bibfnamefont {K.}~\bibnamefont
  {{El-Badry}}}, \bibinfo {author} {\bibfnamefont {E.}~\bibnamefont
  {{Quataert}}}, \bibinfo {author} {\bibfnamefont {D.~R.}\ \bibnamefont
  {{Weisz}}}, \bibinfo {author} {\bibfnamefont {N.}~\bibnamefont {{Choksi}}},\
  and\ \bibinfo {author} {\bibfnamefont {M.}~\bibnamefont {{Boylan-Kolchin}}},\
  }\href {https://doi.org/10.1093/mnras/sty3007} {\bibfield  {journal}
  {\bibinfo  {journal} {\mnras}\ }\textbf {\bibinfo {volume} {482}},\ \bibinfo
  {pages} {4528} (\bibinfo {year} {2019})},\ \Eprint
  {https://arxiv.org/abs/1805.03652} {arXiv:1805.03652 [astro-ph.GA]}
  \BibitemShut {NoStop}%
\bibitem [{\citenamefont {{Tremaine}}\ \emph {et~al.}(1975)\citenamefont
  {{Tremaine}}, \citenamefont {{Ostriker}},\ and\ \citenamefont
  {{Spitzer}}}]{1975ApJ...196..407T}%
  \BibitemOpen
  \bibfield  {author} {\bibinfo {author} {\bibfnamefont {S.~D.}\ \bibnamefont
  {{Tremaine}}}, \bibinfo {author} {\bibfnamefont {J.~P.}\ \bibnamefont
  {{Ostriker}}},\ and\ \bibinfo {author} {\bibfnamefont {L.}~\bibnamefont
  {{Spitzer}}, \bibfnamefont {Jr.}},\ }\href {https://doi.org/10.1086/153422}
  {\bibfield  {journal} {\bibinfo  {journal} {\apj}\ }\textbf {\bibinfo
  {volume} {196}},\ \bibinfo {pages} {407} (\bibinfo {year}
  {1975})}\BibitemShut {NoStop}%
\bibitem [{\citenamefont {{Capuzzo-Dolcetta}}(1993)}]{1993ApJ...415..616C}%
  \BibitemOpen
  \bibfield  {author} {\bibinfo {author} {\bibfnamefont {R.}~\bibnamefont
  {{Capuzzo-Dolcetta}}},\ }\href {https://doi.org/10.1086/173189} {\bibfield
  {journal} {\bibinfo  {journal} {\apj}\ }\textbf {\bibinfo {volume} {415}},\
  \bibinfo {pages} {616} (\bibinfo {year} {1993})},\ \Eprint
  {https://arxiv.org/abs/astro-ph/9301006} {arXiv:astro-ph/9301006 [astro-ph]}
  \BibitemShut {NoStop}%
\bibitem [{\citenamefont {{Gnedin}}\ \emph {et~al.}(2014)\citenamefont
  {{Gnedin}}, \citenamefont {{Ostriker}},\ and\ \citenamefont
  {{Tremaine}}}]{2014ApJ...785...71G}%
  \BibitemOpen
  \bibfield  {author} {\bibinfo {author} {\bibfnamefont {O.~Y.}\ \bibnamefont
  {{Gnedin}}}, \bibinfo {author} {\bibfnamefont {J.~P.}\ \bibnamefont
  {{Ostriker}}},\ and\ \bibinfo {author} {\bibfnamefont {S.}~\bibnamefont
  {{Tremaine}}},\ }\href {https://doi.org/10.1088/0004-637X/785/1/71}
  {\bibfield  {journal} {\bibinfo  {journal} {\apj}\ }\textbf {\bibinfo
  {volume} {785}},\ \bibinfo {eid} {71} (\bibinfo {year} {2014})},\ \Eprint
  {https://arxiv.org/abs/1308.0021} {arXiv:1308.0021 [astro-ph.CO]}
  \BibitemShut {NoStop}%
\bibitem [{\citenamefont {{Arca-Sedda}}\ and\ \citenamefont
  {{Capuzzo-Dolcetta}}(2014)}]{2014MNRAS.444.3738A}%
  \BibitemOpen
  \bibfield  {author} {\bibinfo {author} {\bibfnamefont {M.}~\bibnamefont
  {{Arca-Sedda}}}\ and\ \bibinfo {author} {\bibfnamefont {R.}~\bibnamefont
  {{Capuzzo-Dolcetta}}},\ }\href {https://doi.org/10.1093/mnras/stu1683}
  {\bibfield  {journal} {\bibinfo  {journal} {\mnras}\ }\textbf {\bibinfo
  {volume} {444}},\ \bibinfo {pages} {3738} (\bibinfo {year} {2014})},\ \Eprint
  {https://arxiv.org/abs/1405.7593} {arXiv:1405.7593 [astro-ph.GA]}
  \BibitemShut {NoStop}%
\bibitem [{\citenamefont {{Ma}}\ \emph {et~al.}(2016)\citenamefont {{Ma}},
  \citenamefont {{Hopkins}}, \citenamefont {{Faucher-Gigu{\`e}re}},
  \citenamefont {{Zolman}}, \citenamefont {{Muratov}}, \citenamefont
  {{Kere{\v{s}}}},\ and\ \citenamefont {{Quataert}}}]{2016MNRAS.456.2140M}%
  \BibitemOpen
  \bibfield  {author} {\bibinfo {author} {\bibfnamefont {X.}~\bibnamefont
  {{Ma}}}, \bibinfo {author} {\bibfnamefont {P.~F.}\ \bibnamefont {{Hopkins}}},
  \bibinfo {author} {\bibfnamefont {C.-A.}\ \bibnamefont
  {{Faucher-Gigu{\`e}re}}}, \bibinfo {author} {\bibfnamefont {N.}~\bibnamefont
  {{Zolman}}}, \bibinfo {author} {\bibfnamefont {A.~L.}\ \bibnamefont
  {{Muratov}}}, \bibinfo {author} {\bibfnamefont {D.}~\bibnamefont
  {{Kere{\v{s}}}}},\ and\ \bibinfo {author} {\bibfnamefont {E.}~\bibnamefont
  {{Quataert}}},\ }\href {https://doi.org/10.1093/mnras/stv2659} {\bibfield
  {journal} {\bibinfo  {journal} {\mnras}\ }\textbf {\bibinfo {volume} {456}},\
  \bibinfo {pages} {2140} (\bibinfo {year} {2016})},\ \Eprint
  {https://arxiv.org/abs/1504.02097} {arXiv:1504.02097 [astro-ph.GA]}
  \BibitemShut {NoStop}%
\bibitem [{\citenamefont {{Santoliquido}}\ \emph {et~al.}(2020)\citenamefont
  {{Santoliquido}}, \citenamefont {{Mapelli}}, \citenamefont {{Bouffanais}},
  \citenamefont {{Giacobbo}}, \citenamefont {{Di Carlo}}, \citenamefont
  {{Rastello}}, \citenamefont {{Artale}},\ and\ \citenamefont
  {{Ballone}}}]{2020ApJ...898..152S}%
  \BibitemOpen
  \bibfield  {author} {\bibinfo {author} {\bibfnamefont {F.}~\bibnamefont
  {{Santoliquido}}}, \bibinfo {author} {\bibfnamefont {M.}~\bibnamefont
  {{Mapelli}}}, \bibinfo {author} {\bibfnamefont {Y.}~\bibnamefont
  {{Bouffanais}}}, \bibinfo {author} {\bibfnamefont {N.}~\bibnamefont
  {{Giacobbo}}}, \bibinfo {author} {\bibfnamefont {U.~N.}\ \bibnamefont {{Di
  Carlo}}}, \bibinfo {author} {\bibfnamefont {S.}~\bibnamefont {{Rastello}}},
  \bibinfo {author} {\bibfnamefont {M.~C.}\ \bibnamefont {{Artale}}},\ and\
  \bibinfo {author} {\bibfnamefont {A.}~\bibnamefont {{Ballone}}},\ }\href
  {https://doi.org/10.3847/1538-4357/ab9b78} {\bibfield  {journal} {\bibinfo
  {journal} {\apj}\ }\textbf {\bibinfo {volume} {898}},\ \bibinfo {eid} {152}
  (\bibinfo {year} {2020})},\ \Eprint {https://arxiv.org/abs/2004.09533}
  {arXiv:2004.09533 [astro-ph.HE]} \BibitemShut {NoStop}%
\bibitem [{\citenamefont {{Bouffanais}}\ \emph {et~al.}(2021)\citenamefont
  {{Bouffanais}}, \citenamefont {{Mapelli}}, \citenamefont {{Santoliquido}},
  \citenamefont {{Giacobbo}}, \citenamefont {{Di Carlo}}, \citenamefont
  {{Rastello}}, \citenamefont {{Artale}},\ and\ \citenamefont
  {{Iorio}}}]{2021MNRAS.507.5224B}%
  \BibitemOpen
  \bibfield  {author} {\bibinfo {author} {\bibfnamefont {Y.}~\bibnamefont
  {{Bouffanais}}}, \bibinfo {author} {\bibfnamefont {M.}~\bibnamefont
  {{Mapelli}}}, \bibinfo {author} {\bibfnamefont {F.}~\bibnamefont
  {{Santoliquido}}}, \bibinfo {author} {\bibfnamefont {N.}~\bibnamefont
  {{Giacobbo}}}, \bibinfo {author} {\bibfnamefont {U.~N.}\ \bibnamefont {{Di
  Carlo}}}, \bibinfo {author} {\bibfnamefont {S.}~\bibnamefont {{Rastello}}},
  \bibinfo {author} {\bibfnamefont {M.~C.}\ \bibnamefont {{Artale}}},\ and\
  \bibinfo {author} {\bibfnamefont {G.}~\bibnamefont {{Iorio}}},\ }\href
  {https://doi.org/10.1093/mnras/stab2438} {\bibfield  {journal} {\bibinfo
  {journal} {\mnras}\ }\textbf {\bibinfo {volume} {507}},\ \bibinfo {pages}
  {5224} (\bibinfo {year} {2021})},\ \Eprint {https://arxiv.org/abs/2102.12495}
  {arXiv:2102.12495 [astro-ph.HE]} \BibitemShut {NoStop}%
\bibitem [{\citenamefont {{Planck Collaboration}}\ \emph
  {et~al.}(2020)\citenamefont {{Planck Collaboration}}, \citenamefont
  {{Aghanim}}, \citenamefont {{Akrami}}, \citenamefont {{Ashdown}} \emph
  {et~al.}}]{2020A&A...641A...6P}%
  \BibitemOpen
  \bibfield  {author} {\bibinfo {author} {\bibnamefont {{Planck
  Collaboration}}}, \bibinfo {author} {\bibfnamefont {N.}~\bibnamefont
  {{Aghanim}}}, \bibinfo {author} {\bibfnamefont {Y.}~\bibnamefont {{Akrami}}},
  \bibinfo {author} {\bibfnamefont {M.}~\bibnamefont {{Ashdown}}}, \emph
  {et~al.},\ }\href {https://doi.org/10.1051/0004-6361/201833910} {\bibfield
  {journal} {\bibinfo  {journal} {\aap}\ }\textbf {\bibinfo {volume} {641}},\
  \bibinfo {eid} {A6} (\bibinfo {year} {2020})},\ \Eprint
  {https://arxiv.org/abs/1807.06209} {arXiv:1807.06209 [astro-ph.CO]}
  \BibitemShut {NoStop}%
\bibitem [{\citenamefont {{Fishbach}}\ \emph {et~al.}(2018)\citenamefont
  {{Fishbach}}, \citenamefont {{Holz}},\ and\ \citenamefont
  {{Farr}}}]{2018ApJ...863L..41F}%
  \BibitemOpen
  \bibfield  {author} {\bibinfo {author} {\bibfnamefont {M.}~\bibnamefont
  {{Fishbach}}}, \bibinfo {author} {\bibfnamefont {D.~E.}\ \bibnamefont
  {{Holz}}},\ and\ \bibinfo {author} {\bibfnamefont {W.~M.}\ \bibnamefont
  {{Farr}}},\ }\href {https://doi.org/10.3847/2041-8213/aad800} {\bibfield
  {journal} {\bibinfo  {journal} {\apjl}\ }\textbf {\bibinfo {volume} {863}},\
  \bibinfo {eid} {L41} (\bibinfo {year} {2018})},\ \Eprint
  {https://arxiv.org/abs/1805.10270} {arXiv:1805.10270 [astro-ph.HE]}
  \BibitemShut {NoStop}%
\bibitem [{\citenamefont {{Briel}}\ \emph {et~al.}(2023)\citenamefont
  {{Briel}}, \citenamefont {{Stevance}},\ and\ \citenamefont
  {{Eldridge}}}]{2023MNRAS.520.5724B}%
  \BibitemOpen
  \bibfield  {author} {\bibinfo {author} {\bibfnamefont {M.~M.}\ \bibnamefont
  {{Briel}}}, \bibinfo {author} {\bibfnamefont {H.~F.}\ \bibnamefont
  {{Stevance}}},\ and\ \bibinfo {author} {\bibfnamefont {J.~J.}\ \bibnamefont
  {{Eldridge}}},\ }\href {https://doi.org/10.1093/mnras/stad399} {\bibfield
  {journal} {\bibinfo  {journal} {\mnras}\ }\textbf {\bibinfo {volume} {520}},\
  \bibinfo {pages} {5724} (\bibinfo {year} {2023})},\ \Eprint
  {https://arxiv.org/abs/2206.13842} {arXiv:2206.13842 [astro-ph.HE]}
  \BibitemShut {NoStop}%
\bibitem [{\citenamefont {{Duquennoy}}\ and\ \citenamefont
  {{Mayor}}(1991)}]{1991A&A...248..485D}%
  \BibitemOpen
  \bibfield  {author} {\bibinfo {author} {\bibfnamefont {A.}~\bibnamefont
  {{Duquennoy}}}\ and\ \bibinfo {author} {\bibfnamefont {M.}~\bibnamefont
  {{Mayor}}},\ }\href@noop {} {\bibfield  {journal} {\bibinfo  {journal}
  {\aap}\ }\textbf {\bibinfo {volume} {248}},\ \bibinfo {pages} {485} (\bibinfo
  {year} {1991})}\BibitemShut {NoStop}%
\bibitem [{\citenamefont {{Raghavan}}\ \emph {et~al.}(2010)\citenamefont
  {{Raghavan}}, \citenamefont {{McAlister}}, \citenamefont {{Henry}},
  \citenamefont {{Latham}}, \citenamefont {{Marcy}}, \citenamefont {{Mason}},
  \citenamefont {{Gies}}, \citenamefont {{White}},\ and\ \citenamefont {{ten
  Brummelaar}}}]{2010ApJS..190....1R}%
  \BibitemOpen
  \bibfield  {author} {\bibinfo {author} {\bibfnamefont {D.}~\bibnamefont
  {{Raghavan}}}, \bibinfo {author} {\bibfnamefont {H.~A.}\ \bibnamefont
  {{McAlister}}}, \bibinfo {author} {\bibfnamefont {T.~J.}\ \bibnamefont
  {{Henry}}}, \bibinfo {author} {\bibfnamefont {D.~W.}\ \bibnamefont
  {{Latham}}}, \bibinfo {author} {\bibfnamefont {G.~W.}\ \bibnamefont
  {{Marcy}}}, \bibinfo {author} {\bibfnamefont {B.~D.}\ \bibnamefont
  {{Mason}}}, \bibinfo {author} {\bibfnamefont {D.~R.}\ \bibnamefont {{Gies}}},
  \bibinfo {author} {\bibfnamefont {R.~J.}\ \bibnamefont {{White}}},\ and\
  \bibinfo {author} {\bibfnamefont {T.~A.}\ \bibnamefont {{ten Brummelaar}}},\
  }\href {https://doi.org/10.1088/0067-0049/190/1/1} {\bibfield  {journal}
  {\bibinfo  {journal} {\apjs}\ }\textbf {\bibinfo {volume} {190}},\ \bibinfo
  {pages} {1} (\bibinfo {year} {2010})},\ \Eprint
  {https://arxiv.org/abs/1007.0414} {arXiv:1007.0414 [astro-ph.SR]}
  \BibitemShut {NoStop}%
\bibitem [{\citenamefont {{Sana}}\ \emph {et~al.}(2012)\citenamefont {{Sana}},
  \citenamefont {{de Mink}}, \citenamefont {{de Koter}}, \citenamefont
  {{Langer}}, \citenamefont {{Evans}}, \citenamefont {{Gieles}}, \citenamefont
  {{Gosset}}, \citenamefont {{Izzard}}, \citenamefont {{Le Bouquin}},\ and\
  \citenamefont {{Schneider}}}]{2012Sci...337..444S}%
  \BibitemOpen
  \bibfield  {author} {\bibinfo {author} {\bibfnamefont {H.}~\bibnamefont
  {{Sana}}}, \bibinfo {author} {\bibfnamefont {S.~E.}\ \bibnamefont {{de
  Mink}}}, \bibinfo {author} {\bibfnamefont {A.}~\bibnamefont {{de Koter}}},
  \bibinfo {author} {\bibfnamefont {N.}~\bibnamefont {{Langer}}}, \bibinfo
  {author} {\bibfnamefont {C.~J.}\ \bibnamefont {{Evans}}}, \bibinfo {author}
  {\bibfnamefont {M.}~\bibnamefont {{Gieles}}}, \bibinfo {author}
  {\bibfnamefont {E.}~\bibnamefont {{Gosset}}}, \bibinfo {author}
  {\bibfnamefont {R.~G.}\ \bibnamefont {{Izzard}}}, \bibinfo {author}
  {\bibfnamefont {J.~B.}\ \bibnamefont {{Le Bouquin}}},\ and\ \bibinfo {author}
  {\bibfnamefont {F.~R.~N.}\ \bibnamefont {{Schneider}}},\ }\href
  {https://doi.org/10.1126/science.1223344} {\bibfield  {journal} {\bibinfo
  {journal} {Science}\ }\textbf {\bibinfo {volume} {337}},\ \bibinfo {pages}
  {444} (\bibinfo {year} {2012})},\ \Eprint {https://arxiv.org/abs/1207.6397}
  {arXiv:1207.6397 [astro-ph.SR]} \BibitemShut {NoStop}%
\bibitem [{\citenamefont {{Moe}}\ and\ \citenamefont {{Di
  Stefano}}(2017)}]{2017ApJS..230...15M}%
  \BibitemOpen
  \bibfield  {author} {\bibinfo {author} {\bibfnamefont {M.}~\bibnamefont
  {{Moe}}}\ and\ \bibinfo {author} {\bibfnamefont {R.}~\bibnamefont {{Di
  Stefano}}},\ }\href {https://doi.org/10.3847/1538-4365/aa6fb6} {\bibfield
  {journal} {\bibinfo  {journal} {\apjs}\ }\textbf {\bibinfo {volume} {230}},\
  \bibinfo {eid} {15} (\bibinfo {year} {2017})},\ \Eprint
  {https://arxiv.org/abs/1606.05347} {arXiv:1606.05347 [astro-ph.SR]}
  \BibitemShut {NoStop}%
\bibitem [{\citenamefont {{Tagawa}}\ \emph
  {et~al.}(2020{\natexlab{a}})\citenamefont {{Tagawa}}, \citenamefont
  {{Haiman}},\ and\ \citenamefont {{Kocsis}}}]{2020ApJ...898...25T}%
  \BibitemOpen
  \bibfield  {author} {\bibinfo {author} {\bibfnamefont {H.}~\bibnamefont
  {{Tagawa}}}, \bibinfo {author} {\bibfnamefont {Z.}~\bibnamefont {{Haiman}}},\
  and\ \bibinfo {author} {\bibfnamefont {B.}~\bibnamefont {{Kocsis}}},\ }\href
  {https://doi.org/10.3847/1538-4357/ab9b8c} {\bibfield  {journal} {\bibinfo
  {journal} {\apj}\ }\textbf {\bibinfo {volume} {898}},\ \bibinfo {eid} {25}
  (\bibinfo {year} {2020}{\natexlab{a}})},\ \Eprint
  {https://arxiv.org/abs/1912.08218} {arXiv:1912.08218 [astro-ph.GA]}
  \BibitemShut {NoStop}%
\bibitem [{\citenamefont {{Ford}}\ and\ \citenamefont
  {{McKernan}}(2022)}]{2022MNRAS.517.5827F}%
  \BibitemOpen
  \bibfield  {author} {\bibinfo {author} {\bibfnamefont {K.~E.~S.}\
  \bibnamefont {{Ford}}}\ and\ \bibinfo {author} {\bibfnamefont
  {B.}~\bibnamefont {{McKernan}}},\ }\href
  {https://doi.org/10.1093/mnras/stac2861} {\bibfield  {journal} {\bibinfo
  {journal} {\mnras}\ }\textbf {\bibinfo {volume} {517}},\ \bibinfo {pages}
  {5827} (\bibinfo {year} {2022})},\ \Eprint {https://arxiv.org/abs/2109.03212}
  {arXiv:2109.03212 [astro-ph.HE]} \BibitemShut {NoStop}%
\bibitem [{\citenamefont {{Arca Sedda}}\ \emph
  {et~al.}(2023{\natexlab{c}})\citenamefont {{Arca Sedda}}, \citenamefont
  {{Naoz}},\ and\ \citenamefont {{Kocsis}}}]{2023Univ....9..138A}%
  \BibitemOpen
  \bibfield  {author} {\bibinfo {author} {\bibfnamefont {M.}~\bibnamefont
  {{Arca Sedda}}}, \bibinfo {author} {\bibfnamefont {S.}~\bibnamefont
  {{Naoz}}},\ and\ \bibinfo {author} {\bibfnamefont {B.}~\bibnamefont
  {{Kocsis}}},\ }\href {https://doi.org/10.3390/universe9030138} {\bibfield
  {journal} {\bibinfo  {journal} {Universe}\ }\textbf {\bibinfo {volume} {9}},\
  \bibinfo {eid} {138} (\bibinfo {year} {2023}{\natexlab{c}})},\ \Eprint
  {https://arxiv.org/abs/2302.14071} {arXiv:2302.14071 [astro-ph.GA]}
  \BibitemShut {NoStop}%
\bibitem [{\citenamefont {{Rowan}}\ \emph {et~al.}(2025)\citenamefont
  {{Rowan}}, \citenamefont {{Whitehead}},\ and\ \citenamefont
  {{Kocsis}}}]{2025MNRAS.544.4576R}%
  \BibitemOpen
  \bibfield  {author} {\bibinfo {author} {\bibfnamefont {C.}~\bibnamefont
  {{Rowan}}}, \bibinfo {author} {\bibfnamefont {H.}~\bibnamefont
  {{Whitehead}}},\ and\ \bibinfo {author} {\bibfnamefont {B.}~\bibnamefont
  {{Kocsis}}},\ }\href {https://doi.org/10.1093/mnras/staf1896} {\bibfield
  {journal} {\bibinfo  {journal} {\mnras}\ }\textbf {\bibinfo {volume} {544}},\
  \bibinfo {pages} {4576} (\bibinfo {year} {2025})},\ \Eprint
  {https://arxiv.org/abs/2412.12086} {arXiv:2412.12086 [astro-ph.HE]}
  \BibitemShut {NoStop}%
\bibitem [{\citenamefont {{Branchesi}}\ \emph {et~al.}(2023)\citenamefont
  {{Branchesi}}, \citenamefont {{Maggiore}}, \citenamefont {{Alonso}},
  \citenamefont {{Badger}}, \citenamefont {{Banerjee}} \emph
  {et~al.}}]{2023JCAP...07..068B}%
  \BibitemOpen
  \bibfield  {author} {\bibinfo {author} {\bibfnamefont {M.}~\bibnamefont
  {{Branchesi}}}, \bibinfo {author} {\bibfnamefont {M.}~\bibnamefont
  {{Maggiore}}}, \bibinfo {author} {\bibfnamefont {D.}~\bibnamefont
  {{Alonso}}}, \bibinfo {author} {\bibfnamefont {C.}~\bibnamefont {{Badger}}},
  \bibinfo {author} {\bibfnamefont {B.}~\bibnamefont {{Banerjee}}}, \emph
  {et~al.},\ }\href {https://doi.org/10.1088/1475-7516/2023/07/068} {\bibfield
  {journal} {\bibinfo  {journal} {\jcap}\ }\textbf {\bibinfo {volume} {2023}},\
  \bibinfo {eid} {068} (\bibinfo {year} {2023})},\ \Eprint
  {https://arxiv.org/abs/2303.15923} {arXiv:2303.15923 [gr-qc]} \BibitemShut
  {NoStop}%
\bibitem [{\citenamefont {{Abac}}\ \emph {et~al.}(2025)\citenamefont {{Abac}},
  \citenamefont {{Abramo}}, \citenamefont {{Albanesi}}, \citenamefont
  {{Albertini}}, \citenamefont {{Agapito}}, \citenamefont {{Agathos}},
  \citenamefont {{Albertus}}, \citenamefont {{Andersson}}, \citenamefont
  {{Andrade}}, \citenamefont {{Andreoni}}, \citenamefont {{Angeloni}},
  \citenamefont {{Antonelli}}, \citenamefont {{Antoniadis}}, \citenamefont
  {{Antonini}}, \citenamefont {{Arca Sedda}} \emph
  {et~al.}}]{2025arXiv250312263A}%
  \BibitemOpen
  \bibfield  {author} {\bibinfo {author} {\bibfnamefont {A.}~\bibnamefont
  {{Abac}}}, \bibinfo {author} {\bibfnamefont {R.}~\bibnamefont {{Abramo}}},
  \bibinfo {author} {\bibfnamefont {S.}~\bibnamefont {{Albanesi}}}, \bibinfo
  {author} {\bibfnamefont {A.}~\bibnamefont {{Albertini}}}, \bibinfo {author}
  {\bibfnamefont {A.}~\bibnamefont {{Agapito}}}, \bibinfo {author}
  {\bibfnamefont {M.}~\bibnamefont {{Agathos}}}, \bibinfo {author}
  {\bibfnamefont {C.}~\bibnamefont {{Albertus}}}, \bibinfo {author}
  {\bibfnamefont {N.}~\bibnamefont {{Andersson}}}, \bibinfo {author}
  {\bibfnamefont {T.}~\bibnamefont {{Andrade}}}, \bibinfo {author}
  {\bibfnamefont {I.}~\bibnamefont {{Andreoni}}}, \bibinfo {author}
  {\bibfnamefont {F.}~\bibnamefont {{Angeloni}}}, \bibinfo {author}
  {\bibfnamefont {M.}~\bibnamefont {{Antonelli}}}, \bibinfo {author}
  {\bibfnamefont {J.}~\bibnamefont {{Antoniadis}}}, \bibinfo {author}
  {\bibfnamefont {F.}~\bibnamefont {{Antonini}}}, \bibinfo {author}
  {\bibfnamefont {M.}~\bibnamefont {{Arca Sedda}}}, \emph {et~al.},\ }\href
  {https://doi.org/10.48550/arXiv.2503.12263} {\bibfield  {journal} {\bibinfo
  {journal} {arXiv e-prints}\ ,\ \bibinfo {eid} {arXiv:2503.12263}} (\bibinfo
  {year} {2025})},\ \Eprint {https://arxiv.org/abs/2503.12263}
  {arXiv:2503.12263 [gr-qc]} \BibitemShut {NoStop}%
\bibitem [{\citenamefont {{Baker}}\ \emph {et~al.}(2006)\citenamefont
  {{Baker}}, \citenamefont {{Centrella}}, \citenamefont {{Choi}}, \citenamefont
  {{Koppitz}},\ and\ \citenamefont {{van Meter}}}]{2006PhRvD..73j4002B}%
  \BibitemOpen
  \bibfield  {author} {\bibinfo {author} {\bibfnamefont {J.~G.}\ \bibnamefont
  {{Baker}}}, \bibinfo {author} {\bibfnamefont {J.}~\bibnamefont
  {{Centrella}}}, \bibinfo {author} {\bibfnamefont {D.-I.}\ \bibnamefont
  {{Choi}}}, \bibinfo {author} {\bibfnamefont {M.}~\bibnamefont {{Koppitz}}},\
  and\ \bibinfo {author} {\bibfnamefont {J.}~\bibnamefont {{van Meter}}},\
  }\href {https://doi.org/10.1103/PhysRevD.73.104002} {\bibfield  {journal}
  {\bibinfo  {journal} {\prd}\ }\textbf {\bibinfo {volume} {73}},\ \bibinfo
  {eid} {104002} (\bibinfo {year} {2006})},\ \Eprint
  {https://arxiv.org/abs/gr-qc/0602026} {arXiv:gr-qc/0602026 [gr-qc]}
  \BibitemShut {NoStop}%
\bibitem [{\citenamefont {{Berti}}\ \emph {et~al.}(2007)\citenamefont
  {{Berti}}, \citenamefont {{Cardoso}}, \citenamefont {{Gonzalez}},
  \citenamefont {{Sperhake}}, \citenamefont {{Hannam}}, \citenamefont
  {{Husa}},\ and\ \citenamefont {{Br{\"u}gmann}}}]{2007PhRvD..76f4034B}%
  \BibitemOpen
  \bibfield  {author} {\bibinfo {author} {\bibfnamefont {E.}~\bibnamefont
  {{Berti}}}, \bibinfo {author} {\bibfnamefont {V.}~\bibnamefont {{Cardoso}}},
  \bibinfo {author} {\bibfnamefont {J.~A.}\ \bibnamefont {{Gonzalez}}},
  \bibinfo {author} {\bibfnamefont {U.}~\bibnamefont {{Sperhake}}}, \bibinfo
  {author} {\bibfnamefont {M.}~\bibnamefont {{Hannam}}}, \bibinfo {author}
  {\bibfnamefont {S.}~\bibnamefont {{Husa}}},\ and\ \bibinfo {author}
  {\bibfnamefont {B.}~\bibnamefont {{Br{\"u}gmann}}},\ }\href
  {https://doi.org/10.1103/PhysRevD.76.064034} {\bibfield  {journal} {\bibinfo
  {journal} {\prd}\ }\textbf {\bibinfo {volume} {76}},\ \bibinfo {eid} {064034}
  (\bibinfo {year} {2007})},\ \Eprint {https://arxiv.org/abs/gr-qc/0703053}
  {arXiv:gr-qc/0703053 [gr-qc]} \BibitemShut {NoStop}%
\bibitem [{\citenamefont {{Hinder}}\ \emph {et~al.}(2008)\citenamefont
  {{Hinder}}, \citenamefont {{Vaishnav}}, \citenamefont {{Herrmann}},
  \citenamefont {{Shoemaker}},\ and\ \citenamefont
  {{Laguna}}}]{2008PhRvD..77h1502H}%
  \BibitemOpen
  \bibfield  {author} {\bibinfo {author} {\bibfnamefont {I.}~\bibnamefont
  {{Hinder}}}, \bibinfo {author} {\bibfnamefont {B.}~\bibnamefont
  {{Vaishnav}}}, \bibinfo {author} {\bibfnamefont {F.}~\bibnamefont
  {{Herrmann}}}, \bibinfo {author} {\bibfnamefont {D.~M.}\ \bibnamefont
  {{Shoemaker}}},\ and\ \bibinfo {author} {\bibfnamefont {P.}~\bibnamefont
  {{Laguna}}},\ }\href {https://doi.org/10.1103/PhysRevD.77.081502} {\bibfield
  {journal} {\bibinfo  {journal} {\prd}\ }\textbf {\bibinfo {volume} {77}},\
  \bibinfo {eid} {081502} (\bibinfo {year} {2008})},\ \Eprint
  {https://arxiv.org/abs/0710.5167} {arXiv:0710.5167 [gr-qc]} \BibitemShut
  {NoStop}%
\bibitem [{\citenamefont {{Hofmann}}\ \emph {et~al.}(2016)\citenamefont
  {{Hofmann}}, \citenamefont {{Barausse}},\ and\ \citenamefont
  {{Rezzolla}}}]{2016ApJ...825L..19H}%
  \BibitemOpen
  \bibfield  {author} {\bibinfo {author} {\bibfnamefont {F.}~\bibnamefont
  {{Hofmann}}}, \bibinfo {author} {\bibfnamefont {E.}~\bibnamefont
  {{Barausse}}},\ and\ \bibinfo {author} {\bibfnamefont {L.}~\bibnamefont
  {{Rezzolla}}},\ }\href {https://doi.org/10.3847/2041-8205/825/2/L19}
  {\bibfield  {journal} {\bibinfo  {journal} {\apjl}\ }\textbf {\bibinfo
  {volume} {825}},\ \bibinfo {eid} {L19} (\bibinfo {year} {2016})},\ \Eprint
  {https://arxiv.org/abs/1605.01938} {arXiv:1605.01938 [gr-qc]} \BibitemShut
  {NoStop}%
\bibitem [{\citenamefont {{Fishbach}}\ \emph {et~al.}(2021)\citenamefont
  {{Fishbach}}, \citenamefont {{Doctor}}, \citenamefont {{Callister}},
  \citenamefont {{Edelman}}, \citenamefont {{Ye}}, \citenamefont {{Essick}},
  \citenamefont {{Farr}}, \citenamefont {{Farr}},\ and\ \citenamefont
  {{Holz}}}]{2021ApJ...912...98F}%
  \BibitemOpen
  \bibfield  {author} {\bibinfo {author} {\bibfnamefont {M.}~\bibnamefont
  {{Fishbach}}}, \bibinfo {author} {\bibfnamefont {Z.}~\bibnamefont
  {{Doctor}}}, \bibinfo {author} {\bibfnamefont {T.}~\bibnamefont
  {{Callister}}}, \bibinfo {author} {\bibfnamefont {B.}~\bibnamefont
  {{Edelman}}}, \bibinfo {author} {\bibfnamefont {J.}~\bibnamefont {{Ye}}},
  \bibinfo {author} {\bibfnamefont {R.}~\bibnamefont {{Essick}}}, \bibinfo
  {author} {\bibfnamefont {W.~M.}\ \bibnamefont {{Farr}}}, \bibinfo {author}
  {\bibfnamefont {B.}~\bibnamefont {{Farr}}},\ and\ \bibinfo {author}
  {\bibfnamefont {D.~E.}\ \bibnamefont {{Holz}}},\ }\href
  {https://doi.org/10.3847/1538-4357/abee11} {\bibfield  {journal} {\bibinfo
  {journal} {\apj}\ }\textbf {\bibinfo {volume} {912}},\ \bibinfo {eid} {98}
  (\bibinfo {year} {2021})},\ \Eprint {https://arxiv.org/abs/2101.07699}
  {arXiv:2101.07699 [astro-ph.HE]} \BibitemShut {NoStop}%
\bibitem [{\citenamefont {{Biscoveanu}}\ \emph {et~al.}(2022)\citenamefont
  {{Biscoveanu}}, \citenamefont {{Callister}}, \citenamefont {{Haster}},
  \citenamefont {{Ng}}, \citenamefont {{Vitale}},\ and\ \citenamefont
  {{Farr}}}]{2022ApJ...932L..19B}%
  \BibitemOpen
  \bibfield  {author} {\bibinfo {author} {\bibfnamefont {S.}~\bibnamefont
  {{Biscoveanu}}}, \bibinfo {author} {\bibfnamefont {T.~A.}\ \bibnamefont
  {{Callister}}}, \bibinfo {author} {\bibfnamefont {C.-J.}\ \bibnamefont
  {{Haster}}}, \bibinfo {author} {\bibfnamefont {K.~K.~Y.}\ \bibnamefont
  {{Ng}}}, \bibinfo {author} {\bibfnamefont {S.}~\bibnamefont {{Vitale}}},\
  and\ \bibinfo {author} {\bibfnamefont {W.~M.}\ \bibnamefont {{Farr}}},\
  }\href {https://doi.org/10.3847/2041-8213/ac71a8} {\bibfield  {journal}
  {\bibinfo  {journal} {\apjl}\ }\textbf {\bibinfo {volume} {932}},\ \bibinfo
  {eid} {L19} (\bibinfo {year} {2022})},\ \Eprint
  {https://arxiv.org/abs/2204.01578} {arXiv:2204.01578 [astro-ph.HE]}
  \BibitemShut {NoStop}%
\bibitem [{\citenamefont {{Farah}}\ \emph {et~al.}(2026)\citenamefont
  {{Farah}}, \citenamefont {{Vijaykumar}},\ and\ \citenamefont
  {{Fishbach}}}]{2026arXiv260103456F}%
  \BibitemOpen
  \bibfield  {author} {\bibinfo {author} {\bibfnamefont {A.~M.}\ \bibnamefont
  {{Farah}}}, \bibinfo {author} {\bibfnamefont {A.}~\bibnamefont
  {{Vijaykumar}}},\ and\ \bibinfo {author} {\bibfnamefont {M.}~\bibnamefont
  {{Fishbach}}},\ }\href {https://doi.org/10.48550/arXiv.2601.03456} {\bibfield
   {journal} {\bibinfo  {journal} {arXiv e-prints}\ ,\ \bibinfo {eid}
  {arXiv:2601.03456}} (\bibinfo {year} {2026})},\ \Eprint
  {https://arxiv.org/abs/2601.03456} {arXiv:2601.03456 [astro-ph.HE]}
  \BibitemShut {NoStop}%
\bibitem [{\citenamefont {{Torniamenti}}\ \emph {et~al.}(2024)\citenamefont
  {{Torniamenti}}, \citenamefont {{Mapelli}}, \citenamefont {{P{\'e}rigois}},
  \citenamefont {{Arca Sedda}}, \citenamefont {{Artale}}, \citenamefont
  {{Dall'Amico}},\ and\ \citenamefont {{Vaccaro}}}]{2024A&A...688A.148T}%
  \BibitemOpen
  \bibfield  {author} {\bibinfo {author} {\bibfnamefont {S.}~\bibnamefont
  {{Torniamenti}}}, \bibinfo {author} {\bibfnamefont {M.}~\bibnamefont
  {{Mapelli}}}, \bibinfo {author} {\bibfnamefont {C.}~\bibnamefont
  {{P{\'e}rigois}}}, \bibinfo {author} {\bibfnamefont {M.}~\bibnamefont {{Arca
  Sedda}}}, \bibinfo {author} {\bibfnamefont {M.~C.}\ \bibnamefont {{Artale}}},
  \bibinfo {author} {\bibfnamefont {M.}~\bibnamefont {{Dall'Amico}}},\ and\
  \bibinfo {author} {\bibfnamefont {M.~P.}\ \bibnamefont {{Vaccaro}}},\ }\href
  {https://doi.org/10.1051/0004-6361/202449272} {\bibfield  {journal} {\bibinfo
   {journal} {\aap}\ }\textbf {\bibinfo {volume} {688}},\ \bibinfo {eid} {A148}
  (\bibinfo {year} {2024})},\ \Eprint {https://arxiv.org/abs/2401.14837}
  {arXiv:2401.14837 [astro-ph.HE]} \BibitemShut {NoStop}%
\bibitem [{\citenamefont {{Ye}}\ and\ \citenamefont
  {{Fishbach}}(2024)}]{2024ApJ...967...62Y}%
  \BibitemOpen
  \bibfield  {author} {\bibinfo {author} {\bibfnamefont {C.~S.}\ \bibnamefont
  {{Ye}}}\ and\ \bibinfo {author} {\bibfnamefont {M.}~\bibnamefont
  {{Fishbach}}},\ }\href {https://doi.org/10.3847/1538-4357/ad3ba8} {\bibfield
  {journal} {\bibinfo  {journal} {\apj}\ }\textbf {\bibinfo {volume} {967}},\
  \bibinfo {eid} {62} (\bibinfo {year} {2024})},\ \Eprint
  {https://arxiv.org/abs/2402.12444} {arXiv:2402.12444 [astro-ph.HE]}
  \BibitemShut {NoStop}%
\bibitem [{\citenamefont {{van Son}}\ \emph {et~al.}(2022)\citenamefont {{van
  Son}}, \citenamefont {{de Mink}}, \citenamefont {{Callister}}, \citenamefont
  {{Justham}}, \citenamefont {{Renzo}}, \citenamefont {{Wagg}}, \citenamefont
  {{Broekgaarden}}, \citenamefont {{Kummer}}, \citenamefont {{Pakmor}},\ and\
  \citenamefont {{Mandel}}}]{2022ApJ...931...17V}%
  \BibitemOpen
  \bibfield  {author} {\bibinfo {author} {\bibfnamefont {L.~A.~C.}\
  \bibnamefont {{van Son}}}, \bibinfo {author} {\bibfnamefont {S.~E.}\
  \bibnamefont {{de Mink}}}, \bibinfo {author} {\bibfnamefont {T.}~\bibnamefont
  {{Callister}}}, \bibinfo {author} {\bibfnamefont {S.}~\bibnamefont
  {{Justham}}}, \bibinfo {author} {\bibfnamefont {M.}~\bibnamefont {{Renzo}}},
  \bibinfo {author} {\bibfnamefont {T.}~\bibnamefont {{Wagg}}}, \bibinfo
  {author} {\bibfnamefont {F.~S.}\ \bibnamefont {{Broekgaarden}}}, \bibinfo
  {author} {\bibfnamefont {F.}~\bibnamefont {{Kummer}}}, \bibinfo {author}
  {\bibfnamefont {R.}~\bibnamefont {{Pakmor}}},\ and\ \bibinfo {author}
  {\bibfnamefont {I.}~\bibnamefont {{Mandel}}},\ }\href
  {https://doi.org/10.3847/1538-4357/ac64a3} {\bibfield  {journal} {\bibinfo
  {journal} {\apj}\ }\textbf {\bibinfo {volume} {931}},\ \bibinfo {eid} {17}
  (\bibinfo {year} {2022})},\ \Eprint {https://arxiv.org/abs/2110.01634}
  {arXiv:2110.01634 [astro-ph.HE]} \BibitemShut {NoStop}%
\bibitem [{\citenamefont {{Amaro-Seoane}}\ \emph {et~al.}(2007)\citenamefont
  {{Amaro-Seoane}}, \citenamefont {{Gair}}, \citenamefont {{Freitag}},
  \citenamefont {{Miller}}, \citenamefont {{Mandel}}, \citenamefont
  {{Cutler}},\ and\ \citenamefont {{Babak}}}]{2007CQGra..24R.113A}%
  \BibitemOpen
  \bibfield  {author} {\bibinfo {author} {\bibfnamefont {P.}~\bibnamefont
  {{Amaro-Seoane}}}, \bibinfo {author} {\bibfnamefont {J.~R.}\ \bibnamefont
  {{Gair}}}, \bibinfo {author} {\bibfnamefont {M.}~\bibnamefont {{Freitag}}},
  \bibinfo {author} {\bibfnamefont {M.~C.}\ \bibnamefont {{Miller}}}, \bibinfo
  {author} {\bibfnamefont {I.}~\bibnamefont {{Mandel}}}, \bibinfo {author}
  {\bibfnamefont {C.~J.}\ \bibnamefont {{Cutler}}},\ and\ \bibinfo {author}
  {\bibfnamefont {S.}~\bibnamefont {{Babak}}},\ }\href
  {https://doi.org/10.1088/0264-9381/24/17/R01} {\bibfield  {journal} {\bibinfo
   {journal} {Classical and Quantum Gravity}\ }\textbf {\bibinfo {volume}
  {24}},\ \bibinfo {pages} {R113} (\bibinfo {year} {2007})},\ \Eprint
  {https://arxiv.org/abs/astro-ph/0703495} {arXiv:astro-ph/0703495 [astro-ph]}
  \BibitemShut {NoStop}%
\bibitem [{\citenamefont {{Hughes}}\ and\ \citenamefont
  {{Blandford}}(2003)}]{2003ApJ...585L.101H}%
  \BibitemOpen
  \bibfield  {author} {\bibinfo {author} {\bibfnamefont {S.~A.}\ \bibnamefont
  {{Hughes}}}\ and\ \bibinfo {author} {\bibfnamefont {R.~D.}\ \bibnamefont
  {{Blandford}}},\ }\href {https://doi.org/10.1086/375495} {\bibfield
  {journal} {\bibinfo  {journal} {\apjl}\ }\textbf {\bibinfo {volume} {585}},\
  \bibinfo {pages} {L101} (\bibinfo {year} {2003})},\ \Eprint
  {https://arxiv.org/abs/astro-ph/0208484} {arXiv:astro-ph/0208484 [astro-ph]}
  \BibitemShut {NoStop}%
\bibitem [{\citenamefont {{Banagiri}}\ \emph {et~al.}(2025)\citenamefont
  {{Banagiri}}, \citenamefont {{Thrane}},\ and\ \citenamefont
  {{Lasky}}}]{2025arXiv250915646B}%
  \BibitemOpen
  \bibfield  {author} {\bibinfo {author} {\bibfnamefont {S.}~\bibnamefont
  {{Banagiri}}}, \bibinfo {author} {\bibfnamefont {E.}~\bibnamefont
  {{Thrane}}},\ and\ \bibinfo {author} {\bibfnamefont {P.~D.}\ \bibnamefont
  {{Lasky}}},\ }\href {https://doi.org/10.48550/arXiv.2509.15646} {\bibfield
  {journal} {\bibinfo  {journal} {arXiv e-prints}\ ,\ \bibinfo {eid}
  {arXiv:2509.15646}} (\bibinfo {year} {2025})},\ \Eprint
  {https://arxiv.org/abs/2509.15646} {arXiv:2509.15646 [astro-ph.HE]}
  \BibitemShut {NoStop}%
\bibitem [{\citenamefont {{Tong}}\ \emph {et~al.}(2025)\citenamefont {{Tong}},
  \citenamefont {{Fishbach}}, \citenamefont {{Thrane}}, \citenamefont
  {{Mould}}, \citenamefont {{Callister}}, \citenamefont {{Farah}},
  \citenamefont {{Guttman}}, \citenamefont {{Banagiri}}, \citenamefont
  {{Beltran-Martinez}}, \citenamefont {{Farr}}, \citenamefont {{Galaudage}},
  \citenamefont {{Godfrey}}, \citenamefont {{Heinzel}}, \citenamefont
  {{Kalomenopoulos}}, \citenamefont {{Miller}},\ and\ \citenamefont
  {{Vijaykumar}}}]{2025arXiv250904151T}%
  \BibitemOpen
  \bibfield  {author} {\bibinfo {author} {\bibfnamefont {H.}~\bibnamefont
  {{Tong}}}, \bibinfo {author} {\bibfnamefont {M.}~\bibnamefont {{Fishbach}}},
  \bibinfo {author} {\bibfnamefont {E.}~\bibnamefont {{Thrane}}}, \bibinfo
  {author} {\bibfnamefont {M.}~\bibnamefont {{Mould}}}, \bibinfo {author}
  {\bibfnamefont {T.~A.}\ \bibnamefont {{Callister}}}, \bibinfo {author}
  {\bibfnamefont {A.}~\bibnamefont {{Farah}}}, \bibinfo {author} {\bibfnamefont
  {N.}~\bibnamefont {{Guttman}}}, \bibinfo {author} {\bibfnamefont
  {S.}~\bibnamefont {{Banagiri}}}, \bibinfo {author} {\bibfnamefont
  {D.}~\bibnamefont {{Beltran-Martinez}}}, \bibinfo {author} {\bibfnamefont
  {B.}~\bibnamefont {{Farr}}}, \bibinfo {author} {\bibfnamefont
  {S.}~\bibnamefont {{Galaudage}}}, \bibinfo {author} {\bibfnamefont
  {J.}~\bibnamefont {{Godfrey}}}, \bibinfo {author} {\bibfnamefont
  {J.}~\bibnamefont {{Heinzel}}}, \bibinfo {author} {\bibfnamefont
  {M.}~\bibnamefont {{Kalomenopoulos}}}, \bibinfo {author} {\bibfnamefont
  {S.~J.}\ \bibnamefont {{Miller}}},\ and\ \bibinfo {author} {\bibfnamefont
  {A.}~\bibnamefont {{Vijaykumar}}},\ }\href
  {https://doi.org/10.48550/arXiv.2509.04151} {\bibfield  {journal} {\bibinfo
  {journal} {arXiv e-prints}\ ,\ \bibinfo {eid} {arXiv:2509.04151}} (\bibinfo
  {year} {2025})},\ \Eprint {https://arxiv.org/abs/2509.04151}
  {arXiv:2509.04151 [astro-ph.HE]} \BibitemShut {NoStop}%
\bibitem [{\citenamefont {{Antonini}}\ \emph {et~al.}(2025)\citenamefont
  {{Antonini}}, \citenamefont {{Romero-Shaw}}, \citenamefont {{Callister}},
  \citenamefont {{Dosopoulou}}, \citenamefont {{Chattopadhyay}}, \citenamefont
  {{Gieles}},\ and\ \citenamefont {{Mapelli}}}]{2025arXiv250904637A}%
  \BibitemOpen
  \bibfield  {author} {\bibinfo {author} {\bibfnamefont {F.}~\bibnamefont
  {{Antonini}}}, \bibinfo {author} {\bibfnamefont {I.}~\bibnamefont
  {{Romero-Shaw}}}, \bibinfo {author} {\bibfnamefont {T.}~\bibnamefont
  {{Callister}}}, \bibinfo {author} {\bibfnamefont {F.}~\bibnamefont
  {{Dosopoulou}}}, \bibinfo {author} {\bibfnamefont {D.}~\bibnamefont
  {{Chattopadhyay}}}, \bibinfo {author} {\bibfnamefont {M.}~\bibnamefont
  {{Gieles}}},\ and\ \bibinfo {author} {\bibfnamefont {M.}~\bibnamefont
  {{Mapelli}}},\ }\href {https://doi.org/10.48550/arXiv.2509.04637} {\bibfield
  {journal} {\bibinfo  {journal} {arXiv e-prints}\ ,\ \bibinfo {eid}
  {arXiv:2509.04637}} (\bibinfo {year} {2025})},\ \Eprint
  {https://arxiv.org/abs/2509.04637} {arXiv:2509.04637 [astro-ph.HE]}
  \BibitemShut {NoStop}%
\bibitem [{\citenamefont {{Vijaykumar}}\ \emph {et~al.}(2026)\citenamefont
  {{Vijaykumar}}, \citenamefont {{Farah}},\ and\ \citenamefont
  {{Fishbach}}}]{2026arXiv260103457V}%
  \BibitemOpen
  \bibfield  {author} {\bibinfo {author} {\bibfnamefont {A.}~\bibnamefont
  {{Vijaykumar}}}, \bibinfo {author} {\bibfnamefont {A.~M.}\ \bibnamefont
  {{Farah}}},\ and\ \bibinfo {author} {\bibfnamefont {M.}~\bibnamefont
  {{Fishbach}}},\ }\href {https://doi.org/10.48550/arXiv.2601.03457} {\bibfield
   {journal} {\bibinfo  {journal} {arXiv e-prints}\ ,\ \bibinfo {eid}
  {arXiv:2601.03457}} (\bibinfo {year} {2026})},\ \Eprint
  {https://arxiv.org/abs/2601.03457} {arXiv:2601.03457 [astro-ph.HE]}
  \BibitemShut {NoStop}%
\bibitem [{\citenamefont {{Ray}}\ and\ \citenamefont
  {{Kalogera}}(2025)}]{2025arXiv251018867R}%
  \BibitemOpen
  \bibfield  {author} {\bibinfo {author} {\bibfnamefont {A.}~\bibnamefont
  {{Ray}}}\ and\ \bibinfo {author} {\bibfnamefont {V.}~\bibnamefont
  {{Kalogera}}},\ }\href {https://doi.org/10.48550/arXiv.2510.18867} {\bibfield
   {journal} {\bibinfo  {journal} {arXiv e-prints}\ ,\ \bibinfo {eid}
  {arXiv:2510.18867}} (\bibinfo {year} {2025})},\ \Eprint
  {https://arxiv.org/abs/2510.18867} {arXiv:2510.18867 [astro-ph.HE]}
  \BibitemShut {NoStop}%
\bibitem [{\citenamefont {{Gerosa}}\ and\ \citenamefont
  {{Fishbach}}(2021)}]{2021NatAs...5..749G}%
  \BibitemOpen
  \bibfield  {author} {\bibinfo {author} {\bibfnamefont {D.}~\bibnamefont
  {{Gerosa}}}\ and\ \bibinfo {author} {\bibfnamefont {M.}~\bibnamefont
  {{Fishbach}}},\ }\href {https://doi.org/10.1038/s41550-021-01398-w}
  {\bibfield  {journal} {\bibinfo  {journal} {Nature Astronomy}\ }\textbf
  {\bibinfo {volume} {5}},\ \bibinfo {pages} {749} (\bibinfo {year} {2021})},\
  \Eprint {https://arxiv.org/abs/2105.03439} {arXiv:2105.03439 [astro-ph.HE]}
  \BibitemShut {NoStop}%
\bibitem [{\citenamefont {{Payne}}\ \emph {et~al.}(2024)\citenamefont
  {{Payne}}, \citenamefont {{Kremer}},\ and\ \citenamefont
  {{Zevin}}}]{2024ApJ...966L..16P}%
  \BibitemOpen
  \bibfield  {author} {\bibinfo {author} {\bibfnamefont {E.}~\bibnamefont
  {{Payne}}}, \bibinfo {author} {\bibfnamefont {K.}~\bibnamefont {{Kremer}}},\
  and\ \bibinfo {author} {\bibfnamefont {M.}~\bibnamefont {{Zevin}}},\ }\href
  {https://doi.org/10.3847/2041-8213/ad3e82} {\bibfield  {journal} {\bibinfo
  {journal} {\apjl}\ }\textbf {\bibinfo {volume} {966}},\ \bibinfo {eid} {L16}
  (\bibinfo {year} {2024})},\ \Eprint {https://arxiv.org/abs/2402.15066}
  {arXiv:2402.15066 [gr-qc]} \BibitemShut {NoStop}%
\bibitem [{\citenamefont {{Colpi}}\ \emph {et~al.}(2024)\citenamefont
  {{Colpi}}, \citenamefont {{Danzmann}}, \citenamefont {{Hewitson}},
  \citenamefont {{Holley-Bockelmann}}, \citenamefont {{Jetzer}}, \citenamefont
  {{Nelemans}}, \citenamefont {{Petiteau}} \emph
  {et~al.}}]{2024arXiv240207571C}%
  \BibitemOpen
  \bibfield  {author} {\bibinfo {author} {\bibfnamefont {M.}~\bibnamefont
  {{Colpi}}}, \bibinfo {author} {\bibfnamefont {K.}~\bibnamefont {{Danzmann}}},
  \bibinfo {author} {\bibfnamefont {M.}~\bibnamefont {{Hewitson}}}, \bibinfo
  {author} {\bibfnamefont {K.}~\bibnamefont {{Holley-Bockelmann}}}, \bibinfo
  {author} {\bibfnamefont {P.}~\bibnamefont {{Jetzer}}}, \bibinfo {author}
  {\bibfnamefont {G.}~\bibnamefont {{Nelemans}}}, \bibinfo {author}
  {\bibfnamefont {A.}~\bibnamefont {{Petiteau}}}, \emph {et~al.},\ }\href
  {https://doi.org/10.48550/arXiv.2402.07571} {\bibfield  {journal} {\bibinfo
  {journal} {arXiv e-prints}\ ,\ \bibinfo {eid} {arXiv:2402.07571}} (\bibinfo
  {year} {2024})},\ \Eprint {https://arxiv.org/abs/2402.07571}
  {arXiv:2402.07571 [astro-ph.CO]} \BibitemShut {NoStop}%
\bibitem [{\citenamefont {{Arca Sedda}}\ \emph
  {et~al.}(2020{\natexlab{b}})\citenamefont {{Arca Sedda}}, \citenamefont
  {{Berry}}, \citenamefont {{Jani}}, \citenamefont {{Amaro-Seoane}},
  \citenamefont {{Auclair}}, \citenamefont {{Baird}}, \citenamefont {{Baker}},
  \citenamefont {{Berti}}, \citenamefont {{Breivik}}, \citenamefont
  {{Burrows}}, \citenamefont {{Caprini}}, \citenamefont {{Chen}}, \citenamefont
  {{Doneva}}, \citenamefont {{Ezquiaga}}, \citenamefont {{Saavik Ford}},
  \citenamefont {{Katz}}, \citenamefont {{Kolkowitz}}, \citenamefont
  {{McKernan}}, \citenamefont {{Mueller}}, \citenamefont {{Nardini}},
  \citenamefont {{Pikovski}}, \citenamefont {{Rajendran}}, \citenamefont
  {{Sesana}}, \citenamefont {{Shao}}, \citenamefont {{Tamanini}}, \citenamefont
  {{Vartanyan}}, \citenamefont {{Warburton}}, \citenamefont {{Witek}},
  \citenamefont {{Wong}},\ and\ \citenamefont {{Zevin}}}]{2020CQGra..37u5011A}%
  \BibitemOpen
  \bibfield  {author} {\bibinfo {author} {\bibfnamefont {M.}~\bibnamefont
  {{Arca Sedda}}}, \bibinfo {author} {\bibfnamefont {C.~P.~L.}\ \bibnamefont
  {{Berry}}}, \bibinfo {author} {\bibfnamefont {K.}~\bibnamefont {{Jani}}},
  \bibinfo {author} {\bibfnamefont {P.}~\bibnamefont {{Amaro-Seoane}}},
  \bibinfo {author} {\bibfnamefont {P.}~\bibnamefont {{Auclair}}}, \bibinfo
  {author} {\bibfnamefont {J.}~\bibnamefont {{Baird}}}, \bibinfo {author}
  {\bibfnamefont {T.}~\bibnamefont {{Baker}}}, \bibinfo {author} {\bibfnamefont
  {E.}~\bibnamefont {{Berti}}}, \bibinfo {author} {\bibfnamefont
  {K.}~\bibnamefont {{Breivik}}}, \bibinfo {author} {\bibfnamefont
  {A.}~\bibnamefont {{Burrows}}}, \bibinfo {author} {\bibfnamefont
  {C.}~\bibnamefont {{Caprini}}}, \bibinfo {author} {\bibfnamefont
  {X.}~\bibnamefont {{Chen}}}, \bibinfo {author} {\bibfnamefont
  {D.}~\bibnamefont {{Doneva}}}, \bibinfo {author} {\bibfnamefont {J.~M.}\
  \bibnamefont {{Ezquiaga}}}, \bibinfo {author} {\bibfnamefont {K.~E.}\
  \bibnamefont {{Saavik Ford}}}, \bibinfo {author} {\bibfnamefont {M.~L.}\
  \bibnamefont {{Katz}}}, \bibinfo {author} {\bibfnamefont {S.}~\bibnamefont
  {{Kolkowitz}}}, \bibinfo {author} {\bibfnamefont {B.}~\bibnamefont
  {{McKernan}}}, \bibinfo {author} {\bibfnamefont {G.}~\bibnamefont
  {{Mueller}}}, \bibinfo {author} {\bibfnamefont {G.}~\bibnamefont
  {{Nardini}}}, \bibinfo {author} {\bibfnamefont {I.}~\bibnamefont
  {{Pikovski}}}, \bibinfo {author} {\bibfnamefont {S.}~\bibnamefont
  {{Rajendran}}}, \bibinfo {author} {\bibfnamefont {A.}~\bibnamefont
  {{Sesana}}}, \bibinfo {author} {\bibfnamefont {L.}~\bibnamefont {{Shao}}},
  \bibinfo {author} {\bibfnamefont {N.}~\bibnamefont {{Tamanini}}}, \bibinfo
  {author} {\bibfnamefont {D.}~\bibnamefont {{Vartanyan}}}, \bibinfo {author}
  {\bibfnamefont {N.}~\bibnamefont {{Warburton}}}, \bibinfo {author}
  {\bibfnamefont {H.}~\bibnamefont {{Witek}}}, \bibinfo {author} {\bibfnamefont
  {K.}~\bibnamefont {{Wong}}},\ and\ \bibinfo {author} {\bibfnamefont
  {M.}~\bibnamefont {{Zevin}}},\ }\href
  {https://doi.org/10.1088/1361-6382/abb5c1} {\bibfield  {journal} {\bibinfo
  {journal} {Classical and Quantum Gravity}\ }\textbf {\bibinfo {volume}
  {37}},\ \bibinfo {eid} {215011} (\bibinfo {year} {2020}{\natexlab{b}})},\
  \Eprint {https://arxiv.org/abs/1908.11375} {arXiv:1908.11375 [gr-qc]}
  \BibitemShut {NoStop}%
\bibitem [{\citenamefont {{Ajith}}\ \emph {et~al.}(2025)\citenamefont
  {{Ajith}}, \citenamefont {{Seoane}}, \citenamefont {{Arca Sedda}},
  \citenamefont {{Arcodia}}, \citenamefont {{Badaracco}}, \citenamefont
  {{Banerjee}}, \citenamefont {{Belgacem}}, \citenamefont {{Benetti}},
  \citenamefont {{Benetti}}, \citenamefont {{Bobrick}}, \citenamefont
  {{Bonforte}}, \citenamefont {{Bortolas}} \emph
  {et~al.}}]{2025JCAP...01..108A}%
  \BibitemOpen
  \bibfield  {author} {\bibinfo {author} {\bibfnamefont {P.}~\bibnamefont
  {{Ajith}}}, \bibinfo {author} {\bibfnamefont {P.~A.}\ \bibnamefont
  {{Seoane}}}, \bibinfo {author} {\bibfnamefont {M.}~\bibnamefont {{Arca
  Sedda}}}, \bibinfo {author} {\bibfnamefont {R.}~\bibnamefont {{Arcodia}}},
  \bibinfo {author} {\bibfnamefont {F.}~\bibnamefont {{Badaracco}}}, \bibinfo
  {author} {\bibfnamefont {B.}~\bibnamefont {{Banerjee}}}, \bibinfo {author}
  {\bibfnamefont {E.}~\bibnamefont {{Belgacem}}}, \bibinfo {author}
  {\bibfnamefont {G.}~\bibnamefont {{Benetti}}}, \bibinfo {author}
  {\bibfnamefont {S.}~\bibnamefont {{Benetti}}}, \bibinfo {author}
  {\bibfnamefont {A.}~\bibnamefont {{Bobrick}}}, \bibinfo {author}
  {\bibfnamefont {A.}~\bibnamefont {{Bonforte}}}, \bibinfo {author}
  {\bibfnamefont {E.}~\bibnamefont {{Bortolas}}}, \emph {et~al.},\ }\href
  {https://doi.org/10.1088/1475-7516/2025/01/108} {\bibfield  {journal}
  {\bibinfo  {journal} {\jcap}\ }\textbf {\bibinfo {volume} {2025}},\ \bibinfo
  {eid} {108} (\bibinfo {year} {2025})},\ \Eprint
  {https://arxiv.org/abs/2404.09181} {arXiv:2404.09181 [gr-qc]} \BibitemShut
  {NoStop}%
\bibitem [{\citenamefont {{Rantala}}\ \emph {et~al.}(2025)\citenamefont
  {{Rantala}}, \citenamefont {{Lah{\'e}n}}, \citenamefont {{Naab}},
  \citenamefont {{Escobar}},\ and\ \citenamefont
  {{Iorio}}}]{2025MNRAS.543.2130R}%
  \BibitemOpen
  \bibfield  {author} {\bibinfo {author} {\bibfnamefont {A.}~\bibnamefont
  {{Rantala}}}, \bibinfo {author} {\bibfnamefont {N.}~\bibnamefont
  {{Lah{\'e}n}}}, \bibinfo {author} {\bibfnamefont {T.}~\bibnamefont {{Naab}}},
  \bibinfo {author} {\bibfnamefont {G.~J.}\ \bibnamefont {{Escobar}}},\ and\
  \bibinfo {author} {\bibfnamefont {G.}~\bibnamefont {{Iorio}}},\ }\href
  {https://doi.org/10.1093/mnras/staf1519} {\bibfield  {journal} {\bibinfo
  {journal} {\mnras}\ }\textbf {\bibinfo {volume} {543}},\ \bibinfo {pages}
  {2130} (\bibinfo {year} {2025})},\ \Eprint {https://arxiv.org/abs/2506.04330}
  {arXiv:2506.04330 [astro-ph.GA]} \BibitemShut {NoStop}%
\bibitem [{\citenamefont {{Kalogera}}(2000)}]{2000ApJ...541..319K}%
  \BibitemOpen
  \bibfield  {author} {\bibinfo {author} {\bibfnamefont {V.}~\bibnamefont
  {{Kalogera}}},\ }\href {https://doi.org/10.1086/309400} {\bibfield  {journal}
  {\bibinfo  {journal} {\apj}\ }\textbf {\bibinfo {volume} {541}},\ \bibinfo
  {pages} {319} (\bibinfo {year} {2000})},\ \Eprint
  {https://arxiv.org/abs/astro-ph/9911417} {arXiv:astro-ph/9911417 [astro-ph]}
  \BibitemShut {NoStop}%
\bibitem [{\citenamefont {{Baibhav}}\ and\ \citenamefont
  {{Kalogera}}(2024)}]{2024arXiv241203461B}%
  \BibitemOpen
  \bibfield  {author} {\bibinfo {author} {\bibfnamefont {V.}~\bibnamefont
  {{Baibhav}}}\ and\ \bibinfo {author} {\bibfnamefont {V.}~\bibnamefont
  {{Kalogera}}},\ }\href {https://doi.org/10.48550/arXiv.2412.03461} {\bibfield
   {journal} {\bibinfo  {journal} {arXiv e-prints}\ ,\ \bibinfo {eid}
  {arXiv:2412.03461}} (\bibinfo {year} {2024})},\ \Eprint
  {https://arxiv.org/abs/2412.03461} {arXiv:2412.03461 [astro-ph.HE]}
  \BibitemShut {NoStop}%
\bibitem [{\citenamefont {{Tagawa}}\ \emph
  {et~al.}(2020{\natexlab{b}})\citenamefont {{Tagawa}}, \citenamefont
  {{Haiman}}, \citenamefont {{Bartos}},\ and\ \citenamefont
  {{Kocsis}}}]{2020ApJ...899...26T}%
  \BibitemOpen
  \bibfield  {author} {\bibinfo {author} {\bibfnamefont {H.}~\bibnamefont
  {{Tagawa}}}, \bibinfo {author} {\bibfnamefont {Z.}~\bibnamefont {{Haiman}}},
  \bibinfo {author} {\bibfnamefont {I.}~\bibnamefont {{Bartos}}},\ and\
  \bibinfo {author} {\bibfnamefont {B.}~\bibnamefont {{Kocsis}}},\ }\href
  {https://doi.org/10.3847/1538-4357/aba2cc} {\bibfield  {journal} {\bibinfo
  {journal} {\apj}\ }\textbf {\bibinfo {volume} {899}},\ \bibinfo {eid} {26}
  (\bibinfo {year} {2020}{\natexlab{b}})},\ \Eprint
  {https://arxiv.org/abs/2004.11914} {arXiv:2004.11914 [astro-ph.HE]}
  \BibitemShut {NoStop}%
\bibitem [{\citenamefont {{McKernan}}\ and\ \citenamefont
  {{Ford}}(2024)}]{2024MNRAS.531.3479M}%
  \BibitemOpen
  \bibfield  {author} {\bibinfo {author} {\bibfnamefont {B.}~\bibnamefont
  {{McKernan}}}\ and\ \bibinfo {author} {\bibfnamefont {K.~E.~S.}\ \bibnamefont
  {{Ford}}},\ }\href {https://doi.org/10.1093/mnras/stae1351} {\bibfield
  {journal} {\bibinfo  {journal} {\mnras}\ }\textbf {\bibinfo {volume} {531}},\
  \bibinfo {pages} {3479} (\bibinfo {year} {2024})},\ \Eprint
  {https://arxiv.org/abs/2309.15213} {arXiv:2309.15213 [astro-ph.HE]}
  \BibitemShut {NoStop}%
\bibitem [{\citenamefont {{Fabj}}\ \emph {et~al.}(2025)\citenamefont {{Fabj}},
  \citenamefont {{Tiede}}, \citenamefont {{Rowan}}, \citenamefont {{Pessah}},\
  and\ \citenamefont {{Samsing}}}]{2025arXiv251007952F}%
  \BibitemOpen
  \bibfield  {author} {\bibinfo {author} {\bibfnamefont {G.}~\bibnamefont
  {{Fabj}}}, \bibinfo {author} {\bibfnamefont {C.}~\bibnamefont {{Tiede}}},
  \bibinfo {author} {\bibfnamefont {C.}~\bibnamefont {{Rowan}}}, \bibinfo
  {author} {\bibfnamefont {M.}~\bibnamefont {{Pessah}}},\ and\ \bibinfo
  {author} {\bibfnamefont {J.}~\bibnamefont {{Samsing}}},\ }\href
  {https://doi.org/10.48550/arXiv.2510.07952} {\bibfield  {journal} {\bibinfo
  {journal} {arXiv e-prints}\ ,\ \bibinfo {eid} {arXiv:2510.07952}} (\bibinfo
  {year} {2025})},\ \Eprint {https://arxiv.org/abs/2510.07952}
  {arXiv:2510.07952 [astro-ph.HE]} \BibitemShut {NoStop}%
\bibitem [{\citenamefont {{Loredo}}(2004)}]{2004AIPC..735..195L}%
  \BibitemOpen
  \bibfield  {author} {\bibinfo {author} {\bibfnamefont {T.~J.}\ \bibnamefont
  {{Loredo}}},\ }in\ \href {https://doi.org/10.1063/1.1835214} {\emph {\bibinfo
  {booktitle} {Bayesian Inference and Maximum Entropy Methods in Science and
  Engineering: 24th International Workshop on Bayesian Inference and Maximum
  Entropy Methods in Science and Engineering}}},\ \bibinfo {series} {American
  Institute of Physics Conference Series}, Vol.\ \bibinfo {volume} {735},\
  \bibinfo {editor} {edited by\ \bibinfo {editor} {\bibfnamefont
  {R.}~\bibnamefont {{Fischer}}}, \bibinfo {editor} {\bibfnamefont
  {R.}~\bibnamefont {{Preuss}}},\ and\ \bibinfo {editor} {\bibfnamefont
  {U.~V.}\ \bibnamefont {{Toussaint}}}}\ (\bibinfo  {publisher} {AIP},\
  \bibinfo {year} {2004})\ pp.\ \bibinfo {pages} {195--206},\ \Eprint
  {https://arxiv.org/abs/astro-ph/0409387} {arXiv:astro-ph/0409387 [astro-ph]}
  \BibitemShut {NoStop}%
\bibitem [{\citenamefont {{Mould}}\ \emph {et~al.}(2023)\citenamefont
  {{Mould}}, \citenamefont {{Gerosa}}, \citenamefont {{Dall'Amico}},\ and\
  \citenamefont {{Mapelli}}}]{mould2023}%
  \BibitemOpen
  \bibfield  {author} {\bibinfo {author} {\bibfnamefont {M.}~\bibnamefont
  {{Mould}}}, \bibinfo {author} {\bibfnamefont {D.}~\bibnamefont {{Gerosa}}},
  \bibinfo {author} {\bibfnamefont {M.}~\bibnamefont {{Dall'Amico}}},\ and\
  \bibinfo {author} {\bibfnamefont {M.}~\bibnamefont {{Mapelli}}},\ }\href
  {https://doi.org/10.1093/mnras/stad2502} {\bibfield  {journal} {\bibinfo
  {journal}
  {\href{https://academic.oup.com/mnras/article/525/3/3986/7246896}{\mnras}}\
  }\textbf {\bibinfo {volume} {525}},\ \bibinfo {pages} {3986} (\bibinfo {year}
  {2023})},\ \Eprint {https://arxiv.org/abs/2305.18539} {arXiv:2305.18539
  [astro-ph.HE]} \BibitemShut {NoStop}%
\bibitem [{\citenamefont {{Jeffreys}}(1939)}]{jeffreys1939}%
  \BibitemOpen
  \bibfield  {author} {\bibinfo {author} {\bibfnamefont {H.}~\bibnamefont
  {{Jeffreys}}},\ }\href@noop {} {\emph {\bibinfo {title} {{Theory of
  Probability}}}}\ (\bibinfo {year} {1939})\BibitemShut {NoStop}%
\bibitem [{\citenamefont {{Dupletsa}}\ \emph {et~al.}(2023)\citenamefont
  {{Dupletsa}}, \citenamefont {{Harms}}, \citenamefont {{Banerjee}},
  \citenamefont {{Branchesi}}, \citenamefont {{Goncharov}}, \citenamefont
  {{Maselli}}, \citenamefont {{Oliveira}}, \citenamefont {{Ronchini}},\ and\
  \citenamefont {{Tissino}}}]{2023A&C....4200671D}%
  \BibitemOpen
  \bibfield  {author} {\bibinfo {author} {\bibfnamefont {U.}~\bibnamefont
  {{Dupletsa}}}, \bibinfo {author} {\bibfnamefont {J.}~\bibnamefont {{Harms}}},
  \bibinfo {author} {\bibfnamefont {B.}~\bibnamefont {{Banerjee}}}, \bibinfo
  {author} {\bibfnamefont {M.}~\bibnamefont {{Branchesi}}}, \bibinfo {author}
  {\bibfnamefont {B.}~\bibnamefont {{Goncharov}}}, \bibinfo {author}
  {\bibfnamefont {A.}~\bibnamefont {{Maselli}}}, \bibinfo {author}
  {\bibfnamefont {A.~C.~S.}\ \bibnamefont {{Oliveira}}}, \bibinfo {author}
  {\bibfnamefont {S.}~\bibnamefont {{Ronchini}}},\ and\ \bibinfo {author}
  {\bibfnamefont {J.}~\bibnamefont {{Tissino}}},\ }\href
  {https://doi.org/10.1016/j.ascom.2022.100671} {\bibfield  {journal} {\bibinfo
   {journal} {Astronomy and Computing}\ }\textbf {\bibinfo {volume} {42}},\
  \bibinfo {eid} {100671} (\bibinfo {year} {2023})},\ \Eprint
  {https://arxiv.org/abs/2205.02499} {arXiv:2205.02499 [gr-qc]} \BibitemShut
  {NoStop}%
\bibitem [{\citenamefont {{Thorne}}(1987)}]{1987thyg.book..330T}%
  \BibitemOpen
  \bibfield  {author} {\bibinfo {author} {\bibfnamefont {K.~S.}\ \bibnamefont
  {{Thorne}}},\ }in\ \href@noop {} {\emph {\bibinfo {booktitle} {Three Hundred
  Years of Gravitation}}},\ \bibinfo {editor} {edited by\ \bibinfo {editor}
  {\bibfnamefont {S.~W.}\ \bibnamefont {{Hawking}}}\ and\ \bibinfo {editor}
  {\bibfnamefont {W.}~\bibnamefont {{Israel}}}}\ (\bibinfo {year} {1987})\ pp.\
  \bibinfo {pages} {330--458}\BibitemShut {NoStop}%
\bibitem [{\citenamefont {{Allen}}\ \emph {et~al.}(2012)\citenamefont
  {{Allen}}, \citenamefont {{Anderson}}, \citenamefont {{Brady}}, \citenamefont
  {{Brown}},\ and\ \citenamefont {{Creighton}}}]{2012PhRvD..85l2006A}%
  \BibitemOpen
  \bibfield  {author} {\bibinfo {author} {\bibfnamefont {B.}~\bibnamefont
  {{Allen}}}, \bibinfo {author} {\bibfnamefont {W.~G.}\ \bibnamefont
  {{Anderson}}}, \bibinfo {author} {\bibfnamefont {P.~R.}\ \bibnamefont
  {{Brady}}}, \bibinfo {author} {\bibfnamefont {D.~A.}\ \bibnamefont
  {{Brown}}},\ and\ \bibinfo {author} {\bibfnamefont {J.~D.~E.}\ \bibnamefont
  {{Creighton}}},\ }\href {https://doi.org/10.1103/PhysRevD.85.122006}
  {\bibfield  {journal} {\bibinfo  {journal} {\prd}\ }\textbf {\bibinfo
  {volume} {85}},\ \bibinfo {eid} {122006} (\bibinfo {year} {2012})},\ \Eprint
  {https://arxiv.org/abs/gr-qc/0509116} {arXiv:gr-qc/0509116 [gr-qc]}
  \BibitemShut {NoStop}%
\bibitem [{\citenamefont {Abac}\ \emph {et~al.}(2025)\citenamefont {Abac},
  \citenamefont {Abouelfettouh}, \citenamefont {Acernese}, \citenamefont {The
  LIGO Scientific~Collaboration},\ and\ \citenamefont {the
  KAGRA~Collaboration}}]{Abac_2025}%
  \BibitemOpen
  \bibfield  {author} {\bibinfo {author} {\bibfnamefont {A.~G.}\ \bibnamefont
  {Abac}}, \bibinfo {author} {\bibfnamefont {I.}~\bibnamefont {Abouelfettouh}},
  \bibinfo {author} {\bibfnamefont {F.}~\bibnamefont {Acernese}}, \bibinfo
  {author} {\bibfnamefont {t.~V.~C.}\ \bibnamefont {The LIGO
  Scientific~Collaboration}},\ and\ \bibinfo {author} {\bibnamefont {the
  KAGRA~Collaboration}},\ }\href {https://doi.org/10.3847/2041-8213/ae0d54}
  {\bibfield  {journal} {\bibinfo  {journal} {The Astrophysical Journal
  Letters}\ }\textbf {\bibinfo {volume} {993}},\ \bibinfo {pages} {L21}
  (\bibinfo {year} {2025})}\BibitemShut {NoStop}%
\bibitem [{\citenamefont {{Giacobbo}}\ \emph {et~al.}(2018)\citenamefont
  {{Giacobbo}}, \citenamefont {{Mapelli}},\ and\ \citenamefont
  {{Spera}}}]{2018MNRAS.474.2959G}%
  \BibitemOpen
  \bibfield  {author} {\bibinfo {author} {\bibfnamefont {N.}~\bibnamefont
  {{Giacobbo}}}, \bibinfo {author} {\bibfnamefont {M.}~\bibnamefont
  {{Mapelli}}},\ and\ \bibinfo {author} {\bibfnamefont {M.}~\bibnamefont
  {{Spera}}},\ }\href {https://doi.org/10.1093/mnras/stx2933} {\bibfield
  {journal} {\bibinfo  {journal} {\mnras}\ }\textbf {\bibinfo {volume} {474}},\
  \bibinfo {pages} {2959} (\bibinfo {year} {2018})},\ \Eprint
  {https://arxiv.org/abs/1711.03556} {arXiv:1711.03556 [astro-ph.SR]}
  \BibitemShut {NoStop}%
\bibitem [{\citenamefont {{Giacobbo}}\ and\ \citenamefont
  {{Mapelli}}(2018)}]{2018MNRAS.480.2011G}%
  \BibitemOpen
  \bibfield  {author} {\bibinfo {author} {\bibfnamefont {N.}~\bibnamefont
  {{Giacobbo}}}\ and\ \bibinfo {author} {\bibfnamefont {M.}~\bibnamefont
  {{Mapelli}}},\ }\href {https://doi.org/10.1093/mnras/sty1999} {\bibfield
  {journal} {\bibinfo  {journal} {\mnras}\ }\textbf {\bibinfo {volume} {480}},\
  \bibinfo {pages} {2011} (\bibinfo {year} {2018})},\ \Eprint
  {https://arxiv.org/abs/1806.00001} {arXiv:1806.00001 [astro-ph.HE]}
  \BibitemShut {NoStop}%
\bibitem [{\citenamefont {{Giacobbo}}\ and\ \citenamefont
  {{Mapelli}}(2020)}]{2020ApJ...891..141G}%
  \BibitemOpen
  \bibfield  {author} {\bibinfo {author} {\bibfnamefont {N.}~\bibnamefont
  {{Giacobbo}}}\ and\ \bibinfo {author} {\bibfnamefont {M.}~\bibnamefont
  {{Mapelli}}},\ }\href {https://doi.org/10.3847/1538-4357/ab7335} {\bibfield
  {journal} {\bibinfo  {journal} {\apj}\ }\textbf {\bibinfo {volume} {891}},\
  \bibinfo {eid} {141} (\bibinfo {year} {2020})},\ \Eprint
  {https://arxiv.org/abs/1909.06385} {arXiv:1909.06385 [astro-ph.HE]}
  \BibitemShut {NoStop}%
\bibitem [{\citenamefont {{Fryer}}\ \emph {et~al.}(2012)\citenamefont
  {{Fryer}}, \citenamefont {{Belczynski}}, \citenamefont {{Wiktorowicz}},
  \citenamefont {{Dominik}}, \citenamefont {{Kalogera}},\ and\ \citenamefont
  {{Holz}}}]{2012ApJ...749...91F}%
  \BibitemOpen
  \bibfield  {author} {\bibinfo {author} {\bibfnamefont {C.~L.}\ \bibnamefont
  {{Fryer}}}, \bibinfo {author} {\bibfnamefont {K.}~\bibnamefont
  {{Belczynski}}}, \bibinfo {author} {\bibfnamefont {G.}~\bibnamefont
  {{Wiktorowicz}}}, \bibinfo {author} {\bibfnamefont {M.}~\bibnamefont
  {{Dominik}}}, \bibinfo {author} {\bibfnamefont {V.}~\bibnamefont
  {{Kalogera}}},\ and\ \bibinfo {author} {\bibfnamefont {D.~E.}\ \bibnamefont
  {{Holz}}},\ }\href {https://doi.org/10.1088/0004-637X/749/1/91} {\bibfield
  {journal} {\bibinfo  {journal} {\apj}\ }\textbf {\bibinfo {volume} {749}},\
  \bibinfo {eid} {91} (\bibinfo {year} {2012})},\ \Eprint
  {https://arxiv.org/abs/1110.1726} {arXiv:1110.1726 [astro-ph.SR]}
  \BibitemShut {NoStop}%
\bibitem [{\citenamefont {{Mapelli}}\ \emph {et~al.}(2020)\citenamefont
  {{Mapelli}}, \citenamefont {{Spera}}, \citenamefont {{Montanari}},
  \citenamefont {{Limongi}}, \citenamefont {{Chieffi}}, \citenamefont
  {{Giacobbo}}, \citenamefont {{Bressan}},\ and\ \citenamefont
  {{Bouffanais}}}]{2020ApJ...888...76M}%
  \BibitemOpen
  \bibfield  {author} {\bibinfo {author} {\bibfnamefont {M.}~\bibnamefont
  {{Mapelli}}}, \bibinfo {author} {\bibfnamefont {M.}~\bibnamefont {{Spera}}},
  \bibinfo {author} {\bibfnamefont {E.}~\bibnamefont {{Montanari}}}, \bibinfo
  {author} {\bibfnamefont {M.}~\bibnamefont {{Limongi}}}, \bibinfo {author}
  {\bibfnamefont {A.}~\bibnamefont {{Chieffi}}}, \bibinfo {author}
  {\bibfnamefont {N.}~\bibnamefont {{Giacobbo}}}, \bibinfo {author}
  {\bibfnamefont {A.}~\bibnamefont {{Bressan}}},\ and\ \bibinfo {author}
  {\bibfnamefont {Y.}~\bibnamefont {{Bouffanais}}},\ }\href
  {https://doi.org/10.3847/1538-4357/ab584d} {\bibfield  {journal} {\bibinfo
  {journal} {\apj}\ }\textbf {\bibinfo {volume} {888}},\ \bibinfo {eid} {76}
  (\bibinfo {year} {2020})},\ \Eprint {https://arxiv.org/abs/1909.01371}
  {arXiv:1909.01371 [astro-ph.HE]} \BibitemShut {NoStop}%
\bibitem [{\citenamefont {{Woosley}}(2017)}]{2017ApJ...836..244W}%
  \BibitemOpen
  \bibfield  {author} {\bibinfo {author} {\bibfnamefont {S.~E.}\ \bibnamefont
  {{Woosley}}},\ }\href {https://doi.org/10.3847/1538-4357/836/2/244}
  {\bibfield  {journal} {\bibinfo  {journal} {\apj}\ }\textbf {\bibinfo
  {volume} {836}},\ \bibinfo {eid} {244} (\bibinfo {year} {2017})},\ \Eprint
  {https://arxiv.org/abs/1608.08939} {arXiv:1608.08939 [astro-ph.HE]}
  \BibitemShut {NoStop}%
\bibitem [{\citenamefont {{Qin}}\ \emph {et~al.}(2018)\citenamefont {{Qin}},
  \citenamefont {{Fragos}}, \citenamefont {{Meynet}}, \citenamefont
  {{Andrews}}, \citenamefont {{S{\o}rensen}},\ and\ \citenamefont
  {{Song}}}]{2018A&A...616A..28Q}%
  \BibitemOpen
  \bibfield  {author} {\bibinfo {author} {\bibfnamefont {Y.}~\bibnamefont
  {{Qin}}}, \bibinfo {author} {\bibfnamefont {T.}~\bibnamefont {{Fragos}}},
  \bibinfo {author} {\bibfnamefont {G.}~\bibnamefont {{Meynet}}}, \bibinfo
  {author} {\bibfnamefont {J.}~\bibnamefont {{Andrews}}}, \bibinfo {author}
  {\bibfnamefont {M.}~\bibnamefont {{S{\o}rensen}}},\ and\ \bibinfo {author}
  {\bibfnamefont {H.~F.}\ \bibnamefont {{Song}}},\ }\href
  {https://doi.org/10.1051/0004-6361/201832839} {\bibfield  {journal} {\bibinfo
   {journal} {\aap}\ }\textbf {\bibinfo {volume} {616}},\ \bibinfo {eid} {A28}
  (\bibinfo {year} {2018})},\ \Eprint {https://arxiv.org/abs/1802.05738}
  {arXiv:1802.05738 [astro-ph.SR]} \BibitemShut {NoStop}%
\bibitem [{\citenamefont {{Shibata}}\ and\ \citenamefont
  {{Fujibayashi}}(2025)}]{2025arXiv250915619S}%
  \BibitemOpen
  \bibfield  {author} {\bibinfo {author} {\bibfnamefont {M.}~\bibnamefont
  {{Shibata}}}\ and\ \bibinfo {author} {\bibfnamefont {S.}~\bibnamefont
  {{Fujibayashi}}},\ }\href {https://doi.org/10.48550/arXiv.2509.15619}
  {\bibfield  {journal} {\bibinfo  {journal} {arXiv e-prints}\ ,\ \bibinfo
  {eid} {arXiv:2509.15619}} (\bibinfo {year} {2025})},\ \Eprint
  {https://arxiv.org/abs/2509.15619} {arXiv:2509.15619 [astro-ph.HE]}
  \BibitemShut {NoStop}%
\bibitem [{\citenamefont {{Amaro-Seoane}}\ and\ \citenamefont
  {{Chen}}(2016)}]{2016MNRAS.458.3075A}%
  \BibitemOpen
  \bibfield  {author} {\bibinfo {author} {\bibfnamefont {P.}~\bibnamefont
  {{Amaro-Seoane}}}\ and\ \bibinfo {author} {\bibfnamefont {X.}~\bibnamefont
  {{Chen}}},\ }\href {https://doi.org/10.1093/mnras/stw503} {\bibfield
  {journal} {\bibinfo  {journal} {\mnras}\ }\textbf {\bibinfo {volume} {458}},\
  \bibinfo {pages} {3075} (\bibinfo {year} {2016})},\ \Eprint
  {https://arxiv.org/abs/1512.04897} {arXiv:1512.04897 [astro-ph.CO]}
  \BibitemShut {NoStop}%
\bibitem [{\citenamefont {{Harris}}(2010)}]{2010arXiv1012.3224H}%
  \BibitemOpen
  \bibfield  {author} {\bibinfo {author} {\bibfnamefont {W.~E.}\ \bibnamefont
  {{Harris}}},\ }\href@noop {} {\bibfield  {journal} {\bibinfo  {journal}
  {arXiv e-prints}\ ,\ \bibinfo {eid} {arXiv:1012.3224}} (\bibinfo {year}
  {2010})},\ \Eprint {https://arxiv.org/abs/1012.3224} {arXiv:1012.3224
  [astro-ph.GA]} \BibitemShut {NoStop}%
\bibitem [{\citenamefont {{Georgiev}}\ \emph {et~al.}(2016)\citenamefont
  {{Georgiev}}, \citenamefont {{B{\"o}ker}}, \citenamefont {{Leigh}},
  \citenamefont {{L{\"u}tzgendorf}},\ and\ \citenamefont
  {{Neumayer}}}]{2016MNRAS.457.2122G}%
  \BibitemOpen
  \bibfield  {author} {\bibinfo {author} {\bibfnamefont {I.~Y.}\ \bibnamefont
  {{Georgiev}}}, \bibinfo {author} {\bibfnamefont {T.}~\bibnamefont
  {{B{\"o}ker}}}, \bibinfo {author} {\bibfnamefont {N.}~\bibnamefont
  {{Leigh}}}, \bibinfo {author} {\bibfnamefont {N.}~\bibnamefont
  {{L{\"u}tzgendorf}}},\ and\ \bibinfo {author} {\bibfnamefont
  {N.}~\bibnamefont {{Neumayer}}},\ }\href
  {https://doi.org/10.1093/mnras/stw093} {\bibfield  {journal} {\bibinfo
  {journal} {\mnras}\ }\textbf {\bibinfo {volume} {457}},\ \bibinfo {pages}
  {2122} (\bibinfo {year} {2016})},\ \Eprint {https://arxiv.org/abs/1601.02613}
  {arXiv:1601.02613 [astro-ph.GA]} \BibitemShut {NoStop}%
\bibitem [{\citenamefont {{Heggie}}(1975)}]{1975MNRAS.173..729H}%
  \BibitemOpen
  \bibfield  {author} {\bibinfo {author} {\bibfnamefont {D.~C.}\ \bibnamefont
  {{Heggie}}},\ }\href {https://doi.org/10.1093/mnras/173.3.729} {\bibfield
  {journal} {\bibinfo  {journal} {\mnras}\ }\textbf {\bibinfo {volume} {173}},\
  \bibinfo {pages} {729} (\bibinfo {year} {1975})}\BibitemShut {NoStop}%
\bibitem [{\citenamefont {{Goodman}}\ and\ \citenamefont
  {{Hut}}(1993)}]{1993ApJ...403..271G}%
  \BibitemOpen
  \bibfield  {author} {\bibinfo {author} {\bibfnamefont {J.}~\bibnamefont
  {{Goodman}}}\ and\ \bibinfo {author} {\bibfnamefont {P.}~\bibnamefont
  {{Hut}}},\ }\href {https://doi.org/10.1086/172200} {\bibfield  {journal}
  {\bibinfo  {journal} {\apj}\ }\textbf {\bibinfo {volume} {403}},\ \bibinfo
  {pages} {271} (\bibinfo {year} {1993})}\BibitemShut {NoStop}%
\bibitem [{\citenamefont {{Binney}}\ and\ \citenamefont
  {{Tremaine}}(2008)}]{2008gady.book.....B}%
  \BibitemOpen
  \bibfield  {author} {\bibinfo {author} {\bibfnamefont {J.}~\bibnamefont
  {{Binney}}}\ and\ \bibinfo {author} {\bibfnamefont {S.}~\bibnamefont
  {{Tremaine}}},\ }\href@noop {} {\emph {\bibinfo {title} {{Galactic Dynamics:
  Second Edition}}}}\ (\bibinfo {year} {2008})\BibitemShut {NoStop}%
\bibitem [{\citenamefont {{Peebles}}(1972)}]{1972ApJ...178..371P}%
  \BibitemOpen
  \bibfield  {author} {\bibinfo {author} {\bibfnamefont {P.~J.~E.}\
  \bibnamefont {{Peebles}}},\ }\href {https://doi.org/10.1086/151797}
  {\bibfield  {journal} {\bibinfo  {journal} {\apj}\ }\textbf {\bibinfo
  {volume} {178}},\ \bibinfo {pages} {371} (\bibinfo {year}
  {1972})}\BibitemShut {NoStop}%
\bibitem [{\citenamefont {{Merritt}}(2013)}]{2013degn.book.....M}%
  \BibitemOpen
  \bibfield  {author} {\bibinfo {author} {\bibfnamefont {D.}~\bibnamefont
  {{Merritt}}},\ }\href@noop {} {\emph {\bibinfo {title} {{Dynamics and
  Evolution of Galactic Nuclei}}}}\ (\bibinfo {year} {2013})\BibitemShut
  {NoStop}%
\bibitem [{\citenamefont {{Bahcall}}\ and\ \citenamefont
  {{Wolf}}(1976)}]{1976ApJ...209..214B}%
  \BibitemOpen
  \bibfield  {author} {\bibinfo {author} {\bibfnamefont {J.~N.}\ \bibnamefont
  {{Bahcall}}}\ and\ \bibinfo {author} {\bibfnamefont {R.~A.}\ \bibnamefont
  {{Wolf}}},\ }\href {https://doi.org/10.1086/154711} {\bibfield  {journal}
  {\bibinfo  {journal} {\apj}\ }\textbf {\bibinfo {volume} {209}},\ \bibinfo
  {pages} {214} (\bibinfo {year} {1976})}\BibitemShut {NoStop}%
\bibitem [{\citenamefont {{Brockamp}}\ \emph {et~al.}(2011)\citenamefont
  {{Brockamp}}, \citenamefont {{Baumgardt}},\ and\ \citenamefont
  {{Kroupa}}}]{2011MNRAS.418.1308B}%
  \BibitemOpen
  \bibfield  {author} {\bibinfo {author} {\bibfnamefont {M.}~\bibnamefont
  {{Brockamp}}}, \bibinfo {author} {\bibfnamefont {H.}~\bibnamefont
  {{Baumgardt}}},\ and\ \bibinfo {author} {\bibfnamefont {P.}~\bibnamefont
  {{Kroupa}}},\ }\href {https://doi.org/10.1111/j.1365-2966.2011.19580.x}
  {\bibfield  {journal} {\bibinfo  {journal} {\mnras}\ }\textbf {\bibinfo
  {volume} {418}},\ \bibinfo {pages} {1308} (\bibinfo {year} {2011})},\ \Eprint
  {https://arxiv.org/abs/1108.2270} {arXiv:1108.2270 [astro-ph.GA]}
  \BibitemShut {NoStop}%
\bibitem [{\citenamefont {{Arca Sedda}}\ \emph {et~al.}(2018)\citenamefont
  {{Arca Sedda}}, \citenamefont {{Askar}},\ and\ \citenamefont
  {{Giersz}}}]{2018MNRAS.479.4652A}%
  \BibitemOpen
  \bibfield  {author} {\bibinfo {author} {\bibfnamefont {M.}~\bibnamefont
  {{Arca Sedda}}}, \bibinfo {author} {\bibfnamefont {A.}~\bibnamefont
  {{Askar}}},\ and\ \bibinfo {author} {\bibfnamefont {M.}~\bibnamefont
  {{Giersz}}},\ }\href {https://doi.org/10.1093/mnras/sty1859} {\bibfield
  {journal} {\bibinfo  {journal} {\mnras}\ }\textbf {\bibinfo {volume} {479}},\
  \bibinfo {pages} {4652} (\bibinfo {year} {2018})},\ \Eprint
  {https://arxiv.org/abs/1801.00795} {arXiv:1801.00795 [astro-ph.GA]}
  \BibitemShut {NoStop}%
\bibitem [{\citenamefont {{Chandrasekhar}}(1943)}]{1943ApJ....97..255C}%
  \BibitemOpen
  \bibfield  {author} {\bibinfo {author} {\bibfnamefont {S.}~\bibnamefont
  {{Chandrasekhar}}},\ }\href {https://doi.org/10.1086/144517} {\bibfield
  {journal} {\bibinfo  {journal} {\apj}\ }\textbf {\bibinfo {volume} {97}},\
  \bibinfo {pages} {255} (\bibinfo {year} {1943})}\BibitemShut {NoStop}%
\bibitem [{\citenamefont {{Spitzer}}(1987)}]{1987degc.book.....S}%
  \BibitemOpen
  \bibfield  {author} {\bibinfo {author} {\bibfnamefont {L.}~\bibnamefont
  {{Spitzer}}},\ }\href@noop {} {\emph {\bibinfo {title} {{Dynamical evolution
  of globular clusters}}}}\ (\bibinfo {year} {1987})\BibitemShut {NoStop}%
\bibitem [{\citenamefont {{Miller}}\ and\ \citenamefont
  {{Lauburg}}(2009)}]{2009ApJ...692..917M}%
  \BibitemOpen
  \bibfield  {author} {\bibinfo {author} {\bibfnamefont {M.~C.}\ \bibnamefont
  {{Miller}}}\ and\ \bibinfo {author} {\bibfnamefont {V.~M.}\ \bibnamefont
  {{Lauburg}}},\ }\href {https://doi.org/10.1088/0004-637X/692/1/917}
  {\bibfield  {journal} {\bibinfo  {journal} {\apj}\ }\textbf {\bibinfo
  {volume} {692}},\ \bibinfo {pages} {917} (\bibinfo {year} {2009})},\ \Eprint
  {https://arxiv.org/abs/0804.2783} {arXiv:0804.2783 [astro-ph]} \BibitemShut
  {NoStop}%
\bibitem [{\citenamefont {{Samsing}}\ \emph {et~al.}(2014)\citenamefont
  {{Samsing}}, \citenamefont {{MacLeod}},\ and\ \citenamefont
  {{Ramirez-Ruiz}}}]{2014ApJ...784...71S}%
  \BibitemOpen
  \bibfield  {author} {\bibinfo {author} {\bibfnamefont {J.}~\bibnamefont
  {{Samsing}}}, \bibinfo {author} {\bibfnamefont {M.}~\bibnamefont
  {{MacLeod}}},\ and\ \bibinfo {author} {\bibfnamefont {E.}~\bibnamefont
  {{Ramirez-Ruiz}}},\ }\href {https://doi.org/10.1088/0004-637X/784/1/71}
  {\bibfield  {journal} {\bibinfo  {journal} {\apj}\ }\textbf {\bibinfo
  {volume} {784}},\ \bibinfo {eid} {71} (\bibinfo {year} {2014})},\ \Eprint
  {https://arxiv.org/abs/1308.2964} {arXiv:1308.2964 [astro-ph.HE]}
  \BibitemShut {NoStop}%
\bibitem [{\citenamefont {{Samsing}}\ and\ \citenamefont
  {{Ramirez-Ruiz}}(2017)}]{2017ApJ...840L..14S}%
  \BibitemOpen
  \bibfield  {author} {\bibinfo {author} {\bibfnamefont {J.}~\bibnamefont
  {{Samsing}}}\ and\ \bibinfo {author} {\bibfnamefont {E.}~\bibnamefont
  {{Ramirez-Ruiz}}},\ }\href {https://doi.org/10.3847/2041-8213/aa6f0b}
  {\bibfield  {journal} {\bibinfo  {journal} {\apjl}\ }\textbf {\bibinfo
  {volume} {840}},\ \bibinfo {eid} {L14} (\bibinfo {year} {2017})},\ \Eprint
  {https://arxiv.org/abs/1703.09703} {arXiv:1703.09703 [astro-ph.HE]}
  \BibitemShut {NoStop}%
\bibitem [{\citenamefont {{G{\"u}ltekin}}\ \emph {et~al.}(2004)\citenamefont
  {{G{\"u}ltekin}}, \citenamefont {{Miller}},\ and\ \citenamefont
  {{Hamilton}}}]{2004ApJ...616..221G}%
  \BibitemOpen
  \bibfield  {author} {\bibinfo {author} {\bibfnamefont {K.}~\bibnamefont
  {{G{\"u}ltekin}}}, \bibinfo {author} {\bibfnamefont {M.~C.}\ \bibnamefont
  {{Miller}}},\ and\ \bibinfo {author} {\bibfnamefont {D.~P.}\ \bibnamefont
  {{Hamilton}}},\ }\href {https://doi.org/10.1086/424809} {\bibfield  {journal}
  {\bibinfo  {journal} {\apj}\ }\textbf {\bibinfo {volume} {616}},\ \bibinfo
  {pages} {221} (\bibinfo {year} {2004})},\ \Eprint
  {https://arxiv.org/abs/astro-ph/0402532} {arXiv:astro-ph/0402532 [astro-ph]}
  \BibitemShut {NoStop}%
\bibitem [{\citenamefont {{Larsen}}(2009)}]{2009A&A...494..539L}%
  \BibitemOpen
  \bibfield  {author} {\bibinfo {author} {\bibfnamefont {S.~S.}\ \bibnamefont
  {{Larsen}}},\ }\href {https://doi.org/10.1051/0004-6361:200811212} {\bibfield
   {journal} {\bibinfo  {journal} {\aap}\ }\textbf {\bibinfo {volume} {494}},\
  \bibinfo {pages} {539} (\bibinfo {year} {2009})},\ \Eprint
  {https://arxiv.org/abs/0812.1400} {arXiv:0812.1400 [astro-ph]} \BibitemShut
  {NoStop}%
\bibitem [{\citenamefont {{Peters}}(1964)}]{1964PhRv..136.1224P}%
  \BibitemOpen
  \bibfield  {author} {\bibinfo {author} {\bibfnamefont {P.~C.}\ \bibnamefont
  {{Peters}}},\ }\href {https://doi.org/10.1103/PhysRev.136.B1224} {\bibfield
  {journal} {\bibinfo  {journal} {Physical Review}\ }\textbf {\bibinfo {volume}
  {136}},\ \bibinfo {pages} {1224} (\bibinfo {year} {1964})}\BibitemShut
  {NoStop}%
\bibitem [{\citenamefont {{Banerjee}}(2021)}]{2021MNRAS.500.3002B}%
  \BibitemOpen
  \bibfield  {author} {\bibinfo {author} {\bibfnamefont {S.}~\bibnamefont
  {{Banerjee}}},\ }\href {https://doi.org/10.1093/mnras/staa2392} {\bibfield
  {journal} {\bibinfo  {journal} {\mnras}\ }\textbf {\bibinfo {volume} {500}},\
  \bibinfo {pages} {3002} (\bibinfo {year} {2021})},\ \Eprint
  {https://arxiv.org/abs/2004.07382} {arXiv:2004.07382 [astro-ph.HE]}
  \BibitemShut {NoStop}%
\bibitem [{\citenamefont {{Fragione}}\ \emph {et~al.}(2019)\citenamefont
  {{Fragione}}, \citenamefont {{Grishin}}, \citenamefont {{Leigh}},
  \citenamefont {{Perets}},\ and\ \citenamefont
  {{Perna}}}]{2019MNRAS.488...47F}%
  \BibitemOpen
  \bibfield  {author} {\bibinfo {author} {\bibfnamefont {G.}~\bibnamefont
  {{Fragione}}}, \bibinfo {author} {\bibfnamefont {E.}~\bibnamefont
  {{Grishin}}}, \bibinfo {author} {\bibfnamefont {N.~W.~C.}\ \bibnamefont
  {{Leigh}}}, \bibinfo {author} {\bibfnamefont {H.~B.}\ \bibnamefont
  {{Perets}}},\ and\ \bibinfo {author} {\bibfnamefont {R.}~\bibnamefont
  {{Perna}}},\ }\href {https://doi.org/10.1093/mnras/stz1651} {\bibfield
  {journal} {\bibinfo  {journal} {\mnras}\ }\textbf {\bibinfo {volume} {488}},\
  \bibinfo {pages} {47} (\bibinfo {year} {2019})},\ \Eprint
  {https://arxiv.org/abs/1811.10627} {arXiv:1811.10627 [astro-ph.GA]}
  \BibitemShut {NoStop}%
\bibitem [{\citenamefont {{Netopil}}\ \emph {et~al.}(2016)\citenamefont
  {{Netopil}}, \citenamefont {{Paunzen}}, \citenamefont {{Heiter}},\ and\
  \citenamefont {et~al.}}]{2016A&A...585A.150N}%
  \BibitemOpen
  \bibfield  {author} {\bibinfo {author} {\bibfnamefont {M.}~\bibnamefont
  {{Netopil}}}, \bibinfo {author} {\bibfnamefont {E.}~\bibnamefont
  {{Paunzen}}}, \bibinfo {author} {\bibfnamefont {U.}~\bibnamefont
  {{Heiter}}},\ and\ \bibinfo {author} {\bibnamefont {et~al.}},\ }\href
  {https://doi.org/10.1051/0004-6361/201526370} {\bibfield  {journal} {\bibinfo
   {journal} {\aap}\ }\textbf {\bibinfo {volume} {585}},\ \bibinfo {eid} {A150}
  (\bibinfo {year} {2016})},\ \Eprint {https://arxiv.org/abs/1511.08884}
  {arXiv:1511.08884 [astro-ph.SR]} \BibitemShut {NoStop}%
\bibitem [{\citenamefont {{Milosavljevi{\'c}}}(2004)}]{2004ApJ...605L..13M}%
  \BibitemOpen
  \bibfield  {author} {\bibinfo {author} {\bibfnamefont {M.}~\bibnamefont
  {{Milosavljevi{\'c}}}},\ }\href {https://doi.org/10.1086/420696} {\bibfield
  {journal} {\bibinfo  {journal} {\apjl}\ }\textbf {\bibinfo {volume} {605}},\
  \bibinfo {pages} {L13} (\bibinfo {year} {2004})},\ \Eprint
  {https://arxiv.org/abs/astro-ph/0310574} {arXiv:astro-ph/0310574 [astro-ph]}
  \BibitemShut {NoStop}%
\bibitem [{\citenamefont {{Nayakshin}}\ \emph {et~al.}(2007)\citenamefont
  {{Nayakshin}}, \citenamefont {{Cuadra}},\ and\ \citenamefont
  {{Springel}}}]{2007MNRAS.379...21N}%
  \BibitemOpen
  \bibfield  {author} {\bibinfo {author} {\bibfnamefont {S.}~\bibnamefont
  {{Nayakshin}}}, \bibinfo {author} {\bibfnamefont {J.}~\bibnamefont
  {{Cuadra}}},\ and\ \bibinfo {author} {\bibfnamefont {V.}~\bibnamefont
  {{Springel}}},\ }\href {https://doi.org/10.1111/j.1365-2966.2007.11938.x}
  {\bibfield  {journal} {\bibinfo  {journal} {\mnras}\ }\textbf {\bibinfo
  {volume} {379}},\ \bibinfo {pages} {21} (\bibinfo {year} {2007})},\ \Eprint
  {https://arxiv.org/abs/astro-ph/0701141} {arXiv:astro-ph/0701141 [astro-ph]}
  \BibitemShut {NoStop}%
\bibitem [{\citenamefont {{Antonini}}\ \emph {et~al.}(2012)\citenamefont
  {{Antonini}}, \citenamefont {{Capuzzo-Dolcetta}}, \citenamefont
  {{Mastrobuono-Battisti}},\ and\ \citenamefont
  {{Merritt}}}]{2012ApJ...750..111A}%
  \BibitemOpen
  \bibfield  {author} {\bibinfo {author} {\bibfnamefont {F.}~\bibnamefont
  {{Antonini}}}, \bibinfo {author} {\bibfnamefont {R.}~\bibnamefont
  {{Capuzzo-Dolcetta}}}, \bibinfo {author} {\bibfnamefont {A.}~\bibnamefont
  {{Mastrobuono-Battisti}}},\ and\ \bibinfo {author} {\bibfnamefont
  {D.}~\bibnamefont {{Merritt}}},\ }\href
  {https://doi.org/10.1088/0004-637X/750/2/111} {\bibfield  {journal} {\bibinfo
   {journal} {\apj}\ }\textbf {\bibinfo {volume} {750}},\ \bibinfo {eid} {111}
  (\bibinfo {year} {2012})},\ \Eprint {https://arxiv.org/abs/1110.5937}
  {arXiv:1110.5937 [astro-ph.GA]} \BibitemShut {NoStop}%
\bibitem [{\citenamefont {{Rafelski}}\ \emph {et~al.}(2012)\citenamefont
  {{Rafelski}}, \citenamefont {{Wolfe}}, \citenamefont {{Prochaska}},
  \citenamefont {{Neeleman}},\ and\ \citenamefont
  {{Mendez}}}]{2012ApJ...755...89R}%
  \BibitemOpen
  \bibfield  {author} {\bibinfo {author} {\bibfnamefont {M.}~\bibnamefont
  {{Rafelski}}}, \bibinfo {author} {\bibfnamefont {A.~M.}\ \bibnamefont
  {{Wolfe}}}, \bibinfo {author} {\bibfnamefont {J.~X.}\ \bibnamefont
  {{Prochaska}}}, \bibinfo {author} {\bibfnamefont {M.}~\bibnamefont
  {{Neeleman}}},\ and\ \bibinfo {author} {\bibfnamefont {A.~J.}\ \bibnamefont
  {{Mendez}}},\ }\href {https://doi.org/10.1088/0004-637X/755/2/89} {\bibfield
  {journal} {\bibinfo  {journal} {\apj}\ }\textbf {\bibinfo {volume} {755}},\
  \bibinfo {eid} {89} (\bibinfo {year} {2012})},\ \Eprint
  {https://arxiv.org/abs/1205.5047} {arXiv:1205.5047 [astro-ph.CO]}
  \BibitemShut {NoStop}%
\bibitem [{\citenamefont {{De Cia}}\ \emph {et~al.}(2018)\citenamefont {{De
  Cia}}, \citenamefont {{Gal-Yam}}, \citenamefont {{Rubin}}, \citenamefont
  {{Leloudas}}, \citenamefont {{Vreeswijk}}, \citenamefont {{Perley}},
  \citenamefont {{Quimby}}, \citenamefont {{Yan}}, \citenamefont {{Sullivan}},
  \citenamefont {{Fl{\"o}rs}}, \citenamefont {{Sollerman}}, \citenamefont
  {{Bersier}}, \citenamefont {{Cenko}}, \citenamefont {{Gal-Yam}},
  \citenamefont {{Maguire}}, \citenamefont {{Ofek}}, \citenamefont
  {{Prentice}}, \citenamefont {{Schulze}}, \citenamefont {{Spyromilio}},
  \citenamefont {{Valenti}}, \citenamefont {{Arcavi}}, \citenamefont {{Corsi}},
  \citenamefont {{Howell}}, \citenamefont {{Mazzali}}, \citenamefont
  {{Kasliwal}}, \citenamefont {{Taddia}},\ and\ \citenamefont
  {{Yaron}}}]{2018ApJ...860..100D}%
  \BibitemOpen
  \bibfield  {author} {\bibinfo {author} {\bibfnamefont {A.}~\bibnamefont {{De
  Cia}}}, \bibinfo {author} {\bibfnamefont {A.}~\bibnamefont {{Gal-Yam}}},
  \bibinfo {author} {\bibfnamefont {A.}~\bibnamefont {{Rubin}}}, \bibinfo
  {author} {\bibfnamefont {G.}~\bibnamefont {{Leloudas}}}, \bibinfo {author}
  {\bibfnamefont {P.}~\bibnamefont {{Vreeswijk}}}, \bibinfo {author}
  {\bibfnamefont {D.~A.}\ \bibnamefont {{Perley}}}, \bibinfo {author}
  {\bibfnamefont {R.}~\bibnamefont {{Quimby}}}, \bibinfo {author}
  {\bibfnamefont {L.}~\bibnamefont {{Yan}}}, \bibinfo {author} {\bibfnamefont
  {M.}~\bibnamefont {{Sullivan}}}, \bibinfo {author} {\bibfnamefont
  {A.}~\bibnamefont {{Fl{\"o}rs}}}, \bibinfo {author} {\bibfnamefont
  {J.}~\bibnamefont {{Sollerman}}}, \bibinfo {author} {\bibfnamefont
  {D.}~\bibnamefont {{Bersier}}}, \bibinfo {author} {\bibfnamefont {S.~B.}\
  \bibnamefont {{Cenko}}}, \bibinfo {author} {\bibfnamefont {M.}~\bibnamefont
  {{Gal-Yam}}}, \bibinfo {author} {\bibfnamefont {K.}~\bibnamefont
  {{Maguire}}}, \bibinfo {author} {\bibfnamefont {E.~O.}\ \bibnamefont
  {{Ofek}}}, \bibinfo {author} {\bibfnamefont {S.}~\bibnamefont {{Prentice}}},
  \bibinfo {author} {\bibfnamefont {S.}~\bibnamefont {{Schulze}}}, \bibinfo
  {author} {\bibfnamefont {J.}~\bibnamefont {{Spyromilio}}}, \bibinfo {author}
  {\bibfnamefont {S.}~\bibnamefont {{Valenti}}}, \bibinfo {author}
  {\bibfnamefont {I.}~\bibnamefont {{Arcavi}}}, \bibinfo {author}
  {\bibfnamefont {A.}~\bibnamefont {{Corsi}}}, \bibinfo {author} {\bibfnamefont
  {D.~A.}\ \bibnamefont {{Howell}}}, \bibinfo {author} {\bibfnamefont
  {P.}~\bibnamefont {{Mazzali}}}, \bibinfo {author} {\bibfnamefont {M.~M.}\
  \bibnamefont {{Kasliwal}}}, \bibinfo {author} {\bibfnamefont
  {F.}~\bibnamefont {{Taddia}}},\ and\ \bibinfo {author} {\bibfnamefont
  {O.}~\bibnamefont {{Yaron}}},\ }\href
  {https://doi.org/10.3847/1538-4357/aab9b6} {\bibfield  {journal} {\bibinfo
  {journal} {\apj}\ }\textbf {\bibinfo {volume} {860}},\ \bibinfo {eid} {100}
  (\bibinfo {year} {2018})},\ \Eprint {https://arxiv.org/abs/1708.01623}
  {arXiv:1708.01623 [astro-ph.HE]} \BibitemShut {NoStop}%
\bibitem [{\citenamefont {{Bouffanais}}\ \emph {et~al.}(2019)\citenamefont
  {{Bouffanais}}, \citenamefont {{Mapelli}}, \citenamefont {{Gerosa}},
  \citenamefont {{Di Carlo}}, \citenamefont {{Giacobbo}}, \citenamefont
  {{Berti}},\ and\ \citenamefont {{Baibhav}}}]{2019ApJ...886...25B}%
  \BibitemOpen
  \bibfield  {author} {\bibinfo {author} {\bibfnamefont {Y.}~\bibnamefont
  {{Bouffanais}}}, \bibinfo {author} {\bibfnamefont {M.}~\bibnamefont
  {{Mapelli}}}, \bibinfo {author} {\bibfnamefont {D.}~\bibnamefont {{Gerosa}}},
  \bibinfo {author} {\bibfnamefont {U.~N.}\ \bibnamefont {{Di Carlo}}},
  \bibinfo {author} {\bibfnamefont {N.}~\bibnamefont {{Giacobbo}}}, \bibinfo
  {author} {\bibfnamefont {E.}~\bibnamefont {{Berti}}},\ and\ \bibinfo {author}
  {\bibfnamefont {V.}~\bibnamefont {{Baibhav}}},\ }\href
  {https://doi.org/10.3847/1538-4357/ab4a79} {\bibfield  {journal} {\bibinfo
  {journal} {\apj}\ }\textbf {\bibinfo {volume} {886}},\ \bibinfo {eid} {25}
  (\bibinfo {year} {2019})},\ \Eprint {https://arxiv.org/abs/1905.11054}
  {arXiv:1905.11054 [astro-ph.HE]} \BibitemShut {NoStop}%
\end{thebibliography}%

\end{document}